\documentclass[a4paper,12pt]{article}
\usepackage{epsfig}
\usepackage{color}
\usepackage{latexsym}

\makeatletter

\def\cont{\mathbin{\dimen0=\ht\strutbox \dimen1=\ht\strutbox \divide\dimen0 by 2
\divide\dimen1 by 4
  \hbox{\vbox{\hrule width\dimen1}\hskip-0.4pt\vrule   height\dimen0}}\,}

\def\pback#1{{
\mathchoice{\StemPullBack{#1}{\leftarrowfill}}
     {\StemPullBack{#1}{\leftarrowfill}}
             {\IndxPullBack{#1}{\leftarrowfill}}
         {\IndxPullBack{#1}{\leftarrowfill}}}\vphantom{#1}}

\newcommand{\Leftarrowfill}[0]{$\m@th  \mathord\Leftarrow  \mkern-6mu
\cleaders\hbox{$\mkern-2mu \mathord= \mkern-2mu$}\hfill
\mkern -6mu \mathord=$}
	 
\def\ppback#1{{
\mathchoice{\StemPullBack{#1}{\Leftarrowfill}}
     {\StemPullBack{#1}{\Leftarrowfill}}
             {\IndxPullBack{#1}{\Leftarrowfill}}
         {\IndxPullBack{#1}{\Leftarrowfill}}}\vphantom{#1}}

\newcommand{\StemPullBack}[2]{
  \vtop{\mathsurround=0pt
  \ialign{##\crcr$\textstyle{#1}\strut$\crcr
    \noalign{\kern-0.4ex\nointerlineskip}{\tiny#2}\crcr}}}

\newcommand{\Cyc}[1]{
  \vtop{\mathsurround=0pt
  \ialign{##\crcr$\textstyle{\rm Cyc}\strut$\crcr
    \noalign{\kern-0.4ex\nointerlineskip}{\tiny#1}\crcr}}\ }

\newcommand{\IndxPullBack}[2]{
  \vtop{\mathsurround=0pt
  \ialign{##\crcr\hfil$\scriptstyle{#1}$\hfil\crcr
    \noalign{\kern+0.4ex\nointerlineskip}{\tiny#2}\crcr}}}

\newcommand{\barbox}{\overline{\mbox{\rule{0pt}{2.5mm}$\Box$}}} 

\newcommand{\Real} {{\mbox{\rm$\mbox{I}\!\mbox{R}$}}} 
\newcommand{\dI} {{\mbox{\rm$\mbox{d}\!\mbox{I}$}}}

\newcommand{\calN}{{\cal N}}
\newcommand{\cN}{{\cal N}}
\newcommand{\cA}{{\cal A}}
\newcommand{\cB}{{\cal B}}
\newcommand{\cC}{{\cal C}}
\newcommand{\cD}{{\cal D}}
\newcommand{\cS}{{\cal S}}
 
\newcommand{\cF}{{\cal F}}
\newcommand{\cV}{{\cal V}} 
\newcommand{\cH}{{\cal H}}
\newcommand{\cU}{{\cal U}} 
\newcommand{\ag}{\alpha} 
\newcommand{\bg}{\beta} 
\newcommand{\Dg}{\Delta}
\newcommand{\dg}{\delta} 
\newcommand{\cg}{\gamma} 
\newcommand{\Cg}{\Gamma} 
\newcommand{\eg}{\epsilon}

\newcommand{\lam}{\lambda}
\newcommand{\Lam}{\Lambda}

\newcommand{\vareg}{\varepsilon}
\newcommand{\Sg}{\Sigma} 
\newcommand{\sg}{\sigma}

\newcommand{\di}{\partial} 
\newcommand{\be}{\begin{equation}} 
\newcommand{\ee}{\end{equation}} 
\newcommand{\bearr}{\begin{eqnarray}}
\newcommand{\eearr}{\end{eqnarray}} 
 
\newcommand{\QED}{\rule{1.5mm}{3mm}} 
\newcommand{\dgF}{\dg^{\scriptscriptstyle F}}
\newcommand{\DDg}{\mbox{\boldmath $\Delta$}}
\newcommand{\halfR}{\scriptstyle{\frac{1}{2}}\Real}
\newcommand{\ocirc}[1]{\stackrel{\circ}{#1}}
\newcommand{\Dund}{\underline{\cal D}}

\newtheorem{definition}{Definition}[section]
\newtheorem{theorem}{Theorem}[section]
\newtheorem{proposition}[theorem]{Proposition}
 
\newtheorem{lemma}[theorem]{Lemma} 

\begin{document} 

\title{The symplectic 2-form and Poisson bracket of null canonical gravity}
\author{Michael P. Reisenberger\\
        Instituto de F\a'{\i}sica, Facultad de Ciencias,\\
        Universidad de la Rep\a'ublica Oriental del Uruguay,\\
        Igu\a'a 4225, esq. Mataojo, Montevideo, Uruguay
}
\maketitle

\begin{abstract}
It is well known that free (unconstrained) initial data for the gravitational field in general 
relativity can be identified on an initial hypersurface consisting of two intersecting null 
hypersurfaces. Here the phase space of vacuum general relativity associated with such an initial
data hypersurface is defined; a Poisson bracket is defined, via Peierls' prescription, on 
sufficiently regular functions on this phase space, called ``observables''; and a bracket on 
initial data is defined so that it reproduces the Peierls bracket between observables when these
are expressed in terms of the initial data. The brackets between all elements of a free initial 
data set are calculated explicitly. The bracket on initial data presented here has all the 
characteristics of a Poisson bracket except that it does not satisfy the Jacobi relations 
(even though the brackets between the observables do). 

The initial data set used is closely related to that of Sachs \cite{Sachs}. However, one 
significant difference is that it includes a ``new'' pair of degrees of freedom on the 
intersection of the two null hypersurfaces which are present but quite hidden in Sachs' 
formalism. As a step in the calculation an explicit expression for the symplectic 2-form in 
terms of these free initial data is obtained.
\end{abstract}

\section{Introduction}

Canonical formulations of general relativity are usually couched in terms of initial 
data on spacelike hypersurfaces \cite{ADM}\cite{Sen}\cite{Ashtekar}. However it has been widely 
known since the early 1960's that the initial value problem on piecewise null hypersurfaces 
(hypersurfaces with null normal) is technically simpler than the spacelike initial value 
problem. The constraints reduce to ordinary differential equations, and {\em free} 
initial data, that parametrize the solutions to the constraints, can be identified explicitly 
\cite{Sachs}\cite{Penrose}\cite{Bondi}\cite{Dautcourt}. 

A canonical formalism in terms of the free null initial data would be of interest for many 
reasons. It might offer insight into the properties of classical solutions of general 
relativity, it should provide a powerful framework for the numerical computation of these 
solutions and finally, in the context of quantum gravity, it promises to shed light on black 
hole entropy and the entropy bounds proposed by Beckenstein and Bousso 
\cite{Beckenstein}\cite{Susskind}\cite{Bousso}.

The main missing ingredient of such a formalism is the Poisson bracket on the space of free
initial data. In 1978 Gambini and Restuccia \cite{GR}, using a judiciously chosen coordinate
system, were able to obtain a perturbation series in Newton's gravitational constant for 
the bracket between the main free data. The main data is a specification of the conformal 
equivalence class of the (degenerate) 3-metric on the initial data hypersurface. In Gambini and 
Restuccia's work, and in the present work, this is determined by the ``conformal 2-metric'', a 
field given on all of the initial data hypersurface, in contrast to the remaining data which
lives on the intersection of the two null hypersurfaces only.

Here I will present closed form expressions for the brackets between all elements of a complete
free initial data set for vacuum general relativity. Although the approach taken is 
entirely different from that of Gambini and Restuccia, the bracket between the conformal 
2-metric at distinct points obtained is equal to that obtained by summing their series.

The first step in my approach is to define a Poisson bracket on a sufficiently rich family of
sufficiently nice diffeomorphism invariant functions of the spacetime geometry of the 
Cauchy developments of the the initial data surface, which will be called ``observables''.
These Poisson brackets are defined by Peierls' prescription, which, unlike other definitions of 
the Poisson bracket, involves no arbitrary elements of the classical description of the 
system (such as the choice of time, or the choice of fields), and which is directly related to 
the quantum commutator. The next step is to define a bracket on a suitable free initial data 
set so that it reproduces the Poisson brackets between the observables. This requirement is 
fulfilled if the bracket is a certain generalized inverse to the symplectic 2-form.
It turns out to be possible to calculate such a generalized inverse explicitly,
yielding explicit closed form expressions for the brackets between all data.

Unfortunately (at least from the point of view of a quantization project) the bracket
obtained does not satisfy the Jacobi relations. The requirement that the bracket reproduce the 
Peierls brackets between observables seems absolutely necessary. Nevertheless there still seems
to be some freedom in the choice of bracket. Though the condition used here to determine the 
brackets on initial data is clearly {\em sufficient} to guarantee that the consequent brackets
of observables reproduces the Peierls bracket, it is not clear that it is necessary. Some 
relaxation of this condition seems possible. Moreover, the brackets compatible with this 
condition are almost, but not quite, unique. The author is currently investigating whether a 
true Poisson bracket on initial data which reproduces the Peierls bracket on observables can be 
found. 

Null hypersurfaces are always swept out by congruences of null geodesics, and these generically 
form caustics. It seems preferable, at least at this stage, to work with smooth and 
structurally simple null hypersurfaces, so the hypersurfaces have been truncated such that they 
contain no caustics. This requires us to work with initial data on hypersurfaces with boundary, 
which brings with it new problems. On the other hand, it seems very natural and appropriate to
general relativity to describe it via an ``atlas of phase spaces''\footnote{I thank J-A Zapata
for this evocative phrase.} rather than to attempt to force it into a single phase space 
- thereby imposing global causality which, though comfortingly familiar, is entirely alien to 
the local nature of the theory.

Be that as it may, the presence of boundaries has frustrated all my attempts to obtain the 
brackets by more conventional means, and is the reason that the bracket is a generalized inverse
of the symplectic 2-form instead of a true inverse. Here a generalized inverse of the symplectic
2-form inverts the action of this 2-form on phase space vectors corresponding to perturbations 
of the gravitational field that vanish in a neighbourhood of the boundary of the initial data 
hypersurface. 

A variation of a solution can be represented as a variation of the initial data. However, for 
some initial data variables a variation of the solution metric that vanishes near the boundary 
of the initial data hypersurface may produce variations of the initial data that do not vanish 
near the boundary. This occurs for instance for Sachs' initial data. It depends on how the 
initial data at point is related to the metric field. However, to calculate the generalized 
inverse of the symplectic 2-form it is convenient to use free initial data fields that {\em are}
invariant near the boundary under variations that leave the solution metric undisturbed in a 
neighbourhood of the boundary, or indeed under variations that simply leave the 4-geometry 
undisturbed in such a neighbourhood. This requirement has led me to define a new free initial 
data set, differing somewhat from that of Sachs. It contains a new field on the intersection of 
the two null hypersurfaces. The information in this field is contained in Sachs's data but 
rather implicitly. The inclusion of this field is quite important in the phase space formalism 
since it is essentially the canonical conjugate to another datum on the intersection 2-surface 
(which is included explicitly in the Sachs data).
    
In the following section I define the initial data hypersurface and the associated phase space, 
observables and the Peierls bracket between them, and the auxiliary pre-Poisson bracket on
initial data. The relation of the Peierls bracket and the symplectic 2-form is found, and 
also the sufficient condition ensuring that the bracket on initial data reproduces the Peierls
bracket on observables. The symplectic 2-form corresponding to the Einstein-Hilbert action is 
calculated explicitly in terms of the spacetime metric. Finally, the meaning of gauge invariance
for initial data on a hypersurface with boundary is considered.
In section \ref{free_data_coords} some convenient spacetime coordinate systems adapted
to the geometry are presented and
an initial data set suitable for our purposes is proposed and demonstrated to be free and
complete. This section also contains a treatment of the relation between the variations
of field components referred to different coordinate systems that are adapted to the fields. 
Then, in section \ref{pre_symplectic_2form} the symplectic 2-form is expressed in terms of the 
initial data chosen. Finally, in section \ref{Poisson_bracket} the 
brackets between the free initial data are calculated in closed form. 

I close the paper with some concluding remarks about open issues and potential directions
of research in section \ref{conclusions}.

\section{Covariant canonical theory} \label{covariant_canonical_theory}

\subsection{The problem} \label{problem}

General relativity is a local theory in the sense that initial data in a bounded region of 
space - a bounded achronal hypersurface in spacetime - suffices to determine completely the 
gravitational field in a contiguous spacetime region, the causal domain of dependence\footnote{
Achronal sets and domains of dependence are defined in appendix \ref{definitions}, see also 
\cite{Hawking_Ellis} or \cite{Wald}.}
of the hypersurface. It does not matter what is happening in the rest of the universe. We may 
thus focus attention on this domain of dependence and regard the space of values of the initial 
data on the hypersurface as a space of states, or phase space, of the gravitational field in the
domain of dependence. 

Now in General Relativity diffeomorphic solutions to Einstein's field equations are equivalent, 
i.e. they represent the same predictions concerning measurements. Thus the phase space, which we
shall denote $\Phi$, actually consist of the equivalence classes of initial data sets 
corresponding to diffeomorphic solutions. Equivalently, it may be identified with the space of 
diffeomorphism equivalence classes of maximal Cauchy 
developments\footnote{
The, slightly generalized, definition of Cauchy developments, used here as well as that of  
maximal Cauchy developments, is given in appendix \ref{definitions}. For the sake of simplicity 
we shall only consider $C^\infty$ developments.}
of data on the given hypersurface.\footnote{
$\Phi$ is often called the ``physical" or ``reduced" phase space to distinguish it from the 
larger ``kinematic" phase space $K$, constructed without use of the field equations. 
(see for example \cite{Henneaux_Teitelboim}. A geometrical 
construction of $K$ for asymptotically flat and for compact boundaryless 
spacetimes is given in \cite{Lee_Wald}). The Poisson bracket we are looking 
for - the Poisson bracket on the physical phase space $\Phi$, expressed directly 
in terms coordinates on $\Phi$ - is often called the Dirac bracket.}

The aim of the present work is to find the Poisson bracket on the phase space $\Phi$ 
corresponding to an initial data hypersurface consisting of two intersecting null hypersurfaces.

What will actually be obtained is the Poisson bracket on a particularly 
nice subset of functions on this phase space, which in turn suffices to  
determine, almost uniquely, the brackets between initial data.
The brackets between the initial data obtained define a {\em pre-Poisson bracket} on the 
algebra of phase space functions, that is, a bracket having 
all the properties of a Poisson bracket save that it does not necessarily satisfy 
the Jacobi relation. The lack of a Jacobi relation is of course an important issue 
when considering the quantization of the system, since commutators satisfy this relation 
identically. 

\begin{figure}
\begin{center}
\input{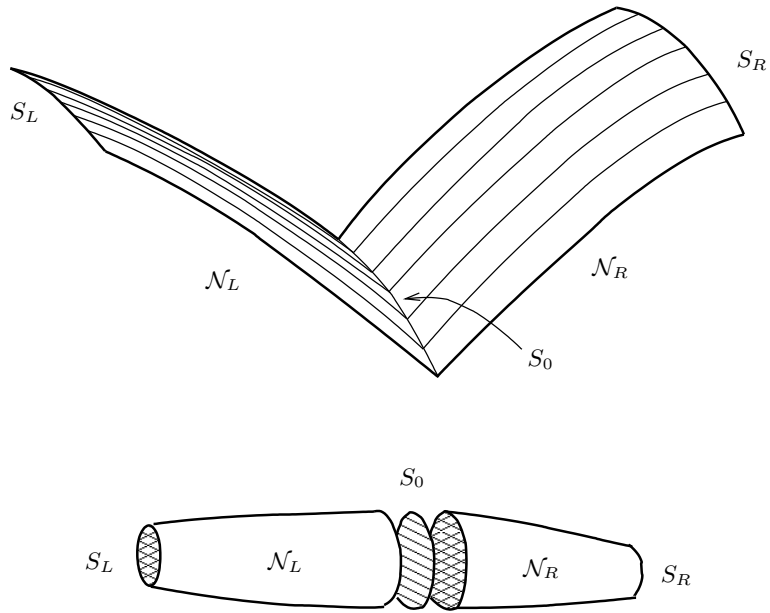}
\caption{The upper diagram shows an initial data surface like $\calN$ embedded in a 2+1 
dimensional spacetime. In 3+1 dimensional spacetimes each of the components $\calN_L$, 
$\calN_R$, and $S_0$, of course have one dimension more than shown in this diagram. This
is shown in the lower diagram, which depicts a three dimensional hypersurface $\cN$ 
corresponding to a $3+1$ spacetime. Unlike the upper diagram, the lower diagram represents 
only the intrinsic differential topology of $\cN$, and does not indicate it's embedding in
spacetime.}
\label{Nfigure}
\end{center}
\end{figure}

Let us define precisely the initial data hypersurface to be considered.\label{cN_def}
It is a compact manifold (with boundary and edges),\footnote{
Manifolds with boundary and edges are defined in appendix \ref{definitions}.}
$\cN$, consisting of two null hypersurface branches $\cN_L$ and $\cN_R$ joined on a smooth 
spacelike 2-surface $S_0$ diffeomorphic to a disk (See Fig. \ref{Nfigure}). $\cN_L$ and $\cN_R$ 
are swept out by the two congruences of future directed null geodesics (called {\em generators})
emerging normally from $S_0$, and are truncated at spacelike disks - $S_L$ and $S_R$ 
respectively - before these generators form caustics.

From the point of view of the initial value problem $\cN$ is simply a compact 3-manifold with 
boundary, consisting of two 3-cylinders ($\cN_L$ and $\cN_R$) joined end to end on the 
2-surface $S_0$. The initial data consists of smooth 3-metrics specified on $\cN_L$ and $\cN_R$,
with a continuity condition at $S_0$, and some further quantities specified on $S_0$ only 
(see subsection \ref{initial_data}). In order that the branches $\cN_L$ and $\cN_R$ 
be null in the solution matching the data, the 3-metric on each branch must be degenerate. That 
is, there must exist at each point a degeneracy vector, a vector tangent to the branch that is 
orthogonal to all tangent vectors, which corresponds to the null direction tangent to the 
embedding of the branch in spacetime. Moreover, the space of such degeneracy vectors must be one
dimensional at any point at which the branch is embedded smoothly (which requires that the 
tangent vectors to the branch are mapped to a three dimensional subspace of the spacetime 
tangent space). In fact the metric on a cross section transverse to the degeneracy vectors must 
be positive definite. Requiring these conditions to hold at all points of $\cN_L$ and $\cN_R$ is
equivalent to requiring that the generators form no caustics on $\cN$. See appendix 
\ref{nullhypersurfaces}. 

Now let us consider the maximal Cauchy developments of data on $\cN$. The representation
of a initial data set, or phase point, by its Cauchy development will play a central role 
in the definition of Poisson brackets for the following reason: Recall that the bracket of two 
phase space functions is again a function on phase space, and therefore takes as its argument a 
phase point. It turns out that the value of the Poisson bracket at a phase point can be very 
naturally defined in terms of the spacetime geometry of the maximal Cauchy development 
corresponding to that phase point. The fundamental definition involves only retarded and 
advanced Green's functions for the linearized Einstein equation on this Cauchy development. 

But how do we know that Cauchy developments of a given set of data on $\cN$ actually exist?
Rendall \cite{Rendall} has proved for a particular set of initial data variables, equivalent 
to that we shall use, that if these data are smooth on $\cN$ they determine a $C^\infty$ Cauchy 
development of at least a neighbourhood $\tilde{\cN} \subset \cN$ of $S_0$ in $\cN$. In fact 
Rendall's proof, which reduces the initial value problem on $\cN$ to an equivalent initial value
problem for data on a spacelike hypersurface, combined with standard results on the spacelike 
initial value problem (\cite{Wald} Theorem 10.2.2), implies that the data on $\tilde{\cN}$ 
determines a smooth maximal Cauchy development uniquely up to diffeomorphisms. 

I shall assume that Rendall's results in fact apply to all of $\cN$, i.e. $\tilde{\cN}$ 
can be taken to be all of $\cN$, assumption which I consider very likely to be correct.
If it is {\em not} then the results are still valid if $\cN$ is replaced by a subset 
$int\tilde{\cN}$ - which is entirely equivalent, by diffeomorphism invariance, to placing 
additional conditions on the initial data on $\cN$. 

This assumption implies that there exists a smooth boundaryless, globally hyperbolic\footnote{
Global hyperbolicity is defined in appendix \ref{definitions}.} 
metric manifold $M$ containing $\cN$ and its Cauchy development isometrically in its 
interior.\footnote{
Rendall embeds $\cN$ in $\Real^4$ and then shows that there exists a smooth Lorentzian metric 
$g'$ on $\Real^4$ such that $g'$ to the past of $\cN$ smoothly matches any Cauchy development 
(on $\Real^4$) of the initial data to the future of $\cN$. $g'$ and a Cauchy development spliced
together in this way defines a smooth metric on an open connected subset $U$ of $\Real^4$ 
excluding $\di\cN$ but including $int \cN$ such that $U - \cN$ is not connected. Now the initial
data used by Rendall is free. The only condition it must satisfy is smoothness. Thus it may be 
extended to a larger hypersurface $\cN^+$, of the same form as $\cN$, but containing $\cN$ in 
its interior. There thus exists a boundaryless metric manifold $U^+$ containing all of $\cN$ 
which extends the Cauchy development of $\cN$. The results of appendix \ref{nullhypersurfaces}, 
which take as a starting point that $\cN$ is embedded achronally in a boundaryless Lorentzian 
metric manifold therefore apply. In that appendix it is shown that $\cN$, and its maximal Cauchy
development, are contained isometrically in a smooth, boundaryless metric manifold $M$ which is 
globally hyperbolic (see prop. \ref{Mhat_good}).}
{}\footnote{
$\di X$ will always denote the boundary component of $X$ viewed as a manifold with boundary. 
See appendix \ref{definitions}. This boundary will in general differ from the 
topological boundary of the embedding of $X$ in another manifold.}

Indeed, in the present work a Cauchy development shall be defined as the domain of dependence 
of the initial data hypersurface embedded achronally in a manifold with a metric that satisfies 
Einstein's equation, at least within the domain of dependence (see def. 
\ref{Cauchy_development}). Note that the domain of dependence $D[\cN]$ of $\cN$ is not in 
general a manifold, not even a manifold with boundary.
On the other hand $\cD \equiv int D[\cN]$, the interior of $D[\cN]$, and $int \cN$ together do 
form a manifold with boundary, $\Dund$. $\Dund$ with its metric is essentially the whole Cauchy 
development. The underline in $\Dund$ is meant to represent the fact that $\Dund$ consists of 
$\cD$ and its past boundary. (Prop. \ref{cN_horizon} indicates that $D[\cN]$ consists entirely 
of the future domain of dependence of $\cN$.) $\Dund$ may also be thought of as $D[\cN]$ minus 
its future Cauchy horizon (prop. \ref{Dund_horizonless}).\footnote{
Cauchy horizons are defined in appendix \ref{definitions}.} 

\begin{figure}
\begin{center}
\input{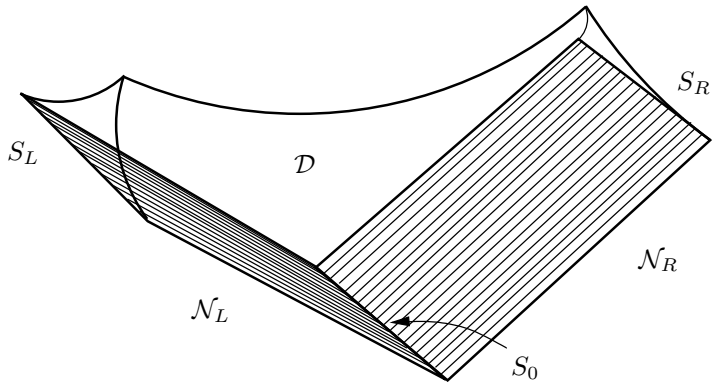}
\caption{The domain of dependence of a pair of plane null rectangles in $2+1$ 
dimensional Minkowski space is shown. The straight lines ruling the rectangles are the 
generators. The domain of dependence is bounded to the future by the light cones of the two 
ends of $S_0$ and the null surfaces orthogonal to $S_R$ and $S_L$.}
\label{Mfig}
\end{center}
\end{figure}

By prop. \ref{Cauchys_diffeo} the subsets $\Dund$ of all Cauchy developments, of distinct data, 
are diffeomorphic as manifolds (though of course not in general isometric).
They are thus all diffeomorphic to a particular such manifold, $\Dund_0$, which we 
may take to be the union of the interior of the hypersurface 
$\cN_0 = \{ t = |x|, |x|\leq 1, y^2 + z^2 \leq 1\}$ and the interior of its domain of dependence
in $\Real^4$ equipped with the Minkowski metric $ds^2 = -dt^2 + dx^2 + dy^2 + dz^2$ - see Fig. 
\ref{Mfig}.

\begin{figure}
\begin{center}
\input{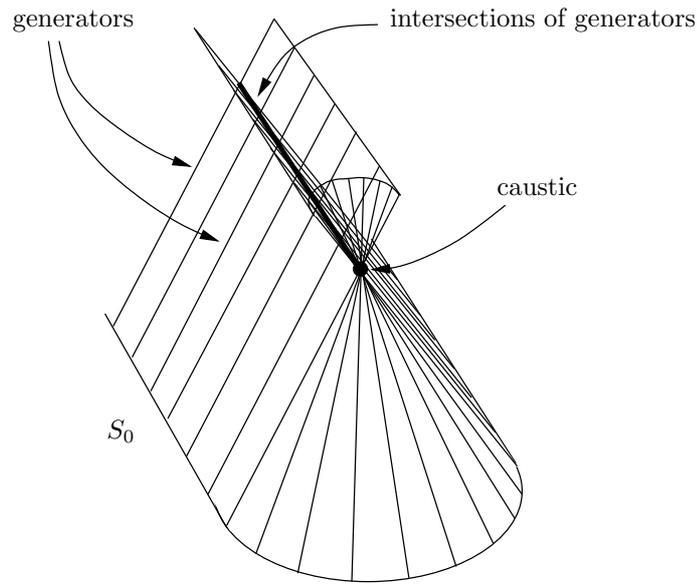}
\caption{An example of caustics and intersections of generators. $S_0$ is a spacelike curve in 
$2 + 1$ dimensional Minkowski spacetime having the shape of a half racetrack) - a semicircle
extended at each end by a tangent straight line. The congruence of null geodesics normal to 
$S_0$ and directed to the future and inward - the generators shown in the diagram - sweep out a
null surface having the form of a ridge roof, terminated by a (half) cone over the semicircle.
The generators from the semicircle form a caustic at the vertex of the cone. There neighbouring
generators intersect. On the other hand generators from the two straight segments of $S_0$ cross
on a line (the ridge of the roof) starting at the caustic, but the generators that cross there 
are not neighbours at $S_0$.}
\label{self_intersection}
\end{center}
\end{figure}

[How can one be certain that $\cN$, equipped with given initial data is indeed achronally 
embedded in the solution space. This question also arises in the spacelike Cauchy problem, 
but it becomes especially urgent when the initial data hypersurface is piecewise null, like 
$\cN$, because the generators will generally cross eventually, spoiling the achronality of the 
hypersurface they sweep out (by the corollary to theorem 8.1.2. of \cite{Wald}). Truncating the 
generators before a caustic is formed does not exclude the possibility that 
{\em non-neighbouring} generators intersect. (See Fig. \ref{self_intersection}). Thus it seems 
one would have to truncate the generators prior to any crossing, and the question arises how 
this requirement can be expressed as a restriction on the admissible initial data. 

It turns out that that is not necessary. Suppose $\cN$ is swept out by the two 
future directed null geodesic congruences normal to a spacelike disk in a spacetime $X$, as per 
its definition, and that it contains no caustics of the generators. (Suitable restrictions on 
the initial data do assure that - see prop. \ref{no_caustics}.) Then if the generators of $\cN$ 
cross, there always exists another spacetime, forming a locally isometric covering of a 
neighbourhood of $\cN$ in $X$, in which the generators do {\em not} cross and the $\cN$ is 
achronal. See appendix \ref{nullhypersurfaces}, and in particular prop. \ref{cN_achronal}.
Thus no restriction on the initial data is necessary to ensure the achronality of $\cN$.
Moreover, the solutions in which $\cN$ has self intersections and is not achronal are
represented in the phase space of Cauchy developments by covering manifolds.

A subtlety arises in connection with the Poisson brackets, which are defined in terms of small 
perturbations of the spacetime geometry. The small perturbations of the geometry of a non simply
connected spacetime $X$ are distinct from the small perturbations of its covering manifold 
$\tilde{X}$. Nevertheless it seems that if one is careful to identify functions of the geometry 
of $X$ with functions of the the geometry of $\tilde{X}$ that are invariant under the action of 
the fundamental group of $X$ on $\tilde{X}$ then the brackets between such functions calculated 
in $X$ are equal to those calculated between the corresponding quantities in $\tilde{X}$.]

%
%

All of the Cauchy developments can therefore be represented up to diffeomorphisms by metrics on 
the single, fixed manifold $\Dund_0$. Mapping the metric on $\Dund$ in each Cauchy development
to $\Dund_0$ via the diffeomorphism of prop. \ref{Cauchys_diffeo} one obtains a space $\cS_0$
of solutions to Einsteins equations representing the Cauchy developments. The diffeomorphism 
equivalence classes of solutions in $\cS_0$ form the phase space $\Phi$.    

The fact that each metric in $\cS_0$ is (diffeomorphic to) a maximal Cauchy development of null
initial data on $\cN$ is equivalent to the requirement that it be a solution such that 
$int \cN_0 \subset \Dund_0$ is null in the solution metric and the interior of $\Dund_0$ is the 
interior of the domain of dependence of $int\cN_0$ in $\Dund_0$, and also in any smooth 
extension of $\Dund_0$ and its metric that satisfies the field equation. (The final condition 
ensures maximality.) 

There is a further condition the metrics in $\cS_0$ must satisfy. As we have seen, our 
assumptions regarding the initial value problem on $\cN$ imply that the metric of the Cauchy 
development can be smoothly extended to a boundaryless manifold $M$ containing $\cN$, and 
in particular $\di\cN$. This gives rise to a restriction on the metrics in $\cS_0$. If 
$\Dund \cup \di\cN$ in $M$ is diffeomorphic to $\Dund_0 \cup \di\cN_0$ in $\Real^4$ 
then the restriction is simply that the solutions in $\cS_0$ must be smoothly extendible to 
$\di\cN_0$ in $\Real^4$. However we have not proved that these sets are diffeomorphic, only that
$\Dund$ and $\Dund_0$ are, so the form this restriction takes in $\cS_0$ is not known with 
certainty at present.\footnote{
What conceivably might happen, though I believe it cannot, is that on some generator from
$\di S_0$ the future boundary of $\Dund$ in $M$ is not only tangential to $\cN$ on the 
generator, as it is in flat spacetime, but it meets $\cN$ so ``softly'' that also the second 
derivatives (of spacetime coordinates) are equal on the two hypersurfaces.}

The requirement of smoothness at $\di\cN$ is relevant to the analysis of gauge transformations 
in subsection \ref{gauge}. In the definition of Poisson brackets, which is our main focus of 
interest, the issue can be sidestepped. The value of a Poisson bracket $\{A,B\}$ at a given 
solution, or phase point, $g$ involves only the variations of the quantities $A$ and $B$ under 
perturbations about the given fiducial solution. It is thus natural to work with metric fields 
living on the spacetime manifold $M_g$ of this fiducial solution, or on the domain 
of dependence of $\cN$ in $M_g$, rather than on Minkowski space or on the domain of 
dependence of $\cN_0$ in Minkowski space. $\cS$ shall be the space of maximal Cauchy 
developments (or ``solutions'' for short) mapped diffeomorphically to $\Dund_g$, in complete 
analogy with $\cS_0$. $\cF$ shall be the vector space of all smooth metric fields on 
$M_g$. 

Of course, basing the description on $\Dund_g$ and $M_g$, rather than on $\Dund_0$ and 
Minkowski space, does not by itself resolve the issue of boundary conditions at $\di\cN$ for 
metrics other than the fiducial solution metric. One is just replacing one preferred solution, 
Minkowski space, by another, $g$. However, it does allow the issue to be resolved for the small 
perturbations relevant to the definition of the bracket. We shall see that the Poisson bracket 
$\{A, B\}$ of real valued functions $A$ and $B$ on $\cF$ involves the variations of $A$ and $B$ 
under perturbations of the metric in the space, $L_g$, of smooth solutions on $M_g$ to the
field equation linearized about $g$.\footnote{
The linearized field equation has well defined solutions on all of $M_g$ even though the
metric $g$ need be a solution to the (vacuum) Einstein equation only on $D[\cN]$.}
{}\footnote{ 
Note that if $g$ is linearization stable then the space, $L_g|\Dund$, of solutions to the 
linearized field equations in $L_g$, restricted to $\Dund$, is identical with the space $T_g$ 
of tangents to the space of solutions to the full field equations on $\Dund$. 
At a solution that is not linearization stable $L_g|\Dund$ is larger than $T_g$. 

$T_g$ is essentially the tangent space to $\cS$ at $g \in \cS$. The solutions in $\cS$ are 
restricted by the requirement that the metric on $\Dund$ be a maximal Cauchy development of null
data on $\cN$, but in fact any variation of the solution about $g \in \cS$ (i.e. any variation 
in $T_g$) can be made to respect these restrictions by adding suitable diffeomorphism 
generators, so the tangent space to $\cS$ is $T_g$ with a certain ``diffeomorphism gauge 
fixing''.
 
It seems likely that most Cauchy developments of data on $\cN$ are linearization stable. 
Moncrief \cite{Moncrief} has shown that Cauchy developments of compact spacelike Cauchy 
surfaces without boundaries are linearization stable iff they have no Killing vectors.}
(see (\ref{Peierls2}) or (\ref{PB2})).


In the sequel the only $M$ and $\Dund$ considered will be those of the fiducial solution 
(whichever solution this is), so the subscript $g$ will be dropped.

It should be emphasized that while diffeomorphic metric fields may be considered physically
equivalent they are not the {\em same} in our mathematical model. A manifold $X$ consists of 
distinct {\it a priori} identifiable points (see the definition of manifolds in, for example 
\cite{Wald}) and it makes sense mathematically to speak of the value of the metric at the same
point for different metric fields. This means that one may also distinguish adapted coordinate
systems, such as Riemann normal coordinates, which depend on the metric placed on $X$, from 
fixed coordinates, which do not and are so to speak ``painted on the points of $X$". It also 
means that the variation $\dg g_{\mu\nu}(p)$ of the metric at a point $p$ is meaningful, a fact
which we shall use implicitly when formulating general relativity in terms of a variational 
principle and in the associated definitions of the symplectic structure and the Poisson bracket 
of the theory. 

Free initial data on $\cN$ provide a convenient coordinate system on the phase space. By 
definition each valuation of such data corresponds to a solution - that is, the data are not 
subject to constraints - and this solution is uniquely determined, up to diffeomorphisms, on 
$D[\cN]$. 
The algebra of functions on the phase space $\Phi$ is thus represented by the algebra of 
functions of the initial data,\footnote{
Some distinct data sets define diffeomorphic solutions, so the initial data will be multivalued 
as functions on $\Phi$. A function of initial data that represents a function on $\Phi$ must of 
course have the same value on all data sets that correspond to a given diffeomorphism 
equivalence class of solutions, that is, on all {\em equivalent} initial data sets. When using 
the ADM spacelike initial data \cite{ADM} or the Sen-Ashtekar variables \cite{Sen}
\cite{Ashtekar} to represent functions on the physical phase space this requirement - the 
requirement that the functions be gauge invariant - is highly non-trivial, since finding the 
class of initial data sets equivalent to a given set requires solving Einstein's field 
equations. Here the freedom in the data corresponding to a given 
diffeomorphism equivalence class of solutions amounts only to the freedom to change coordinates 
on the two spacelike disks $S_L$ and $S_R$, and in this way relabel the generators of $\cN_L$ 
and $\cN_R$. The action of these transformations on the initial data is simple and explicitly 
known - it does not require the solution of the field equations - so the multivaluedness of our 
initial data should not represent a problem. It is an interesting and open question whether this
freedom in our data should be understood as a gauge freedom (generated by some sort of 
constraint?) or rather as a global symmetry.}
and the (pre-)Poisson bracket can be specified explicitly by giving the brackets between the 
initial data.

Free initial data on piecewise null hypersurfaces have been found by several authors 
\cite{Sachs}\cite{Penrose}\cite{Bondi}\cite{Dautcourt}. Sachs \cite{Sachs} in particular has 
proposed a set of free initial data suitable for $\cN$, and has argued (somewhat heuristically) 
that the values of these data determine the solution in a neighbourhood of $\cN$ in $\Dund$ 
uniquely
up to diffeomorphisms. Rendall \cite{Rendall} has proved that these data, if $C^\infty$ on each 
of the null branches $\cN_L$ and $\cN_R$ of $\cN$ and continuous in a suitable sense at $S_0$, 
indeed determine a $C^\infty$ solution uniquely up to diffeomorphisms in a neighbourhood of 
$S_0$ in $\Dund$. This is somewhat weaker than what is needed, but it does support Sachs' 
original claim. Here, as already stated earlier in other words, it will be assumed that Sachs' 
data determine the solution, modulo diffeomorphisms, on all of $\Dund$. The brackets will be 
calculated between the elements of an initial data set, defined in subsection 
\ref{initial_data}, that is equivalent to, but slightly different from Sachs' data.

To be exact, in the present work the pre-Poisson bracket on initial data will be evaluated at 
$C^\infty$ solutions without Killing vectors such that the generators of a given branch, $\cN_L$
or $\cN_R$, of $\cN$ are either everywhere converging or everywhere diverging. The last 
condition ensures that the cross sectional area of a bundle of neighbouring generators is either
monotonically increasing or monotonically decreasing, which makes it possible to use the so 
called area distance, a quantity proportional to the square root of this cross sectional area, 
to parametrize the generators.

The corresponding limitation on the applicability of our formulae for the bracket on initial 
data is far weaker than one might think. Causality requires that the Poisson bracket between 
two observables is non-zero only if these depend on fields at causally related points (points 
connected by a causal curve), and it is natural (as will be discussed further on) to require the
same for the brackets between initial data. On $\cN$ causally related points must lie on the 
same generator (by prop. \ref{domain_of_influence_on_N}), so the only non-trivial brackets 
between 
data on $\cN$ are those between data on the same generator. Now suppose that the monotonicity 
conditions on the cross sectional area of bundles of generators does {\em not} hold for $\cN$, 
and suppose $p \in int S_0$ is the origin of a generator on which we wish to evaluate a bracket.
If the generators of both $\cN_L$ and $\cN_R$ have non-zero expansion rates at $p$, which 
generically is the case, then we can find a disk $S_0' \subset S_0$ containing $p$ in its 
interior on which these expansion rates are everywhere non-zero and of uniform sign. Our results
then apply to a hypersurface $\cN'$ swept out by the generators emerging from $S_0'$, for if the
expansion rate at $p$ is negative then the Raychaudhuri equation guarantees that it remains 
negative along the generators until a caustic, or conjugate point, is reached, while if the 
expansion rate is positive the generators may be cut off before they begin to reconverge.
Thus our results give the bracket between the data except when the expansion rate
of the generators of $\cN_L$ or $\cN_R$ vanish at $p$, or when one of the initial data in the 
bracket pertains to a point which lies beyond the cutoff of $\cN'$ because the neighbouring 
generators, initially diverging at $S_0$, have begun to reconverge. In fact, in the case of 
re converging generators the formulae we will find for the brackets of initial data are still 
meaningful and can be used to assign values to these. If the expansion rate of the generators 
vanishes at $p$ the expressions obtained for the brackets of some of the data are still 
meaningfully but other brackets are singular. Whether any of these extrapolations are correct 
remains to be determined since these cases are not treated in the present derivation of 
the bracket. 

As the fundamental definition of the physical Poisson bracket an adaptation to general 
relativity of the Peierls bracket \cite{Peierls} will be adopted. Let $g$ be a solution to the 
field equations and let $A$ and $B$ be diffeomorphism invariant functionals of the metric, and 
suppose that the functional gradient $\dg A/\dg g^{\mu\nu}$ of $A$ has compact support in 
spacetime. [The support of $\dg A/\dg g^{\mu\nu}$ will be called the {\em domain of sensitivity}
of $A$, and will be denoted by $s_A$.] The value of the Peierls bracket $\{A,B\}$ at $g$ is 
determined by the first order perturbation of the solution, and thus of the value of $B$, 
caused by the addition of a term proportional to $A$ to the action, $I$. Of course the 
perturbation of the 
metric is not determined uniquely by the change in the action. Aside from the freedom to apply 
diffeomorphisms, one may choose a retarded perturbation, an advanced perturbation, or a 
combination of these. Let $g^+$ be a solution to the field equations corresponding to 
the modified action $I_\lam = I + \lambda A$ (with $\lam$ a real constant) which equals $g$ 
outside the causal future (domain of influence) $J^+[s_A]$ of $s_A$ and $g^-$ a solution that 
agrees with $g$ outside the causal past $J^-[s_A]$ of $s_A$.\footnote{
The causal past and future are defined in appendix \ref{definitions}.}
(That is, $g^+$ is the solution with a retarded perturbation, while $g^-$ incorporates an 
advanced perturbation.)
Then the Peierls bracket is
    \be    \label{PB1}
                \{A,B\} = \Dg_A B = \frac{d}{d\lambda} B(g_\lambda^+)|_{\lambda = 0} 
                                  - \frac{d}{d\lambda} B(g_\lambda^-)|_{\lambda = 0}. 
    \ee

As the action for gravity we will use the Einstein-Hilbert action,\footnote{
The normalization of the action is chosen so that when a particle of mass $m$ is coupled to the 
gravitational field with it's action normalized so that the canonical momentum of the particle 
is $m\: dx^\mu/d\tau$, with $\tau$ the proper time of the particle, then Newton's law of 
gravitation (with G = Newton's constant) is recovered in the appropriate limit.
The sign conventions for the curvature tensor and scalar are those of \cite{Wald}, that is, 
$R = R_{\mu\nu}{}^{\mu\nu}$ with 
$[\nabla_\mu,\nabla_\nu]\omega_\sg = R_{\mu\nu\sg}{}^\rho \omega_\rho$ for any 1-form $\omega$.}
        \be		\label{EH}
                I = \frac{1}{16\pi G}\int_Q R \sqrt{-g}\: d^4 x,
        \ee
where $Q$ is the spacetime domain of integration of the action, $R$ the scalar curvature defined
by the spacetime metric $g_{\mu\nu}$, $g$ the determinant of this metric, and $d^4x$ the 
coordinate volume form in a coordinate system $x$ fixed in the manifold $Q$. 

No boundary terms need be added to the action because boundary terms do not affect the Poisson
bracket implicit in the action. (See \cite{Lee_Wald} for a discussion of the effect of boundary
terms on the presymplectic 2-form, and sub-subsection \ref{Peierls_symplectic} for the relation 
between the presymplectic 2-form and the Peierls bracket.)   

This definition of the Poisson bracket has the virtue of being manifestly covariant and, 
moreover, is very directly related to the commutators of quantum theory. 
Peierls showed that for field theories in Minkowski space his bracket is equal
to that computed from the conventional Poisson bracket on initial data.
Further on I will show that this bracket has all the properties of a Poisson bracket, including
antisymmetry and the Jacobi relation, which are not obvious from the expression (\ref{PB1}).
In fact it will be argued that the Peierls bracket between diffeomorphism invariant (and 
sufficiently smooth) functions of the spacetime geometry is equal to the conventional Poisson 
bracket between these functions implied by the Poisson brackets between the canonical ADM 
variables on a spacelike Cauchy surface \cite{ADM}.  

Unfortunately the initial data on $\cN$ used in the present paper are too singular as functions 
of the spacetime geometry to allow an unambiguous, direct computation of the Peierls bracket 
between them. My approach is therefore to use Peierls' definition to define the bracket on a set
of sufficiently regular diffeomorphism invariant functions of the metric on $\Dund$, which will 
be called ``observables'', and then to look for a bracket on the initial data which reproduces
the Peierls bracket on the observables when these observables are expressed as functions of the 
initial data. 

I shall call a functional $A$ of the metric on $\Dund$ an {\em observable} if it is 
diffeomorphism invariant and its functional gradient 
$\ag_{\mu\nu} = \frac{1}{\sqrt{-g}} \frac{\dg A}{\dg g^{\mu\nu}}$ is $C^\infty$ and of compact 
support contained in the interior of $\Dund$. The support of $\ag$ is of course the 
domain of sensitivity of $A$.

If all measurable functions of the metric are observables in this sense then our approach is 
certainly justified because the only predictive content of a field theory is what it has to say 
about measurable quantities. The predictive content of a classical field theory 
consists of the relations between measurable functions of the fields implied by the field 
equations. In addition if the field theory is derived from a local action, like general 
relativity is, then the action defines a Poisson bracket between the measurable functions of the
classical fields which reflects the commutation relations of the corresponding quantum 
observables.\footnote{
The same classical theory can sometimes be derived from distinct actions 
that define distinct Poisson brackets, for instance in the case of the harmonic oscillator. This
presumably means that there are also distinct quantum systems which behave according to this 
classical theory in the classical limit. The Poisson structure is merely a convenience at the 
classical level which allows one to put theories into a universal (canonical) form. It assumes
physical significance only at the quantum level.} 
A classical field theory contains no further information about the classical world it models or 
the quantum world which might underlie this classical world. In particular the initial data and 
their brackets are no more than a means to represent the measurable quantities and their 
brackets.

Thus, in a mathematical formulation of a field theory in terms of observables these should be 
nice enough, as functions of the field, that the Poisson bracket between them can be defined, 
and yet rich enough so that all measurable quantities are represented by observables, or at 
least the limits of sequences of observables. Furthermore it is at least convenient and probably
necessary that the Poisson brackets of observables are themselves observables, so that the 
observables form a closed Poisson algebra.

It is not clear that the class of observables we have defined above meet these 
criteria,\footnote{
It is not hard to see that the domain of sensitivity of the Peierls bracket of two of our
observables is compact and contained in the interior of $\Dund$. However it is not obvious 
that the functional gradient of the bracket must be $C^\infty$ in spacetime.} 
but it is reasonable to expect that our observables form at least a subset of an algebra of 
observables of general relativity that does, and, as we shall see, there are enough of them so 
that their Peierls brackets, together with a reasonable causality condition, determine almost 
unique brackets between our null initial data.

Do observables as defined above in fact exist? An example shows that they do: For a generic 
$C^\infty$ metric $g$ on $\Dund$, and a point $p$ in the interior of $\Dund$, $C^\infty$ 
coordinates consisting of scalars formed from the curvature can be found which are non-singular 
on a neighbourhood of $p$. Choose a compact domain $c_0 \subset \cD$, with non-empty interior,
in this neighbourhood and let $[c] \subset \Real^4$ be the coordinate range corresponding to 
$c_0$. Then for all metrics in a suitable neighbourhood $U \subset \cF$ of $g$, in which all 
components of the metric with respect to the curvature scalar coordinates are nearly equal to 
those of $g$ throughout the coordinate range $[c]$, the same curvature scalars continue to be 
good spacetime coordinates on the pre-image $c$ of $[c]$ and this pre-image will be a compact 
subset of $\cD$ diffeomorphic to $c_0$.

Now note that the metric components taken in these coordinates are diffeomorphism invariants. 
Suitable functionals of these metric components on $[c]$ define observables. The domain of 
sensitivity of such a functional is necessarily a closed subset of $c$, and thus a compact
subset of the interior of $\Dund$. [The functional gradient of such a functional, $A$, vanishes 
outside $c$ because a smooth infinitesimal variation of the metric which vanishes on $c$ leaves 
the curvature scalars invariant on $c$ and changes them only smoothly and infinitesimally 
outside $c$. This implies that $A$ is unchanged because the preimage of $[c]$ remains unchanged 
(i.e. equal to $c$) and the metric components with respect to the curvature scalar coordinates 
are unchanged in $[c]$.] It is not difficult to construct explicitly a functional $A$ of this 
form so that its functional gradient exists, for metrics in $U$, and is $C^\infty$ in spacetime.
For instance, if $x^\mu$ are manifold fixed coordinates, $z^\ag$ are curvature scalar 
coordinates, and $f$ is a smooth function from $\Real^4$ to $\Real$ that vanishes outside $[c]$ 
then 
\be
A = \int_{[c]} f(z) \sqrt{-det[g_{\ag\bg}](z)} d^4 z 
= \int_c f(z(x)) \sqrt{-det[g_{\mu\nu}](x)} d^4 x
\ee
satisfies all requirements. (It is less obvious how to construct observables that are well 
defined on all of $\cF$, or $\cS$, and still satisfy our requirements, but that is not necessary
here.)

For non-generic metrics $g$ which have Killing vectors the situation seems to be different.
The construction of functionally differentiable observables which distinguish 
between metrics with isometries (Killing vectors) seems to be problematic. Perhaps this should 
not surprise us since the quotient manifold of the space of metric fields $\cF$ by the group of 
diffeomorphisms of $\Dund$ is singular at metrics with isometries. Notice also that observables 
that are actually measured are always defined using some sort of inhomogeneity or asymmetry, if 
not in the metric then in the matter fields, to single out points or regions of spacetime. 
Recall that metrics with isometries are explicitly excluded from the scope of the present work. 

The definition (\ref{PB1}) of the Peierls bracket has one serious problem: It is in general 
not obvious that the exact stationary points $g^+$ and $g^-$ of the modified action 
$I + \lam A$ actually exist. Fortunately this problem is easily overcome.
The exact stationary points $g^+$ and $g^-$ are not really necessary in order to define
the Peierls bracket. The first order (in $\lam$) approximations to these are sufficient.
The Peierls bracket is a creature of first order perturbation theory and is well defined 
whenever this theory is:
Exact stationarity of the modified action requires 
\be             \label{exact_stationarity}
G_{\mu\nu} + 16\pi G\lam \ag_{\mu\nu} = 0
\ee
where $G_{\mu\nu} = R_{\mu\nu} - 1/2 g_{\mu\nu} R$ is the Einstein tensor. This is just
Einstein's field equation with stress energy tensor $-2\lam \ag_{\mu\nu}$. [The diffeomorphism
invariance of $A$ implies that $\ag_{\mu\nu}$ is divergenceless, as the stress energy tensor in 
Einstein's equation must be: For any $C^\infty$ vector field $\chi$
\be             \label{div_alpha_0}
0 = {\pounds}_\chi A = -2\int_\cD \ag_{\mu\nu} \nabla^\mu \chi^\nu \sqrt{-g}\: d^4 x = 
 2\int_\cD [\nabla^\mu \ag_{\mu\nu}] \chi^\nu \sqrt{-g}\: d^4 x
\ee
and thus $\nabla^\mu \ag_{\mu\nu} = 0$.\footnote{Conversely, if $A$ is defined on a 
diffeomorphism invariant domain of metric fields in $\cF$ and $\ag$ is divergenceless on all of 
this domain, A is diffeomorphism invariant.}]

On the other hand, if $g$ is a solution to Einstein's vacuum equation $G_{\mu\nu} = 0$, then the
modified action is stationary to first order in $\lam$ at $g + \lam \dg_A g$ provided
\be             \label{lin_field_eq}
\dg_A G_{\mu\nu} + 16\pi G \ag_{\mu\nu} = 0.
\ee
Here $\dg_A$ acting on a functional of the metric is the linear order variation of the 
functional due to the variation $\dg_A g$ of the metric.

The Peierls bracket of two observables $A$ and $B$ may be defined by
\be     \label{Peierls2}
\{A, B\} = \Dg_A B
\ee
with
\be             \label{Dg_def}
\Dg_A = \dg^+_A - \dg^-_A
\ee
where $\dg^+_A g$ and $\dg^-_A g$ are solutions to (\ref{lin_field_eq}) that vanish outside 
$J^+[s_A]$ and $J^-[s_A]$ respectively.

When the exact solutions $g^+$ and $g^-$ exist then $dg^\pm/d\lam$ satisfies the conditions
defining $\dg_A^\pm g$, so the definition (\ref{PB1}) of $\{A, B\}$ is consistent with
(\ref{Peierls2}) and (\ref{Dg_def}). The advantage of the definition via (\ref{Peierls2}) and 
(\ref{Dg_def}) is that one can show that $\dg^+_A g$ and $\dg^-_A g$ always exist, provided $A$ 
is an observable ($\ag$ is $C^\infty$, divergenceless, and of compact support contained in the 
interior of $\Dund$), and is unique up to diffeomorphisms, assuring the existence and uniqueness
of $\{A, B\}$ for any observable $B$. All that is needed are Green's functions for linearized 
general relativity, so that the advanced ($-$) and retarded (+) perturbations of the metric 
$\dg^\pm_A g$ can be expressed as\footnote{
As already noted $-2\lam \ag$ acts as a stress energy tensor generating the perturbation
of the metric. The first order (in $\lam$) expression given in (\ref{pert1}) for the retarded 
perturbation should be compared with formulae for the 
gravitational radiation produced by a given stress energy. (see for example \cite{Wald}).}
        \be             \label{pert1}
\dg_A^\pm g_{\sg\tau}(y) 
	= -16\pi G\int \ag^{\mu\nu}(x) G^{\pm}_{\mu\nu\,\sg\tau} (x, y) \sqrt{-g}\, d^4 x,
        \ee
where $G^-$ and $G^+$ are respectively the advanced and the retarded Green's functions for 
the linearized Einstein equation in the gauge used. In terms of these Green's functions the 
Peierls bracket is
        \be             \label{PB2}
            \{A,B\} = 16\pi G\int \ag^{\mu\nu}(x) \Delta_{\mu\nu\,\sg\tau}(x,y) \bg^{\sg\tau}(y)
	    \sqrt{-g}\,d^4 x\:\sqrt{-g}\,d^4 y,
        \ee
with $\Delta = G^+ - G^-$ and $\bg_{\mu\nu} = \frac{1}{\sqrt{-g}} \frac{\dg B}{\dg g^{\mu\nu}}$ 
the functional gradient of $B$. Note that (\ref{PB2}) not only provides a quite explicit 
definition of the Peierls bracket, but also allows the bracket to be extended to a much wider 
class of functionals than our observables.

To define the desired Green's function it is necessary to fix the freedom to change solutions
to (\ref{lin_field_eq}) by linearized diffeomorphisms,
$\dg_A g_{\mu\nu} \rightarrow \dg_A g_{\mu\nu} + 2\nabla_{(\mu}\xi_{\nu)}$. 
We shall adopt the ``transverse gauge'' in which it is required that the trace reverse, 
$\bar{\cg}_{\mu\nu} = \cg_{\mu\nu} - \frac{1}{2} g_{\mu\nu} (g^{\sg\rho}\cg_{\sg\rho})$, 
of the perturbation $\cg_{\mu\nu} = \dg g_{\mu\nu}$ of the metric satisfies the condition 
\be             \label{transverse_gauge}
        \chi_\nu \equiv \nabla^\mu \bar{\cg}_{\mu\nu} = 0.
\ee
The linearized Einstein tensor may be written as
\be             \label{dgG}
\dg G_{\mu\nu} = \nabla_{(\mu}\chi_{\nu)} - 1/2 g_{\mu\nu}\nabla_\sg\chi^\sg
- 1/2 (\nabla^\sg\nabla_\sg \bar{\cg}_{\mu\nu} - 2R^\sg{}_{\mu\nu}{}^\tau \bar{\cg}_{\sg\tau}),
\ee
so if (\ref{transverse_gauge}) holds for $\dg_A g$ then (\ref{lin_field_eq}) is equivalent to 
the linear, diagonal, second order hyperbolic system
\be             \label{reduced_linearized}
\nabla^\sg\nabla_\sg \bar{\cg}_{\mu\nu} - 2 R^\sg{}_{\mu\nu}{}^\rho \bar{\cg}_{\sg\rho} 
= 32\pi G \ag_{\mu\nu}
\ee
(see \cite{Wald} section 7.5.)
This equation defines unique advanced and retarded Green's functions \cite{three_muses}, with 
which one may calculate the unique advanced and retarded solutions, $\bar{\cg}^\pm$, to
(\ref{reduced_linearized}). Moreover, these Green's functions are regular in the sense that 
the solutions obtained by integrating them against a $C^\infty$ source distribution $\ag$ are
also $C^\infty$. 

To show that the $\bar{\cg}^\pm$ correspond to solutions of the linearized Einstein equations
(\ref{lin_field_eq}) 
it is sufficient to show that they satisfy the transverse gauge condition. Taking the divergence
of (\ref{reduced_linearized}) and using the Bianchi identity and the fact that the unperturbed 
metric $g_{\mu\nu}$ satisfies Einstein's vacuum field equation one obtains
\be             \label{transverse_gauge_preservation}
\nabla^\sg\nabla_\sg \nabla^\mu \bar{\cg}_{\mu\nu} = 32\pi G \nabla^\mu \ag_{\mu\nu},
\ee
a linear, diagonal, second order hyperbolic system for $\nabla^\mu \bar{\cg}_{\mu\nu}$, with
the divergence of $\ag$ as the source. Since this source term vanishes, because $A$ is 
diffeomorphism invariant, and $\bar{\cg}^\pm$ vanishes either to the past or to the future
of the support of $\ag$, $\nabla^\mu \bar{\cg}^\pm_{\mu\nu} = 0$ everywhere. The corresponding 
metric perturbations, $\dg_A^\pm g_{\mu\nu} = \bar{\cg}^\pm_{\mu\nu} - 1/2 g_{\mu\nu}g^{\sg\tau}
\bar{\cg}^\pm_{\sg\tau}$, which may be expressed in the form (\ref{pert1}), thus {\em are}
transverse gauge solutions to the linearized Einstein equations with source $\ag$, 
(\ref{lin_field_eq}). They may therefore be used to evaluate the variations $\dg_A^\pm B$ of 
$B$. 

Any variation of the metric may be transformed to one satisfying the transverse gauge condition 
by a suitable linearized diffeomorphism - see \cite{Wald}. The unicity of the advanced and 
retarded solutions to (\ref{reduced_linearized}) therefore implies that all retarded and 
advanced perturbations satisfying (\ref{lin_field_eq}) are equal, modulo a linearized 
diffeomorphism, to $\dg_A^+ g_{\mu\nu}$ and $\dg_A^- g_{\mu\nu}$ respectively. Since $B$ is 
diffeomorphism invariant, they all define the same variations of $B$, $\dg_A^\pm B$, as these. 
$\{A, B\}$ is thus well defined and unique.

The difficulty in applying Peierls' definition of the bracket directly to null initial data can 
now be understood. The basic problem is that the Green's functions $G^\pm(p,q)$ and 
$\Delta(p,q)$ are discontinuous in $q$ on the light cone of $p$ (the boundary of the causal 
domain of influence $J^+[p]\cup J^-[p]$ of $p$ in $M$). On a flat spacetime $G^+$ and 
$G^-$ are distributions supported entirely on the light cone itself. On curved spacetimes 
additional contributions should appear that are smooth functions on the interior of $J^+[p]$ 
but discontinuous on the boundary.\footnote{
The retarded Green's function in curved spacetime for the massless Klein-Gordon equation, which 
is very similar to the transverse gauge linearized Einstein equation, can be estimated in the 
limit in which the two points, $p$ and $q$, are close using an adiabatic expansion (see Birrel 
and Davies \cite{Birrel_Davies} p. 74). In normal coordinates about $p$ the zero order term is 
the Minkowski spacetime Green's function, and the lowest order correction (in a vacuum Einstein 
spacetime) is
\be
\frac{1}{1440\pi} R^{\lam\rho\kappa}{}_\mu R_{\lam\rho\kappa\nu} (q-p)^\mu (q-p)^\nu 
\theta(p,q),
\ee
where $\theta(p,q)$ is $+1$ when
$q$ is in the causal future of $p$, and zero otherwise.}

This means that the Peierls bracket (\ref{PB2}) between null initial data is ambiguous: As will 
be evident from the definitions of our initial data, the domain of sensitivity of an initial 
datum associated with a point $p \in \cN$ consists of (a portion of) the generator, or 
generators, through $p$. The Peierls brackets between initial data at causally related points of
$\cN$, which necessarily lie on the same generator (by prop. \ref{domain_of_influence_on_N} in 
appendix \ref{nullhypersurfaces}), is thus ambiguous because the integral (\ref{PB2}) is 
ambiguous in this case. It would be interesting to find out whether the bracket on null initial 
data found in the present work can be obtained from a simple disambiguation of (\ref{PB2}).

\subsection{The Peierls bracket, the presymplectic 2-form, and the bracket on initial data}

We have defined the Peierls bracket between observables, our aim is now to find a bracket 
$\{\cdot,\cdot\}_*$ on the null initial data which reproduces the Peierls bracket on observables
when these are expressed in terms of this data. In the following I will show that the definition
(\ref{Peierls2}, \ref{Dg_def}) of the Peierls bracket in terms of first order perturbations 
governed by equation (\ref{lin_field_eq}) implies that it is, in a certain sense, inverse to the
presymplectic 2-form on the space $L_g$ of solutions to the linearized field equations, an 
object which can be computed explicitly in terms of initial data from the 
action. We shall see, moreover, that to ensure that the bracket on the initial data on $\cN$ 
reproduces the Peierls bracket on observables it is sufficient that this bracket is 
also inverse (in a slightly different sense) to the same presymplectic 2-form. This condition 
will be used in section \ref{Poisson_bracket} to calculate the bracket on initial data.

\subsubsection{The Peierls bracket and the presymplectic 2-form}\label{Peierls_symplectic}

The relation between the Peierls bracket and the presymplectic 2-form is valid for any local,
Lagrangean field theory - that is, for any theory derived from an action which is the spacetime
integral of a Lagrangean density which is a function, at each point, only of the fields and 
their derivatives up to a finite order (see \cite{Lee_Wald} for a precise definition of this 
class of theories). The specific features of general relativity are therefore not needed in the 
demonstration of this relation, and a notation which pushes these features into the background 
will be adopted. The key result, equation (\ref{inverse2}), may also obtained less abstractly 
working with the explicit form of the linearized Einstein equations, as is shown in appendix 
\ref{another_approach}, where (\ref{inverse2}) is obtained from a Kirchhoff type formula for the
solution to the linearized field equations in terms of initial data. 

Our first step will be to demonstrate a relation between variations of an observable $A$
and the perturbation $\Dg_A$ produced by that observable. Specifically it will be shown that the
contraction of $\Dg_A$\footnote{Like any smooth variation of the metric field, $\Dg_A$ can be 
regarded as a tangent vector to $\cF$.} with the presymplectic 2-form is equal to the gradient 
of $A$ when both are pulled back to the space $L_g$ of perturbations about a solution $g$
which satisfy the linearized field equations.\footnote{
What this says is essentially that $\Dg_A$ is (minus) the Hamiltonian vector field, $v_A$ 
generated by the Hamiltonian $A$ and the presymplectic 2-form $\Omega$. In general $v_A$ 
is defined by the condition that $\dg A = - \Omega[v_A, \dg]$ $\forall \dg$ tangent to the 
phase space.}

We shall work with the action on a domain $Q$ bounded to the past by $\cN$ and 
to the future by a smooth closed Cauchy surface $\Sg_+$\footnote{
Cauchy surfaces are defined in appendix \ref{definitions}. The existence of Cauchy surfaces
of $\cD$ is guaranteed by the global hyperbolicity of $\cD$ (prop. 
\ref{closed_cauchy_unnecessary} or \cite{Hawking_Ellis} prop. 6.6.3.) and Geroch's 
theorem, prop. 6.6.8. of \cite{Hawking_Ellis}. That these may be taken to be smooth follows 
from the result of \cite{Bernal_Sanchez0}.}
of $\cD$, the interior of which is contained in $\cD$. Specifically $Q$ is the closure in 
$M$ of the open set $I^+[\cN]\cap I^-[\Sg_+] \subset \cD$.\footnote{
The chronological past, $I^-[S]$, and future, $I^+[S]$, of a subset $S$ of spacetime are defined
in appendix \ref{definitions}.}
$\Sg_+$ is chosen to lie to the future of the domain of sensitivity, $s_A$, of $A$ so that 
$s_A$ is contained in the interior of $Q$. (See Fig. \ref{Qfigure}.) 

By prop. \ref{cauchy_D} $\Sg_+$ is compact, so prop. 6.6.1 of \cite{Hawking_Ellis} and its 
corollary imply that, in the globally hyperbolic spacetime $M$, 
$J^+[\cN]\cap J^-[\Sg_+] = \overline{I^+[\cN]}\cap \overline{I^-[\Sg_+]} \supset Q$ is compact.
Since $Q$ is closed it is also compact. Integrals over $Q$ are therefore defined and the Stokes 
theorem applies to $Q$ and $\di Q$. 

$\cN$ and $\Sg_+$ form the entire boundary of $Q$ in $M$: Suppose $p \in Q$, then, since $Q$ is 
the closure of a subset of $\cD$, $p \in \overline{\cD}$. Since 
$\cD \equiv int D[\cN] = int D[\Sg_+]$ \cite{Hawking_Ellis} prop. 6.5.1 requires that every 
inextendible timelike curve $\cg$ through $p$ crosses both $\cN$ and $\Sg_+$. $\cg$
must cross $\cN$ at $p$ or at a point $r$ to the past of $p$, because 
$Q \subset \overline{I^+[\cN]}$ which is disjoint from $I^-[\cN]$ since $\cN$ is achronal,
and similarly $\cg$ must cross $\Sg_+$ at $p$ or at a point $s$ to the future of $p$.
Thus if $p$ does not lie in either $\cN$ or $\Sg_+$ then $p \in I^+(r)\cap I^-(s)$, and 
therefore $p$ lies in the interior of $Q$. (This is of course consistent with prop. 
\ref{cauchy_D} which shows that $\di\Sg_+ = \di\cN$.) 

For later convenience $\Sg_+$ and $\cN$ will {\em both} be oriented with the positive side to 
the future.\footnote{
The space of n-forms at a point of an $n$ dimensional manifold is one dimensional. Thus there 
are two equivalence classes of such n-forms under multiplication by
positive real numbers. The orientation of a manifold is a continuous choice of one of these 
equivalence classes as the ``positive" one. A vector transverse to an $n-1$ dimensional 
submanifold emerges on the ``positive side" of the submanifold if the contraction of the vector 
with a positive $n$-form is a positive $(n-1)$-form of the submanifold.}
Thus, as oriented manifolds, $\di Q = \Sg_+ \cup -\cN$ and $\di\Sg_+ = \di\cN$. 
 
\begin{figure}

\begin{center}
\input{nullfig4.pstex_t}
\caption{The diagram shows schematically (or literally in $1 + 1$ dimensional spacetime) the 
initial data hypersurface $\cN$, its maximal Cauchy development (minus future Cauchy horizon) 
$\Dund$, a boundaryless manifold
$M$ containing $\Dund$ and $\cN$, a Cauchy surface $\Sg_+$ of the interior of $\Dund$, the
domain $Q$ between $\cN$ and $\Sg_+$, and the domain of sensitivity $s_A$ of an observable $A$.
The support in $\Dund$ of the retarded and advanced perturbations of the metric generated by $A$
are also indicated - as white regions where the vertical or diagonal hatching is absent.}  
\end{center}
\label{Qfigure}
\end{figure}

At an exact stationary point of the modified action $I + \lam A$ the variation of this action is
a boundary term:\footnote{By a stationary point of an action we shall always mean a field 
configuration at which the action is stationary with respect to all variations {\em that vanish
in a neighbourhood of the boundary}. That is, we mean a solution to the field equations 
corresponding to the action.}
        \be             \label{stationary}
                \dg I + \lam \dg A = \phi[\dg],
        \ee
where $\phi$ is the integral of a local functional of the fields and their variations over the 
boundary of $Q$.\footnote{
We will always take the view that $\dg$ represents the derivative along an unspecified 
one parameter family of field configurations, and is thus a vector in the tangent space to the 
space of fields being considered. Another point of view, adopted in \cite{Crnkovic_Witten}, is 
that $\dg$ is the exterior derivative operator on the space of fields, and it's contraction with
a vector is the derivative along the vector. Though this second point of view is not adopted in 
the discussion (and we denote the exterior derivative operator on the space of fields by $\dI$) 
our formulae, with slight modifications, may be interpreted in this way, if one takes $\dg$ 
without subscripts (and $\dI$) to be the exterior derivative, and $\dg$s with subscripts to be 
directional derivatives along vector fields.}

As we have seen already, whether or not exact stationary points of $I + \lam A$ exist, 
approximate stationary points where the action $I + \lam A$ is stationary up
to first order in $\lam$ can be constructed by adding to any exact stationary metric $g$ of the
unperturbed action $I$ a suitable correction $\lam \dg_A g$ linear in $\lam$: At the 
perturbed metric $g + \lam \dg_A g$,
        \be             \label{stationary_order1}
                \dg I + \lam \dg A = \phi[\dg] + O(\lam^{1+}),
        \ee
where $O(\lam^{1+})$ vanishes faster than $\lam$ as $\lam \rightarrow 0$. 

Note that $A$ does not contribute to $\phi$, because its domain of sensitivity $s_A$ does 
not intersect the boundary of $Q$, and consequently $\phi$ is independent of $\lam$.
When $I$ is the Einstein-Hilbert action (\ref{EH}) on $Q$: 
        \bearr  
    \phi[\dg] & = & -\frac{1}{8\pi G}\int_Q [\nabla_{[\mu} \dg\Gamma^{\sg}_{\sg]\nu} g^{\mu\nu}]
                                \sqrt{-g}\:d^4x\\
              & = & -\frac{1}{8\pi G}\int_{\di Q} \dg\Gamma^{[\sg}_{\sg\nu} g^{\mu]\nu}
                                \sqrt{-g}\:d\Sigma_\mu, \label{phiEH}
        \eearr
where $\nabla$ is the metric compatible covariant derivative, $\Gamma$ is the Christoffel 
symbol, and $d\Sigma_\mu = 1/3!\:\eg_{\mu\nu\rho\tau}\:dx^\nu\wedge dx^\rho \wedge dx^\tau$ is 
the coordinate volume 3-form on $\di Q$.\footnote{%
The exterior, or wedge, product of a sequence of forms, of valence $n_1$, $n_2$, .... adding up 
to a 
total valence of $n$, is defined to be $n!/(n_1! n_2! ...)$ times the antisymmetric component of 
their tensor product. For example if $a$ is a 1-form and $b$ a 2-form, then
        \be
                [a\wedge b]_{\mu\nu\sg} = 3\,a_{[\mu}b_{\nu\sg]}.
        \ee
The integral of an $n$-form $\ag$ over an $n$-manifold $X$ is defined by
        \be
            \int_X \ag = \frac{1}{n!}\int_X \eg^{\mu_1 ... \mu_n}\ag_{\mu_1 ... \mu_n}\:d^n x,
        \ee
with $\eg$ antisymmetric and $\eg^{1 2 ... n} = 1$.}
  
The linearized field equation that governs $\dg_A g$ is obtained by differentiating 
(\ref{stationary_order1}) with respect to $\lam$ and setting $\lam = 0$. Note that the 
derivative receives contributions both from the explicit $\lam$ dependence 
in (\ref{stationary_order1}) and from the $\lam$ dependence of the metric $g + \lam\dg_A g$. 
The result is
        \be             \label{linearized}
                \dg_A \dg I + \dg A = \dg_A \phi[\dg],
        \ee
where, as in (\ref{lin_field_eq}), $\dg_A$ acting on a functional of the metric is the linear
variation of the functional due to the variation $\dg_A g$ in the metric.\footnote{
Note that $\dg$ and $\dg_A$ in 
(\ref{linearized}) may be viewed as vector fields on the space $\cF_Q$ of field configurations 
on $Q$, but they may equally well be taken to be vector fields on the space $\cF$ of field 
configurations on all of $M$ since no term in (\ref{linearized}) depends on their action on 
the fields outside $Q$. We shall take the latter point of view.}
The contribution from the term $O(\lam^{1+})$ vanishes because $d O(\lam^{1+})/d\lam|_{\lam=0} =
\lim_{\lam \rightarrow 0} O(\lam^{1+})/\lam = 0$.

In general relativity the validity of (\ref{linearized}) for all $\dg$ with support contained in
the interior of $Q$ is equivalent to the validity of (\ref{lin_field_eq}) there. Of course 
(\ref{linearized}) also has a boundary term, probed by $\dg$s supported on the boundary. This 
term provides additional information, not contained in (\ref{lin_field_eq}), which is the focus 
of our present interest because, as we shall see, it determines the symplectic structure and 
Peierls bracket of the theory.

Perturbations $g + \tau\dg_0 g$ of $g$ which are first order (in $\tau$) stationary points
of the {\em unperturbed} action $I$ obey the same equation, but without the source term
$\dg A$:
                \be             \label{linearized_vacuum}
                \dg_0 \dg I = \dg_0 \phi[\dg].
        \ee
In general relativity the bulk term of this equation corresponds to the linearized vacuum 
Einstein equation on the interior of $Q$.

Setting $\dg = \dg_0$ in (\ref{linearized}) and $\dg = \dg_A$ in (\ref{linearized_vacuum})
one obtains
                \bearr
                        \dg_0 A & = & \dg_A \phi[\dg_0] - \dg_A\dg_0 I  \\
                                & = & \dg_A \phi[\dg_0] - \dg_0 \phi[\dg_A] - [\dg_A,\dg_0] I
                \eearr
The commutator $[\dg_A,\dg_0]$ of the two vector fields $\dg_A$ and $\dg_0$ on $\cF$
is again a vector field (the Lie bracket of the two), so, since $g$ is an exact stationary point
of $I$, $[\dg_A, \dg_0] I = \phi[[\dg_A,\dg_0]]$, and
\bearr          
        \dg_0 A & = & \dg_A \phi[\dg_0] - \dg_0 \phi[\dg_A] - \phi[[\dg_A,\dg_0]] 
\label{inverse0-}\\
                & = & \dI \wedge \phi [\dg_A,\dg_0].    \label{inverse0} 
\eearr
Here $\dI$ is the exterior derivative operator on $\cF$, so the right hand side of 
(\ref{inverse0}) is the curl of $\phi$ evaluated on the pair of vector fields $\dg_A$ and 
$\dg_0$.\footnote{
In general, if $\bg$ is a 1-form and $v_1$ and $v_2$ are vector fields then
\be
        d \wedge \bg (v_1, v_2) = d_{v_1} \bg (v_2) - d_{v_2} \bg (v_1) - \bg ([v_1,v_2])
\ee
where $[v_1,v_2]$ is the Lie bracket of the vectors defined by 
$[v_1,v_2] = d_{v_1} v_2 - d_{v_2} v_1$ or equivalently by the requirement that it's action 
(on a scalar) is equal to the commutator of the actions of $v_1$ and $v_2$: 
$d_{[v_1,v_2]} = [d_{v_1}, d_{v_2}]$.}
Its value at $g$ depends only on the values of these vector fields at $g$ and not on how they
vary away from $g$. This is because the commutator term in (\ref{inverse0}) cancels 
the contributions to the first two terms coming from the action of $\dg_A$ on $\dg_0$ and 
{\em vice versa}.\footnote{\label{commuting_delta}
If the reader prefers, the action of $\dg_A$ on $\dg_0$ and the action of $\dg_0$ on $\dg_A$, 
and thus also the commutator $[\dg_A,\dg_0]$, can be set to zero by taking the variations 
$\dg_A g_{\mu\nu}$ and $\dg_0 g_{\mu\nu}$ to be independent of the metric on which they are 
evaluated, that is, by taking $\dg_A$ and $\dg_0$ to be constant vector fields on the vector 
space $\cF$.}

Now consider the case in which $\dg_A$ is the retarded perturbation $\dg^+_A$, which vanishes 
outside $J^+[s_A]$ in the globally hyperbolic spacetime $M$.  The achronality of $\cN$ 
requires $J^+[s_A]$ to be disjoint from $\cN$ (otherwise a future directed timelike curve from 
$\cN$ to a point in $s_A$ and a future directed causal curve from that point to $\cN$ could be 
deformed to a timelike curve from $\cN$ to $\cN$). Furthermore, the facts that $s_A$ is compact 
and $M$ globally hyperbolic imply that $J^+[s_A]$ is closed (\cite{Hawking_Ellis} 
prop. 6.6.1.) so it is disjoint from an open neighbourhood of $\cN$. The boundary integral 
$\phi$ in (\ref{inverse0}) may therefore be replaced by its restriction to $\Sg_+$, denoted 
$\Theta_{\Sg_+}$. Then  
        \be             \label{inverse+}
                \dg_0 A = \dI \wedge \Theta_{\Sg_+}[\dg^+_A, \dg_0].
        \ee
(Since $\phi$ is an integral of a {\em local} functional of the fields - see (\ref{phiEH}), 
its functional derivatives are local functionals of the fields, so $\dI \wedge \phi[\dg^+_A, 
\dg_0]$ receives no contribution from the parts of $\di Q$ outside the (closed) support of 
$\dg^+_A$.) In principle the 
restriction of the boundary integral $\phi$ to just the portion $\Sg_+$ of the boundary is not 
unambiguous. The integrand of $\phi$ is defined only up to total derivatives, that integrate to 
zero because $\di\di Q = \emptyset$, but which can contribute boundary terms to $\Theta_{\Sg_+}$
(see \cite{Lee_Wald}). However, these terms do not affect (\ref{inverse+})
because $\dg_A^+$ vanishes in a neighbourhood of $\di\Sg_+ = \di\cN \subset \cN$.

Now, we also have - for the same reasons -  that the {\em advanced} perturbation $\dg_A^-$ 
vanishes on an open neighbourhood of the future boundary $\Sg_+$. Thus, on $\Sg_+$, the variation
$\dg_A^+$ may be replaced by
        \be             \label{DAdef2}
                \Dg_A = \dg_A^+ - \dg_A^-.
        \ee
Equation (\ref{inverse+}) then becomes
        \be             \label{inverse1}
                \dg_0 A = \dI \wedge \Theta_{\Sg_+}[\Dg_A, \dg_0]  
                \equiv \Omega_{\Sg_+}[\Dg_A,\dg_0],
        \ee
where $\Omega_{\Sg_+}$ is the {\em presymplectic 2-form} on $\cF$ associated with the 
hypersurface $\Sg_+$. For any pair of variations, $\dg_1$ and $\dg_2$, 
        \be             \label{Omega_def}
                \Omega_{\Sg_+}[\dg_1,\dg_2] = \dI \wedge \Theta_{\Sg_+}[\dg_1, \dg_2] =  
                \dg_1\Theta_{\Sg_+}[\dg_2] - \dg_2\Theta_{\Sg_+}[\dg_1] 
		- \Theta_{\Sg_+}[[\dg_1,\dg_2]].
        \ee
(\ref{inverse1}) holds for any $\dg_0$ in the space of solutions to the unperturbed linearized 
field equation on $Q$, (\ref{linearized_vacuum}). All the more so it holds for a variation 
$\dg_0 \in L_g$ which satisfies the linearized field equation on all of $M$. 

Notice that the variation $\Dg_A g$ also satisfies the linearized field equation 
(\ref{linearized_vacuum}): Subtraction of equation (\ref{linearized}) with 
$\dg_A = \dg_A^-$ from the same equation with $\dg_A = \dg_A^+$ yields $\Dg_A\dg I = 
\Dg_A\phi[\dg]$ for all $\dg$, which is just (\ref{linearized_vacuum}). Indeed, since this
is true also when $Q$ is replaced by any compact subset of $M$ with smooth boundary, $\Dg_A 
\in L_g$.\footnote{
If $g$ is linearization stable this means that $\Dg_A$, with a suitable ``diffeomorphism gauge 
fixing'', is tangent to $\cS$. In \ref{inverse+} and the following developments $\dg^\pm_A$ can 
be any retarded/advanced perturbation to the metric generated by $A$ (i.e. by the source 
$16\pi G \ag$ in the linearized field equation). It need not satisfy any further restriction, 
such as the transverse gauge condition, and so neither does $\Dg_A$. Diffeomorphism generators 
with support contained in the causal domain of influence of $s_A$ may be freely added to 
$\Dg_A$. If this freedom is exploited to ensure that $\Dg_A$, viewed as a perturbation of the 
metric, does not move the boundaries of the domain of dependence of $\cN$ in $M$ then it 
is tangent to $\cS$ provided $g$ is linearization stable.}
(This last result also follows directly from (\ref{lin_field_eq}).) Indeed, since 
$\Dg_A B = \{A, B\}$ for any observable $B$, $\Dg_A$ must be the Hamiltonian vector field of 
$-A$.

Thus, if $\bar{\Omega}$ is defined to be the restriction of the presymplectic 2-form 
$\Omega_{\Sg_+}$ to $L_g$, then (\ref{inverse1}) implies that\footnote{
Restricting $\dg_0$ in (\ref{inverse1}) to lie in $L_g$ implies no loss of generality
since any solution of the linearized field equations on $Q$ can be extended to a solution
on all of $M$.}
                \be                             \label{inverse2}
                \dg A = \bar{\Omega}[\Dg_A,\dg]\ \ \ \forall\: \dg \in L_g.
        \ee
(\ref{inverse2}) is the equation we were looking for. It is the key equation relating the 
Peierls bracket (defined by $\Dg_A$) to the presymplectic 2-form. 

(\ref{inverse2}) determines $\Dg_A$ up to degeneracy 
vectors of $\bar{\Omega}$, that is, up to vectors $\cg$ such that\footnote{
In fact the ambiguity in $\Dg_A$ is further restricted by the definitions of $\dg_A^+$ and 
$\dg_A^-$, which require that $\Dg_A$ vanishes outside of the causal domains of influence 
$J^+[s_A]$ and $J^-[s_A]$ of $s_A$.} 
                \be             \label{degeneracy}
                \bar{\Omega}[\cg, \dg] = 0\ \ \ \forall\: \dg \in L_g.
        \ee
This is sufficient to determine uniquely the Peierls bracket, $\{B,A\} = \Dg_B A$, of $A$ with 
any observable $B$ with domain of sensitivity contained in $Q$: (\ref{inverse2}) holds for any 
$\dg \in L_g$. It therefore holds in particular when $\dg = \Dg_B$. Thus 
\be             \label{pb3}
\{B,A\} = \Dg_B A = \bar{\Omega}[\Dg_A,\Dg_B].
\ee
Clearly the right side of this equation is unaffected by the addition of degeneracy vectors of
$\bar{\Omega}$ to $\Dg_A$ and $\Dg_B$.

$\bar{\Omega}$ need not be defined using $\Sg_+$. One obtains the same 2-form on 
$L_g$ if one defines it using $\cN$ in place of $\Sg_+$.

\noindent{\em Proof}: $\phi = \Theta_{\Sg_+} - \Theta_{\cN}$.
Thus, if $\dg_1$ and $\dg_2$ belong to $L_g$, then\footnote{
Recall that the commutator of the actions, as derivative operators, of two vector fields is 
equal to the action of their Lie bracket, which is itself a vector field.}
        \bearr
                \bar{\Omega}_{\Sg_+}[\dg_1,\dg_2] - \bar{\Omega}_{\cN}[\dg_1,\dg_2] 
                 & = & \dg_1 \phi[\dg_2] - \dg_2 \phi[\dg_1] - \phi[[\dg_1,\dg_2]]\\
                 & = & \dg_1\dg_2 I - \dg_2\dg_1 I - [\dg_1,\dg_2] I = 0\ \ \QED.
							\label{hypersurface_independence}
        \eearr 
Actually, since $\Theta_{\Sg_+}[\dg]$ does not depend on the shape of the remainder, 
$\di Q - \Sg_+$, of the boundary of $Q$, any piecewise differentiable hypersurface $\Sg$ which 
together with $\Sg_+$ encloses a compact spacetime region on which Einstein's field equations 
hold can be used instead of $\cN$ or $\Sg_+$ to evaluate $\bar{\Omega}[\dg_1,\dg_2]$, and gives 
the same result. In particular {\em any} smooth Cauchy surface of $\cD$ will do. [This does not 
exhaust the possibilities. In principle the hypersurface need not even be achronal, or contained
in $\overline{\cD}$. On the other hand our proof only guarantees the validity of 
(\ref{inverse2}) if $s_A \subset \cD$.]

From (\ref{phiEH}) one immediately obtains $\Omega$ for the Einstein-Hilbert action:
        \bearr  
                \Omega_\Sg[\delta_1,\delta_2] & = & \delta_1 \Theta_\Sg[\delta_2] 
		- \delta_2 \Theta_\Sg[\delta_1] 
                - \Theta_\Sg[[\dg_1,\dg_2]]\\
                        & = & -\frac{1}{8\pi G} \int_\Sg \delta_2 
\Gamma^{[\sg}_{\sg\nu} \delta_1 [g^{\mu]\nu}\sqrt{-g}] d\Sg_\mu - (1 \leftrightarrow 2) 
\label{presymp0}
        \eearr
This agrees (up to an over all factor) with the expression proposed by Crnkovic and Witten 
\cite{Crnkovic_Witten}. Moreover, when $\Sg$ is spacelike (\ref{presymp0}) reduces to 
the standard 
expression in terms of ADM variables \cite{ADM}\cite{Lee_Wald}: In Gaussian normal coordinates 
$r^\ag$ based on $\Sg$ $r^0 = 0$ on $\Sg$ and $\di/\di r^0$ is the unit future directed
tangent to the congruence of geodesics normal to $\Sg$. Consequently $g^{00} = g_{00} = -1$,
$g^{0i} = g_{0i} = 0$, and $g_{ij} = h_{ij}$ where indices $i,j,k,...$ range over $\{1,2,3\}$ 
and $h$ is the induced 3-metric on $\Sg$. Moreover $\Cg^0_{00} = \Cg^i_{00} = 0$ and 
$\Cg^0_{ij} = K_{ij}$, $\Cg^i_{j0} = h^{ik}K_{kj}$ where $K_{ij} = 1/2\, \di_0 h_{ij}$ is the 
extrinsic curvature of $\Sg$. The presymplectic 2-form is thus
\be
	\Omega_\Sg[\dg_1,\dg_2] =  \frac{1}{16\pi G} \int_\Sg \dg_2 K \dg_1 \sqrt{h} 
	+ \dg_2 K_{ij}\dg_1[h^{ij}\sqrt{h}] d^3 r - (1 \leftrightarrow 2),
\ee
where $K = K_{ij} h^{ij}$ and $h = det[h_{ij}]$. In terms of $h_{ij}$ and the ADM momentum 
variable\footnote{
The normalization of $\pi$ corresponds to the normalization of the action (\ref{EH}) used here.}
\be		\label{ADM_momentum_variable}
\pi^{ij} = \sqrt{h}[K^{ij} - K h^{ij}]/16\pi G
\ee
the presymplectic 2-form becomes
\be		\label{spacelike_presymp}
 	\Omega_\Sg[\dg_1,\dg_2] = \int_\Sg \dg_1 \pi^{ij} \dg_2 h_{ij} d^3 r 
	- (1 \leftrightarrow 2).
\ee

\subsubsection{The Peierls bracket is a Poisson bracket} \label{Peierls_is_Poisson}

(\ref{pb3}) allows us to verify that the Peierls bracket $\{\cdot,\cdot\}$ really is a Poisson 
bracket on the observables. In order that $\{\cdot,\cdot\}$ be a Poisson 
bracket it must \cite{Marsden} 
\begin{itemize}

\item[i] be skew symmetric,

\item[ii] be linear in each argument (under addition and multiplication by a constant),

\item[iii] satisfy the Leibniz rule $\{A, BC\} = \{A, B\} C + B \{A, C\}$,

and

\item[iv] satisfy the Jacobi relation,
\be             \label{Jacobi_id}
\{A,\{B,C\}\} + \{B,\{C,A\}\} + \{C,\{A,B\}\} = 0
\ee
\end{itemize}

(\ref{pb3}) immediately shows that i, ii and iii are satisfied. The last requirement, the Jacobi
relation, follows from the fact that $\Omega$, being the curl of $\Theta$, is itself 
a closed 2-form. For any 2-form $\ag$ and triplet of vector fields $u, v, w$
\be  \label{curl_id}
\Cyc{u,v,w} d_u \ag(v,w) = d\wedge \ag(u,v,w) + \Cyc{u,v,w} \ag([u,v],w)
\ee
(where $\Cyc{u,v,w}$ denotes the sum over cyclic permutations of the sequence $u,v,w$). From 
(\ref{pb3}), (\ref{curl_id}) and the fact that $\dI \wedge \Omega = 0$ it follows that\footnote{
Actually it is not necessary to invoke a differentiable structure on $\cF$ to obtain the Jacobi 
relation. The crucial relation 
\be
\Cyc{1,2,3} \dg_1\bar{\Omega}[\dg_2,\dg_3] = \Cyc{1,2,3} \bar{\Omega}[[\dg_1,\dg_2],\dg_3],
\ee
that we have obtained from (\ref{curl_id}) and $\dI \wedge \Omega = 0$, follows algebraically 
from (\ref{Omega_def}) - 
\be
\bar{\Omega}[\dg_1,\dg_2] = \dg_1\Theta_{\Sg_+}[\dg_2] - \dg_2\Theta_{\Sg_+}[\dg_1] 
- \Theta_{\Sg_+}[[\dg_1,\dg_2]]\ \ \ \forall \dg_1, \dg_2 \in L_g.
\ee}
\be
X \equiv \Cyc{A,B,C} \{A,\{ B, C \}\} 
= - \Cyc{A,B,C} \Dg_A \bar{\Omega}[\Dg_B,\Dg_C]
= - \Cyc{A,B,C} \bar{\Omega}[[\Dg_A,\Dg_B],\Dg_C]
\ee
and therefore
\bearr
X & = &  \Cyc{A,B,C} [\Dg_A,\Dg_B] C 
= \Cyc{A,B,C} \{A,\{ B, C \}\} - \{B,\{ A, C \}\}\\
&& = \Cyc{A,B,C} (X - \{C,\{ A, B \}\})
= 3 X - X = 2X
\eearr
Thus $X = 0$, which is precisely the Jacobi relation (\ref{Jacobi_id}).

\subsubsection{Definition of the bracket on initial data}\label{def_bracket_initial_data}

We turn now to the problem of defining a bracket on the initial data on $\cN$,
to be called the {\em auxiliary bracket} and denoted by $\{\cdot,\cdot\}_*$, which reproduces 
the Peierls bracket on observables when these are expressed in terms of the initial data.

A skew symmetric bracket, $\{\varphi_I,\varphi_J\}_*$, defined on all pairs $\varphi_I$, 
$\varphi_J$ of initial data determine a bracket 
\be             \label{auxbracketdef1}
\{F,G\}_* = \sum_{I,J} \frac{\di F}{\di \varphi_I} \{\varphi_I,\varphi_J\}_* 
\frac{\di G}{\di \varphi_J}.
\ee
on all pairs $F$, $G$ of differentiable functions of the initial data.
Here the indices $I$, $J$ comprehend all continuous and discrete labels of the initial data 
components (including position in $\cN$), $\sum_{I,J}$ represents the integral and/or sum 
over the set of such labels and $\di/\di\varphi_I$ represents ordinary
partial differentiation or functional differentiation as appropriate.
$\{\cdot,\cdot\}_*$ as defined by (\ref{auxbracketdef1}) is skew symmetric, linear in its 
arguments, and satisfies the Leibniz rule. This makes it a {\em pre-Poisson} bracket - 
a Poisson bracket save that it may not satisfy the Jacobi relation.  

This bracket is defined on observables only if they are differentiable functionals of the 
initial data. That they are can be demonstrated if it is assumed, as we have done elsewhere, 
that Rendall's \cite{Rendall} results on the Cauchy development of null initial data apply 
to all of $\cN$, and not just to a neighbourhood of $S_0$ in $\cN$. Rendall shows that in 
the Cauchy developments corresponding to a differentiable one parameter family of Sachs initial 
data the metric at a given harmonic coordinate point is also differentiable in the family 
parameter. The same is then easily seen to be the case for the closely related initial data we
shall use (defined in subsection \ref{coordinates}). 

The derivative, $\dg g_{\mu\nu}$, of the metric field in the family parameter is a solution to 
the linearized field equation, so the corresponding derivative of an observable $A$ satisfies
$\dg A = \Omega_\cN[\Dg_A,\dg]$. The linear operator $\Omega_\cN[\Dg_A,\cdot]$ is a distribution
on the derivative $\dg\varphi_I$ of the initial data. Indeed it can be seen at once from the 
explicit form (\ref{Omega_R}) of $\Omega_\cN$ in terms of our initial data, and the smoothness 
of $\Dg_A$, that $\Omega_\cN[\Dg_A,\cdot]$ is a regular distribution, equivalent to integration 
of the variations of the initial data against smooth functions. Since $\Omega_\cN[\Dg_A,\cdot]$ 
depends only on the definition of $A$ and on the unperturbed metric, and not on which one 
parameter family of initial data is being considered, it is in fact the functional derivative 
of $A$ by the initial data. $\di A/\di \varphi_I$ thus exists and is a smooth function of the
the continuous position label in the index $I$. 

Note that $\Dg_A$ vanishes in a spacetime neighbourhood of $\di\cN$ (because both $\dg_A^-$ and 
$\dg_A^+$ do so). We will take care in the definition of our initial data to make sure that
variations of the metric that vanish near $\di\cN$ also leave the initial data undisturbed near 
$\di\cN$, so $\Dg_A \varphi_I = 0$ for data in some neighbourhood of $\di\cN$. Since 
$\Omega_\cN[\Dg_A,\dg]$ is an integral over the initial data and their variations under $\Dg_A$
and $\dg$ on $\cN$ (see (\ref{Omega_R})) the variations $\dg\varphi_I$ corresponding to points 
sufficiently close to $\di\cN$ will not contribute. That is, the derivative 
$\di A/\di \varphi_I$ vanishes for $I$ corresponding to points in a neighbourhood of $\di\cN$. 

A sufficient condition ensuring that $\{\cdot,\cdot\}_*$ reproduces the Peierls bracket on 
observables is that the motion $\{A,\cdot\}_*$ generated by any observable $A$ via the bracket 
satisfies\footnote{
$\{A,\cdot\}_*$ is a variation of the initial data. That it necessarily corresponds to 
a solution to the linearized Einstein equation of course follows from the extensions of 
Rendall's results \cite{Rendall} that we have assumed, but it can also be seen directly from 
the explicit formula for the solution (\ref{Kirch2}) obtained in appendix 
\ref{another_approach}.}
\be
        \dg A = \bar{\Omega}[\{A,\cdot\}_*,\dg]\ \ \ \forall\:\dg \in L_g^0,
        \label{auxbracketdef0}
\ee
where $L_g^0$ is the space of variations satisfying the linearized field equations and the 
requirement that $\dg g_{\mu\nu}$ vanishes in some neighbourhood of $\di\cN$ in $M$. 
Then 
\be             \label{brackets_equal}
\{A,B\} = \Dg_A B = \bar{\Omega}[\{B,\cdot\}_*,\Dg_A] = -\bar{\Omega}[\Dg_A, \{B,\cdot\}_*] 
= - \{B, A\}_* = \{A,B\}_*.
\ee
for all observables $A$ and $B$. The first equality is just the definition of the Peierls 
bracket; the second follows from (\ref{auxbracketdef0}), using the fact that $\Dg_A$ vanishes
in a neighbourhood of $\di\cN$; the third from the antisymmetry of $\bar{\Omega}$; the fourth 
from (\ref{inverse2}); and the last from the assumed antisymmetry of $\{\cdot,\cdot\}_*$.
 
Condition (\ref{auxbracketdef0}) is satisfied if
\be                     \label{auxbracketdef}
        \dg \varphi_I = \Omega_\cN[\{\varphi_I,\cdot\}_*,\dg]\ \ \ \forall\:\dg \in L_g^0
\ee
for all initial data $\varphi_I$ living on the interior of $\cN$. Indeed it is sufficient that 
(\ref{auxbracketdef}) holds when both sides are integrated against smooth functions
that vanish in a neighbourhood of $\di\cN$, for (\ref{auxbracketdef0}) is obtained from 
(\ref{auxbracketdef}) by contracting both sides of (\ref{auxbracketdef}) with 
$\di A/\di \varphi_I$. 

(\ref{auxbracketdef}) even implies that $\{A,\cdot\}_* = \Dg_A$
distributionally on initial data on the interior of $\cN$. If $\theta$ is initial data
smeared with a smooth function supported in $int\cN$ then $\{\theta,\cdot\}_*$, like
$\{A,\cdot\}_*$, corresponds to an element of $L_g$. One may therefore repeat the arguments
of equation (\ref{brackets_equal}), from the second equality on, with $\theta$ in place of $B$, 
and conclude that $\Dg_A \theta = \{A,\theta\}_*$. (This result also follows from 
(\ref{Kirch1}) and (\ref{auxbracketdef}).)

Notice that both (\ref{auxbracketdef0}) and (\ref{auxbracketdef}), by ensuring that 
$\{\cdot,\cdot\}_*$ reproduces the Peierls bracket on observables, also ensure that the Jacobi 
relation holds for $\{\cdot,\cdot\}_*$ on observables. Nevertheless neither 
(\ref{auxbracketdef0}) nor (\ref{auxbracketdef}) imply that the Jacobi relation holds for 
brackets between arbitrary functions of the data.\footnote{
One might think that the Jacobi relation follows from (\ref{auxbracketdef}), which 
essentially requires that $\{\cdot,\cdot\}_*$ be inverse to $\bar{\Omega}$. As we saw in 
sub-subsection \ref{Peierls_is_Poisson}, the analogous 
condition (\ref{inverse2}) does imply the Jacobi relation for the Peierls bracket between 
observables. However (\ref{auxbracketdef}) applies only to variations $\dg$ vanishing in some 
neighbourhood of $\di\cN$, so $\{\cdot,\cdot\}_*$ is only almost inverse to $\bar{\Omega}$, 
which turns out not to be sufficient.}
Indeed the bracket solving
(\ref{auxbracketdef0}) that we find does not satisfy the Jacobi relation. As long as one is only
interested in reproducing the Peierls Poisson algebra of the observables the validity of 
the Jacobi relation for brackets of the initial data is evidently not necessary, but if one 
wishes to quantize the system by replacing these brackets by commutators this relation becomes 
indispensable, since it holds identically for commutators. 

Our task will therefore be to find a skew symmetric bracket on initial data, 
$\{\varphi_I,\varphi_J\}_*$, satisfying (\ref{auxbracketdef}). Two further conditions will be 
imposed on this bracket. The first of these relates to constraints. We shall calculate the
bracket on a set of initial data which are almost, but not entirely, free. They are subject to 
one constraint at the intersection 2-surface $S_0$ of the two null branches of $\cN$. 
The brackets on these almost free data then allow us to evaluate the brackets of a second set 
of data, which are completely free. 
As is always the case for constrained data the bracket on the almost free data defines an
equivalent bracket on the completely free data iff it respects the constraints.
That is to say, we must require that the functions of the data that are constrained to vanish 
also have vanishing brackets with all data. 

The second, and last, condition is a requirement of causality. We require that the auxiliary 
bracket vanishes between data at points of $\cN$ that are not connected by a causal curve.
Since disjoint generators are not connected by causal curves (prop. 
\ref{domain_of_influence_on_N}) it follows that the only non-zero brackets are those between 
data on the same generator, which facilitates considerably our calculations. 

Whether the causality requirement is independent of (\ref{auxbracketdef}) or implicit in it is 
unclear at present.  
The causality requirement seems quite reasonable in view of the causality property of the 
Peierls bracket - namely that the Peierls bracket between observables with domains of 
sensitivity unconnected by any causal curve vanishes. As will be clear from the definitions
of our data, the domain of sensitivity of a datum at a point $p \in \cN$, as a functional of 
the spacetime metric,\footnote{
Quite generally the domain of sensitivity of a functional $F$ of the metric will be defined to 
be the support in spacetime of its functional gradient $\dg F/\dg g^{\mu\nu}$.}
is contained in the generator through $p$, or the pair of generators through $p$ in the case 
that $p \in S_0$. Thus the Peierls bracket between data on distinct generators, if it can
be defined, should vanish. Indeed this can be done. (\ref{PB2}) allows us to define the Peierls 
bracket between initial data that do not live on the same generator, provided a suitable fixing 
of the diffeomorphism freedom in the data and the Green's functions is adopted, and this Peierls
bracket must vanish because the advanced and retarded Green's functions that enter (\ref{PB2}) 
vanish between points that cannot be connected by causal curves. 
That the causality requirement is at least compatible with the other requirements follows 
{\em a posteriori} from the fact that a bracket satisfying all the requirements is found.

\subsubsection{Degeneracy vectors of the presymplectic 2-form and the consistency of the 
definition of the auxiliary pre-Poisson bracket}

The consistency of (\ref{auxbracketdef}) will be fully established {\em post hoc} by the fact 
that a solution to this condition is found. Nevertheless it is interesting to attempt to verify
consistency already at this stage, before entering into the details of a particular 
representation of the solutions (i.e. of the initial data). We shall succeed in doing so in
the generic case in which a neighbourhood of each point can be coordinatized by scalars defined
by the geometry. 

The consistency of (\ref{auxbracketdef}) requires that all data $\varphi_I$ be invariant under 
the action of degeneracy vectors $\cg$ of $\bar{\Omega}$ such that the corresponding variation 
of the metric, $\cg g$, vanishes in a spacetime neighbourhood of $\di\cN$. At least on generic 
solutions these turn out to be diffeomorphism generators, i.e $\cg = {\pounds}_\xi$ for some 
vector field $\xi$ on $\Dund$.\footnote{
The tangent to a one parameter family of diffeomorphisms is the Lie derivative ${\pounds}_\xi$ 
along the vector field $\xi$ of tangents to the orbits of the manifold points.}
In fact the following stronger result holds

\begin{proposition}  \label{Degenerate_is_diff}
In the generic case, in which there exist at each point of $\Dund$ four smooth functions of the 
metric $g$ and a finite number of its derivatives, having smooth, linearly independent 
spacetime gradients, any variation $\cg \in L_g$ that satisfies the generalized degeneracy 
condition
\be		\label{generalized_degeneracy}
\bar{\Omega}(\cg,\dg) = 0 \ \ \ \forall \dg \in L_g^0
\ee
is a diffeomorphism generator on $\Dund$. 
\end{proposition}  
\noindent {\em Proof}: Any divergenceless, $C^\infty$, symmetric, bivalent tensor field $\ag$ 
of compact support contained in $\cD$ defines, via (\ref{Dg_def}) and 
(\ref{pert1}) a vector $\Dg_A \in L_g^0$ which, by (\ref{inverse2}), must satisfy 
\be
\int \dg g_{\mu\nu}\:\ag^{\mu\nu}\sqrt{-g}\:d^4 x = \bar{\Omega}[\Dg_A,\dg]\ \ 
\forall\: \dg \in L_g.
\ee
If $\dg$ is a generalized degeneracy vector $\cg$, satisfying (\ref{generalized_degeneracy}) 
then the right side vanishes and we have
\be
\int \cg g_{\mu\nu}\:\ag^{\mu\nu}\sqrt{-g}\:d^4 x = 0
\ee	\label{int_gamma_g_alpha}
for all $\ag$ satisfying our conditions. 

Let $p$ be a point of $\Dund$, and $z^\ag, \ag = 0,...,3$ four smooth functionals of the metric 
with linearly independent gradients $dz^\ag$. By the inverse function theorem these form
a smooth chart on a neighbourhood $U$ of $p$ in $M\supset \Dund$. The components of the 
metric in this chart are diffeomorphism invariant, so differentiating any functional of these 
components by the components of the metric in a manifold fixed chart yields a symmetric, 
divergenceless tensor. Generalizing this observation we shall consider linear functionals of 
the form
\be
F[\dg g] = \int \dg g_{\ag\bg} \bar{f}^{\ag\bg}\sqrt{-g_z}\,d^4 z
\ee
where $g_{\ag\bg}$ are the components of the metric in the $z$ chart and $\bar{f}^{\ag\bg}$ 
is an arbitrary symmetric matrix of functions with support contained in $U\cap \cD$. The 
variation of $g_{\ag\bg}$ is related to the variations of the components, $g_{\mu\nu}$, of the 
metric in a manifold fixed chart $x$ by
\be
\dg g_{\ag\bg} = \frac{\di x^\mu}{\di z^\ag} \frac{\di x^\nu}{\di z^\bg}[\dg g_{\mu\nu} - 
{\pounds}_{\xi[\dg]} g_{\mu\nu}]
\ee
where
\be
\xi[\dg]^\mu(q) = \frac{\di x^\mu}{\di z^\ag}(q) 
\int \dg g_{\nu\rho}(r) \frac{\dg z^\ag(q)}{\dg g_{\nu\rho}(r)}\,d^4 x(r)
\ee
at each $q \in U$. ($q$ is fixed in the spacetime manifold, $z^\ag(q)$ thus varies when the 
metric varies.) Thus
\bearr
F[\dg g] & = & \int [\dg g_{\mu\nu} - 2 \nabla_{(\mu} \xi[\dg]_{\nu)}] \bar{f}^{\mu\nu}
      \sqrt{-g_x}\,d^4 x\\
& = & \int \{\dg g_{\mu\nu}\bar{f}^{\mu\nu} + 2 \xi[\dg]^\rho g_{\rho\nu}\nabla_{\mu} 
      \bar{f}^{\mu\nu}\} \sqrt{-g_x}\,d^4 x\\
& = & \int \dg g_{\mu\nu} f^{\mu\nu} \sqrt{-g_x}\,d^4 x
\eearr
with
\be
\bar{f}^{\mu\nu} = \frac{\di x^\mu}{\di z^\ag}\frac{\di x^\nu}{\di z^\bg} \bar{f}^{\ag\bg}
\ee
and
\be
f^{\mu\nu}(r) = \bar{f}^{\mu\nu}(r) + \frac{2}{\sqrt{-g_x}}\int  
\frac{\dg z^\ag}{\dg g_{\mu\nu}(r)} \frac{\di x^\rho}{\di z^\ag} g_{\rho\tau}\nabla_{\sg} 
\bar{f}^{\sg\tau}\sqrt{-g_x}\,d^4 x
\ee
This tensor is smooth because the functional derivative is a sum with smooth coefficients of 
derivatives, up to finite order, of delta distributions, and this distribution is convolved
with a smooth function. Clearly the support of $f$ is contained in that of $\bar{f}$, which in
turn is contained in $U\cap \cD$. Finally, $f$ is divergenceless. This can be verified by 
letting $\dg = {\pounds}\zeta$ with $\zeta$ smooth and of compact support contained in $U$:
\be \label{divergenceless_proof}
F[{\pounds}\zeta] = - \int \zeta_{\nu} \nabla_\mu f^{\mu\nu}\,\sqrt{-g_x}\,d^4 x 
\ee
vanishes because $\xi[{\pounds}\zeta] = \zeta$, since $z^\ag$ are functionals of the metric 
only, and thus the variation of the $z$ components of the metric vanishes. 
((\ref{divergenceless_proof}) can also be verified by direct calculation.) 
(\ref{divergenceless_proof}) for all smooth, compactly supported $\zeta$ implies 
$\nabla_\mu f^{\mu\nu} = 0$. 


Thus $F[\cg] = 0$ for all $\bar{f}^{\ag\bg}$, (which are smooth and have compact support 
contained in $U\cap \cD$) which implies that 
\be
0 = \cg g_{\ag\bg} = \cg g_{\mu\nu} + {\pounds}_{\xi[\cg]} g_{\mu\nu}.
\ee
That is $\cg = {\pounds}_\xi$, a diffeomorphism generator, on $U\cap \cD$. Clearly this 
implies that $\cg$ is a diffeomorphism generator throughout $\cD$. 

In fact $\cg$ is a diffeomorphism generator also on $int \cN$. By the Whitney extension theorem
(see \cite{Abraham_Robbin}) $\cg g$ can be extended smoothly to a neighbourhood of $int \cN$ in 
$M$. $\xi[\cg]$ is defined on the intersection of $U$ with this neighbourhood. Therefore, 
since both $\cg$ and ${\pounds}_{\xi[\cg]}$ define smooth variations of the metric, 
$\cg g = {\pounds}_{\xi[\cg]} g$ on $U \cap int \cN$ which is a subset of the boundary of $\cD$ 
in the topology of $U$.
\QED
\newline

The genericity assumption does not actually seem to be necessary. In particular, the result 
holds in a flat spacetime, which does not satisfy this assumption. 

In any case, we may conclude that in the generic case that a generalized degeneracy vector is a 
diffeomorphism generator on $\Dund$. Conversely, a 
direct calculation (which will be useful for another purpose further on)\footnote{
A more general argument is given in subsection \ref{gauge}.}
shows that 

\begin{proposition} \label{Diff_is_degenerate}
Any diffeomorphism generator is a generalized degeneracy vector of $\bar{\Omega}$.
\end{proposition}
\noindent {\em Proof}:
Suppose $\xi$ is a $C^\infty$ vector field on $\Dund$, $\dg$ is a variation satisfying the 
linearized vacuum field equation, and $\Sg$ is a smooth Cauchy surface of $\cD$, contained
in $\cD$. Note that a smooth Cauchy surface exists by \cite{Bernal_Sanchez0}, and that by 
prop. \ref{cauchy_D} the closure $\bar{\Sg}$ of $\Sg$ in $M$ is an embedded compact 
3-manifold with boundary $\di\cN$. 
For the sake of simplicity (it makes no difference to the result) suppose $\xi$ is independent 
of the metric, and thus $\dg \xi = 0$. Then
	\be	\label{omega_diffeo}
		\bar{\Omega}[{\pounds_\xi},\dg] = {\pounds_\xi} \Theta_\Sg[\dg] 
		- \dg \Theta_\Sg[{\pounds_\xi}].
	\ee
Now 
	\bearr
		\Theta_\Sg[\dg] & = & -\frac{1}{8\pi G}\int_{\Sg} \dg\Gamma^{[\sg}_{\sg\nu} 
		                      g^{\mu]\nu}\sqrt{-g}\:d\Sigma_\mu \\
				& = & \int_{\Sg} \ag
	\eearr
where $\ag$ is the 3-form $-1/(3!\,8\pi G)\:\dg\Gamma^{[\sg}_{\sg\nu} g^{\mu]\nu}
\eg_{\mu\lam\rho\tau}\sqrt{-g}\:dx^\lam\wedge dx^\rho \wedge dx^\tau$. Thus
	\be
		{\pounds_\xi} \Theta_\Sg[\dg] = \int_\Sg {\pounds_\xi} \ag = \int_\Sg \xi\cont 
		[d\wedge \ag] + \int_\Sg d \wedge [\xi\cont\ag].
	\ee
But $d \wedge \ag = 1/(16\pi G) \dg  R \sqrt{-g}\: dx^0 \wedge dx^1 \wedge dx^2 \wedge dx^3$,
the divergence term in the variation of the Einstein-Hilbert Lagrangean density, which vanishes 
because $g \in \cS$ and $\dg \in L_g$ and thus $R = \dg R = 0$. Therefore
	\bearr
		{\pounds_\xi} \Theta_\Sg[\dg] & = & \int_{\di\bar{\Sg}} \xi\cont\ag \\
		& = & \frac{1}{8\pi G}\int_{\di\bar{\Sg}} \xi^\mu 
		\dg\Gamma^{[\sg}_{\sg\tau} g^{\nu]\tau} \sqrt{-g}\:d\Sigma_{\mu\nu} \\
		& = & \frac{1}{8\pi G}\int_{\di\bar{\Sg}} \xi^\mu 
(\nabla^\nu \dg \sqrt{-g} + \frac{1}{2}\nabla_\sg \dg g^{\nu\sg} \sqrt{-g})\:d\Sigma_{\mu\nu}.
\label{L_xi_Theta_delta}
	\eearr						
Here $d\Sg_{\mu\nu} = 1/2\,\eg_{\mu\nu\rho\tau}\:dx^\rho\wedge dx^\tau$ 
is the coordinate area 2-form, and in the last line the identity 
\be \label{dGamma_id}
\dg\Cg^\sg_{\mu\nu} = \frac{1}{2}g^{\sg\tau}\{\nabla_\nu g_{\tau\mu} + \nabla_\mu g_{\tau\nu} 
- \nabla_\tau g_{\mu\nu}\}.
\ee
has been used.  

The second term in (\ref{omega_diffeo}) is the $\dg$ variation of
	\be
		\Theta_\Sg[{\pounds_\xi}] = -\frac{1}{8\pi G}
		\int_{\Sg} {\pounds_\xi}\Gamma^{[\sg}_{\sg\nu} g^{\mu]\nu}\sqrt{-g}\:d\Sigma_\mu
 	\ee
But (\ref{dGamma_id}) and Einstein's field equation, which is satisfied by $g$, imply that
	\bearr
		{\pounds_\xi}\Gamma^{[\sg}_{\sg\nu} g^{\mu]\nu}\sqrt{-g} 
	& = & (\nabla^\mu \nabla_\sg \xi^\sg -\frac{1}{2}\nabla_\sg \nabla^\sg \xi^\mu 
		-\frac{1}{2}\nabla_\sg \nabla^\mu \xi^\sg) \sqrt{-g} \\
		& = & \nabla_\sg \nabla^{[\mu} \xi^{\sg]} \sqrt{-g}	\\
		& = & \di_\sg [\nabla^{[\mu} \xi^{\sg]} \sqrt{-g}].
	\eearr
Taking the integral one obtains
	\be		\label{diffTheta1}
		\Theta_\Sg[{\pounds_\xi}] = -\frac{1}{16\pi G}
				\int_{\di\bar{\Sg}} \nabla^{[\mu} \xi^{\nu]}\sqrt{-g}\:
				d\Sigma_{\mu\nu}. 
	\ee

Summing (\ref{L_xi_Theta_delta}) and minus the $\dg$ variation of (\ref{diffTheta1}) one obtains
\bearr
\bar{\Omega}[\pounds_\xi,\dg] 
& = & \frac{1}{16\pi G} \int_{\di\bar{\Sg}} \xi^\mu\{2\nabla^\nu \dg\sqrt{-g}
      + \nabla_\sg \dg g^{\nu\sg}\sqrt{-g}\} + \dg \{\nabla^\mu \xi^\nu \sqrt{-g}\} 
        d\Sg_{\mu\nu},  \label{boundary_diff2}	\\
& = & \frac{1}{16\pi G} \int_{\di\bar{\Sg}} 3\xi^{[\mu}\dg\Gamma^\sg_{\sg\tau} g^{\nu]\tau}
	\sqrt{-g}
      + \dg[\sqrt{-g} g^{\sg\mu}]\nabla_\sg\xi^\nu d\Sg_{\mu\nu},     \label{boundary_diff}
\eearr
Since this is a boundary integral it certainly vanishes if $\dg$ vanishes in a neighbourhood of 
$\di\bar{\Sg} = \di\cN$, showing that any diffeomorphism generator is a generalized degeneracy 
vector of $\bar{\Omega}$. \QED
\newline 

In order that ${\pounds}_\xi$ be a {\em true} degeneracy vector the boundary integral 
(\ref{boundary_diff}) must vanish for all $\dg \in L_g$. A sufficient condition is that $\xi$ 
and it's gradient vanish on $\di\Sg$. It is not clear to the author whether this is a 
necessary condition.

According to prop. \ref{Degenerate_is_diff}, when the metric $g$ on $\Dund$ is generic
a degeneracy vector of $\bar{\Omega}$ which preserves the metric in a neighbourhood of $\di\cN$
is a diffeomorphism generator which vanishes in this neighbourhood. Our initial data will 
be invariant under all diffeomorphisms which preserve $\di\cN$, so they are certainly invariant 
under such degeneracy vectors, implying that (\ref{auxbracketdef}) is consistent in this case. 

On the other hand on any solution, even one with a non-generic metric, (\ref{auxbracketdef}) 
only defines $\{\varphi_I,\cdot\}_*$ up to a generalized degeneracy vector. Thus, according to 
prop. \ref{Diff_is_degenerate}, one is free to add a diffeomorphism generator. This ambiguity 
is to be expected if $\{\cdot,\cdot\}_*$ is determined only by the requirement that it 
reproduces the Peierls bracket on observables, because observables are diffeomorphism invariant.


The consequent ambiguity of the brackets between our initial data is not very great however, 
because the data are invariant under a large subgroup of diffeomorphisms. Nevertheless,
the freedom to add diffeomorphism generators to the solutions of (\ref{auxbracketdef}) will be 
used to simplify calculations further on.

Finally, the fact that any diffeomorphism generator that vanishes in a neighbourhood of $\di\cN$
is a true degeneracy vector of $\bar{\Omega}$ implies that we may, without changing the
consequences of (\ref{auxbracketdef}), add a suitable diffeomorphism generator to each variation
in $L_g^0$ such that the total variation preserves the null character of the branches of $\cN$.
This partial ``gauge fixing'' of $L_g^0$, which will be adopted from here on, simplifies our 
subsequent considerations, principally because the fixed hypersurface $\cN$ then carries null 
initial data for the varied solutions and the variation of this data is simply a function of 
the variation of the metric (and of its derivatives) on $\cN$. 

It is tempting to try to go further and restrict the variations so that they leave the 
generators of $\cN$ fixed (i.e. the null generators for a varied solution are the same as in the
fiducial solution). This {\em can} be done by adding suitable diffeomorphism generators 
${\pounds}_\xi$ to the variations in $L_g^0$, but in general $\xi$ cannot be chosen to vanish 
in a neighbourhood of $\di\cN$, so the resulting variation will not lie in $L_g^0$: Imagine 
following the generators back in time from the boundary $\di\cN$ to $S_0$, that is, from the 
disks $S_L$ and $S_R$ where the generators of $\cN_L$ and $\cN_R$, respectively, cross the 
boundary. (See Fig. \ref{Nfigure}).
A perturbation which vanishes in a neighbourhood of $\di\cN$ will leave the generators fixed in 
that neighbourhood. However, it will in general affect which generator from $S_L$ meets a given 
generator from $S_R$ at $S_0$. So no diffeomorphism acting only in the interior of $\cN$ can 
restore the generators to their unperturbed positions. Aside from the exceptional cases in which
the the geometry near $\di\cN$ admits non-trivial isometries, these are the only diffeomorphisms
that leave the metric invariant near $\di\cN$. Thus, while a perturbation of the metric that 
preserves the geometry near $\di\cN$ can always be reduced by the addition of a suitable 
diffeomorphism to a perturbation that vanishes (i.e. preserves the metric tensor) in a 
neighbourhood of $\di\cN$, it is not in general possible to add a diffeomorphism so that the 
total perturbation leaves both the metric invariant near $\di\cN$ {\em and} the generators 
invariant on all of $\cN$. 

We shall continue to use variations $\dg$ which vanish near $\di\cN$ in 
(\ref{auxbracketdef}), and the generators will not in general be fixed under under these 
variations, even though the hypersurface $\cN$ they collectively sweep out is. 

\subsubsection{The auxiliary bracket for spacelike initial data}

Our conditions also serve to define a bracket on spacelike initial data, which
turns out to be the same as the Poisson bracket used in the ADM formulation of canonical GR 
\cite{ADM}. This implies that the Peierls bracket between a pair of observables of the domain 
of dependence of a Cauchy surface $\Sg$ is equal to the bracket between these in the ADM theory.
Our theory is thus equivalent to the ADM theory.

The auxiliary bracket is a solution to (\ref{auxbracketdef}), but it is convenient to first
solve the simpler, but analogous condition
\be                     \label{kinematic_bracket_def}
\dg \varphi_I = \Omega_\Sg[\{\varphi_I,\cdot\}_\Dg,\dg]\ \ \ \forall\:\dg \ 
\mbox{vanishing near}\:\di\Sg.
\ee
This condition is stronger than (\ref{auxbracketdef}) because it must hold for all $\dg$ 
vanishing in a neighbourhood of the boundary, not just those satisfying the linearized field 
equations. On the other hand $\{\varphi_I,\cdot\}_\Dg$ will not be required to satisfy the 
linearized field equations either (for this reason $\Omega$ must be used instead of 
$\bar{\Omega}$).\footnote{
$\{\cdot,\cdot\}_\Dg$ is (a generalization of) the kinematic Poisson bracket defined in 
\cite{Lee_Wald}.}
When $\Sg$ is a spacelike Cauchy surface of $\Dund$ one may use the ADM variables, consisting of
the induced 3-metric $h$ and the momentum variable $\pi$ defined in 
(\ref{ADM_momentum_variable}), as the initial data. 
In terms of these variables the presymplectic 2-form is given by the simple expression 
(\ref{spacelike_presymp}). It follows immediately that the unique solution to 
(\ref{kinematic_bracket_def}) is the standard ADM kinematic Poisson bracket:
\bearr
&&\{h_{ij}(x),\pi^{kl}(y)\}_\Dg = \dg^k_{(i}\dg^l_{j)}\dg^3(x,y) \\
&&\{h_{ij}(x),h_{kl}(y)\}_\Dg = \{\pi^{ij}(x),\pi^{kl}(y)\}_\Dg = 0
\eearr
for all $x$ and $y$ in the interior of $\Sg$. 

Now let's return to the auxiliary bracket $\{\cdot,\cdot\}_*$, which satisfies 
(\ref{auxbracketdef}). (\ref{auxbracketdef}) requires that $\dg \varphi_I 
= \Omega_\Sg[\{\varphi_I,\cdot\}_*,\dg]$
holds for all $\dg$ that vanish near $\di\Sg$ and satisfy the linearized field equations (that 
is $\forall \dg \in L_g^0$). Moreover $\{\varphi_I,\cdot\}_*$ is also required to satisfy the 
linearized field equations. These conditions are equivalent to requiring that the actions of 
$\dg$ and $\{\varphi_I,\cdot\}_*$ annihilate the constraints on the data on $\Sg$ (for if a 
perturbation of the initial data satisfies the linearized constraints then there always exists 
a solution to the linearized field equations matching the perturbation).\footnote{
{\em Proof:} The sourceless transverse gauge linearized field equation 
\be\label{transverse_lin}
\nabla^\sg\nabla_\sg \bar{\cg}_{\mu\nu} - 2R^\sg{}_{\mu\nu}{}^\tau \bar{\cg}_{\sg\tau} = 0
\ee
for the trace reversed metric perturbation $\bar{\cg}_{\mu\nu} = \dg g_{\mu\nu} 
- \frac{1}{2} g_{\mu\nu} (g^{\sg\rho}\dg g_{\sg\rho})$ determines a unique solution on the 
entire domain of dependence of $\Sg$ given any smooth symmetric tensor fields 
$\bar{\cg}_{\mu\nu}$ and $\nabla_n \bar{\cg}_{\mu\nu}$ on $\Sg$, where 
$n$ is the unit normal to $\Sg$. (see (\ref{reduced_linearized}).)
If these data satisfy the transverse gauge condition 
\be
\nabla^\mu \bar{\cg}_{\mu\nu} \equiv \chi_\nu = 0
\ee
on $\Sg$ and the linearized constraints
\be
0 = n^\mu\dg G_{\mu\nu} = n^\mu[\nabla_{(\mu}\chi_{\nu)} - 1/2 g_{\mu\nu}\nabla_\sg\chi^\sg
- 1/2 (\nabla^\sg\nabla_\sg \bar{\cg}_{\mu\nu} - 2R^\sg{}_{\mu\nu}{}^\tau \bar{\cg}_{\sg\tau})]
\ee
then in the corresponding solutions to (\ref{transverse_lin}) also the normal derivative of 
the transverse gauge condition holds: $\nabla_n \chi_\nu = 0$. But taking the divergence of 
(\ref{transverse_lin}) we find that this equation implies $\nabla^\sg\nabla_\sg \chi_\nu = 0$. 
The fact that $\chi$ and its normal derivative vanish 
on $\Sg$ thus implies that it vanishes on all of the domain of dependence. But then 
(\ref{transverse_lin}) is equivalent to the linearized Einstein vacuum field 
equation. $\dg g_{\mu\nu} = \cg_{\mu\nu} = \bar{\cg}_{\mu\nu} - 1/2 g_{\mu\nu}g^{\sg\tau}
\bar{\cg}_{\sg\tau}$ thus solves this equation.}
If $\dg$ annihilates the constraints 
then it follows from (\ref{kinematic_bracket_def}) that for any constraint $C$
\be
\Omega_\Sg[\{C,\cdot\}_\Dg,\dg] = 0.
\ee
Thus, if we could add some combination of actions of constraints $\{C_i,\cdot\}_\Dg$ to
$\{\varphi_I,\cdot\}_\Dg$ so that the sum annihilates the constraints then we would have a 
solution $\{\varphi_I,\cdot\}_*$ to (\ref{auxbracketdef}). This can be done, at least formally:
A gauge fixing is chosen that completely fixes the gauge transformations generated by the first 
class constraints, and $\dg$ is required to respect this gauge - that is, to annihilate the 
gauge fixing constraints. Then the {\em Dirac bracket},
\be\label{Dirac_bracket}
\{\cdot,\cdot\}_* = \{\cdot,\cdot\}_\Dg - \sum_{ij}\{\cdot,C_i\}_\Dg X^{ij}\{C_j,\cdot\}_\Dg,
\ee
fulfills our requirements. Here the set $\{C_i\}$ includes all constraints, also the gauge 
fixing constraints, and $X$ is the inverse of the matrix $\{C_i, C_j\}_\Dg$. 
The Dirac bracket is clearly a pre-Poisson bracket. In fact it also satisfies the Jacobi 
relations,\footnote{
To verify the Jacobi relations for the Dirac bracket note that the kinematic bracket 
$\{\cdot,\cdot\}_\Dg$ does satisfy the Jacobi relations, and that $\{A,B\}_* = \{A^*,B^*\}_\Dg$ 
where $A^* = A - \sum_{ij}\{A,C_i\}_\Dg X^{ij} C_j$ and $B^*$ is defined similarly.}
Note however that several issues arise when this construction is attempted for general 
relativity, which posses an infinite number of constraints. (A finite set of constraints 
at each point of $\Sg$). An inverse $X$ to $\{C_i, C_j\}_\Dg$ may not exist even if the
latter matrix is non-degenerate - i.e. has no degeneracy vectors; the sum in 
(\ref{Dirac_bracket}) may not converge; the sums in the proof of the Jacobi relations may not 
converge.

All the constraints on the ADM variables $h$ and $\pi$ are first class and generate, via the 
kinematic bracket $\{\cdot,\cdot\}_\Dg$, diffeomorphisms of the Cauchy surface $\Sg$ in the 
solution spacetime \cite{ADM}. The effect of such a diffeomorphism of $\Sg$ on the the data on 
$\Sg$ is of course equivalent to that of leaving $\Sg$ fixed and applying a suitable 
diffeomorphisms to the spacetime metric field. Since the observables are invariant under such 
diffeomorphisms the kinematic bracket between an observable and a constraint must vanish. 
Since $X$ has non-zero matrix elements only between the original first class constraints and 
the corresponding gauge fixing constraints it follows that the kinematic bracket between two 
observables, is equal to their Dirac bracket, which in turn, by (\ref{brackets_equal}), is 
equal to their Peierls bracket. In other words the Peierls bracket of the observables is equal 
to the standard ADM bracket between them.

\subsection{Gauge transformations}\label{gauge}


Degeneracy vectors of the presymplectic 2-form are normally associated with gauge 
transformations. Though it is not strictly necessary for our purpose of obtaining the 
auxiliary pre-Poisson bracket between initial data it is interesting, and ultimately important 
for the theory, to see how this works out in the situation we are considering, in which the 
initial data is specified on a hypersurface with a finite boundary. Here a definition of gauge 
transformations is proposed and some of its consequences are worked out.

Gauge transformations are transformations of the solutions that affect only those degrees of 
freedom of the field which are in principle unpredictable. A more precise definition (inspired 
by that of \cite{Lee_Wald}) would be the following: 

A map $\theta$ from the space of solutions to itself is a gauge transformation iff for any pair 
$C_1$, $C_2$ of disjoint closed sets in spacetime there exists another map $\theta'$ from the 
space of solutions to itself that reduces to $\theta$ on the field on $C_1$ and to the identity 
on the field on $C_2$. 

Put another way, the gauge transformed solution on any closed portion $C_1$ of spacetime can 
always be joined to the untransformed solution on any disjoint closed region $C_2$, forming 
another solution. Thus knowledge of the field on all of $C_2$ does not suffice to determine 
whether or not the field is transformed by $\theta$ on $C_1$.

Here a slightly different definition will be proposed. As discussed earlier, the developments
of smooth initial data are expected to be smooth at $\di\cN$. However disjoint 
subsets of $\Dund$ which are closed in $\Dund$ may have closures in $M$ which are not 
disjoint, containing common points on $\di\cN$ which is outside $\Dund$. The requirement that
gauge related solutions can be spliced on all pairs of disjoint sets that are closed in $\Dund$
combined with the requirement that the resulting spliced solutions be smooth at $\di\cN$ 
puts restrictions on the gauge transformations near $\di\cN$. Whether gauge transformations 
should be restricted in this way or not depends on the usefulness of the notion of gauge 
obtained. Here I opt to remove these restrictions on the transformations by placing further 
restrictions on the pairs of subsets $C_1, C_2$. The resulting set of gauge transformations is 
certainly simpler as a consequence.

\begin{definition}\label{gauge_def2}
A map $\theta$ from the space of solutions $\cS$ to itself is a gauge transformation iff for 
any pair $C_1$, $C_2$ of subsets of $\Dund$ having disjoint closures in $M$ there exists 
another map $\theta': \cS \rightarrow \cS$ that reduces to $\theta$ on the field on $C_1$ and 
to the identity on the field on $C_2$. 
\end{definition}

Note that the manifold $\Dund$ here is the union of $int \cN$ and the interior of its domain
of dependence in some, arbitrarily chosen solution which we take as fiducial. (For instance 
$\Dund_0$ formed from $\cN_0$ and the interior of its domain of dependence in Minkowski space 
could be used - see subsection \ref{problem}.) As in subsection \ref{problem} $\cS$ consists of 
the metric fields of the maximal Cauchy developments, mapped via the diffeomorphism of prop. 
\ref{Cauchys_diffeo} to $\Dund$. 

According to the definition def. \ref{gauge_def2} (and also the previous one) the gauge 
transformations of general relativity are diffeomorphisms, as one would expect:

\begin{proposition}
A gauge transformations of general relativity acts on any given solution as a 
diffeomorphism of $\Dund$ to $\Dund$.\footnote{
A gauge transformation may act as one diffeomorphism on one solution and a different 
diffeomorphism on another. Thus as a map from $\cS$ to $\cS$ it is not necessarily equivalent
to any one diffeomorphism.}
\end{proposition}

{\em Proof:} Consider a particular solution $g$ to Einsteins field equations on $\Dund$ and a 
gauge transformation $\theta$. Let $C_1$ be a compact set in the interior $\cD$ of $\Dund$ and 
$C_2$ the causal past in $\cD$ of a spacelike Cauchy surface $\Sg_2$ of $\cD$ to the past of 
$C_1$. $C_2$ is closed in $\cD$ since it is the complement of the open set $I^+[\Sg_2,\cD]$ in 
$\cD$. Its closure in $M$ thus intersects $\cD$ only on $C_2$ itself, and thus does not
intersect $C_1$, which, being compact, is already closed in $M$.

The well known uniqueness theorems for the Cauchy problem of general relativity (see 
\cite{Hawking_Ellis}) then show that the solution on $C_2$ determines that on $C_1$ up to a 
diffeomorphism, which necessarily maps $\Dund - C_2$ to itself (since $\cS$ contains only 
solutions in which $\Dund - C_2$ is the spacetime of the maximal future Cauchy development 
of $\Sg_2$). It follows that on $C_1$, on the given solution, $\theta$ is equivalent to a 
diffeomorphism that maps $C_1$ to a compact subset of $\cD$. 

In fact $\theta$ is equivalent on all of $\cD$ to a diffeomorphism $\phi$ from $\cD$ 
to itself: If $C_1'$ is a compact subset of $\cD$ which contains $C_1$ then the diffeomorphism 
on $C_1$ corresponding to $\theta$ may be chosen to be the restriction to $C_1$ of 
the diffeomorphism corresponding to $\theta$ on $C_1'$. If the metric on $C_1$ has no 
isometries this follows from the fact that the mapping $\theta$ of the metric determines a 
unique diffeomorphism. When there are isometries we are still free to {\em impose} this 
condition. Now, because $\cD$ is paracompact it may be covered by a sequence of 
such nested compact sets. If there are no isometries of all of $\cD$ then some compact set in 
the sequence, and all following it, will also be free of isometries so $\theta$ 
defines a unique diffeomorphism on each set in the sequence, such that the diffeomorphisms
on the larger sets are extensions of the diffeomorphisms on the smaller sets. $\theta$ 
is therefore equivalent to a diffeomorphism $\phi: \cD \rightarrow \cD$. If $\cD$ does
have isometries a compact set in the sequence may be chosen that has only the isometries
of $\cD$ as a whole. The diffeomorphisms equivalent to $\theta$ on the smaller sets in the 
sequence as well as on the larger sets may then be chosen to be consistent with the 
diffeomorphism on this set, and again $\theta$ is equivalent to a diffeomorphism 
$\phi: \cD \rightarrow \cD$, although in this case it is not unique.

This result extends to the bounding hypersurface $int\cN$. That is, on all of $\Dund$, 
including $int\cN$, $\theta$ is equivalent to a diffeomorphism from $\Dund$ to $\Dund$. 
To show this we shall demonstrate, using the fact that both $g$ and $\theta g$ are smooth on 
$\Dund$, and thus can be smoothly extended to a boundaryless manifold $Z \subset M$ 
containing $\Dund \supset int\cN$, that a suitable extension of the diffeomorphism $\phi$ to 
$int\cN$ can be constructed.

Let $p$ be a point of $int\cN$ and let $U$ be a convex normal neighbourhood of $p$ in $Z$ 
according to a smooth extension $\hat{g}$ of the metric $g$ to $Z$. Let $q$ be a point of 
$U \cap \cD$ from which $p$ can be reached via a past directed timelike geodesic $\ell$, and 
suppose that a Riemann $\hat{g}$-normal coordinate chart, $y$, covering $U$ and based at $q$ is
set up. The map $\phi$ takes $q$ to $q' \in \cD$ and the restriction of $y$ to $U \cap \cD$ 
to a new chart $y'$ with origin at $q'$ which is normal with respect to the metric $\theta g$. 
An extension of the chart $y'$ to a slightly larger domain defines, together with the chart $y$,
the extension of $\phi$ that is needed.

First we need to understand a little more about the image under the diffeomorphism $\phi$
of the domain $U\cap \cD$. Let $\ell^\circ$ be the segment of the geodesic $\ell$ which lies 
in $\cD$, and let ${\ell^\circ}' = \phi(\ell^\circ)$. ${\ell^\circ}'$ is timelike according 
to $\theta g$ and past directed, if the time orientation on $\Dund$ is always chosen so that 
$int\cN$ lies to the past of $\cD$, because with this convention $\phi$ preserves the time 
orientation of timelike curves. Since $q'$ lies in $\cD$, and thus in the future domain of 
dependence, according to $\theta g$, of $int\cN$ in $\Dund$, but ${\ell^\circ}'$ does not cross 
$int\cN$ ${\ell^\circ}'$ must be an extendible curve - it must have a past endpoint, $p'$, in 
$\Dund$. $p'$ cannot lie in $\cD$ since $\ell^\circ$ would then have a past endpoint in $\cD$. 
$p'$ thus lies in $\int\cN$. ($p'$ will turn out to be the 
image of $p$ under the extension of $\phi$.) 

The same argument applies when $p$ is replaced by any other point $\tilde{p}$ in the open 
subset of $int\cN\cap U$ that lies in the chronological past $I^-[q]$ of $q$. That is, the 
image under $\phi$ of the geodesic segment from $q$ to $\tilde{p}$ (not including $\tilde{p}$ 
itself) has a past endpoint on $int\cN$.

Let $\hat{g}'$ be a smooth extension of $\theta g$ from $\Dund$ to $Z$, then the past 
directed, timelike geodesics from $q$ that define the normal coordinates $y'$ near $p'$ can 
be extended smoothly through their past end points on $int\cN \subset Z$. These extended 
geodesics can then be used to extend the domain of the normal coordinates $y'$. The geodesics
do not form any caustics on $int\cN$ near $p'$: In terms of the metric components
\be
\hat{g}'_{\ag\bg} = \hat{g}'(\frac{\di}{\di {y'}^\ag},\frac{\di}{\di {y'}^\bg})
\ee
in normal coordinates a caustic occurs at a point if and only if $det[\hat{g}'_{\ag\bg}]$
approaches zero as the point is approached. But the $y'$ components of the metric 
$\theta g$ at a point $r \in \cD$ are the same as the $y$ components of the metric $g$ at
$\phi^{-1}(r) \in \cD$. There can thus be no caustic at $p'$ since the $y$ chart does 
{\em not} break down at $p$, being good on all of $U$. The inverse function theorem then 
implies that the $y'$ coordinates constructed from the extended timelike geodesics form a 
good chart of a neighbourhood of $p'$ in $Z$. Since the diffeomorphism ${y'}^{-1}\circ y$ 
coincides with $\phi$ on the portion of its domain $V$ lying in $\cD$, and furthermore it 
maps no point outside $\cD$ into $\cD$ and $y'(p') = y(p)$, the definition 
$\phi = {y'}^{-1}\circ y$ on $V$ extends $\phi$ as a diffeomorphism to a domain including a 
neighbourhood of $p$ in $Z$.

Of course $\phi(p) = p'$. Moreover, the limiting values of the $y$ components of 
$\hat{g}$ as $p$ is approached from $\cD$ are the same as the limiting values of the $y'$ 
components of $\hat{g}'$ as $p'$ is approached from $\cD$, which implies that the actual 
values of the $y$ components of $\hat{g}$ at $p$ and the $y'$ components of $\hat{g}'$ at $p'$ 
are equal, because the smoothness of the metrics $\hat{g}$ and $\hat{g}'$ imply that the charts 
$y$ and $y'$, and the corresponding metric components, are smooth. Thus $\theta g = \phi^* g$
in a neighbourhood of $p$ in $int\cN$.

$\phi$ can be extended in this way to all of $int\cN$. If $\phi_1$ and $\phi_2$ are two 
extensions of $\phi$ such that a point $t \in int\cN$ lies in the intersection of their domains 
of definition then, because $\phi_1 = \phi_2$ on $\cD$ and both are continuous 
$\phi_1(t) = \phi_2(t)$. 
\QED
\\

Now consider linearized gauge transformations. The tangent $\cg \in L_g$ at a solution $g$ to a 
one parameter family of gauge transforms of $g$ clearly satisfies the following condition: For 
all pairs $C_1, C_2 \subset \Dund$ having disjoint closures in $M$ there exists 
$\cg' \in L_g$, such that $\cg' = \cg$ on $C_1$ and $\cg' = 0$ on $C_2$. I shall call any 
$\cg \in L_g$ that satisfies this condition a {\em linearized gauge transformation}. These can 
be characterized as follows:

\begin{proposition}
Linearized gauge transformations are generalized degeneracy vectors of the presymplectic 2-form.
\end{proposition}
\noindent {\em Proof:} The hypersurface independence of $\bar{\Omega}$ implies that any 
linearized gauge transformation $\cg_0$ which vanishes in a neighbourhood of $\di\cN$ is a 
degeneracy vector of $\bar{\Omega}$: Let $\Sg_1, \Sg_2 \subset \cD$ be smooth Cauchy surfaces 
of $\cD$, such that\footnote{
Time ordered sequences of Cauchy surfaces exist by Geroch's theorem, prop. 6.6.8. of 
\cite{Hawking_Ellis}, and may be taken to be smooth by the result of \cite{Bernal_Sanchez0}.}
\be \label{Sg_1_Sg_2_ordering}
\Sg_2 \subset I^+[\Sg_1;\cD],
\ee  
and let $C'_1 = J^-[\Sg_1;\cD]$ and $C_2 = J^+[\Sg_2;\cD]$. These are disjoint closed subsets of
$\cD$. However their closures, $\overline{C}'_1$, $\overline{C}_2$ in $M$ intersect on
$\di\cN$. A point $p \in \overline{C}'_1 \cap \overline{C}_2$ lies in 
$\overline{D[\cN;M]} = \overline{D[\overline{\Sg}_1;M]} 
= \overline{D[\overline{\Sg}_2;M]}$, and thus, by prop. 8.3.2. of \cite{Wald}
any inextendible timelike curve through $p$ must cross $\overline{\Sg}_1$, at $p$ or later,
and $\overline{\Sg}_2$, at $p$ or earlier. But $\overline{\Sg}_2 \subset 
\overline{I^+[\Sg_1;M]}$ so the crossing of $\overline{\Sg}_2$ must be after or 
simultaneous with the crossing of $\overline{\Sg}_1$. The only consistent possibility is that
$p \in \overline{\Sg}_1 \cap \overline{\Sg}_2$, which by (\ref{Sg_1_Sg_2_ordering}) and
prop. \ref{cauchy_D} implies that $p \in \di\cN$.

Let $C_1$ be the support of $\cg_0$ in the causal past of $\Sg_1$. The closure $\overline{C}_1$
of this set in $M$ is a subset of $\overline{C}'_1$ which excludes a neighbourhood of 
$\di\cN$. $\overline{C}_1$ therefore does not intersect $\overline{C}_2$. By the definition of 
linearized gauge transformations there must thus exist $\cg'_0 \in L_g$ such that 
$\cg'_0 = \cg_0$ on $C_1$ and $\cg'_0 = 0$ on $C_2$. Therefore
\be	\label{gauge_degeneracy1}
\bar{\Omega}[\cg_0,\cdot] = \bar{\Omega}_\Sg[\cg'_0,\cdot] = \bar{\Omega}_{\Sg'}[\cg'_0,\cdot] 
= 0.
\ee

Now consider an arbitrary linearized gauge transformation $\cg$ and a solution $\dg$ to the 
linearized field equations vanishing in a neighbourhood $U \subset M$ of $\di\cN$. By 
definition there 
exists another linearized gauge transformation $\cg'$ which is equal to $\cg$ on $\cD - U$ 
(which contains the support of $\dg$) and which vanishes on a closed subset of $U$ which 
contains $\di\cN$ in its interior. Thus 
\be
\bar{\Omega}[\cg,\dg] = \bar{\Omega}[\cg',\dg]
\ee
and, since $\cg'$ vanishes in a neighbourhood of $\di\cN$, this vanishes by 
(\ref{gauge_degeneracy1}). This is valid for any $\dg \in L_g^0$ so $\cg$ is a generalized 
degeneracy vector of $\bar{\Omega}$. \QED
\\

This last result and prop. \ref{Degenerate_is_diff} imply that, at least when the metric on 
$\Dund$ is generic (in the sense of prop. \ref{Degenerate_is_diff}), a linearized gauge 
transformation must be a smooth 
diffeomorphism generator - that is, a Lie derivative ${\pounds}_\xi$ along a smooth vector 
field $\xi$. This can in fact be demonstrated directly, without the genericity assumption, by 
an argument analogous to that used above to characterize the gauge transformations of full 
general relativity, from the fact that data on a Cauchy surface of $\cD$ determines the solution
to the linearized field equations up to the addition of terms of the form 
${\pounds}_\xi g_{\mu\nu} = 2 \nabla_{(\mu} \xi_{\nu)}$ (see Appendix \ref{another_approach}).
Conversely, it is easy to see that any Lie derivative ${\pounds}_\xi$ along a smooth vector
field $\xi$ is a linearized gauge transformation according to our definition. 

The linearized 
gauge transformations are thus precisely the diffeomorphism generators, they are always
generalized degeneracy vectors of $\bar{\Omega}$, and, at least for a generic metric on $\Dund$
all degeneracy vectors are diffeomorphism generators and thus linearized gauge transformations.

It would perhaps be natural to impose a further condition on the linearized gauge 
transformations that all tangents to full gauge transformations satisfy, namely that the 
integral curves of $\xi$ do not leave $\Dund$. This would restrict the linearized gauge 
transformations to only a subset of the generalized degeneracy vectors of $\bar{\Omega}$, 
although it does not require them to be true degeneracy vectors.

The best - most natural and most useful - choice of definition of gauge transformations in the 
context of a spacetime with boundaries is not clear to the author. In this section some possible
definitions have been presented. Note however that nothing in the remainder of the paper depends
on the choice of these definitions. 
Indeed the general notion of gauge transformations is not used in the paper, precisely because
no clearly preferred definition was found. Instead the classes of transformations considered are
identified explicitly in each instance (e.g. diffeomorphisms that preserve a neighbourhood of 
$\di\cN$).

\section{Free initial data and adapted coordinates} \label{free_data_coords}

\subsection{The area parameter}\label{area_parameter}

The tangents $n_A$ to the generators that sweep out the branch $\cN_A$\footnote{
Here and from here on subscripts $A,B,...$ represent indices ranging over $L$ and $R$.}
of $\cN$ are also normal to $\cN_A$ (prop. \ref{tangent_is_normal}). As a consequence 
$\cN_A$ is a null hypersurface (its normals are null vectors) and the 3-metric
induced on $\cN_A$ by the spacetime metric is degenerate, as there is a vector in 
the tangent space - namely $n_A$ - which is orthogonal to all tangent vectors.
Put another way, if one uses a tangent to the generators as the first vector in a basis 
of the tangent space of $\cN_A$ then the 3-metric takes the form
\be		\label{degenerate_standard_form}
\left[\begin{array}{cc}  0 & \begin{array}{cc} 0 & 0 \end{array} \\
                \begin{array}{c} 0 \\ 0 \end{array} & h_{ij} \end{array}\right],
\ee
where $h_{ij}$ is a symmetric $2\times 2$ matrix.

The degeneracy of the 3-metric implies that the distance between neighbouring points $p$ and $q$ on 
$\cN_A$ does not depend on the component
of the (infinitesimal) displacement vector $\vec{pq}$ parallel to the generators. Similarly the area,
\be
\cA = \sqrt{t_1\cdot t_1\,t_2\cdot t_2 - (t_1\cdot t_2)^2},
\ee
of an infinitesimal parallelogram in $\cN_A$ spanned by vectors $t_1$ and $t_2$ based at 
$p \in \cN_A$ does not depend on the components of $t_1$ and $t_2$ along the generators. 
It depends on $p$ and on the set of generators cut by the parallelogram, but not on the 
``angle" at which it cuts these, that is, on the tangent plane of the parallelogram. This 
fact makes possible a very useful (non-affine) parametrization of the generators. It implies 
that the cross sectional area of an infinitesimal bundle of generators neighbouring a given 
generator $\cg$, and terminating on a surface of area $\bar{\cA}$ in $S_A$, depends only on the 
point $p \in \cg$ at which the cross section is taken, and (linearly) on $\bar{\cA}$. In other 
words, the factor $\cA(p)/\bar{\cA}$ by which the area expands following the generators from 
$S_A$ to $p$ depends only on $p$. The expansion factor can therefore be used to parametrize 
$\cg$, provided it increases or decreases monotonically along $\cg$.

We shall use the square root of the expansion factor, 
\be             \label{r_def}
r = \sqrt{\cA/\bar{\cA}},
\ee
to parametrize the generators. The cross sectional area of a bundle of neighbouring generators 
is then proportional to $r^2$, which means that $r$ is an {\em area parameter}. That is, $r(p)$ 
is proportional to the distance from $p$ to the caustic point, where neighbouring generators 
focus and $\cA = 0$, inferred from the area subtended at the caustic by a standard disk at $p$ 
by assuming that this area falls as $1/\mbox{\em distance}^2$ as in flat spacetime.
\footnote{
In vacuum general relativity a bundle of neighbouring null geodesics is either incomplete
or has at least one caustic, except in the special case that its cross sectional area is 
strictly constant along the bundle. (This follows from the Raychaudhuri equation discussed in 
the next paragraph.) Such constant cross section bundles do not exist in generic vacuum 
solutions (see prop. 4.4.5. of \cite{Hawking_Ellis} and the discussion which follows it).
If a bundle of null geodesics has several caustics, it is most natural to $r$ to be a measure 
of distance to the nearest caustic in the direction along the bundle in which $r$ is decreasing.
Of course this is just an issue of interpretation. (\ref{r_def}) defines $r$ regardless of the 
interpretation adopted.}

As mentioned in subsection \ref{problem} we shall consider only initial data such that
the generators of a given branch, $\cN_L$ or $\cN_R$, of $\cN$ are either everywhere diverging 
($\cA$ increasing) or everywhere converging toward the future ($\cA$ decreasing).
This ensures that $r$ is a good parameter on each generator of the branch.

Parametrizing the generators, and the initial data on them, by $r$ makes it easy to incorporate
the condition that $\cN_L$ and $\cN_R$ are truncated before their generators reach a caustic.
One simply requires that each branch corresponds to an interval of positive, non-zero, values of $r$.
\footnote{
This does not in itself guarantee that the generators do not cross in spacetime, just that
{\em neighbouring} generators do not (more precisely, that the separation of neighbouring 
generators does not become zero {\em to first order} in their separation on $S_0$). But 
recall that we are considering Cauchy developments, and in a Cauchy development of initial 
data on $\cN$, the generators of $\cN$ never intersect (except of course at their starting 
points on $S_0$). [If $\cN$ intersects itself in a solution to Einstein's field equations then 
(a portion of) the domain of dependence of $\cN$ in this solution is represented in the 
catalogue of Cauchy developments of $\cN$ by a covering manifold (see appendix 
\ref{nullhypersurfaces} prop. \ref{cN_achronal}).]

If the reader prefers not to represent spacetime by a covering manifold an alternative argument 
is that in order to calculate the brackets of the data along a given generator $\cg$ 
one may use instead of $\cN$ an initial data surface $\cN'$ swept out by the generators 
emerging from a sufficiently small disk $S_0' \subset S_0$ containing the base point of $\cg$ 
in $S_0$ such that these generators do {\em not} cross.}
The use of the area parameter also simplifies the mathematics of the null initial value 
problem. This was noted and exploited by Gambini and Restuccia \cite{GR} to obtain an
expansion in powers of Newton's constant for the brackets between the main null
initial data.

\subsection{A geometrical description of the initial data}\label{initial_data}

We now turn to the initial data, beginning with a coordinate free description. All the 
data have an interpretation in terms of the geometry of the corresponding solution, 
but first let us consider $\cN$ and the initial data on their own, prior to embedding 
in the solution spacetime.

The set of initial data that we will consider consists of a 3-metric on $\cN$ and 
further data specified on $S_0$ only. The 3-metric is specified by two degenerate, 
positive semi-definite 3-metrics $g_L$ and $g_R$ defined on $\cN_L$ and $\cN_R$ respectively.
$g_L$ and $g_R$ are required to be smooth and to match 
at $S_0$ in the sense that they induce the same 2-metric on $S_0$. The degeneracy subspace 
$N_p^A$ of $g_A$ at every point $p \in \cN_A$ is one dimensional, and the integral curves of 
these subspaces, which will be called the ``generators" of $\cN_A$, form a congruence that 
threads the cylinder $\cN_A$ along its axis. That is, each integral curve originates on $S_0$ 
and terminates on $S_A$. The congruence as a whole defines a (trivial) fibration of $\cN_A$ 
with the generators being the fibres and $S_0$ the base space.

The data on $S_0$ are most easily described in terms of two fields, $\chi$ and $\omega$, that 
depend on a choice of parametrization of the generators. We shall therefore suppose that a 
good parametrization has been (arbitrarily) chosen on the generators. The parametrization 
defines fields of tangent vectors on $\cN_L$ and $\cN_R$ which will be denoted, like the 
tangent vectors to the parametrized generators in the spacetime context, by $n_L$ and $n_R$ 
respectively.

$\chi$ is a smooth scalar field on $S_0$. $\omega$ is a smooth 1-form field on $S_0$, 
co-tangent to $S_0$. Under a change of parametrization such that $n_L$ and $n_R$ are 
rescaled by factors $\ag_L$ and $\ag_R$
\bearr
\chi   & \rightarrow & \ag_L\ag_R \chi      \label{chitransform}\\
\omega & \rightarrow & \omega + \ppback{d} \log \ag_L - \ppback{d} \log \ag_R,
\label{omegatransform}
\eearr
where a $\Leftarrow$ beneath a form indicates that the pullback of the form to $S_0$ is 
to be taken.

In the corresponding solution to Einstein's vacuum equations the 3-metric on $\cN$ matches the 
3-metric induced on the embedding of $\cN$ by the spacetime geometry (this means of course that 
the ``generators" defined as the integral curves of the degeneracy subspaces of the 3-metric 
coincide with the generators of the embedding of $\cN$) and $\chi$ and $\omega$ are related to 
the spacetime metric and connection via
\be     \label{chiinterp}
\chi = n_L\cdot n_R
\ee
and 
\be \label{omegainterp}
\omega = \frac{n_L^\mu \ppback{\nabla} n_{R\,\mu} - n_R^\mu \ppback{\nabla} n_{L\,\mu}}{n_L\cdot n_R}.
\ee  

$\chi$ thus measures the component of the spacetime metric not captured by the 3-metrics 
$g_L$ and $g_R$ of $\cN$ at $S_0$, while $\omega$ is somewhat akin to an extrinsic curvature,
since it measures an aspect of the derivatives along $S_0$ of its two null normals, $n_L$ and $n_R$.

Note that the sign of $\chi$ is determined once the sense of $n_L$ and of $n_R$ are chosen. If both
point away from $S_0$, or both toward $S_0$, then $\chi$ must be negative, for then $n_L$ and 
$n_R$ are both null future directed or both null past directed when $\cN$ is embedded in a solution
spacetime, so their inner product must be negative.

The initial data $\chi$ and $\omega$ have been postulated to transform under a change of 
parametrization of the generators in the same way as the solution spacetime quantities 
that they correspond to, so the correspondence does not depend on the parametrization chosen.

An interesting alternative expression for $\omega$ in terms of the spacetime geometry 
can be obtained if one extends both $n_L$ and $n_R$ to all of $\cN$ in such a way that both are 
orthogonal to the cross sections of $\cN$ cutting the generators at a constant parameter value. 
Then 
\be		\label{normalized_twist}
\omega = \frac{\ppback{[n_L, n_R]}}{n_L\cdot n_R},
\ee
where the Lie bracket, a vector, is mapped to a 1-form using the spacetime metric.
(The right side is called the {\em normalized twist} \cite{Epp} but is not to be confused
with the ``twist" of a single congruence \cite{Wald} which always vanishes for the generators
of a null hypersurface.)

To see this suppose a tangent vector $s$ of $S_0$ is Lie dragged along $n_A$ to all of the branch 
$\cN_A$. This defines a vector field $s$ orthogonal to both $n_L$ and $n_R$, since $s$ is everywhere 
tangent to the isoparametric cross sections. The Lie derivative of the orthogonality relation
$s^\mu n_{B\,\mu} = 0$ together with the fact that $s$ satisfies $\pounds_{n_A} s = 0$ implies
\be
0 = s^\mu {\pounds}_{n_A} n_{B\,\mu} = s^\mu \nabla_{n_A} n_{B\,\mu} + n_{B\,\mu}\nabla_s n_A^\mu.
\ee
Thus for all $s$ tangent to $S_0$, $\omega \cdot s$ takes the same value whether computed using
the expression (\ref{omegainterp}) or the projection of the normalized twist 
(\ref{normalized_twist}), proving their equivalence.

From (\ref{normalized_twist}) it is clear that $\omega$ measures the non-integrability at $S_0$ 
of the normal planes to the isoparametric sections of $\cN$. Integrability cannot in general be 
achieved by reparametrizing the generators because the transformation law (\ref{omegatransform})
implies that the curl of $\omega$ on $S_0$ is independent of the parametrization. 

Sachs \cite{Sachs} and Rendall \cite{Rendall} have given arguments that imply that any choice
of these data (satisfying certain inequalities)\footnote{\label{field_eq_restriction}
The vacuum Einstein equations imply, via the Raychaudhuri focusing equation 
(\cite{Wald} eq. 9.2.32), that if the cross sectional area of an infinitesimal bundle of 
neighbouring generators is decreasing in one direction along the bundle at one point then it 
must continue to decrease in the same direction until it reaches zero. That is to say,if the 
derivative of the area by any differentiable parameter of the generators is negative at a point
then it must remain negative until the area vanishes. This imposes inequalities on the 3-metric 
on $\cN$, although it does not imply any restriction on the variables Sachs used to describe 
the initial data.}
on $\cN$ define a solution to Einstein's field equations that is unique up to diffeomorphisms. 
More precisely, Sachs considered these data expressed in a special coordinate system adapted to 
the data - in which the generators lie along coordinate axes and each generator is parametrized
by a coordinate which becomes an affine parameter in the corresponding solution spacetime - and 
argued that the solution metric exists and is unique in a particular coordinate system adapted 
to the solution which extends his coordinates on $\cN$ to spacetime. Rendall then proved that 
Sachs' claim is true, at least in a neighbourhood of $S_0$ in the causal domain of dependence 
$D[\cN]$ of $\cN$.

The chief advantage of using coordinates adapted to the data, such as Sachs' coordinates, is 
that it makes it easier to identify freely specifiable data. The data $g_L$, $g_R$, $\chi$ and 
$\omega$ are not themselves freely specifiable since $g_L$ and $g_R$ are constrained to be 
degenerate, with restrictions on the degeneracy subspaces. Taking the generators as coordinate 
axes reduces this constraint to the requirement that certain metric components vanish 
(see (\ref{degenerate_standard_form})).
The use of a coordinate adapted to the data to parametrize the generators further reduces the 
freely specifiable data. Sachs normalizes the parameters on the generators so that $\chi = -1$ 
always. Moreover, the requirement that the parameters become affine parameters in the matching 
solution implies a constraint on the parameter dependence of the 3-metric on $\cN$. The cross 
sectional area of an infinitesimal bundle of neighbouring generators must satisfy the 
Raychaudhuri focusing equation corresponding to an affine parametrization (\cite{Wald} eq. 
9.2.32) with the Ricci curvature term set to zero, which substantially reduces the 
independently specifiable metric data.
 
Another way to view the constraint on the metric components just mentioned is the following: 
Given {\em any} 3-metric on $\cN$ satisfying the conditions of smoothness and degeneracy
we have required of the initial data, and satisfying the inequalities 
explained in footnote \ref{field_eq_restriction}, there is a parametrization of each generator, 
unique up to an affine transformation, such that the metric satisfies the constraint. 
If this parameter, which I will term a ``proto affine parameter'' is adopted the metric must 
satisfy the constraint, but only because of the choice of parameter, and not because of 
Einstein's field equations. The field equations imply that in the matching solution 
the proto affine parameter becomes an affine parameter of the spacetime geometry.
This argument will be developed in detail in subsection \ref{freeness}.

We will define both an almost freely specifiable data set and a completely free data set using 
charts similar to those of Sachs but with the proto affine parameter replaced by the area 
parameter $r$ defined in (\ref{area_parameter}). The area parameter provides a good 
parametrization of the 
generators of a branch of $\cN$ iff the cross sectional area of infinitesimal
bundles of neighbouring generators in the branch is either everywhere increasing or everywhere 
decreasing toward the future. This is not the case in all solutions so our formalism cannot be
applied to all initial data compatible with the field equations, but only to data satisfying
this restriction on the 3-metric on $\cN$.\footnote{
This restriction is not a {\em constraint} because it implies only that the 3-metric satisfies
inequalities, and not equations.}
However, as explained in subsection \ref{problem}, this leads to relatively little loss
of generality in our results. The causality requirement implies that only data living on 
the same generator have non-zero brackets. In the calculation of the brackets on the 
generators through a point $p \in S_0$ we may therefore replace $\cN$ by $\cN' \subset \cN$
swept out by the generators through a compact subset $S'_0$ of $S_0$ containing $p$ in its 
interior. If the expansion rate of the congruence of generators of $\cN_L$ and that of the 
generators of $\cN_R$ are both non-zero at $p$ (the generic case) then $S'_0$ may be chosen
so that these expansion rates are non-zero and of uniform sign on $S'_0$. If the expansion 
rate at $S'_0$ of the generators of a branch is negative then by the field equations and the 
Raychaudhuri equation it will be negative on the whole branch, and $r$ will decrease 
monotonically on the generators. If the expansion rate is positive at $S'_0$ then $r$ increases
monotonically along each generator until it leaves the branch or until $r$ reaches a maximum. 
In the latter case our results apply until the maximum is reached. Only if the expansion rate 
of the generators at $p$ of one of the branches of $\cN$ is zero are our methods inapplicable. 

In the following subsection the coordinates that will be used to define our almost free and 
completely free initial data will be defined precisely. The relation between these data and 
Sachs' data is treated in detail in subsection \ref{freeness}.

\subsection{Coordinate systems and free initial data}
\label{coordinates}

Three special coordinate charts will be used to represent the different elements of the initial 
data. The $a_R$ chart, with coordinates $a_R^\ag = (u_R, r_R, y_R^1, y_R^2)$, covers $\cN_R$ 
and a spacetime neighbourhood of $int \cN_R$, is adapted to the generators of $\cN_R$, and is 
fixed in the manifold (i.e. solution independent) on $S_R \subset \di\cN$. The $a_L$ 
chart, which will be defined in strict analogy with the $a_R$ chart, covers $\cN_L$ and a 
spacetime neighbourhood of $int \cN_L$, is adapted to the generators of $\cN_L$, and is fixed to 
the manifold on $S_L \subset \di\cN$. Finally, the $b$ chart, with coordinates $b^\mu = 
(v^+,v^-,\theta^1,\theta^2)$, covers all of $\cN$ and a spacetime neighbourhood of $int \cN$ and 
is adapted to the generators of both branches of $\cN$, but is not {\em a priori} fixed 
anywhere.

Greek lowercase indices $\ag,\bg, ...$ from the beginning of the alphabet will always refer
to an $a$ coordinate basis (of the branch $\cN_A$ under consideration). Greek uppercase indices
$\Lambda, \Xi ...$ refer to the $b$ coordinate basis. 
Lowercase Latin indices $i,j,...$ from the latter part of the alphabet range over $\{1,2\}$ and 
will be used to identify the two adapted coordinates $y^1_A$ and $y^2_A$ and their 
corresponding coordinate basis vectors, while lower case Latin indices $a, b, ...$ from the 
beginning of the alphabet, which also range over $\{1, 2\}$, identify the coordinates 
$\theta^1$ and $\theta^2$ and their basis vectors.

The chart $a^\ag_A = (u_A, r_A, y_A^1, y_A^2)$ adapted to $\cN_A$ is defined in three steps.
First smooth, fixed (not solution dependent), coordinates $y_A^1$ and $y_A^2$ are chosen on 
$S_A$. Next this coordinate system is extended to a chart on all of $\cN_A$ by convecting the 
$y_A^i$ along the generators, i.e. setting them to be constant along the generators, and 
introducing the coordinate $r_A$, taken to be the area parameter along the generators 
defined in subsection
\ref{area_parameter}. Thus on $\cN_A$ the two $y$ coordinates identify the generators and
$r_A$ identifies the points on each generator. Finally coordinates are defined on a spacetime 
domain containing $int \cN_A$ in its interior: Each constant $r_A$ section of $\cN$ is a two 
dimensional spacelike disk and thus posses two null normals at each point, and two 
corresponding normal null geodesics,
One is the generator of $\cN_A$ through the point and the other is transverse to $\cN_A$.
To extend the coordinate system to a neighbourhood of $\cN_A$ the three coordinates $y^1_A$, 
$y^2_A$, and $r_A$ are convected along the transverse normal null geodesics and a fourth 
coordinate, $u_A$, which parametrizes these geodesics 
is defined by the requirements $u_A = 0$ on $\cN_A$ and $\di_{u_A} \cdot \di_{r_A} = -1$.

To lighten the notation the subscript (``$A$'') identifying the branch $\cN_A$ to which the 
coordinates $a_A$ are adapted will be dropped from here on when there is little risk of 
confusion.

The components of the metric on a branch $\cN_A$ take a very restricted form in the $a$ 
coordinates adapted to that branch. The coordinate basis vectors $\di_r$ and $\di_u$ are null, 
and have inner product $\di_r\cdot\di_u = -1$. Moreover both are orthogonal to the constant 
$(u, r)$ surfaces. Thus the line element takes the very simple form
\be             \label{a_line_element}
ds^2 = -2du dr + h_{ij}dy^i dy^j
\ee
on $\cN_A$, where $h_{ij}$ is a symmetric $2\times 2$ matrix which must be positive definite
in order that the spacetime metric be Lorentzian with signature ${}-+++$. The induced 3-metric 
on $\cN_A$ is seen to be 
degenerate and takes precisely the form (\ref{degenerate_standard_form}) obtained in
subsection \ref{area_parameter}. The line element on {\em any} 2-surface cutting a cross 
section of the congruence of generators is $ds^2 = h_{ij}dy^i dy^j$. This means in particular 
that area density of the cross section is,
\be
        \rho = \sqrt{det[h_{ij}]},
\ee
the root of the determinant of the induced 2-metric. (As expected this depends only on
position in $\cN_A$, and not on the tangent plane of the cross section.) 
The value of the coordinate $r$, the area parameter, at a point $p \in \cN_A$ is thus
\be
r(p) = \sqrt{\rho(p)/\bar{\rho}(\cg)}
\ee
where $\bar{\rho}(\cg)$ is the value of $\rho$ at the endpoint on $S_A$ of the generator $\cg$ 
through $p$. Equivalently
\be             \label{rho_form}
\rho(r, y^i) = r^2 \bar{\rho}(y^i).
\ee 
The only freedom in $\rho$ is thus its boundary value $\bar{\rho}$ on $S_A$.

The remaining freedom in $h_{ij}$ resides in the conformal\footnote{
It is invariant under rescaling of the metric, hence the name.}
2-metric, a unimodular (determinant equals 1) $2 \times 2$ matrix defined by
        \be
                e_{ij} = h_{ij}/\rho.
        \ee 

Information about the metric geometry of $\cN_A$ is also found in the range of values the $a$ 
coordinates take on $\cN_A$. $(y^1, y^2)$ ranges over a fixed (solution independent) domain 
diffeomorphic to a disk in $\Real^2$. $r$ on the, other hand, ranges on each generator 
from $1$ on $S_A$ to a value $r_0(y)$ on $S_0$ (depending on the generator). The function 
$r_0(y)$ depends on the solution metric, since a variation 
of the metric will in general change the rate at which the generators are focused, contracting 
or expanding cross sections of bundles of neighbouring generators, and thus affects $r_0$.
[$r_0$ may be either greater than 1, 
(generators converging toward the future) or less than 1 (generators diverging), and though
$r_0$ typically varies over $S_0$, it never passes through $1$ because of the restrictions 
that have been imposed on the initial data to ensure that the area parameter 
is good on every generator.]

Further information is contained in the map $\phi$ from the coordinates $y_R$ to the 
coordinates $y_L$ on $S_0$. A perturbation of the solution metric, for instance a 
gravitational wave pulse passing through $\cN_R$ but not $\cN_L$, will in general affect which
generator from $S_R$ intersects a given generator from $S_L$, and thus which values 
$y_R$ and of $y_L$ correspond to the same point on $S_0$. The map $\phi$ is thus 
solution dependent.

$\phi$ and the functions $r_0(y)$, $\bar{\rho}(y)$, and $e_{ij}(r,y)$ on each branch determine
the 3-metric on $\cN$ modulo the action of a group $\cal D_\cN^\circ$ of homeomorphisms 
$\cN \rightarrow \cN$ which map each branch diffeomorphically to itself and reduce to the 
identity on $S_R$ and $S_L$:\footnote{
These homeomorphisms could perhaps be called diffeomorphisms since they respect all the natural
differentiable structure of $\cN$. Furthermore, by the Whitney extension theorem the 
homeomorphisms in $\cal D_\cN^\circ$ can always be extended to diffeomorphisms in a spacetime 
neighbourhood of $\cN$ taking each branch of $\cN$ to itself and leaving $S_L$ and $S_R$ 
invariant.} 
The functions $r_{L0}(y_R)$, $r_{L0}(y_L)$ and $\phi$ alone, without further specification of 
the 3-metric on $\cN$, determine the $a_R$ and $a_L$ coordinates on $\cN$ up to the action of 
$\cal D_\cN^\circ$. To see this consider the transformation of the $a$ charts due to a change 
of the metric, within the class of metrics we allow on $\cN$, which preserves
$r_{L0}(y_R)$, $r_{L0}(y_L)$ and $\phi$. (Recall that once the $y$ coordinates are fixed on 
$S_L$ and $S_R$ the $a$ charts are completely determined by the metric, so this type of 
transformation is the only freedom that remains in these charts once $r_{L0}(y_R)$, 
$r_{L0}(y_L)$ and $\phi$ are given.) The displacement of the $a$ coordinate points induced 
on each branch by this change of metric leave $S_L$ and $S_R$ invariant, since the coordinates 
are fixed there. Furthermore, the invariance of $r_0(y)$ on each branch (and the smoothness of 
the metric) implies that these displacements define a diffeomorphism of the branch to itself,
with $S_0$ mapped to $S_0$. Finally, the invariance of $\phi$ implies that the diffeomorphisms
on the two branches induce the {\em same} diffeomorphism on $S_0$, assuring that the 
transformation of the $a_L$ and $a_R$ charts is an element of $\cal D_\cN^\circ$. 

If in addition to $r_{L0}(y_R)$, $r_{L0}(y_L)$ and $\phi$ the non-zero $a$ components of the 
3-metric (i.e. $h_{ij} = r^2\bar{\rho}(y)e_{ij}(r,y)$) are given on each branch, then the 
3-metric is determined up to homeomorphisms in $\cal D_\cN^\circ$. This is the most one could 
expect, since all these quantities are invariant under the action of $\cal D_\cN^\circ$ on the 
3-metric.

In the calculation of the brackets we shall use essentially these functions to represent the 
3-metric part of
the initial data. In particular the conformal 2-metric on $\cN$, specified by the functions 
$e_{R\,ij}(r_R,y_R)$ and $e_{L\,ij}(r_L,y_L)$, will represent the bulk of the initial data. 
The remaining data will ultimately be represented by fields living on $S_0$ - fields equivalent
to $\phi$, $r_{A\,0}$, and $\bar{\rho}_A$, and $\chi$ and $\omega$ corresponding to a 
particular parametrization of the generators.

These quantities are not entirely unconstrained. We require that the induced 3-metric be 
consistently defined at $S_0$ in the sense that the same 2-metric is induced on $S_0$ from the 
3-metrics of either branch. Taking into account the transformation from $y_L$ to $y_R$ 
coordinates this leads to the conditions
\be		\label{rho_consistency}
\bar{\rho}_R r^2_{R\,0} = |det \frac{\di y_L}{\di y_R}| \bar{\rho}_L r^2_{L\,0} 
\ee
and 
\be		\label{e_consistency}
e_{R\,ij} = \frac{\di y^m_L}{\di y^i_R}\frac{\di y^n_L}{\di y^j_R}|det \frac{\di y_R}{\di y_L}| 
e_{L\,mn},
\ee
and of course $e_{ij}$ is required to be symmetric and unimodular.

The scheme of two overlapping adapted charts with solution dependent transformations between 
them is rather complicated. Why not base our adapted coordinates at $S_0$, instead of $S_R$ or
$S_L$? The coordinates $b^\mu = (v^+,v^-,\theta^1,\theta^2)$ define such a chart: $\theta^1$ and 
$\theta^2$ are coordinates on $S_0$ which are convected along the generators to all of $\cN$.
$v^-$ and $v^+$ are area parameters on the generators of $\cN_R$ and $\cN_L$ respectively 
that are normalized to $1$ on $S_0$. $v^+$, $\theta^1$, and $\theta^2$ are extended to 
a spacetime region by convecting them from $\cN_L$ along the congruence of null 
geodesics that are normal to the constant $v^+$ sections of $\cN_L$ and transverse to $\cN_L$, 
and $v^-$ is similarly extended by convecting it in the same way from $\cN_R$.

The $b$ coordinates will be used to represent data living on $S_0$, and to facilitate comparison
with Sachs' initial data. It is tempting to express all the data in terms of these coordinates.
In fact the brackets will ultimately be stated for data referred to this chart.
However we will not use such data to {\em calculate} the brackets. To solve 
(\ref{auxbracketdef}) for the auxiliary pre-Poisson bracket we need to be able to identify
those variations of the metric that vanish (in a fixed chart) in a neighbourhood of $\di\cN$ 
from their actions on the initial data. But even variations that vanish near $\di\cN$ do in 
general deflect the 
generator of $\cN_A$ emerging from a given point of $S_0$, changing its course all the way to 
$S_A$, and thus would produce a variation in the $b$ coordinates at $S_A$.\footnote{
The $b$ coordinates can be held fixed near the end $S_A$ of one branch $\cN_A$ under variations 
of the metric vanishing near the boundary by giving the $\theta$ coordinates fixed values on 
$S_A$ and then convecting them to $S_0$ along the generators, and from there to the other 
branch. The $\theta$ would then be given by a solution independent function of the $y_A$ 
coordinates, or at least by one that is invariant under the variations being considered. 
However, since it is easily shown that the map from the $y_L$ to the $y_R$ coordinates varies 
under variations of the solution, even variations that vanish near $\di\cN$, these $\theta$ 
cannot be invariant at the end of the other branch. Thus there is no choice of $\theta$ which 
makes the $b$ coordinates invariant near all of $\di\cN$ under variations of the metric that 
vanish there.}
Therefore data that are local functions of the metric components and their derivatives in the 
$b$ chart generally vary at $\di\cN$ under such variations, complicating their 
identification. It would be simpler if the variations that vanish near $\di\cN$ also left the 
initial data associated with points near $\di\cN$ invariant.

This is the case for data constructed from the $a$ coordinate components of the metric. The 
variations that 
leave the metric invariant near $\di\cN$ also leave the $a$ coordinate 
components of the metric invariant near $\di\cN$, because such a variation 
leaves the $a$ coordinates themselves invariant in a neighbourhood of 
$\di\cN$: The coordinates $y_A$ are {\em fixed} on $S_A$, so a variation 
$\dg$ that vanishes in a spacetime neighbourhood $W$ of $\di\cN$ clearly leaves the 
generator corresponding to given $y_A^i$ invariant in a neighbourhood of $S_A$.
Furthermore any generator originating within a sufficiently small neighbourhood $U$ of 
$\di S_0$ in $S_0$ remains inside $W$ until it leaves $\cN$ 
(See prop. \ref{generators_near_boundary}). Such 
generators are entirely undisturbed by $\dg$. It follows that there 
exists a neighbourhood $V$ of $\di\cN$ in $\cN$, possibly smaller than 
$W\cap\cN$, such that $\dg a_A = 0$ on $V\cap\cN_A$. The construction of the 
$a_A$ coordinates off $\cN_A$ then implies that these are in fact invariant in
a spacetime neighbourhood of $\di\cN_A\cap\cN_A$. 

The chief condition defining the bracket on initial data is that the equation 
(\ref{auxbracketdef}),
\be	\label{auxbracketdef2}
\dg \varphi = \bar{\Omega}[\{\varphi,\cdot\}_*,\dg],
\ee
holds for all initial data $\varphi$ and all variations $\dg \in L_g^0$ - which is to say all 
$\dg$ satisfying the linearized field equations and vanishing in a spacetime neighbourhood of 
$\di\cN$. Under such variations 
\be	\label{dge_boundary_condition}
	\dg e_{A\,ij} = 0\ \ \ \mbox{in a neighbourhood of}\ \di\cN
\ee
and 
\be	\label{dg_bar_rho}
	\dg\bar{\rho}_A = 0.
\ee 
I shall solve (\ref{auxbracketdef2}) for the auxiliary bracket imposing only the restrictions 
(\ref{dge_boundary_condition}) and (\ref{dg_bar_rho}) on the variations $\dg$. This is 
{\em a priori} a weaker restriction than $\dg \in L_g^0$, which requires that
$\dg$ vanishes in a whole spacetime neighbourhood of $\di\cN$. But certainly if 
(\ref{auxbracketdef2}) holds for our larger class of $\dg$s then {\em a fortiori} 
it holds for $\dg \in L_g^0$.

On $\cN$ the $a$ and $b$ coordinates are related in a fairly simple manner.
On $\cN_R$ the coordinates $b^\mu = (v^+,v^-,\theta^1,\theta^2)$ can be expressed 
in terms of the $a_R$ chart as
\be	\label{b-a_R}
v^+ = 1\ \ \ v^- = r_R/r_{R0}(y_R)\ \ \ \theta^a = f_R^a(y_R),
\ee
where $f_R$ is a diffeomorphism from $\Real^2$ to $\Real^2$.
Similarly on $\cN_L$ the $b^\Lambda$ are related to the $a_L^\ag$ via
\be	\label{b-a_L}
v^+ = r_L/r_{L0}(y_L)\ \ \ v^- = 1\ \ \ \theta^a = f_L(y_L).
\ee
Of course $f_R = f_L \circ \phi$ where $\phi$ is the (solution dependent) diffeomorphism 
from $y_R$ to $y_L$ coordinates on $S_0$. 

One could take $f_L$ to be the identity map (i.e. $\theta = y_L$) and $f_R = \phi$. However it 
will be more convenient not to assume any particular relationship between the $\theta$ and $y$
charts. It will be assumed that the $\theta$ coordinates of a point in $S_0$ depends only
on the $y_L$ and $y_R$ coordinates of that point and possibly on the initial data on the 
generators through the point. That way the $b$ coordinates, like the $a$ coordinates, move 
along with the initial data when these are acted on by a homeomorphism in $\cal D_\cN^\circ$, 
and the components of the data in the $b$ chart are invariant under such homeomorphisms.

The parameters $v^+$ and $v^-$ on the generators of $\cN_L$ and $\cN_R$ respectively define
tangent vectors to these generators:
\bearr
&&n_- = \di_{v^-}\ \ \ \ \mbox{on}\ \cN_R	\\
&&n_+ = \di_{v^+}\ \ \ \ \mbox{on}\ \cN_L
\eearr
and define corresponding initial data $\chi_v$ and $\omega_v$ which, in a solution matching the
initial data, are given by
\bearr
\chi_v & = & n_+\cdot n_-	\label{chi_v_def}	\\
\omega_{v\,a} & = & \frac{n_+\cdot\nabla_a n_- - n_-\cdot\nabla_a n_+}{n_+\cdot n_-}.
\label{omega_v_def}
\eearr

Another datum, which will be used in place of $\bar{\rho}_L$ and $\bar{\rho}_R$, is the area 
density on $S_0$ in $\theta$ coordinates:
\be    \label{rho0_rhoR_rhoL}
\rho_0 = r_{R\,0}^2 \bar{\rho}_{y\,R} |\det[\frac{\di y_R}{\di\theta}]| =
r_{L\,0}^2 \bar{\rho}_{y\,L} |\det[\frac{\di y_L}{\di\theta}]|.
\ee
The two data $\bar{\rho}_L$ and $\bar{\rho}_R$ can be replaced by a single datum because 
of the constraint (\ref{rho_consistency}) between them, which is expressed by the second 
equality of (\ref{rho0_rhoR_rhoL}).

The preceding considerations show that a possible initial 
data set for general relativity consists of

\begin{itemize}

\item $e_{L\,ij}(r_L,y_L)$ on $\cN_L$ and $e_{R\,ij}(r_R,y_R)$ on $\cN_R$, 

\end{itemize}
and on $S_0$
\begin{itemize}

\item $r_{L\,0}(y_L)$ and $r_{R\,0}(y_R)$,

\item the maps $f_L: y_L \rightarrow \theta$ and $f_R: y_R \rightarrow \theta$,

\item $\rho_0(\theta)$,

\item $\chi_v(\theta)$,

\item $\omega_{v\,a}(\theta)$.

\end{itemize}

These will called the {\em boundary fixed coordinate data} (BFC data) since they are referred
to charts that are fixed in the manifold on portions of the boundary $\di\cN$ of the initial 
data hypersurface under the variations of the solution being considered.\footnote{
Since the variations of the metric under consideration do not affect generators originating 
sufficiently close to $\di S_0$ even the $b$ chart is fixed in a sufficiently small 
neighbourhood of $\di S_0$ in $S_0$, which is where the data referred to the $b$ chart live.}
They are subject to the one constraint (\ref{e_consistency}) requiring the consistency of the 
definition of the conformal 2-metric at $S_0$ (and of course $e$ is also required to be 
symmetric and unimodular).

When using the BFC data we shall drop the subscript $v$ on $\chi$ and $\omega$. $\chi$ and 
$\omega$ corresponding to other parametrizations of the generators will have no role to play.
It will also turn out to be convenient to use instead of $f_L$ and $f_R$ their inverses, $s_L$ 
and $s_R$. Finally, $\chi$, will often be represented by $\lam = -\log |\chi|$. The sign of 
$\chi$ is uniform on $S_0$ for the solutions and the $\cN$ we admit, and does not change under 
variations, so taking the absolute value represents no real loss of information.

The domains of sensitivity\footnote{
Recall that we have defined the domain of sensitivity a functional $F$ of the spacetime metric 
to be the support of the functional gradient $\dg F/\dg g_{\mu\nu}$ of $F$. Thus perturbations 
of the metric in this domain, and only in this domain, affect the value of $F$ to first order.}
of all these data are such that the causality condition on the auxiliary pre-Poisson bracket 
implies that data living on different generators have vanishing bracket.
These domains are confined either to only one generator or to the 
pair of generators emerging from a single point of $S_0$: The domain of sensitivity of $e$ at 
a point $p$ on a generator $\cg$ is the segment of $\cg$ from $p$ to $\di\cN$, that of 
$r_{A\,0}$ at a point $p_0 \in S_0$ is the whole generator of $\cN_A$ originating at $p_0$, that
of $\theta^a(y_R) = f^a_R(y_R)$ is by assumption confined to the pair of generators originating 
in the point of $S_0$ specified by $\theta$. The domains of sensitivity of all the remaining 
data at the same values of $\theta^a$ are also contained in the union of these generators.

We shall calculate the brackets between the BFC data, but ultimately the brackets will be 
expressed in terms of a completely free data set consisting of
\begin{itemize}
\item $e_{ab}(v,\theta)$, the conformal 2-metric on constant $v^+$, $v^-$ surfaces in $b$ 
coordinates, specified on all of $\cN$
\end{itemize}
and
\begin{itemize}
\item $r_{L\,0}$, $r_{R\,0}$, $s_L$, $s_R$, $\rho_0$, $\lam$, $\omega_a$ specified on $S_0$.
\end{itemize}

The data $r_{A\,0}$ and $f_A$ (or $s_A$) define the transformation from the $a_A$ chart to the 
$b$ chart on $\cN_A$, thus once these data are given specifying $e_{ab}$ is equivalent to 
specifying $e_{A\,ij}$. Recall that because of the degeneracy of the 3-metric of $\cN$ the 
induced metric at a point $p$ on a two dimensional spacelike cross section of $\cN$ depends 
only on $p \in \cN$, and not on the tangent plane of the cross section. Thus 
\be
e_{ab} = \di_a y_A^i\di_b y_A^j |\det \frac{\di \theta}{\di y_A}|e_{A\,ij}
\ee
in complete analogy with (\ref{e_consistency}).

The only constraint on the BFC data, (\ref{e_consistency}), simply requires that $e_{ab}$ 
obtained from $e_L$, $r_{L\,0}$ and $s_L$ agrees with $e_{ab}$ obtained from $e_R$, $r_{R\,0}$ 
and $s_R$ at $S_0$. It does not limit $e_{ab}$ itself. For this reason we shall refer to this 
new data set as the completely free (CF) data.

The data $r_{A\,0}$, will turn out to play a somewhat secondary role. Their variations can be 
eliminated from the presymplectic 2-form. This seems natural as they essentially specify where 
the generators of the branches of $\cN$ have been cut off, something that we are free
to change without changing the solution.     

\subsection{Variations of fields in the $a$ and $b$ charts} \label{variations}

The notion of variations of fields, i.e. the derivative of the fields along a one parameter 
family of field configurations, is fundamental to the variational principle, and the symplectic
potential and presymplectic 2-form constructed from it. We have defined the variation of 
a field on a manifold in terms of a chart fixed to the manifold: The components of the variation
in the fixed chart are the variations of the corresponding components of the field in this 
chart. The transformation of the components of the variation from one fixed chart to another is 
straightforward since the map from the old to the new chart is field independent. Thus the 
variations of tensor components transform as do the original tensor components, 
defining a tensor field, and the variation of a connection also forms a tensor field.

The expression (\ref{presymp0}) for the presymplectic 2-form therefore has a straightforward 
interpretation if it is supposed that the field components are referred to a chart fixed in the 
manifold. 
However we wish to express the presymplectic 2-form in terms of the BFC data, which consists
of metric and connection components in field dependent coordinate systems. How are variations 
in such a moving chart related to variations in a fixed chart? It turns out that they 
differ by a Lie derivative. 

Consider first the simplest case - the variation of a scalar field $\varphi$. It is clear 
that the variation $\dg^w \varphi$ of the field at a given grid point of a chart $w$, 
is related to its variation $\dgF \varphi$ at the corresponding grid point of a coordinate
system $x^\mu$ that is fixed in the manifold via
\be
\dg^w \varphi = \dgF \varphi + \dg^w x^\mu \di_\mu \varphi.
\ee
The variations of the $x$ coordinates at a $w$ chart grid point defines a vector 
$v = \dg^w x^\mu \di_\mu$ which may be thought of as the variation of the position
of the $w$ grid point in the manifold. Thus
\be
\dg^w \varphi = \dgF\varphi + d_v \varphi.
\ee
A similar relation holds for the variations of tensor fields, such as the metric $g$, in 
the $w$ chart:
\be		\label{w_component_variation}
\dg^w g = \dgF g + {\pounds}_v g.
\ee
Here $\dg^w g$ is the tensor field with $w$ components equal to the variation $\dg$
of the $w$ components of $g$ at fixed values of the $w$ coordinates. That is 
\be
[\dg^w g]_w = \dg [g]_w
\ee
where $[g]_w$ indicates the $w$ components of $g$ expressed as functions of the $w$ 
coordinates. Notice that if $w'$ is a chart which moves together with $w$, i.e. the 
mapping between the charts is field independent, then $\dg^{w'} = \dg^w$ on tensors.
In other words, what is important is not the chart but how it moves when the field
is changed. 

These results generalize to any field $X$ that has well defined components in any $C^\infty$
chart. In the possibly field dependent chart $w$ the components of $\dg^w X$ are just the 
variations of the $w$ components $[X]_w$ of $X$. To define the components of $\dg^w X$ in 
another we impose the requirement that, just as for tensors, the variation $\dg^u X$ in a 
co moving chart $u$ be equal to $\dg^w X$. Specifically, suppose $w_0$ is the $w$ chart
corresponding to the fiducial field configuration we are varying about, and $u_0$ is another 
chart. Then $u_0 \circ w_0^{-1}$ is the (field independent) diffeomorphism that maps the $w_0$ 
chart to the $u_0$ chart, $u = u_0 \circ w_0^{-1} \circ w$ is a chart co-moving with $w$ 
that coincides with $u_0$ at the fiducial field configuration, and 
\be  \label{dgwdef}
[\dg^w X]_{u_0} = [\dg^u X]_{u_0} = \dg [X]_u.
\ee

With this definition $\dg^w = \dgF + {\pounds}_v$ quite generally.\footnote{
The general formalism is presented mostly for the sake of completeness. The reader may find
it easier to deduce the form of $\dg^w X$ for each field $X$ that is needed directly from the 
definition (\ref{dgwdef}). The fields of interest, aside from four dimensional tensors, will
be Christoffel symbols and the BFC data.}
Suppose $\dg$ is the 
tangent $d/d\ag$ to a family of configurations of the gravitational field parametrized by 
$\ag \in \Real$, with $\ag = 0$ at the fiducial solution. Let $X_\ag$ be the corresponding 
family of $X$ fields, $w_\ag$ be the corresponding $w$ charts, $\phi_\ag$ the family of 
diffeomorphisms mapping $w_0$ to $w_\ag$, $v$ the vector field generating $\phi_\ag$, and 
$\phi^0_\ag$ the mapping induced by $\phi_\ag$ on the components of $X$, i.e. 
$[X]_{w_\ag} = \phi^0_\ag[[X]_{w_0}]$. Then
\bearr
[\dg^w X]_{w_0} & \equiv & \frac{d}{d\ag}[X_\ag]_{w_\ag}|_{\ag = 0} \\
& = &\frac{d}{d\ag}\phi^0_\ag[[X_\ag]_{w_0}]|_{\ag = 0} 
= [\dgF X]_{w_0} + \frac{d}{d\ag}\phi^0_\ag[[X_0]_{w_0}]|_{\ag = 0}
= [[\dgF + {\pounds_v}] X]_{w_0}.    
\eearr
The equation $d\, \phi^\circ_\ag [X_0]_{w_0}/d\ag|_{\ag = 0} = 
[{\pounds_v} X]_{w_0}$ is the natural generalization of the definition of the Lie derivative 
for tensors (see \cite{Wald} Appendix C, but note that $\phi^*$ there is the inverse of our 
$\phi^\circ$). It defines $\pounds_v$ to be the generator of the 
action of the family of diffeomorphisms $\phi_\ag$ on $X$.

The relation between the variations of components in differently moving charts is entirely 
analogous to that between variations in fixed and moving charts. For instance the variations 
of $a$ and $b$ components are related by
\be	\label{dgb_a}
\dg^b = \dg^a + {\pounds}_\xi,
\ee
where $\xi$ is the difference of the $v$ fields associated with the $a$ and $b$ charts. $\xi$
can be evaluated in the $a$ chart by applying (\ref{dgb_a}) to the $a$ coordinates themselves:
\be
\dg^b a^\ag = 0 + {\pounds}_\xi a^\ag = \xi^\ag,
\ee
and in the $b$ chart by applying it to the $b$ coordinates:
\be
\xi^\mu = - \dg^a b^\mu.
\ee

To lighten the notation the superscript indicating the chart relative to which a 
variation is taken will often be dropped. When no chart is indicated the variation of a field 
is to be taken with respect to its ``natural chart''. The natural chart of a field component
(of a tensor or any other multi-component field) is the chart that the component is referred to.
For instance $\dg g_{\mu\nu}$, with $g_{\mu\nu}$ the components of the metric in a fixed chart, 
is $\dgF g_{\mu\nu}$. The BFC data, and the variants we have discussed, each have a 
corresponding natural chart: $e_{A\,ij}$, $r_{A\,0}$, and $f_A^a$ all have natural chart $a_A$.
$\rho_0$, $\chi$, $\omega_a$, and $s_A^i$ have natural chart $b$. Note that $s_A^i$ is not
an $a_A$ vector components, but a scalar function of the $\theta^a$. Similarly $f_A^a$ is
a scalar function of the $y_A$. The natural chart of $e_{ab}$ is of course $b$. Finally, we 
adopt the convention that the natural chart of a coordinate is the chart it forms part of.
Thus $\dg a^\ag \equiv \dg^a a^\ag = 0$ and $\dg b^\Lambda \equiv \dg^b b^\Lambda = 0$.

Although the $b$ chart is clearly a moving chart, being adapted to the generators of 
$\cN$, one may, when solving (\ref{auxbracketdef}) for the auxiliary pre-Poisson bracket,
restrict attention to variations of the solution that leave the $b^\mu$ fixed in a neighbourhood
of $S_0$, where the data referred to the $b$ chart live:
(\ref{auxbracketdef}) requires that for any initial datum $\varphi$ on the interior of $\cN$
\be		\label{poisson_symplectic3}
\dg \theta = \bar{\Omega}[\{\varphi,\cdot\}_*,\dg]
\ee
for all $\dg$ which satisfy the linearized field equations and vanish in a neighbourhood of 
$\di\cN$. The right side of this equation is unchanged by the addition to either argument 
of $\bar{\Omega}$ of a diffeomorphism generator $\pounds_\tau$ vanishing near $\di\cN$ since 
such a generator is a degeneracy vector of $\bar{\Omega}$. Moreover, if $\varphi$ is a BFC 
datum the left side is also invariant under the addition of such 
a term to $\dg$. Now, if $\varphi$ is a BFC datum living at a point $p \in int\cN$ then 
$\{\varphi,\cdot\}_*$ acts only on the data on the generator through $p$, thus leaving the 
$b^\mu$ invariant in a neighbourhood of $\di\cN - S_L - S_R$. As explained in subsection 
\ref{coordinates}, $\dg$ also leaves the $b^\mu$ invariant in a neighbourhood of 
$\di\cN - S_L - S_R$. The (linearized) diffeomorphisms induced by $\{\varphi,\cdot\}_*$ and 
$\dg$ can thus be cancelled near $S_0$ by diffeomorphism generators that vanish in a 
neighbourhood of $\di\cN$.

Note that the pair of charts $a_R$ and $a_L$ cannot both be held fixed on $S_0$ in the same 
manner as the $b$ chart because the transformation from the $a_R$ chart to the $a_L$ chart 
depends on the fields.

\subsection{Freeness and completeness of the BFC and CF data}	\label{freeness}

To show that the BFC data of section \ref{free_data_coords} is free 
(apart from the constraint (\ref{e_consistency})) and complete it is sufficient
to show that any BFC data determines a set of Sachs data such that a solution matches the 
CF data iff it matches the Sachs data. Since we assume that any Sachs data determines a unique 
solution, modulo diffeomorphisms, the same would also be true for the BFC data.
Furthermore, since the CF data are equivalent to the BFC data, this would establish that the 
CF data are also complete and completely free. 

The BFC data, the CF data, and the Sachs data are all explicit coordinate representations of 
the geometrical initial data defined in subsection \ref{initial_data} - the 3-metric on $\cN$ 
and $\chi$ and $\omega$ on $S_0$ - but referred to different coordinate systems. The BFC data is

in part referred to the $b$ chart and in part to the $a$ charts, the CF data refers to the $b$ 
chart, and the Sachs data refers to essentially the $b$ chart, but with affine parameters 
of the generators in the solution spacetime replacing the area parameters $v^\pm$.\footnote{
Sachs also uses a special chart, adapted to the geometry, on $S_0$ - essentially the $\psi$ 
chart defined in subsection \ref{coordinates}. But this specialization of the chart on $S_0$
plays no role in the results of Sachs and Rendall on the existence and uniqueness of 
solutions, so we may as well use the $\theta$ chart on $S_0$ to define Sachs' data.}
The key point to be demonstrated is thus that the different data sets contain sufficient 
information to define the map to the coordinate system of the other data sets. 

The BFC and the CF data each define the transformation (\ref{b-a_R},\ref{b-a_L}) from the $a$ 
chart on a branch of $\cN$ to the $b$ chart on that branch. In particular they define the map 
from the coordinates $y^i$ on each branch to the coordinates $\theta^a$. This allows one to 
compute the CF data corresponding to a spacetime metric from the BFC data, and {\em vice versa},
establishing the equivalence of the these two data sets.

It remains to show that the BFC data define the map from the $b$ chart to Sachs' coordinates,
in particular the map from the area parameter $r$ to an affine parameter $\eta$ 
(and thus also from $v = r/r_0$ to $\eta$). In a non degenerate Lorentzian solution matching 
the BFC data the generators of $\cN$ are geodesics: The degeneracy vectors of the degenerate 
3-metric defined by the BFC data are orthogonal to all tangents to $\cN$ and null, so $\cN$ is 
a null hypersurface 
and the degeneracy vectors are its normals - from which it follows that the integral curves of 
the degeneracy vectors are null geodesics (see prop. \ref{null_normals_geodesics} of appendix 
\ref{nullhypersurfaces}). Now if $\eta$ is an affine parameter of a generator then the 
corresponding tangent vector $k = \di_\eta$ is parallel transported along the generator: 
$\nabla_k k = 0$. In $a$ coordinates this equation is equivalent to
\be  \label{Gamma_rrr_eq}
    \Gamma^r_{rr} = \frac{d}{dr} \log \frac{d\eta}{dr}.
\ee
But on $\cN$ the field equation $G_{rr} = 0$ is equivalent to
\be     \label{Grr}
    \Gamma^r_{rr} = -\frac{r}{8}\di_r e^{ij} \di_r e_{ij}.
\ee
(see Appendix \ref{Grrzero}.)
Combining these equations one obtains the relation
\be		\label{affine_r}
\frac{d}{dr} \log \frac{d\eta}{dr} = -\frac{r}{8}\di_r e^{ij} \di_r e_{ij}, 
\ee
which may be integrated using BFC data to obtain $\eta$ as a function of $r$ up to an additive 
constant and a constant factor on each generator in concordance with the freedom to change 
affine parameters by affine transformations. (The ``constants'' are independent of $r$, but 
may depend on $y$.) Of course $\eta$ obtained in this way is not actually 
an affine parameter. Only when the field equation $G_{rr} = 0$ holds does (\ref{affine_r}) 
imply (\ref{Gamma_rrr_eq}), making $\eta$ is an affine parameter of the spacetime geometry. 

Sachs sets the affine parameters, $\eta_L$ and $\eta_R$, of the two congruences of generators to
zero on $S_0$. To partly fix the normalization of the affine parameters he also demands that 
$\di_{\eta_L} \cdot \di_{\eta_R} = -1$ on $S_0$. The remaining freedom in the normalization of 
the generators can be fixed in our case by requiring 
\be	\label{normalization_eta_R}
	d\eta_R/dv^- = 1
\ee
on $S_0$. Then, by (\ref{chi_v_def}),
\be	\label{normalization_eta_L}
	d\eta_L/dv^+ = -\chi_v.
\ee

The BFC data determine the $a$ coordinate components of the 3-metric on $\cN$, and also, 
as we have just seen, the transformation to Sachs' coordinates on $\cN$. They therefore 
determine the 3-metric of $\cN$ in Sachs' coordinates. This determines most of the Sachs data. 

The Sachs data consists of the conformal 2-metric, $e_{ab}(\eta,\theta)$, as a function of
Sachs' coordinates on each branch of $\cN$ and four fields on $S_0$, referred to the  
coordinates $\theta^a$: the area density $\rho(\theta)$ and its derivatives $\di_{\eta_L} \rho$ 
and $\di_{\eta_R} \rho$, and $\omega_\eta$, the normalized twist $\omega$ corresponding to the 
parametrizations $\eta_L$ and $\eta_R$ of the generators.\footnote{
Sachs actually takes as his final data a pair of quantities he writes as $C_{A,1}\ \ \ A = 1,2$.
However these are just the $\theta$ coordinate components of the 1-form $-\omega_\eta$. This can
be seen most easily from his equation (19), which when a forgotten factor of $1/2$ is restored 
and it is rewritten in our notation reads
\be
\frac{1}{2} C_{a,1} = \di_{\eta_L} \cdot \nabla_a \di_{\eta_R}.
\ee
This equation, the normalization $\di_{\eta_L} \cdot \di_{\eta_R} = -1$ on $S_0$, and the 
expression (\ref{omegainterp}) imply the claim.}
All but the last of these are subsumed
in the 3-metric of $\cN$.\footnote{
The explicit relations of these data with the BFC data are not complicated. Since the 3-metric
on $\cN$ is degenerate the induced 2-metric at a point on a constant $\eta$ section of $\cN$
is the same as the induced 2-metric on the constant $r$ section at the same point. Thus the 
$e_{ab}$ at a point are determined by the $e_{ij}$ at that point and the standard tensor density
transformation associated with the transformation from $y$ to $\theta$ coordinates. 
$\rho(\theta)$ is both a Sachs and BFC datum, and the derivatives of $\rho$ are fixed by the 
normalization conditions (\ref{normalization_eta_L}) and (\ref{normalization_eta_R}) on the 
$\eta$ parameters:  
\bearr
\di_{\eta_L} \rho & = & 2 \rho\, dv^+/d\eta_L = - 2 \rho/\chi_v \\
\di_{\eta_R} \rho & = & 2 \rho\, dv^-/d\eta_R = 2 \rho.
\eearr}
\footnote{
In the coordinate independent description of the data in subsection
\ref{initial_data} a further datum, $\chi$, is specified on $S_0$, but this is 
identically equal to $-1$ for the parametrization $\eta_L$, $\eta_R$ of the generators.}

The final Sachs datum, $\omega_\eta$, is related to, $\omega_v$, the normalized twist 
corresponding to the parametrizations $v^+$ and $v^-$ of the generators via
\be
\omega_{\eta\,a}(\theta) = \omega_{v\,a}(\theta) - \di_a \log \chi_v(\theta).
\ee
(See (\ref{omegatransform}), (\ref{normalization_eta_L}) and 
(\ref{normalization_eta_R}).)

Any solution matching the BFC data necessarily also matches these Sachs data. Since the solution
is uniquely determined by the Sachs data it is uniquely determined by the BFC data.\footnote{
It has been proved that the Sachs data determine the spacetime metric uniquely up to 
diffeomorphisms in a neighbourhood of $S_0$ in $\Dund$ \cite{Rendall}, and we are assuming 
that they in fact determine a maximal Cauchy development of $\cN$. This implies that 
the components of the metric in Sachs' spacetime coordinates are uniquely determined. 
Sachs' spacetime coordinates are obtained from the $b$ coordinates by replacing the 
coordinates $v^\pm$ by $\eta_L$ and $\eta_R$, everywhere given by the same functions 
of $v^+$ and $v^-$ respectively as on $\cN$. Thus the Sachs data define the solution 
in $b$ coordinates uniquely as well.}
Thus the BFC data are complete. 

The BFC data are also free, aside from the one constraint (\ref{e_consistency}). This follows 
from the fact that, firstly, any BFC data that satisfies (\ref{e_consistency}) defines Sachs 
data, and secondly, that the solution corresponding to this Sachs data really matches all the 
original BFC data.
To verify the second claim consider first the area parameter $r'$, defined like $r$ but
from the 3-metric induced on $\cN$ by the solution. $r'$, like $r$, must satisfy 
(\ref{affine_r}) which may be re-expressed as
\be		\label{affine_r2}
\frac{d^2 r}{d\eta^2} = r \kappa(\eta), 
\ee
with $\kappa = \frac{1}{8}\di_\eta e^{ij} \di_\eta e_{ij}$ a function of $\eta$ that is 
determined by the Sachs data and is thus the same for the original 3-metric as for the 
3-metric induced by the solution. The solution to this linear second order equation is 
completely determined by the value of $d\log r/d\eta$ at $S_0$ and the value of $r$ at the other
end of the generator. At $S_A$ $r'_A = r_A = 1$ by definition, and at $S_0$ $d\log r_A/d\eta_A
= d\log r'_A/d\eta_A = 1/2 d\log \rho/d\eta_A$ since this is part of the Sachs data defined by 
the original BFC data. Thus $r'_A = r_A$, and, since $r(\eta)$ and the Sachs data determine
the 3-metric completely (without use of the field equations), the 3-metric induced from the
solution is equal to the original 3-metric. The 3-metric BFC data induced by the solution 
are thus the same as the original ones. Since $v' \equiv r'/r'_0 = v$ it is also clear that
the remaining BFC data, $\chi_v$ and $\omega_v$, induced by the solution are equal to the
ones originally specified.  

\section{The presymplectic 2-form}\label{pre_symplectic_2form}

\subsection{Preliminaries}\label{preliminaries}

The auxiliary pre-Poisson bracket $\{\cdot,\cdot\}_*$ on functions of the initial data is 
defined by the requirements that it should reproduce the Peierls bracket on observables, be 
antisymmetric, that it be causal in the sense that the bracket vanishes
between data at points that cannot be connected by any causal curve, and that it respects
any constraints that the data must satisfy. As shown in section 
\ref{covariant_canonical_theory}, $\{\cdot,\cdot\}_*$ reproduces the Peierls bracket on 
observables iff for any initial datum $\theta$ on $int \cN$
\be		\label{poisson_symplectic2}
\dg_0 \theta = \bar{\Omega}[\{\theta,\cdot\}_*,\dg_0]
\ee
for all $\dg_0$ which vanish in a neighbourhood of $\di\cN$. 

In this and the following section $\bar{\Omega}[\dg_1,\dg_2]$ will be evaluated explicitly in 
terms of the BFC initial data defined in subsection \ref{coordinates} for variations $\dg_1$ and
$\dg_2$ in a class that is wide enough to include all $\dg_0$ allowed in 
(\ref{poisson_symplectic2}) and 
a solution $\{\theta,\cdot\}_*$ to (\ref{poisson_symplectic2}). Specifically 
$\bar{\Omega}[\dg_1,\dg_2]$ will be calculated for $\dg_1$ and $\dg_2$ satisfying the following 
conditions on each branch $\cN_A$ of the initial data hypersurface:  
\begin{itemize}

\item[a)] The variation $v_A^\ag = -\dgF a_A^\ag$ of the adapted coordinates $a_A^\ag$ 
corresponding to $\cN_A$ vanishes in a {\em neighbourhood} of $\di\cN \cap \cN_A$.

\item[b)] The variation $\dg \bar{\rho}_A$ of the area density $\bar{\rho}_A$
on the truncating 2-surface $S_A$ (where the generators of $\cN_A$ cross the boundary $\di\cN$) 
vanishes.

\end{itemize}

These conditions are clearly satisfied by the variations $\dg_0$ in (\ref{poisson_symplectic2}) 
because they vanish near $\di\cN$. On the other hand, the variation $\{\theta,\cdot\}_*$ 
generated by an initial datum $\theta$ is not in general expected to vanish on all of $\di\cN$. 
However, recall that (\ref{poisson_symplectic2}) determines $\{\theta,\cdot\}_*$ only up to the 
addition of an arbitrary diffeomorphism generator ${\pounds}_\eta$ (a generalized degeneracy 
vector of $\bar{\Omega}$). It turns out that for any bracket satisfying 
(\ref{poisson_symplectic2}), and the causality condition, there exist suitable diffeomorphism 
generators such that the modified bracket obtained by adding these to the original bracket 
satisfies a) and b).

Let us verify this in detail. The requirement that the auxiliary pre-Poisson bracket be causal 
implies that the variation $\{\phi(p),\cdot\}_*$ generated by the value of an initial data field
$\phi$ at the point $p \in \cN$ can be non-zero only on the generators passing through $p$. 
Since all our initial data live on the interior of $\cN$ this implies that the variation 
generated by any initial datum vanishes in some neighbourhood of $\di\cN - S_R - S_L$. But it 
would seem that the conditions a) and b) could be violated at some points in the interior of 
$S_L$ and/or $S_R$.
 
In order to work with smooth variations let us smear the initial data fields with smooth test 
functions. Thus, instead of taking $\theta$ to be the value of an initial data field at an 
interior point of $\cN$ we take $\theta$ to be the integral, weighted by a smooth test 
function with support contained in $int \cN$, of such an initial data field. Imposing 
(\ref{poisson_symplectic2}) for all such test functions implies that it holds distributionally 
for the initial data themselves, which is all we will ultimately require.

Now suppose that $\{\theta,\cdot\}'_*$ is a solution to (\ref{poisson_symplectic2}). 
By hypothesis, the area density $\rho$ transverse to a generator always increases or 
decreases along the generator - it is never stationary - in the spacetime geometry and 
for the $\cN$ we are considering. Therefore if an infinitesimal variation 
of the geometry changes the area density at a point this change can always be cancelled by a 
suitable infinitesimal displacement along the generator. By adding the diffeomorphism generator 
${\pounds}_\eta$ to $\{\theta,\cdot\}'_*$ with $\eta$ taken parallel to the generators and with 

magnitude such that $d_\eta \rho = - \{\theta,\rho\}'_*$ on $S_L$ and $S_R$ one may ensure that 
the area density on these surfaces is invariant under the combined variation. That is, condition
b) is satisfied.

What about condition a)? The modified action $\{\theta,\cdot\}'_* + {\pounds}_\eta$ of $\theta$ 
will in general affect the generator passing through a given (interior) point of $S_A$. Since 
the generator is identified by its point of intersection with $S_A$ this point remains fixed by 
definition, but all other points on the generator will in general be displaced, inducing 
non-zero changes in the adapted coordinates $a^\ag$ arbitrarily close to $S_A$ in violation of 
requirement a). These changes in the $a^\ag$ can clearly be undone in neighbourhoods of $S_L$ 
and $S_R$ by adding a second diffeomorphism generator, ${\pounds}_{\eta'}$, with $\eta'$ chosen 
to vanish on $S_L$ and $S_R$ so it does not affect $\bar{\rho}$ and b) remains satisfied. 
Furthermore, the original variation of the generators must vanish outside the causal domain of 
influence of the support of the smearing function of $\theta$, and thus in a neighbourhood of 
$\di\cN - S_L - S_R$, which implies 
that the action of $\theta$, with the two added diffeomorphism generators, leaves the adapted 
coordinates invariant in a neighbourhood of all of $\di\cN$. That is, $\{\theta,\cdot\}'_* 
+ {\pounds}_\eta + {\pounds}_{\eta'}$ is a solution to (\ref{poisson_symplectic2}) satisfying 
conditions a) and b) above.
We may therefore solve (\ref{poisson_symplectic2}) for $\{\theta,\cdot\}_*$ assuming a) and b) 
hold (and then recover all the remaining solutions by adding diffeomorphism generators to the 
first solution). 

One consequence of condition b) is worth noting. $r_{0\,L}$, $r_{0\,R}$, and $\rho_0$ are 
{\em a priori} independent BFC data but, because $\bar{\rho}_A(y) = \rho_0(f_A(y)) r_0(y)^{-2}
\det[\di_i f_A^a]$ is invariant under the class of variations being considered, the variations 
$\dg r_{A\,0}$ may always be reduced to variations of other BFC data. We shall do this and 
variations of $r_{0\,A}$ will not appear in the expression for the presymplectic 2-form finally 
used to solve for the auxiliary bracket between the data. This does not mean that the brackets 
of the $r_{0\,A}$ are undetermined. Since the  variation $\{\theta,\cdot\}_*$ satisfies b) by
the very argument we have just given $\{\theta,r_{A\,0}\}_*$ can be expressed in terms of the
the brackets of $\theta$ with other data. It is not difficult to do so explicitly and the reader
may do so using the brackets between the remaining BFC data given further on.

In the next subsection the symplectic potential $\Theta[\dg]$ will be evaluated
in terms of the BFC data for $\dg$ satisfying conditions a) and b).
Then, in the following subsection, the expression for $\Theta$ obtained will be used to compute 
$\bar{\Omega}$ in terms of the same data for variations satisfying a) and b). The solution of 
the resulting explicit form of equation (\ref{poisson_symplectic2}), which yields the auxiliary 
pre-Poisson brackets between the initial data, is left to section \ref{Poisson_bracket}.

\subsection{Calculation of the symplectic potential in terms of the almost free BFC initial 
data}\label{potential_calc}

We shall calculate explicitly only the part of the symplectic potential corresponding to the 
branch $\cN_R$, denoted $\Theta_R$. Since the two branches of $\cN$ enter the problem on an 
exactly equal footing, $\cN_L$ contributes an entirely analogous expression. 

$\Theta_R$ is just the contribution of $\cN_R$ to the boundary term (\ref{phiEH}) in the 
variation of the action integral (on a region of which $\cN_R$ forms part of the boundary). 
That is
\be             \label{Theta+}
\Theta_R[\dg] = -\frac{1}{8\pi G}\int_{\cN_R} \dg\Gamma^{[\sg}_{\sg\nu} g^{\mu]\nu}
                                \sqrt{-g}\:d\Sigma_\mu,
\ee
in terms of field components and their variations referred to manifold fixed coordinates.

Recall that the BFC data on $\cN_R$ consist of bulk data, namely the conformal 2-metric 
components in the $a$ coordinates, specified on all of $\cN_R$, and surface data associated with
the 2-surface $S_0$. 
The contributions to $\Theta_R[\dg]$ of the variations of the bulk data $e_{ij}$ and of the 
surface data on $S_0$ can be neatly separated by expressing the variation of the metric as 
the sum of a contribution due to the variation of its components with respect to the $a$ 
coordinates corresponding to to the $R$ branch, and another due to the variation of these $a$ 
coordinates themselves (with respect to fixed coordinates).

That is, we decompose $\Theta_R$ according to
\be             \label{decompTheta}
\Theta_R[\dg] = \Theta_R[\dg^a] - \Theta_R[{\pounds_v}],
\ee
where $v$ is the vector field $-\dgF a^\ag \di_\ag$. The first term can be written immediately 
in terms of $a$ coordinate components, and reduces to an integral over $\cN_R$ of a functional 
of $e_{ij}$ and its variation. The second term turns out to be a surface integral over $S_0$ 
and depends on the data on $S_0$ and their variations (as well as on the derivative of $e_{ij}$ 
along the generators at $S_0$).

In direct analogy with (\ref{diffTheta1}) 
\be
\Theta_R[{\pounds_v}] = -\frac{1}{16\pi G}\int_{\di\cN_R} \nabla^\mu v^\nu \sqrt{-g}\, 
d\Sg_{\mu\nu}.
\ee
By condition a) on the variation we are considering $v$ vanishes in a neighbourhood of 
$\di \cN_R - S_0 = \di \cN \cap \cN_R$, so 
\be             \label{S0initial}
\Theta_R[{\pounds}_v] = \frac{1}{16\pi G}\int_{S_0} \nabla^\mu v^\nu \sqrt{-g}\, 
d\Sg_{\mu\nu}.
\ee
(The sign reflects the orientation of $S_0$ which is taken to be opposite to that of 
$\di\cN_R$.)

As discussed in subsection \ref{variations} one may restrict attention to variations such that 
the $b$ chart is fixed in the manifold in a neighbourhood of $S_0$ in $\cN$. Then $v = - \xi$ on
$S_0$, where $\xi$ is defined by $\dg^{a} = \dg^b - {\pounds}_{\xi}$. That the substitution
of $v$ by $-\xi$ does not affect the symplectic potential may be verified directly using 
(\ref{S0initial}): The entire symplectic potential, obtained by summing the contributions of the
two branches of $\cN$, is
\be             
\Theta[\dg] = \Theta_R[\dg^{a_R}] + \Theta_L[\dg^{a_L}] + \frac{1}{16\pi G}\int_{S_0} 
\nabla^\mu [v_L - v_R]^\nu \sqrt{-g}\, d\Sg_{\mu\nu}.
\ee
As in (\ref{S0initial}) the orientation of $S_0$ is opposite to that of $\di\cN_R$, and thus 
equal to that of $\di\cN_L$. Now since $\dg^b = \dg + {\pounds}_{v_b}$ and
$\dg^{a_A} = \dg + {\pounds}_{v_A}$, $\xi_A = - v_A + v_b$. That is, $\xi_A$ 
differs from $-v_A$ by a vector field $v_b$ that does not depend on the branch $A$. It follows
that $v_L - v_R = - (\xi_L - \xi_R)$. We will therefore take 
\be
\Theta_R[\dg] = \Theta_R[\dg^a] + \Theta_R[{\pounds}_\xi] = \Theta_R[\dg^a]
+\frac{1}{16\pi G}\int_{S_0} \nabla^\mu \xi^\nu \sqrt{-g}\, d\Sg_{\mu\nu}.
\ee
as the contribution of $\cN_R$ to $\Theta[\dg]$. We will also use the conceptual crutch 
of supposing the $b$ chart to be fixed near $S_0$ under the variations being considered, so
$\dg^b = \dg$ on fields there. This condition does not contradict conditions 
a) and b) on the variations, since these conditions refer only to what happens near $S_L$ 
and $S_R$, whereas the diffeomorphism generator that must be added to the variations to keep 
the $b$ chart fixed near $S_0$ may be chosen to have support in a closed 
neighbourhood\footnote{The closure of an open neighbourhood of $S_0$.} of $S_0$ excluding $S_L$ 
and $S_R$.

In terms of the $b$ coordinates the pullback to $S_0$ of $\sqrt{-g}\: d\Sg_{\Lambda\Upsilon}$ is
\bearr
\sqrt{-g}\: \ppback{d\Sg}_{\Lambda\Upsilon} & \equiv & \frac{1}{2}\sqrt{-g}\,
\vareg_{\Lambda\Upsilon\Xi\Pi}\:
\ppback{db}^\Xi\wedge \ppback{db}^\Pi \\
& = & - 2\, \rho_0 \chi\: \di_{[\Lambda} v^+\,\di_{\Upsilon]} v^-\,\ppback{d\theta}^1\wedge 
\ppback{d\theta}^2.
\eearr
Since $v^+$ and $v^-$ are constant on $\cN_R$ and $\cN_L$ respectively $dv^+$ and $dv^-$ must be
proportional to the normals $n_-$ and $n_+$ respectively.\footnote{
Recall that the tangents to the generators $n_- \equiv \di_{v^-}$ are normal to $\cN$
by prop. \ref{tangent_is_normal}.}
In fact, since 
$d_{n_+} v^+ = d_{n_-} v^- = 1$ it follows that $n_\pm = \chi dv^\mp$. Hence
\be
\sqrt{-g}\: \ppback{d\Sg}_{\Lambda\Upsilon} = -\frac{2}{\chi}\: n_{-\,[\Lambda}\,
n_{+\,\Upsilon]}\: \eg
\ee
where $\eg = \rho_0\, d\theta^1 \wedge d\theta^2$ is the area 2-form on $S_0$.

Substituting this form into (\ref{S0initial}) yields
\be     \label{ThetaR0}
\Theta_R[{\pounds_\xi}] = \frac{1}{16\pi G}\int_{S_0} \eg\, \frac{1}{\chi} 
\{n_- \cdot\nabla_{n_+} \xi - n_+ \cdot \nabla_{n_-} \xi\}. 
\ee
The derivative along $n_+$ may be eliminated in favour of a variation of $\chi$:
\bearr
\dg\chi = \dg g_{+-} & = & [\dg^a g]_{+-} + {\pounds_\xi} g_{+-} \\
& = & \di_{v^+} a^\ag \di_{v^-} a^\bg \dg g_{\ag\bg} + 2 n_+^\mu n_-^\nu \nabla_{(\mu} 
\xi_{\nu)}.
\eearr
But $\di_{v^-} a^\bg = r_0 \dg_r^\bg$, and (by (\ref{a_line_element})) $g_{\ag r} = -\dg^u_\ag$,
which is constant, so $[\dg^a g]_{+-} = 0$ and
\be
\dg\chi = n_+\cdot\nabla_{n_-}\xi + n_-\cdot\nabla_{n_+}\xi. 
\ee
The integrand of (\ref{ThetaR0}) is thus equal to
\be             \label{integrandThetaR0}
\eg\, \frac{1}{\chi}[ \dg\chi - 2 n_+\cdot\nabla_{n_-}\xi ].
\ee
Under the variations being considered $u$ remains zero on $\cN_R$.\footnote{
In the definition of the space of solutions $\cS$ and the space of solutions to the 
linearized field equations it is required that $\cN$ always remains null.}
$\xi$ is thus tangent to $\cN_R$ and can be expanded as
\be
\xi = \xi_\perp + \xi^- n_-,
\ee
where $\xi^-$ is the $v^-$ component of $\xi$ and $\xi_\perp$ is tangent to the $v^- = constant$
cross sections of $\cN_R$. 


$\xi_\perp$ generates the variations of the $y$ coordinates at fixed $\theta$ on the 
$v^- = constant$ surfaces induced by the variation of the metric. That is, 
${\pounds_{\xi_\perp}} y^i = \dg^b y^i$. 
Now by definition the curves $y^i = constant$ and $\theta^a = constant$, always coincide with 
the generators, a fact which the displacement of the $a$ chart with respect to the $b$ chart 
induced by a variation of the metric must respect. That is, the constant $y^i$ curves must 
always remain tangent to the vector field $\di_{v^-} = n_-$. Thus 
${\pounds_{n_-}} y^i = \di_{v^-} y^i = 0$ and ${\pounds_{n_-}} \dg^b y^i = \dg^b \di_{v^-} y^i 
= 0$. It follows that
\be
{\pounds_{[n_-,\xi_\perp]}} y^i = [{\pounds_{n_-}}{\pounds_{\xi_\perp}} 
- {\pounds_{\xi_\perp}}{\pounds_{n_-}}] y^i = 0.
\ee
Clearly also ${\pounds_{[n_-,\xi_\perp]}} v^\pm = 0$, so
\be
[n_-,\xi_\perp] \equiv {\pounds}_{n_-}\xi_\perp = 0.
\ee

The second term in (\ref{integrandThetaR0}) may therefore be expanded as
\be     \label{integrandTheta1}
2 n_+ \cdot \nabla_{n_-} \xi = 2 n_+ \cdot \nabla_{n_-}[\xi_\perp + \xi^- n_-] 
= 2\chi d_{n_-} \xi^- + 2 \xi^- n_+ \cdot \nabla_{n_-} n_- + 2 n_+\cdot \nabla_{\xi_\perp} n_-
\ee
Let us consider each term in turn, beginning with the last. From the definition 
(\ref{omega_v_def}) of $\omega$ 
\be     \label{third_term}
2 n_+\cdot \nabla_{\xi_\perp} n_- = d_{\xi_\perp} \chi + \chi \xi_\perp^a \omega_a.
\ee
The middle term is a little more subtle. Since $n_-$ is tangent to a geodesic and
$n_- = \di_{v^-} = r_0\di_r$
\be
n_+\cdot\nabla_{n_-} n_- = n_+ \cdot r_0^2 \nabla_{\di_r} \di_r = n_+\cdot r_0 \Gamma^r_{rr} n_-
= \chi r_0 \Gamma^r_{rr}.
\ee
$\Gamma^r_{rr}$ can be expressed in terms of the BFC data by means of one of the field equation
(we are working in the space of solutions) - see (\ref{Grr}) and Appendix \ref{Grrzero}.
However this will not be necessary since the term we are considering, together with another
that we will find further on, constitute a total variation which may be subtracted off from 
$\Theta$ without altering the presymplectic 2-form.


%

It remains to express $\xi^-$ and $d_{n_-}\xi^-$ in terms of the BFC data and their 
variations. $\xi^-$ turns out to measure the variation of $r_0(y)$. Define
\be
\dg^y = \dg^b - \pounds_{\xi_\perp}.
\ee 
(Since $\dg^b y^i = \xi_\perp^i$ this is the variation associated with the hybrid moving 
coordinate system $(v^+, v^-, y^1, y^2)$.) On $\cN_R$ the variation of $r$ at fixed $v^-$ and 
$y^i$ is then
\be
\dg^y r \equiv \dg^b r - \pounds_{\xi_\perp} r = \dg^a r + \pounds_{\xi^- n_-} r 
= \xi^- \di_{v^-} r = \xi^- r_0.
\ee
Hence $\xi^- = \dg^y r/r_0 = v^- \dg^y \log r_0$. 
On $S_0$ therefore
\be
\xi^- = d_{n_-}\xi^- = \dg^y \log r_0 = \frac{1}{2}\dg^y \log\rho_y,
\ee
or equivalently
\be             \label{xi_minus-eval}
\eg\xi^- = \eg d_{n_-}\xi^- = \frac{1}{2}\dg^y \eg,
\ee
where $\dg^y \bar{\rho} = 0$, which follows from condition b) on $\dg$, has been used. 

Substituting all these results into (\ref{ThetaR0}) we find
\be
\Theta_R[{\pounds_\xi}] = -\frac{1}{16\pi G}\int_{S_0} \eg (\dg^y \lam + \xi^a_{\perp} \omega_a)
+ \dg^y \eg (1 + r_0\Gamma^r_{rr}),
\label{ThetaRsurf}
\ee
where $\lam = - \log |\chi|$.

Now let us turn to the first term in (\ref{decompTheta}), $\Theta_R[\dg^a]$. In working with 
this quantity it is convenient to use the $a$ coordinates, because in these coordinates 
the $a$ variation, $\dg^a$, of a field reduces to simply the variation, at fixed $a^\ag$, of 
the $a$ coordinate components of the field. For instance
\be
[\dg^a\Gamma]^\ag_{\bg\cg}(a) = \dg\Gamma^\ag_{\bg\cg}(a)
\ee
Thus we may write 
\be             \label{Theta_bulk}
\Theta_R[\dg^a] = -\frac{1}{8\pi G}\int_{\cN_R} \dg\Gamma^{[\ag}_{\ag\cg} g^{\bg]\cg}
                                \sqrt{-g}\:d\Sigma_\bg.
\ee

In $a$ coordinates the components of the metric on $\cN_R$ are (by (\ref{a_line_element}))
        \be
\begin{array}{cc} g_{\ag\bg} = \left[\begin{array}{cc} \begin{array}{rr} 0 & -1 \\ -1 & 0 
\end{array} & \begin{array}{rr} {} & {} \\ {} & {} \end{array}\\ 
\begin{array}{rr} {} & {} \\ {} & {} \end{array} & h_{ij} \end{array}\right],     &
                  g^{\ag\bg} = \left[\begin{array}{cc} \begin{array}{rr} 0 & -1 \\ -1 & 0 
\end{array} & \begin{array}{rr} {} & {} \\ {} & {} \end{array}\\ 
\begin{array}{rr} {} & {} \\ {} & {} \end{array} & h^{ij} \end{array}\right],     \end{array}
        \ee        
with $h^{ij}$ the inverse of $h_{ij}$. Thus $\sqrt{-g} = \sqrt{det[h_{ij}]} \equiv \rho_y$ and 
the pull back to $\cN_R$ of $\sqrt{-g}\,d\Sg_\bg$ is
\be
\sqrt{-g}\,\pback{d\Sg}_\bg = \rho_y\dg^u_\bg\,\pback{dy}^1\wedge \pback{dy}^2
\wedge \pback{dr} = \dg^u_\bg\,\pback{\eg}\wedge \pback{dr}.
\ee
( A $\leftarrow$ beneath a form indicates that the form is pulled back to $\cN$.)

The integrand in (\ref{Theta_bulk}) thus reduces to
        \bearr
\dg\Gamma^{[\ag}_{\ag\cg} g^{u]\cg} \eg\wedge dr 
& = & \frac{1}{2}\{ - \dg\Gamma^{\ag}_{\ag r} + g^{\ag\cg} \dg\Gamma_{r\ag\cg}\}\eg\wedge dr\\
& = & - \frac{1}{2}\{ \dg\di_r \log \sqrt{-g} + 2 \dg\Gamma_{rur}
- h^{ij} \dg\Gamma_{rij}\}\eg\wedge dr\\
& = & - \frac{1}{2}\{ \dg\di_r \log \rho_y + \dg\di_u g_{rr} 
+ \frac{1}{2}h^{ij} \dg\di_r h_{ij}\}\eg\wedge dr    \label{integrand_Theta_bulk}
                \eearr
Now recall that $\rho_y(y,r) = r^2\bar{\rho}(y)$, where $\bar{\rho}$ is the area density on 
$S_R$. Thus for the variations we are considering, which leave $\bar{\rho}(y)$ invariant, 
$\dg\rho_y = 0$. As a consequence 
\be     \label{dirrho0}
\dg\di_r\log\rho_y = 0
\ee
and
\be     \label{dirhh}
\frac{1}{2}\dg[\di_r h_{ij}]h^{ij} 
= \dg\di_r \log\rho_y - \frac{1}{2}\di_r[\rho_y\, e_{ij}] \dg\frac{e^{ij}}{\rho_y} 
= - \frac{1}{2}[\di_r e_{ij}] \dg e^{ij}. 
\ee
(Here $e^{ij}$ is the inverse of $e_{ij}$ and the fact that 
$e_{ij}\dg e^{ij} = -\dg \log \det[e_{ij}] = 0$, has been used.)

The remaining, middle, term in (\ref{integrand_Theta_bulk}) is proportional to 
$\dg \Gamma^r_{rr}$, since on $\cN_R$
\be     \label{dga_diu_grr}
        \di_u g_{rr} = 2 \Gamma^r_{rr}.
\ee


Substituting (\ref{dirrho0}), (\ref{dirhh}) and (\ref{dga_diu_grr}) into 
(\ref{integrand_Theta_bulk}) yields an expression for $\Theta_R[\dg^a]$ 
in terms of the BFC data and $\Gamma^r_{rr}$, and their variations. Adding to this the
expression (\ref{ThetaRsurf}) for $\Theta_R[{\pounds_\xi}]$ we obtain 
%
%
\bearr
\Theta_R[\dg] & = & -\frac{1}{16\pi G}\bigg\{ \int_{S_R} \eg \int_{r_0}^1 
\frac{1}{2} r^2 \di_r e_{ij}\dg e^{ij} - 2 r^2 \dg\Gamma^r_{rr} dr \nonumber \\
&& + \int_{S_0} \eg (\dg^y \lam + \xi^a_{\perp}\omega_a) 
+ \dg^y \eg (1 + r_0\Gamma^r_{rr})\bigg\} \label{symp_potential3}
\eearr
Here the identity
\be
\int_{\cN_R} f \eg\wedge dr = \int_{S_R} \eg \int_{r_0}^1  r^2 f dr,
\ee
valid for any function $f$ on $\cN_R$, has been used to turn the integral over $\cN_R$ 
into an integral over each generator followed by an integration of the result over
the set of generators.

The dependence on $\Gamma^r_{rr}$ can be eliminated.
Adding the variation of a functional of the data (an exact form on $\cF$) to the symplectic 
potential does not affect its curl, the presymplectic 2-form. Thus we are free to subtract
from $\Theta_R[\dg]$ the variation
\bearr
\lefteqn{\dg \left[ \frac{1}{8\pi G} \int_{S_R} \eg \int_{r_0}^1 
[r + r^2\Gamma^r_{rr}] dr \right]} \nonumber \\
& = & -\frac{1}{16\pi G} \left\{ \int_{S_R} \eg \int_{r_0}^1 
-2r^2 \dg\Gamma^r_{rr} dr + \int_{S_0} \dg^y \eg (1 + r_0\Gamma^r_{rr})\right\}.
\eearr
To obtain the last equality note firstly that $\dg r_0$ is $\dg^y r$ evaluated at $S_0$ (i.e. 
$v^- = 1$), and secondly, that $\eg_{S_R} r_0 \dg^y r_0 = \frac{1}{2}\bar{\rho}(y)\, 
dy^1\wedge dy^2\:\dg^y r_0^2 = \frac{1}{2} \dg^y \eg_{S_0}$.\footnote{Strictly speaking the 
three 2-forms in the equation live in different spaces. But they are equal when they are all 
pulled back to the cotangent space of $\Real^2$ using the $y$ chart.}
We will therefore take as the symplectic potential contributed 
by $\cN_R$
\be	\label{symp_potential4}
\Theta'_R[\dg] = -\frac{1}{16\pi G}\left\{ \int_{S_R} \eg \int_{r_0}^1 
\frac{1}{2} r^2 \di_r e_{ij} \dg e^{ij}  dr
+ \int_{S_0} \eg (\dg^y \lam + \xi^a_{\perp}\omega_a ) \right\}. 
\ee
Adding the contribution of $\cN_L$ we obtain the entire symplectic potential:
\bearr
\Theta'[\dg] = -\frac{1}{16\pi G}\bigg\{\!\!\!\!\!\!\!\! &&\int_{S_R} \eg \int_{r_0}^1 
\frac{1}{2} r^2 \di_r e_{ij}\dg e^{ij}\,dr \nonumber \\
& + &\!\int_{S_L} \eg \int_{r_0}^1 \frac{1}{2} r^2 \di_r e_{ij}\dg e^{ij}\,dr \nonumber \\
& + &\!\int_{S_0} \eg (\dg^{y_R} \lam + \dg^{y_L} \lam 
+ \xi^a_{\perp\,R}\omega_a - \xi^a_{\perp\,L}\omega_a)\bigg\}. 
\label{symp_potential5}
\eearr
All quantities appearing in the integral over $\cN_R$ are referred to the $a_R$ coordinates and,
similarly, those appearing in the integral over $\cN_L$ are referred to the $a_L$ coordinates.
In the integral over $S_0$ the term $\xi^a_{\perp\,L}\omega_a$ appears with a minus sign
because under the interchange of $L$ and $R$, which interchanges $n_+$ and $n_-$, 
the field $\omega_a = [n_+\cdot\nabla_a n_- - n_-\cdot\nabla_a n_+]/n_-\cdot n_+$ is replaced 
by $-\omega_a$. 

In (\ref{symp_potential5}) there appear, aside from the BFC data and their variations, only
the two area forms $\eg_{S_A} = \bar{\rho}_A dy_A^1\wedge dy_A^2$ and the two vector fields
$\xi_{\perp\,A}$. The area forms we know how to express in terms of BFC data (see the end of 
subsection \ref{preliminaries}), and they will be left as they are. The vector $\xi_{\perp\, A}$
is the variation of the map $s_A$ that gives the $y_A$ coordinates of a generator 
as a function of its $\theta$ coordinates: $\xi^i_{\perp\,A} = \dg^b y_A^i = \dg s_A^i$ or, 
transforming to $\theta$ components,
\be
\xi^a_{\perp\,A} = \frac{\di\theta^a}{\di y_A^i} \dg s_A^i.
\ee
$s_A$ is the inverse of the BFC datum $f_A$ it may be taken as an element of the BFC data, 
in place of $f_A$. (Notice that the $\theta^a$ are the natural coordinates of $s_A$.)

It turns out to be helpful to make explicit the dependence of the $S_0$ integral in 
(\ref{symp_potential5}) on the variations of the $s_A$:
\be
\dg^y \lam = \dg \lam - \xi^a_{\perp} \di_a \lam,
\ee
so
\be
\dg^{y_R} \lam + \dg^{y_L} \lam + \xi^a_{\perp\,R}\omega_a - \xi^a_{\perp\,L}\omega_a
 =  2\dg \lam - \xi^a_{\perp\,R}\omega_{-\,a} - \xi^a_{\perp\,L}\omega_{+\,a},
\ee
where $\omega_+ = \omega + d\lam$ and $\omega_- = - \omega + d\lam$.\footnote{These quantities 
were introduced by Epp in \cite{Epp}.} 
With the definitions 
\be
\hat{\omega}_{R\,i} = \rho_0\omega_{-\,a}\frac{\di\theta^a}{\di y_R^i}\ \ \mbox{and}\ \ 
\hat{\omega}_{L\,m} = \rho_0\omega_{+\,a}\frac{\di\theta^a}{\di y_L^m},
\ee
the contribution of $\cN_R$ to the symplectic potential may be written as
\bearr
\Theta_R'[\dg] = -\frac{1}{16\pi G}\bigg\{\!\!\!\!\! &&\!\!\!\int_{S_R} \eg \int_{r_0}^1 
\frac{1}{2} r^2 \di_r e_{ij}\dg e^{ij}\,dr \nonumber \\
\!\!& + &\!\!\int_{S_0} [\rho_0 \dg\lam - \hat{\omega}_{R\,i}\dg s_R^i]
\: d\theta^1\wedge d\theta^2 \bigg\}. 
\label{symp_potential6}
\eearr
and the contribution $\Theta_L'[\dg]$ of $\cN_L$ is given by a completely analogous expression.

\subsection{Calculation of the presymplectic 2-form in terms of almost free initial data} 
\label{presymp_calculation}

The first term in (\ref{symp_potential6}) is a ``bulk" term, an integral over $\cN_R$, 
while the second term, an integral over $S_0$, is a surface term. There are no contributions 
from the boundary of $\cN_R$ except that from $S_0$.

The contribution of the hypersurface $\cN_R$ to the presymplectic 2-form is 
$\bar{\Omega}_R[\dg_1,\dg_2] = \dg_1 \Theta'_R[\dg_2] - (1 \leftrightarrow 2)$. [For convenience
the commutator $[\dg_1,\dg_2]$ of the vector fields $\dg_1$ and $\dg_2$ on the space of field 
configurations $\cF$ has been set to zero. As discussed in footnote \ref{commuting_delta}, 
this may always be done without restricting the action of $\dg_1$ and $\dg_2$ on the fields 
at the fiducial solution, which is a single point in $\cF$.] The bulk 
term in $\bar{\Omega}_R$ is obtained by varying the bulk term in $\Theta'_R$ with $r_0$ held 
fixed. It is
\be     
\frac{1}{16\pi G}\left\{ \int_{S_R} \eg \int_{r_0}^1 \frac{1}{2} r^2 \dg_1 e^{ij}
\di_r \dg_2 e_{ij}\: dr - (1 \leftrightarrow 2)\right\}.     \label{Omega_bulk_term}
\ee 
The surface term at $S_0$ comes both from the surface term in $\Theta'_R$ and the variation of 
$r_0$ in the bulk term of $\Theta'_R$. The latter contribution is
\bearr
\lefteqn{\frac{1}{16\pi G}\left\{ \int_{S_R} \eg \dg_1^y r_0 [ \frac{1}{2} r^2 \di_r e_{ij}
\dg^a_2 e^{ij} ]_{r_0} - (1 \leftrightarrow 2)\right\}}		\nonumber \\ 
& = & \frac{1}{16\pi G}\left\{ \int_{S_0} \dg_1^y \eg \frac{1}{4} r_0 \di_r e_{ij}
\dg^a_2 e^{ij} - (1 \leftrightarrow 2)\right\}.        \label{r0_variation}
\eearr
(The subscript $r_0$ on the bracket indicates that the bracketed expression is to be evaluated 
at $r = r_0$. That is, the integral ranges over $S_R$ with the integrand at each point of $S_R$ 
given by an expression evaluated at $r = r_0$ on the generator through that point.)

In (\ref{r0_variation}) the derivatives by $r$ along the generators may be replaced by 
derivatives by $v^- = r/r_0$ using $r_0 \di_r = \di_{v^-}$, and the variation $\dg^a$ may be 
replaced by $\dg^y$:
\bearr
\lefteqn{\dg_1^y r_0 \:\dg_2^a e^{ij} - (1 \leftrightarrow 2) }\\
&&  = \dg_1^y r_0 \:\dg_2^y e^{ij} - \dg_1^y r_0 \:\dg_2^y r_0 \di_r e^{ij} 
- (1 \leftrightarrow 2)\\
&&  = \dg_1^y r_0 \:\dg_2^y e^{ij} - (1 \leftrightarrow 2).
\eearr
The first equality is evidently correct if $\dg_2^y e^{ij}$ is interpreted as the variation, 
at fixed $(v^+, v^-, y^1, y^2)$, of the $a$ components $e^{ij}$. The result is stronger 
than that however. Because the 3-metric on $\cN$ is degenerate along the generators it takes 
the form $h_{ij} dy^i \otimes dy^j$ (see \ref{a_line_element}). Therefore the metric induced on 
{\em any} two dimensional cross section of $\cN$ will have $y$ components $h_{ij}$. The $e^{ij}$
are both the $a$ components of the inverse conformal 2-metric, $e^{-1}$, on a constant $(r,u)$ 
surface as well as the components of $e^{-1}$ on a constant $(v^+, v^-)$ surface in the 
hybrid chart $(v^+, v^-, y^1, y^2)$. As a consequence the $\dg^y e^{ij}$ may also be interpreted
as the $(v^+, v^-, y^1, y^2)$ components, or the $a$ components, of $\dg^y e^{-1}$.

The contribution (\ref{r0_variation}) to $\bar{\Omega}_R$ can thus be written as
\be
\frac{1}{16\pi G}\left\{ \int_{S_0} \dg_1^y \eg \frac{1}{4} \di_{v^-} e_{ij}
\dg^y_2 e^{ij} - (1 \leftrightarrow 2)\right\}.        \label{r0_variation2}
\ee
The contribution to $\bar{\Omega}_R$ of the surface term in $\Theta'_R[\dg]$ is 
\be
-\frac{1}{16\pi G}\left\{ \int_{S_0}
\dg_1\rho_0 \dg_2\lam - \dg_1\hat{\omega}_{R\,i}\dg_2 s_R^i\:d\theta^1\wedge d\theta^2
- (1 \leftrightarrow 2)\right\}. \label{variation_S0_term}
\ee
Summing (\ref{Omega_bulk_term}), (\ref{r0_variation2}), and (\ref{variation_S0_term})
one obtains
\bearr
\bar{\Omega}_R[\dg_1,\dg_2] = \frac{1}{16\pi G}\bigg\{ \!\!&\!\! \frac{1}{2} &
\!\!\int_{S_R} \eg \int_{r_0}^1 r^2 
\dg_1 e^{ij}\di_r \dg_2 e_{ij} dr \ \ \ \ \ \ \ \ \ \ \ \ \ \ \ \ \ \ \ \ \nonumber\\
\!\!&\! + \frac{1}{4} & \!\!\int_{S_0} \dg_1^y \eg \di_{v^-} e_{ij} \dg^y_2 e^{ij}\nonumber\\
\!&\!- &\!\! \int_{S_0}\dg_1\rho_0 \dg_2\lam - \dg_1\hat{\omega}_{R\,i}\dg_2 s_R^i\:
d\theta^1\wedge d\theta^2 - (1 \leftrightarrow 2)\bigg\}.
\label{Omega_R}
\eearr  
This is the needed expression for the presymplectic 2-form in terms of the BFC data.

\section{The pre-Poisson bracket on initial data}\label{Poisson_bracket}

With the expression (\ref{Omega_R}) for the presymplectic 2-form in terms of BFC data in hand
we are ready to solve for the auxiliary pre-Poisson bracket satisfying (\ref{auxbracketdef}):
\be
\dg_0 \varphi = \bar{\Omega}[\{\varphi,\cdot\}_*,\dg_0]\ \ \ \forall \dg_0 \in L_g^0.
\ee
Actually, in order to preserve the symmetry between the $L$ and $R$ branches of $\cN$, 
we will solve for an equivalent bracket, $\{\cdot,\cdot\}_\circ$, on a slightly larger 
``extended" phase space, from which the $*$ brackets between the BFC data may be read off.
In subsection \ref{brackets} the resulting bracket is stated explicitly and discussed. The 
calculation of the bracket is given in subsection (\ref{bracket_calc}).

Neither the $*$ nor the $\circ$ bracket satisfies the Jacobi relation. (For this reason they 
are called pre-Poisson brackets and not Poisson brackets.) But recall that (\ref{auxbracketdef})
and the skew symmetry, linearity and Leibniz rule properties of the $*$ bracket suffice to 
ensure that the $*$ brackets of observables constructed from the initial 
data are equal to the Peierls brackets of these and thus satisfy the Jacobi relation. 

\subsection{Preliminaries}

In the following the extended phase space and the $\circ$ bracket will be defined. In addition
a useful parametrization of the degrees of freedom encoded in the conformal 2-metric $e_{ij}$ 
in terms of a single complex valued field, $\mu$, will be presented.

\subsubsection{A helpful extension of the phase space} \label{phase_space_extension}

We have introduced the $b$ chart in order to make it easier to treat the two branches of
$\cN$ democratically - quantities associated with the 2-surface $S_0$ common to the two branches
are referred to the $b$ chart rather than the $a$ chart adapted to either of the branches.
However, to be able to solve (\ref{auxbracketdef}) it is necessary to fix some specific relation
between the coordinates $\theta^a$ and the $y_R^i$ and the $y_L^j$. One could attempt to find a 
suitably symmetrical relation but this does not seem to be simple. The simplest choice is to 
equate the $\theta^a$ with either the $y_R^i$ or the $y_L^j$, which would of course break the 
symmetry between $L$ and $R$. 

A way to maintain the symmetry is to maintain the $\theta$ coordinates independent of the
$y$ coordinates and use a somewhat different bracket. Maintaining the $\theta$ coordinates 
independent introduces a new gauge degree of freedom, namely the freedom to change the $\theta$ 
chart by a diffeomorphism. The value of the presymplectic 2-form on a given pair of variations

does not depend on the $\theta$ chart that has been chosen. Indeed reference to the $\theta$ 
chart can be eliminated altogether by choosing $\theta = y_L$ (for instance) and thus expressing
$\bar{\Omega}[\dg_1,\dg_2]$ entirely in terms of the variations of the fields with respect to
the $a_L$ and $a_R$ charts. Thus a variation of the BFC data which corresponds solely to a 
diffeomorphism of the $\theta$ coordinates, leaving the relationship between the $a_L$
and $a_R$ coordinates and the fields referred to these invariant, is necessarily mapped to zero
by $\bar{\Omega}$. For this reason it is impossible to invert $\bar{\Omega}$ to obtain a 
bracket on a phase space of initial data including fields referred to the $\theta$ chart, 
unless either the $\theta^a$ are taken to be fixed functions of the $y_R^i$ and $y_L^m$ or the 
phase space is extended by including momenta conjugate to the $\theta^a$. 

We shall take essentially the second option, although we will not actually specify variables 
conjugate to the $\theta^a$. What is really necessary is that suitable functions on the extended
phase space generate the $\theta$ diffeomorphisms. The original phase space is then recovered 
as the submanifold on which these generators vanish. 

Such an extension can be realized in a quite natural manner.
In the expression (\ref{Omega_R}) for the $R$ contribution to the presymplectic 2-form and the 
corresponding formula for the $L$ contribution the field $\omega$ appears in the guise of two 
distinct quantities, $\hat{\omega}_{R\,i}$ and $\hat{\omega}_{L\,m}$, derived from $\omega$.
In the required extended phase space these two are independent fields on $S_0$. In the original
phase space they are of course not independent. They are related by the equation 
\be	\label{omega_constraint}
\hat{\omega}_{R\,i}\frac{\di s_R^i}{\di \theta^a} 
+ \hat{\omega}_{L\,m}\frac{\di s_L^m}{\di \theta^a}
= \rho[\omega_{-a} + \omega_{+a}] = 2\rho \frac{\di}{\di\theta^a} \lam,
\ee
which serves as a constraint that defines the original phase space of initial data as a 
submanifold of the extended phase space. As we shall see shortly this constraint also generates 
diffeomorphisms of the $\theta$ chart. 

The symplectic potential on the extended phase space, denoted $\hat{\Theta}$, is taken to be 
equal to the expression (\ref{symp_potential6}) for $\Theta'$ but with $\hat{\omega}_{R\,i}$ 
and $\hat{\omega}_{L\,j}$ independent fields. The presymplectic 2-form $\hat{\Omega}$ is the 
curl of $\hat{\Theta}$. Explicitly $\hat{\Omega}$ is the sum of a contribution from 
$\cN_R$, $\hat{\Omega}_R = A_R + B_R + C_R$ with
\bearr
A_R[\dg_1,\dg_2] & = & -\frac{1}{16\pi G}\int_{S_0}\dg_1\rho_0 \dg_2\lam 
- \dg_1\hat{\omega}_{R\,i}\dg_2 s_R^i\: d\theta^1\wedge d\theta^2 - (1 \leftrightarrow 2) 
\label{AR_def}\\
B_R[\dg_1,\dg_2] & = & \frac{1}{64\pi G}\int_{S_0} \dg_1^y \eg \di_{v^-} e_{ij} \dg^y_2 e^{ij}
- (1 \leftrightarrow 2) \label{BR_def}\\
C_R[\dg_1,\dg_2] & = & \frac{1}{32\pi G}\int_{S_R} \eg \int_{r_0}^1 r^2 
\dg_1 e^{ij}\di_r \dg_2 e_{ij} dr - (1 \leftrightarrow 2), \label{CR_def}
\eearr
and an entirely analogous contribution $\hat{\Omega}_L$ from $\cN_L$ (obtained by 
substituting $L$ for $R$, and $v^+$ for $v^-$, in $\hat{\Omega}_R$).

The auxiliary bracket on the extended phase space, $\{\cdot,\cdot\}_\circ$, is defined by the 
requirement that
\be	\label{auxbracket2def}
\dg \varphi = \hat{\Omega}[\{\varphi,\cdot\}_\circ,\dg]
\ee
for all extended phase space initial data $\varphi$ on the interior of $\cN$ and all variations 
$\dg$ of the data compatible with the metric consistency constraint (\ref{e_consistency}), and 
the condition that $\dg e_{ij} = 0$ at the two ends $S_R$ and $S_L$ of $\cN$. These variations 
$\dg$ will be termed ``admissible". 

Note that we have weakened the condition on the variation $\dg e_{ij}$ of the $a$ chart 
components of $e$. We are not requiring it to vanish in a neighbourhood of $\di\cN$, only on 
$\di\cN$ itself. This does not lead to any actual change of the implications of 
(\ref{auxbracket2def}) for the specific $\hat{\Omega}$ we are using. Restricting the admissible 
$\dg$ to ones for which $\dg e_{ij}$ vanishes in some neighbourhood of $\di\cN$ does not enlarge
the set of brackets solving (\ref{auxbracket2def}), as the reader may easily verify when this 
condition is solved for $\{\cdot,\cdot\}_\circ$ in subsection \ref{bracket_calc}. But a simpler 
definition of the admissible $\dg$ seems preferable, and may facilitate understanding of the 
results, especially in connection with the Jacobi relations. In any case it is clear that 
enlarging the class of admissible $\dg$ does not weaken (\ref{auxbracket2def}), and so our key 
result, that the brackets satisfying this relation reproduce the Peierls brackets of observables
(when the constraint (\ref{omega_constraint}) is imposed) continues to hold.

Let us return to the constraint (\ref{omega_constraint}). 
Imposing this constraint on the interior of $S_0$ is equivalent to requiring that
\bearr
\kappa[v] & = & \frac{1}{16\pi G}\int_{S_0} v^a [2\rho \di_a \lam 
- \hat{\omega}_{R\,i}\di_a s_R^i - \hat{\omega}_{L\,m}\di_a s_L^m] d\theta^1 \wedge d\theta^2 
\label{theta_diff_gen}\\ 
& = & -\hat{\Theta}[\pounds_v],
\eearr
vanishes for all smooth vector fields $v$ tangent to $S_0$ and vanishing on $\di S_0$.
$\kappa[v]$ acts as a diffeomorphism generator, it produces a 
variation $\dg^y \theta^a = -v^a$ in the $\theta$ coordinates at fixed $y_R$ and $y_L$, and no 
change in the relation between the $a_R$ and $a_L$ charts or in the fields as functions of 
either of those charts. Equivalently it acts as $\pounds_v$ on the fields as functions of the 
$\theta^a$. 

To see this note that 
\be    \label{Omega_Lv}
\hat{\Omega}[{\pounds_v},\dg] = {\pounds_v}\hat{\Theta}[\dg] - \dg \hat{\Theta}[\pounds_v],
\ee
where $\dg v$ has been set to zero, so that $[\dg,\pounds_v] = \pounds_{\dg v} = 0$.
$\pounds_v$ acts only on the $S_0$ surface term in $\hat{\Theta}[\dg]$ since the 
bulk term, in which no fields referred to the $\theta^a$ intervene, is invariant under 
diffeomorphisms of the $\theta$ chart. But since the integrand of the $S_0$ term is a scalar 
density under diffeomorphisms of the $\theta$ and $v$ vanishes on $\di S_0$ the $S_0$ 
contribution to 
$\hat{\Theta}[\dg]$ is also invariant under the diffeomorphisms generated by $v$. Thus 
${\pounds_v}\hat{\Theta}[\dg] = 0$ and
\be		\label{diff_generator_proof}
\dg \kappa[v] = \hat{\Omega}[{\pounds_v},\dg].
\ee
Note that $\dg$ need not be an admissible variation. Admissibility is not assumed in 
(\ref{Omega_Lv}) nor in obtaining the explicit expression (\ref{AR_def} - \ref{CR_def}) for 
$\hat{\Omega}$ from $\hat{\Theta}$, via the calculation of subsection 
\ref{presymp_calculation}. (Indeed, in \ref{presymp_calculation} the variations are only 
assumed to satisfy the conditions a) and b) of subsection \ref{preliminaries}.)  
We may therefore take $\dg$ to be the action $\{\varphi,\cdot\}_\circ$ of an initial datum 
$\varphi$, in which case
\be 
\{\kappa[v],\varphi\}_\circ \equiv -\dg\kappa[v] = \hat{\Omega}[\{\varphi,\cdot\}_\circ, 
{\pounds}_v].
\ee
Since ${\pounds_v}$ is an admissible variation (in fact, it leaves $e_{ij}$ invariant)
equation (\ref{auxbracket2def}) implies that
\be
\{\kappa[v],\varphi\}_\circ =  {\pounds}_v\varphi.
\ee
Here ${\pounds}_v$ acts only on the $\theta$ dependence of the fields. That is, it is 
equivalent to a generator of diffeomorphisms of the $\theta$ coordinates.

Because it preserves the symmetry between the branches of $\cN$ the bracket 
$\{\cdot,\cdot\}_\circ$ rather than $\{\cdot,\cdot\}_*$ will be given explicitly here. 
Note however that the bracket $\{\cdot,\cdot\}_*$ on the original phase space corresponding to 
the choice $\theta^1 = y_L^1$, $\theta^2 = y_L^2$ (i.e. $s_L = \mbox{id}$) can be recovered 
easily from $\{\cdot,\cdot\}_\circ$ as a Dirac bracket:
\be  \label{Dirac_b_kappa}
\{\cdot,\cdot\}_* = \{\cdot,\cdot\}_\circ 
- \int_{S_0} \frac{\di\theta^a}{\di y_L^l} \,[
  \{\cdot, s_L^l\}_\circ\{\kappa_a,\cdot\}_\circ 
- \{\cdot, \kappa_a\}_\circ\{s_L^l,\cdot\}_\circ]\,
\:d\theta^1\wedge d\theta^2,  
\ee
where $\kappa_a \equiv \kappa[\di_a] = \frac{1}{16\pi G}[ 2\rho \di_a \lam 
- \hat{\omega}_{R\,i}\di_a s_R^i - \hat{\omega}_{L\,m}\di_a s_L^m]$, and 
$\frac{\di\theta^a}{\di y_L^l}$, with the fixing of $\theta$ adopted is a Kronecker delta. 
Of course the $*$ bracket may also be obtained by solving the original equation 
(\ref{auxbracketdef}) with this fixing of the $\theta$ chart. 

The explicit form of the $\circ$ brackets, given in subsection \ref{brackets}, shows that the 
only datum that has non-zero $\circ$ bracket with $s_L$ is $\hat{\omega}_L$ 
(see (\ref{omega_y_L_brack}) and the subsequent discussion). $\{\cdot,\cdot\}_*$ thus differs 
from $\{\cdot,\cdot\}_\circ$ only when one of the arguments is $\hat{\omega}_L$. 
This datum is not independent on the constraint manifold so the $*$ brackets of a complete set
of phase space coordinates on the gauge fixed constraint manifold, namely the data we have been 
considering excluding $s_L$ and $\hat{\omega}_L$, are identical to the corresponding $\circ$
brackets. The $*$ brackets involving $\hat{\omega}_L$ can be obtained from brackets between 
these other data by substituting the expression for $\hat{\omega}_L$ in terms of these provided 
by the constraint (\ref{omega_constraint}) and the gauge condition $s_L = \mbox{id}$. The $*$ 
brackets of $s_L$ of course vanish. 

\subsubsection{A useful parametrization of the conformal 2-metric degrees of freedom}

We turn now to the definition of a useful parametrization of the conformal 2-metric $e$.
The unimodular symmetric matrix field $e_{ab}$ is free initial data in the sense that it is 
unrestricted by the field equations, but of course the components of $e$ are algebraically 
restricted by the requirements of symmetry and unimodularity. It is thus convenient to 
parametrize the two degrees of freedom of $e$ by an unconstrained complex number. We shall do 
this as follows: The degenerate line element on $\cN$ can always be expressed in terms the 
complex coordinate $z = \theta^1 + i\theta^2$ as
   \be             \label{mu_parametrization}
       ds^2 = h_{ab}d\theta^a d\theta^b 
			= \rho (1 - |\mu|^2)^{-1}[dz + \mu d\bar{z}][d\bar{z} + \bar{\mu}dz].
   \ee
with $\mu$ a complex number valued field of modulus less than 1. $\mu$ provides the desired 
parametrization of $e$. In component form (with respect to the $\theta$ coordinate basis)
   \be             \label{mu_parametrization2}
       e_{ab} = \frac{1}{1-|\mu|^2}\left[\begin{array}{cc} |1+\mu|^2 & 2\mbox{Im} \mu \\
                                         2\mbox{Im} \mu & |1-\mu|^2 \end{array}\right].
   \ee

Under a transformation of coordinates $\theta \rightarrow \theta'$ in which the $\theta'$ 
coordinates, like the $\theta$ coordinates, are constant on the generators of $\cN$, the
components $e_{ab}$ are transformed to
\be		\label{e_transformation}
e'_{cd} = \frac{\di \theta^a}{\di \theta'^c}\frac{\di \theta^b}{\di \theta'^d}
|det \frac{\di \theta'}{\di \theta}| e_{ab}
\ee
(a fact that was used in the definition of the constraint (\ref{e_consistency})).
The corresponding transformation of $\mu$ is 
\be \label{mu_transformation}
\mu \rightarrow \mu' = [\frac{\di z}{\di\bar{z}'} + \mu\frac{\di\bar{z}}{\di\bar{z}'}]/
		[\frac{\di z}{\di z'} + \mu\frac{\di\bar{z}}{\di z'}].
\ee

The $\mu$ field referred to the $y_R$ chart will be denoted by $\mu_R$ and that referred to the
$y_L$ chart by $\mu_L$. 

\subsection{The brackets}	\label{brackets}

In this subsection the $\circ$ brackets will be given between the data 
$\hat{\omega}_{R\,i}$, $\hat{\omega}_{L\,m}$, $s_R^j$, $s_L^n$, $\lam$, $\rho_0$, and
$\mu$ and $\bar{\mu}$ (or equivalently $e_{ab}$), where the conformal 2-metric 
data $\mu$ and $\bar{\mu}$ are, like the other data, {\em referred to the $b$ chart}. 
A subset of these form a {\em free and complete} data set. Moreover, the $\circ$ brackets
among the data in this subset are equal to their $*$ brackets.

As pointed out earlier, the constraint (\ref{e_consistency}) that the $a_R$ and $a_L$ 
components of $e$ must satisfy at $S_0$ is equivalent to the requirement that $e_{ab}$ at $S_0$ 
obtained by transforming $e_{R\,ij}$ from the $a_R$ chart to the $b$ chart is the same as 
obtained by transforming $e_{L\,mn}$ from the $a_L$ chart, i.e. it is equivalent to requiring 
that $e_{ab}$ can be consistently defined. The constraint (\ref{e_consistency}) thus disappears 
if the $b$ components, $e_{ab}$, are used to specify the conformal 2-metric throughout $\cN$, in
place of $e_{R\,ij}$ on $\cN_R$ and $e_{L\,mn}$ on $\cN_L$. 

The other constraint, (\ref{omega_constraint}), is still present, and so is, necessarily, a 
gauge fixing condition defining the $\theta$ coordinates. However, as shown in subsection 
\ref{phase_space_extension}, if the simple gauge fixing $s_L = \mbox{id}$ is adopted then 
(\ref{omega_constraint}) 
can always be solved for $\hat{\omega}_L$, and the remaining data 
$e_{ab}$, $\hat{\omega}_{R\,i}$, $s_R^j$, $\lam$, and $\rho_0$ form a free and complete set.
Using of the explicit expressions for the $\circ$ brackets given here it is easy to check that 
the $*$ (\ref{Dirac_b_kappa}) coincides with the $\circ$ bracket between these data. To maintain
the symmetry between $L$ and $R$ in the formalism, and in order to not commit ourselves to any 
specific gauge for the $\theta$ coordinates, the $\circ$ brackets between all the data, 
including $\hat{\omega}_L$ and $s_L$ will be given.

For $1$ and $2$ points on $S_0$ (or more precisely values of the $b$ coordinates corresponding 
to points on $S_0$)
\bearr
\{\lam(1),\rho_0(2)\}_\circ 
& = & - 8\pi G \dg^2(\theta(2) - \theta(1)),	
\label{lam_rho_bracket}\\
\{\hat{\omega}_{R\,i}(1),s_R^j(2)\}_\circ 
& = & - 16\pi G \dg_i^j\dg^2(\theta(2) - \theta(1)),	
\label{omega_y_R_bracket}\\
\{\hat{\omega}_R[f],\lam(1)\}_\circ & = & \pi G [(\di_{v^-} e^{ab} - \di_{v^+} e^{ab}) 
{\pounds}_f e_{ab}]_1 \label{lam_omega_R_bracket}\\
\{\hat{\omega}_R[f_1],\hat{\omega}_R[f_2]\}_\circ & = & - 2\pi G
\int_{S_0} {\pounds}_{f_1} e_{ab} \,\di_{v^-} e^{ab}{\pounds}_{f_2}\eg - (1 \leftrightarrow 2) 
\label{omega_omega_R_bracket} \\
\{\hat{\omega}_R[f],\hat{\omega}_L[g]\}_\circ & = & 2\pi G
\int_{S_0} [{\pounds}_f e_{ab} \di_{v^+} e^{ab}{\pounds}_g \eg
			- {\pounds}_g e_{ab} \di_{v^-} e^{ab}{\pounds}_f \eg],
\label{omega_R_omega_L_bracket}
\eearr
where $\hat{\omega}_R[f] = \int_{S_0} \hat{\omega}_{R\,i} f^i d\theta^1\wedge d\theta^2$ with 
$f$ a test vector field tangent to $S_0$, and $\hat{\omega}_L[g]$ and $g$ are defined similarly.
Note that {\em the $y_R$ components} $f^i$ of $f$, and similarly the $y_L$ 
components $g^m$ of $g$, are taken to be independent of the dynamical fields and thus $\circ$ 
commute with everything. This is important to remember in order to interpret the given 
expressions for the brackets correctly, as can be clearly seen when the smearing fields are 
eliminated. 

%
%

Expressions analogous to (\ref{omega_y_R_bracket}), (\ref{lam_omega_R_bracket}), and 
(\ref{omega_omega_R_bracket}) give the brackets $\{\hat{\omega}_{L\,m}(1),s_L^n(2)\}_\circ$,
$\{\lam(1),\hat{\omega}_L[g]\}_\circ$, and $\{\hat{\omega}_L[g_1],\hat{\omega}_L[g_2]\}_\circ$ 
respectively.

For $1$ and $2$ points on the same branch of $\cN$
\bearr	\label{mu_mubar_bracket}
\{\mu(1),\bar{\mu}(2)\}_\circ 
& = & 2\pi G\, \dg^2(\theta(2) - \theta(1))\, sgn(1,2) \nonumber \\
&& [\frac{1 - |\mu|^2}{\sqrt{\rho}}]_1 [\frac{1 - |\mu|^2}{\sqrt{\rho}}]_2\:
e^{\int_1^2 \frac{1}{1 - |\mu|^2}[\bar{\mu} d\mu - \mu d\bar{\mu}]},
\eearr
where for $1$ and $2$ lying on the same generator (i.e. $\theta(1) = \theta(2)$) 
\be
sgn(1,2) = \left\{\begin{array}{cl} 
+1 & \mbox{if $2$ is further from $S_0$ along the generator than $1$} \\
-1 & \mbox{if $2$ is closer to $S_0$ along the generator than $1$} \end{array}\right.
\ee 
and the integral in the exponential is evaluated along the segment of the generator from $1$
to $2$. Note that this defines the bracket between $\mu$ and $\bar{\mu}$ smeared with smooth
test functions on $\cN$.

When $1$ and $2$ both lie on $S_0$ $\{\mu(1),\bar{\mu}(2)\}_\circ = 0$. This is not 
unambiguously implicit in (\ref{mu_mubar_bracket}). It is an additional specification of the 
bracket compatible with the defining equation (\ref{auxbracket2def}).


For $1$ and $2$ any two points in $int \cN$ 
\be   \label{mu_mu_brack}
0  =  \{\mu(1),\mu(2)\}_\circ = \{\bar{\mu}(1),\bar{\mu}(2)\}_\circ,	 
\ee
Notice that when $1$ and $2$ both lie on $S_0$ (\ref{mu_mu_brack}), together with 
the vanishing of $\{\mu(1),\bar{\mu}(2)\}_\circ$ in this case, implies 
that $\{e^{cd}(1), e_{ab}(2)\}_\circ = 0$.

These brackets are not uniquely determined by (\ref{auxbracket2def}).
In particular $\dg^2(\theta(2) - \theta(1))\, sgn(1,2)$ in (\ref{mu_mubar_bracket}) may be 
replaced by $\dg^2(\theta(2) - \theta(1))\, sgn(1,2) + i \Upsilon(\theta(1),\theta(2))$, where 
$\Upsilon$ is a real distribution on $\theta$, and possibly a function of the data, resembling 
$\dg^2(\theta(2) - \theta(1))$ in that it has support only at $\theta(2) = \theta(1)$ and that 
it transforms like $\dg^2(\theta(2) - \theta(1))$ under diffeomorphisms of the $\theta$ chart. 

This indeterminacy arises for the following reason: Denote the $a$ chart $\mu$ and $\bar{\mu}$ 
fields on the truncating surface $S_A \subset \di\cN$ of a given branch by $\dot{\mu}$ and 
$\dot{\bar{\mu}}$. The bracket is only defined on data on the {\em interior} of $\cN$, but
we may define $\{\dot{\mu}(y),\cdot\}_\circ$ and $\{\dot{\bar{\mu}}(y),\cdot\}_\circ$ 
to be the limits of $\{\mu(p),\cdot\}_\circ$ and $\{\bar{\mu}(p),\cdot\}_\circ$, respectively, 
as $p$ tends to $S_A$ along the generator specified by $y$. 
For any admissible variation $\dg$ (which we always take to be smooth) 
\be
\hat{\Omega}[\{\dot{\mu}(y),\cdot\}_\circ,\dg] 
= \lim_{p \rightarrow S_A} \hat{\Omega}[\{\mu(p),\cdot\}_\circ,\dg]
= \lim_{p \rightarrow S_A} \dg \mu(p) = 0,
\ee
and similarly for $\{\dot{\bar{\mu}}(y),\cdot\}_\circ$. (\ref{auxbracket2def}) thus permits

the addition of  multiples of $\{\dot{\mu}(y),\cdot\}_\circ$ and 
$\{\dot{\bar{\mu}}(y),\cdot\}_\circ$ to any solution $\{\varphi_I,\cdot\}_\circ$. This 
freedom in the brackets is greatly reduced by the requirements of causality and antisymmetry, 
but not entirely eliminated. 

Another possible indeterminacy is associated with the bracket $\{e^{cd}(1), e_{ab}(2)\}_\circ$ 
for $1, 2 \in S_0$. The vanishing of this bracket is shown to be consistent with 
(\ref{auxbracket2def}), but it is not shown to be necessary. On the other hand, no candidate for
an alternative solution for $\{e^{cd}(1), e_{ab}(2)\}_\circ$ was found. In other words, this 
bracket has not been demonstrated to be uniquely determined, but it may be so.

The freedom in the bracket $\{\mu(1),\bar{\mu}(2)\}_\circ$ propagates to some other brackets.
However, if it is assumed that $\{\mu(1),\bar{\mu}(2)\}_\circ$ is given by the expression 
(\ref{mu_mubar_bracket}) and $\{e^{cd}(1), e_{ab}(2)\}_\circ = 0$ for $1, 2 \in S_0$ all other 
brackets are uniquely determined by (\ref{auxbracket2def}). Here I will make this choice, as 
it is the simplest and provides the most natural seeming solution to (\ref{auxbracket2def}).

For $1 \in S_0$ and $2 \in \cN_R - S_0$
\bearr
\{\lam(1),e_{ab}(2)\}_\circ 
& = & \frac{1}{4} [\di_{v^+} e_{cd}]_1\{e^{cd}(1), e_{ab}(2)\}_\circ
- 4\pi G [\frac{v^-}{\rho_0} \di_{v^-} e_{ab}]_2 \dg^2(\theta(2) - \theta(1)), \nonumber\\
&&	\label{lam_e_bracket}\\
\{\hat{\omega}_R[f],e_{ab}(2)\}_\circ & = & -\frac{1}{2}\int_{S_0} \eg\,{\pounds}_f e_{cd}
\{e^{cd},e_{ab}(2)\}_\circ \nonumber \\
&& - 16\pi G[{\pounds}_f e_{ab} - \frac{1}{2}\di_{v^-} e_{ab}
\frac{{\pounds}_f \rho_0}{\rho_0}]_2,
\label{omegaR_e_bracket}\\
\{\hat{\omega}_L[g],e_{ab}(2)\}_\circ & = & \frac{1}{2}\int_{S_0} [\eg\, {\pounds}_g e_{cd} -
{\pounds}_g \eg\: \di_{v^+} e_{cd}]\{e^{cd},e_{ab}(2)\}_\circ. \nonumber \\
\label{omegaL_e_bracket}
\eearr
Of course analogous expressions with the roles of $\cN_R$ and $\cN_L$ reversed give 
the corresponding brackets when $2 \in \cN_L - S_0$.

When $2$ lies on $S_0$ 
\bearr
\{\lam(1),e_{ab}(2)\}_\circ 
& = & - 2\pi G \frac{1}{\rho_0}[\di_{v^-} e_{ab} + \di_{v^+} e_{ab}] \dg^2(\theta(2) 
- \theta(1)),\label{lam_e_0_bracket}\\
\{\hat{\omega}_R[f],e_{ab}(2)\}_\circ 
& = & - 8\pi G [{\pounds}_f e_{ab} - \frac{1}{2}\di_{v^-} e_{ab}
\frac{{\pounds}_f \rho_0}{\rho_0}]_2.
\label{omega_R_e_0_bracket}\\
\{\hat{\omega}_L[g],e_{ab}(2)\}_\circ 
& = & - 8\pi G [{\pounds}_g e_{ab} - \frac{1}{2}\di_{v^+} e_{ab}
\frac{{\pounds}_g \rho_0}{\rho_0}]_2.
\label{omega_L_e_0_bracket}
\eearr
These brackets for $2 \in S_0$ are the averages of the expressions for the same brackets valid 
in the cases $2 \in \cN_R - S_0$ and $2 \in \cN_L - S_0$. (Recall that 
$\{e^{cd}(1), e_{ab}(2)\}_\circ = 0$ when both $1$ and $2$ lie on $S_0$.)

All other brackets vanish. In particular 
\be   \label{lam_lam_brack}
0 = \{\lam(1),\lam(2)\}_\circ.	
\ee
and $s_R$, $s_L$ and $\rho_0$ commute (that is, they have vanishing $\circ$ brackets) with 
everything except $\hat{\omega}_R$, $\hat{\omega}_L$, and $\lam$ respectively. 

The requirement that $\kappa$ generates diffeomorphisms of the $\theta$ chart - specifically 
that $\{\kappa[v],\varphi\}_\circ =  -{\pounds}_v\varphi$ for all data $\varphi$ on $int\cN$ 
- provides a non-trivial check of the brackets. The reader may easily verify that the 
expressions for the brackets given here satisfy this requirement.

Notice that by (\ref{mu_mubar_bracket}) and (\ref{mu_mu_brack}) the variation 
$\{\mu(1),\cdot\}_\circ$ does not preserve the reality of the metric. This corresponds to the 
fact that $\mu$ is complex. The action of a real functional of $\mu$ and $\bar{\mu}$ does 
preserve the reality of the metric, as can be seen easily from the relations
\bearr	
\overline{\{\mu(1),\bar{\mu}(2)\} _\circ} & = & \{\bar{\mu}(1),\mu(2)\}_\circ 
\label{bracket_reality1}, \\
\overline{\{\mu(1),\mu(2)\} _\circ} & = & \{\bar{\mu}(1),\bar{\mu}(2)\}_\circ 
\label{bracket_reality2},
\eearr 
which follow from (\ref{mu_mubar_bracket}) and (\ref{mu_mu_brack}). (\ref{bracket_reality1}) and
 
(\ref{bracket_reality1}) can be interpreted as asserting that the bracket in itself, 
$\{\cdot,\cdot\}_\circ$, is real, so that complex conjugation acts only on it's arguments.

Since the coordinate systems $y_R$, $y_L$, and $\theta$ labelling the generators of $\cN$
are essentially arbitrary it is important to verify that the brackets are invariant under
diffeomorphisms of these coordinates. All the data except $\mu$ and $\bar{\mu}$ transform
as tensor or tensor density fields under such transformations. For all the brackets except 
those involving $\mu$ and $\bar{\mu}$ it is therefore easy to check
that the brackets and the expressions proposed for them transform in the same way under
the various possible coordinate transformations. Moreover, the transformation law 
(\ref{mu_transformation}) of $\mu$ shows immediately that $\{\mu(1),\mu(2)\}_\circ = 0$ 
is a coordinate invariant statement. The only non-trivial case is therefore the bracket 
(\ref{mu_mubar_bracket}).

To verify the invariance under coordinate transformations of the equation 
(\ref{mu_mubar_bracket}) note first that $\dg^2(\theta(2) - \theta(1))\, sgn(1,2)
\frac{1}{\sqrt{\rho_1}}\frac{1}{\sqrt{\rho_2}}$ is invariant. From (\ref{mu_transformation})
one shows that a change of coordinates transforms the ratio of a variation $\Dg \mu$ and 
$1 - |\mu|^2$ by a phase factor: 
\be	\label{dmu_transformation}
\frac{\Dg \mu'}{1 - |\mu'|^2} = \frac{\overline{\ag + \mu\bg}}{\ag + \mu\bg}
\frac{\Dg \mu}{1 - |\mu|^2},
\ee
where $\ag = \frac{\di z}{\di z'}$ and $\bg = \frac{\di\bar{z}}{\di z'}$. Thus 
\bearr
\lefteqn{[\frac{1}{1 - |\mu'|^2}]_1\{\mu'(1),\bar{\mu}'(2)\}_\circ[\frac{1}{1 - |\mu'|^2}]_2 =} 
\nonumber \\
&& [\frac{\overline{\ag + \mu\bg}}{\ag + \mu\bg}]_1
[\frac{\ag + \mu\bg}{\overline{\ag + \mu\bg}}]_2
[\frac{1}{1 - |\mu|^2}]_1\{\mu(1),\bar{\mu}(2)\}_\circ[\frac{1}{1 - |\mu|^2}]_2.
\eearr
To complete the demonstration of the invariance of (\ref{mu_mubar_bracket}) we only have to show
that the phase $\exp(\int_1^2 \frac{1}{1 - |\mu|^2}
[\bar{\mu}d \mu - \mu d\bar{\mu}])$ transforms in the same way.

(\ref{dmu_transformation}) and (\ref{mu_transformation}) show that
\bearr
\frac{\bar{\mu}'d\mu'}{1 - |\mu'|^2} & = & \frac{\bg + \bar{\mu}\ag}{\ag + \mu\bg}
\frac{d\mu}{1 - |\mu|^2}	\\
& = & \frac{\bg d\mu}{\ag + \mu\bg} + \frac{\bar{\mu} d\mu}{1 - |\mu|^2}\\
& = & d\log(\ag + \mu\bg) + \frac{\bar{\mu}d\mu}{1 - |\mu|^2}.
\eearr
It follows that indeed
\be
e^{\int_1^2 \frac{1}{1 - |\mu'|^2}[\bar{\mu}' d\mu' - \mu' d\bar{\mu}']}
= [\frac{\overline{\ag + \mu\bg}}{\ag + \mu\bg}]_1
[\frac{\ag + \mu\bg}{\overline{\ag + \mu\bg}}]_2\:
e^{\int_1^2 \frac{1}{1 - |\mu|^2}[\bar{\mu}d\mu - \mu d\bar{\mu}]}.
\ee
The equation (\ref{mu_mubar_bracket}) is therefore invariant under transformations of the 
$\theta$ chart.

(\ref{mu_mubar_bracket}) has a further invariance. It is invariant under reparametrization
of the generators. In fact (\ref{mu_mubar_bracket}) does not even refer to a particular 
parametrization of the generators. This makes it easy to take the flat spacetime limit of this 
bracket, and it suggests the conjecture that $\{\mu,\bar{\mu}\}_\circ$ is given by the 
expression (\ref{mu_mubar_bracket}) even when the area distance is not a good parameter along 
the generators. For instance, in flat spacetime with $\cN$ swept out by parallel generators 
(neither diverging nor converging), or in a generic spacetime along a generator with neighbours 
which initially diverge and then re converge. Checking this conjecture is left to future work.

Invariance of the expressions for the brackets under transformations of the $\theta$ coordinates
is necessary because these coordinates are arbitrary, but this is not so for the parametrization
of the generators, because $v = r/r_0$ is not arbitrary. It is completely determined
by the geometry of $\cN$. Indeed the expressions for several of the brackets, involving 
$\di_{v^-} e$ and $\di_{v^+} e$ at $S_0$, are not invariant under such reparametrizations of the
generators. Of course one may still re express these brackets in terms of another 
parametrization, $\lam$, of the generators by replacing $\di_v e_{ab}$ at $S_0$ by 
\be
\frac{\di\lam}{\di v} \di_\lam e_{ab} = 2\frac{\rho_0}{\di_\lam \rho} \di_\lam e_{ab} 
\ee
One sees that the brackets cannot all be straightforwardly extended to hypersurfaces $\cN$ with 
vanishing expansion of the generators at $S_0$. The area distance is not a good parameter at 
$S_0$ in this case, and when it is replaced by a good parameter, such as an affine
parameter, several of the brackets become undefined. The comparison of the present theory with 
canonical formulations of general relativity linearized about Minkowski space 
\cite{Aragone_Gambini} in terms of data on null hyperplanes is thus a delicate matter which is 
also left to future investigations.

\subsection{Calculation of the brackets}	\label{bracket_calc}

How were these brackets obtained from (\ref{auxbracket2def})? In the following 
(\ref{auxbracket2def}) will be solved step by step, yielding the brackets given above and 
at the same time demonstrating that $\{\cdot,\cdot\}_\circ$ defined by these brackets really
satisfies (\ref{auxbracket2def}).

We begin by considering (\ref{auxbracket2def}) for variations $\dg$ such that
$0 = \dg e = \dg s_L = \dg s_R = \dg \rho_0$. Then
\bearr
\lefteqn{\hat{\Omega}[\{\varphi(1),\cdot\}_\circ,\dg] = A[\{\varphi(1),\cdot\}_\circ,\dg] = }
\nonumber \\
&& - \frac{1}{16\pi G}\int_{S_0}
2 \{\varphi(1),\rho_0\}_\circ \dg\lam + \{\varphi(1),s_R^i\}_\circ \dg \hat{\omega}_{R\,i} 
+ \{\varphi(1),s_L^m\}_\circ \dg \hat{\omega}_{L\,m}\: d\theta^1 \wedge d\theta^2.
\eearr
Substituting for $\varphi$ each of the fields coordinatizing the extended phase space one 
finds that the only non-zero bracket involving $\rho_0$ is (\ref{lam_rho_bracket}),
\be
\{\lam(1),\rho_0(2)\}_\circ = - 8\pi G \dg^2(\theta(2) - \theta(1)),	
\ee
the only non-zero bracket with $s_R$ is (\ref{omega_y_R_bracket}),
\be
\{\hat{\omega}_{R\,i}(1),s_R^j(2)\}_\circ = - 16\pi G \dg_i^j\dg^2(\theta(2) - \theta(1)),	
\ee
and the only non-zero bracket with $s_L$ is
\be		\label{omega_y_L_brack}
\{\hat{\omega}_{L\,m}(1),s_L^n(2)\}_\circ = - 16\pi G \dg_m^n\dg^2(\theta(2) - \theta(1)).	
\ee
Moreover, it is easy to see that, given these non-zero brackets and the fact that all other 
brackets involving $\rho_0$, $s_R$, and $s_L$ vanish, (\ref{auxbracket2def}) is satisfied
for $\varphi = \rho_0$, $s_R^i$ or $s_L^m$ and {\em all} $\dg$.

Now let us consider (\ref{auxbracket2def}) with $\varphi = \mu_R$ evaluated at fixed values of 
the $a_R$ coordinates corresponding to a point $1 \in \cN_R - S_0$, and admissible variations 
$\dg$ such that $0 = \dg s_L = \dg s_R = \dg \rho_0$ but $\dg e$ not necessarily zero (except 
of course on $\di\cN$). This will give us the $\circ$ brackets of $\mu$ with $\mu$ and 
$\bar{\mu}$.\footnote{%
To lighten the notation the subscripts $R$ and $L$ indicating the branch to which variables are 
referred will be dropped when the context makes this information redundant. In particular $\mu$
will be used to denote variously $\mu_R$, $\mu_L$, $\mu$ referred to the $\theta$ chart, or 
several of these indistinctly, as the context indicates. For example, in the following 
paragraphs (until equation (\ref{muR_muL_brack})), concerned with the brackets of 
$\mu_R(1)$ for $1 \in \cN_R - S_0$, $\mu(1)$ will denote $\mu_R(1)$ and plain $\mu$ denotes 
$\mu_R$, $\mu_L$, or either depending on the context.}

To solve this condition we first express $\hat{\Omega}[\{\mu(1),\cdot\}_\circ,\dg]$ in terms of
$\mu$, $\bar{\mu}$, and their variations. Since we are only considering variations $\dg$
that satisfy $0 = \dg s_R = \dg \rho_0$ and since $0 = \{\mu, s_R^i\}_\circ 
= \{\mu, \rho_0\}_\circ$ 
\be
\hat{\Omega}_R[\{\mu(1),\cdot\}_\circ,\dg] = C_R[\{\mu(1),\cdot\}_\circ,\dg] 
= \frac{1}{32\pi G} \int_{S_R} \eg \int_{r_0}^1 r^2
\dg_1 e^{ij} \di_r \dg_2 e_{ij} dr - (1 \leftrightarrow 2),
\ee
with $\dg_1 = \{\mu(1), \cdot\}_\circ$ and $\dg_2 = \dg$. To this we must add an entirely 
analogous expression contributed by $\cN_L$.

[Note that while $\dg_2$ is smooth, admissible, and real in the sense that $\dg_2\bar{\mu}$ is 
the complex conjugate of $\dg_2\mu$, $\dg_1 = \{\mu(1), \cdot\}_\circ$ need not have any of 
these properties (and indeed, it turns out to have none of them). In particular we shall see 
that $\dg_1\bar{\mu}$ is {\em not} the complex conjugate of $\dg_1\mu$, so this variation takes 
us outside the space of real metric fields. All the same there is no ambiguity in the 
definition of $\dg_1 e_{ij}$. Even for real variations, such as $\dg$, $\dg\mu$ and 
$\dg\bar{\mu}$ are {\em linearly} independent, that is, no non-zero pair of constants $a, b$ 
exist such that at a given point $a\dg \mu + b\dg \bar{\mu} = 0$ for all possible real $\dg$.
$e_{ij}$ thus has uniquely defined partial derivatives by $\mu$ and $\bar{\mu}$, which in turn
define $\dg_1 e_{ij}$ unambiguously, even when $\dg_1\bar{\mu} \neq \overline{\dg_1\mu}$.] 

In terms of $\mu$
        \bearr
                \lefteqn{\delta_1 e^{ij} \di_r \delta_2 e_{ij} - (1 \leftrightarrow 2)}	
				\nonumber\\ 
& = & -4\frac{1}{(1 - |\mu|^2)^2}\{\delta_1 \mu \di_r \delta_2 \bar{\mu}
		+ \delta_1 \bar{\mu} \di_r \delta_2 \mu \}	\nonumber\\
&& -8\frac{1}{(1 - |\mu|^2)^3}[\mu\di_r \bar{\mu} - \bar{\mu}\di_r \mu] 
                \delta_1\mu\delta_2 \bar{\mu} - (1 \leftrightarrow 2). \label{mu_integrand1}
		\eearr

[This can be obtained as follows: By a direct calculation from (\ref{mu_parametrization2}) one 
sees that 
        \be     \label{Dg_e_sq}
                \Dg e_{ij} \Dg e^{ij} = -8 \frac{1}{(1 - |\mu|^2)^2}\Dg \mu \Dg \bar{\mu},
        \ee
for any variation $\Dg$. (This is valid even when $\Dg\bar{\mu}$ is not the complex conjugate of
$\Dg\mu$.) Now, substituting $\Dg_1 + \Dg_2$ for $\Dg$ in (\ref{Dg_e_sq}) and 
taking the part linear in $\Dg_1$ one obtains
        \be     \label{Dg_1_Dg_2}
                \Dg_1 e_{ij} \Dg_2 e^{ij} = -4 \frac{1}{(1 - |\mu|^2)^2}
                \{\Dg_1 \mu \Dg_2 \bar{\mu} + \Dg_1 \bar{\mu} \Dg_2 \mu \}.       
        \ee
In particular, with $\Dg_1 = \di_r$ and $\Dg_2 = \dg_1$
        \be     \label{scndterm}
                \di_r e_{ij} \delta_1 e^{ij} = -4 \frac{1}{(1 - |\mu|^2)^2}
                \{\di_r \mu \delta_1 \bar{\mu} + \di_r \bar{\mu} \delta_1 \mu \}.       
        \ee
Finally, by acting with $\delta_2$ on (\ref{scndterm}) and anti-symmetrizing
with respect to interchange of $\delta_1$ and $\delta_2$ one obtains (\ref{mu_integrand1}).]

If we let         
\be   \label{alpha_def}
                \ag(r, y) = \int_{r_0}^r  \frac{1}{1 - |\mu|^2}
                                        [\bar{\mu}\di_r \mu - \mu\di_r \bar{\mu}]\ dr,
        \ee
where the integral is taken at constant $y^i$, that is, along a 
generator of $\cN$, then
        \bearr
                \lefteqn{\delta_1 e^{ij} \di_r \delta_2 e_{ij} - (1 \leftrightarrow 2)}	
				\nonumber \\ 
& = & -4\{\frac{e^\ag\delta_1 \mu}{1 - |\mu|^2} 
				\di_r [\frac{e^{-\ag}\delta_2 \bar{\mu}}{1 - |\mu|^2}]
      + \frac{e^{-\ag}\delta_1 \bar{\mu}}{1 - |\mu|^2} 
				\di_r [\frac{e^\ag\delta_2 \mu}{1 - |\mu|^2}]\}
        - (1 \leftrightarrow 2).
        \eearr
(Note that $\ag$ is pure imaginary, so $e^\ag$ is a phase).

This invites us to define, for any variation $\Dg$,
        \be		\label{Box_def}
                \Box = \frac{\sqrt{\rho_y} e^\ag}{1 - |\mu|^2}\Delta \mu\ \ \ \
 \ \barbox = \frac{\sqrt{\rho_y} e^{-\ag}}{1 - |\mu|^2}\Delta \bar{\mu}.
        \ee
$\barbox$ is a {\em formal} complex conjugate of $\Box$, obtained from $\Box$ 
via the substitutions $\mu \leftrightarrow \bar{\mu}$ and $\Dg\mu \leftrightarrow \Dg\bar{\mu}$.
When $\Dg$ is a real variation, so that $\Dg\bar{\mu} = \overline{\Dg\mu}$, $\barbox$ really
is the complex conjugate of $\Box$, but for variations, such as $\dg_1$, which are not real
this is not the case.

With these definitions
        \bearr
C[\{\mu(1),\cdot\}_\circ,\dg] =
           -\frac{1}{8\pi G}[\!\!\!\!\!&& \int_{S_R} dy^1_R \wedge dy^2_R  \int_{r_{R\,0}}^1 
			\Box_{R\,1} \di_r \barbox_{R\,2} + \barbox_{R\,1}\di_r\Box_{R\,2} dr 
				\nonumber \\
			& +\!\! & \int_{S_L} dy^1_L \wedge dy^2_L  \int_{r_{L\,0}}^1 
			\Box_{L\,1} \di_r \barbox_{L\,2} + \barbox_{L\,1} \di_r \Box_{L\,2} dr 
				\nonumber \\
                & -\!\! & (1 \leftrightarrow 2)].        \label{Omega_Box}
        \eearr
($\Box_L$ and $\Box_R$ signify $\Box$ formed according to (\ref{Box_def}) from fields on 
$\cN_L$, referred to the $a_L$ chart, and fields on $\cN_R$, referred to the $a_R$ chart, 
respectively.)

The contribution of either branch to (\ref{Omega_Box}) may be integrated by parts in two 
different ways:
\bearr
\int_{r_0}^1 \Box_1 \di_r \barbox_2 + \barbox_1 \di_r \Box_2 
- \Box_2 \di_r \barbox_1 - \barbox_2 \di_r \Box_1 dr
& = & 2 \int_{r_0}^1 \Box_1 \di_r \barbox_2 + \barbox_1 \di_r \Box_2 dr \nonumber \\
&& + [\Box_1 \barbox_2 + \barbox_1 \Box_2]_{r_0} \label{parts_int1}, \\
\mbox{or} && \nonumber \\
 & = & -2 \int_{r_0}^1 \Box_2 \di_r \barbox_1 + \barbox_2 \di_r \Box_1 dr	\nonumber \\ 
 && -  [\Box_1 \barbox_2 + \barbox_1 \Box_2]_{r_0}.
\label{parts_int2}
\eearr
In both these expressions the boundary term at $r = 1$ (i.e. at $\di\cN$) vanishes
because $\dg e_{ij}$ and thus $\Box_2$ and $\barbox_2$ vanish on $\di\cN$.

$C[\{\mu(1),\cdot\}_\circ,\dg]$ can be put in a form that facilitates the solution
of (\ref{auxbracket2def}) by integrating the contribution of $\cN_R$ by parts one way, according
to (\ref{parts_int2}), and the contribution of $\cN_L$ the other way, according to 
(\ref{parts_int1}). For variations $\dg$ satisfying $0 = \dg \rho_0 = \dg s_R = \dg s_L$ the 
surface terms of the two contributions then cancel: By (\ref{Box_def}) and (\ref{Dg_1_Dg_2})
\bearr
\Box_1 \barbox_2 + \barbox_1 \Box_2 
& = & \frac{\rho_y}{(1 - |\mu|^2)^2}[\dg_1 \mu \dg_2 \bar{\mu} + \dg_1 \bar{\mu} \dg_2 \mu]\\
& = & -\frac{1}{4} \rho_y \dg_1 e_{ij} \dg_2 e^{ij}. \label{surface_term_e_form}
\eearr
Since both $\dg_1 = \{\mu(1),\cdot\}_\circ$ and $\dg_2 = \dg$ leave $\rho_0$, $y_R$ and $y_L$ 
invariant $\dg_1^{a_R} = \dg^b_1 = \dg_1^{a_L}$ and $\dg_2^{a_R} = \dg^b_2 = \dg_2^{a_L}$. The 
fact that $\dg_1$ and $\dg_2$ both respect the constraint (\ref{e_consistency}) thus implies 
that 
\be
\int_{S_0} \rho_{y\,R}\: \dg_1^{a_R} e_{R\,ij}\, \dg_2^{a_R} e_R{}^{ij}\:dy_R^1\wedge dy_R^2
= \int_{S_0} \rho_{y\,L}\: \dg_1^{a_L} e_{L\,mn}\, \dg_2^{a_L} e_L{}^{mn}\:dy_L^1\wedge dy_L^2.
\ee
That is, the surface terms cancel and
        \bearr
C[\{\mu(1),\cdot\}_\circ,\dg] =
        \frac{1}{4\pi G}[\!\!\!\!\! && + \int_{S_R} dy^1_R \wedge dy^2_R  \int_{r_0}^1 
	\Box_{R\,2} \di_r \barbox_{R\,1} + \barbox_{R\,2} \di_r \Box_{R\,1} dr	\nonumber \\
		&& - \int_{S_L} dy^1_L \wedge dy^2_L  \int_{r_0}^1 
		\Box_{L\,1} \di_r \barbox_{L\,2} + \barbox_{L\,1} \di_r \Box_{L\,2} dr ].
        \label{Omega_Box2}
        \eearr
(Further on we shall need to evaluate $C[\{\mu(1),\cdot\}_\circ,\dg]$ for arbitrary admissible 
$\dg$. It is therefore worth noting that (\ref{Omega_Box}), (\ref{parts_int1}), 
(\ref{parts_int2}), and (\ref{surface_term_e_form}) hold for {\em all} admissible $\dg$.)

We are now ready to solve 
\be		\label{mu_brack_eq}
\dg\mu(1) = \hat{\Omega}[\{\mu(1),\cdot\}_\circ,\dg]
\ee
for $\{\mu(1),\mu(2)\}_\circ$ and $\{\mu(1),\bar{\mu}(2)\}_\circ$.
Since $1 \in \cN_R - S_0$ our causality requirement implies that both these brackets vanish 
when $2 \in \cN_L - S_0$, because $2$ is then outside the causal domain of influence, $J[1]$, of
$1$. (See appendix \ref{nullhypersurfaces} prop. \ref{domain_of_influence_on_N}). Thus in the 
second term of (\ref{Omega_Box2}), the contribution of 
$\cN_L$, $\Box_{L\,1}$ and $\barbox_{L\,1}$ can have support only on $S_0$ itself.
Could they be singular distributions supported there? $\dg \mu_L$ and $\dg \bar{\mu}_L$ (and
therefore $\di_r\Box_{L\,2}$ and $\di_r\barbox_{L\,2}$) can be varied freely in a 
neighbourhood of $S_0$ in $\cN_L$.\footnote{
$\dg \mu_L$ and $\dg \bar{\mu}_L$ are not independent since one is the complex conjugate of the 
other, but they are {\em linearly} independent, that is no non-zero pair of constants $a, b$ 
exist such that at a given point $a\dg \mu_L + b\dg \bar{\mu}_L = 0$ for all possible 
variations. This is sufficient for our purposes.}
Moreover, both are independent of $\dg \mu(1)$. It follows
that $\Box_{L\,1}$ and $\barbox_{L\,1}$ must vanish as distributions on $\cN_L-\di\cN$, and that
the contribution of $\cN_L$ to (\ref{Omega_Box2}) vanishes. 

On $\cN_R$ $\Box_{R\,2}$ and $\barbox_{R\,2}$ are freely variable except that they must vanish 
on $\di\cN$. Thus (\ref{mu_brack_eq}) requires that 
\be		\label{mu_mu_brack_0}
\di_r \Box_{R\,1} = 0
\ee
and
\be		\label{mu_mubar_brack_0}
\frac{\sqrt{\rho_y}\, e^\ag}{1 - |\mu|^2} \di_r \barbox_{R\,1} =  4\pi G \dg^2(y - y(1))
\dg (r - r(1)) \sg
\ee
as distributions on $\cN_R - \di\cN$. Here $\sg$ is $+1$ if $r$ increases away from $S_0$
(i.e. if $r_0 < 1$) and $-1$ if $r$ decreases away from $S_0$ ($r_0 > 1$). This factor
must be inserted because the integral over an interval $[a, b]$ of a delta distribution with 
support between $a$ and $b$ is $+1$ if $b > a$ and $-1$ if $b < a$.

(\ref{mu_mu_brack_0}) and (\ref{mu_mubar_brack_0}) are satisfied by
\bearr
\{\mu(1),\mu(2)\}_\circ & = & 0  \label{mu_mu_brack_1}\\
\{\mu(1),\bar{\mu}(2)\}_\circ 
& = & 2\pi G\, \dg^2(y(2) - y(1))\, sgn(r(2) - r(1)) \sg \nonumber \\
&&[\frac{1 - |\mu|^2}{\sqrt{\rho_y}}]_1 [\frac{1 - |\mu|^2}{\sqrt{\rho_y}}]_2
\,e^{\int_{r(1)}^{r(2)} \frac{1}{1 - |\mu|^2}[\bar{\mu}\di_r \mu - \mu\di_r \bar{\mu}] dr} 
\label{mu_mubar_brack_1} 
\eearr
for $1 \in \cN_R - S_0$ and $2 \in \cN_R$, where
\be
sgn(r(2) - r(1)) = \left\{\begin{array}{cl} 
+1 & \mbox{if $r(2) > r(1)$} \\
-1 & \mbox{if $r(2) < r(1)$} \end{array}\right.
\ee 

Notice that when $1$ and $2$ lie on the same generator $sgn(r(2) - r(1)) \sg = sgn(1,2)$ and 
that $\int_{r(1)}^{r(2)} \frac{1}{1 - |\mu|^2}[\bar{\mu}\di_r \mu - \mu\di_r \bar{\mu}] dr
= \int_1^2 \frac{1}{1 - |\mu|^2}[\bar{\mu} d\mu - \mu d\bar{\mu}]$, so (\ref{mu_mubar_brack_1})
differs from (\ref{mu_mubar_bracket}) only in that the $y_R$ chart is used to label
the generators instead of the $\theta$ chart.

These brackets are not the only possible solutions to (\ref{mu_brack_eq}). (\ref{mu_mu_brack_0})
and (\ref{mu_mubar_brack_0}) allow the addition of functions independent of $r$ to 
$\Box_{R\,1}$ and to $\barbox_{R\,1}$. The requirement that the brackets be antisymmetric 
reduces this freedom. In the case of $\{\mu(1),\mu(2)\}_\circ$ it eliminates the ambiguity all 
together - (\ref{mu_mu_brack_1}) is the only possibility. In the case of 
$\{\mu(1),\bar{\mu}(2)\}_\circ$ one is still free to replace 
$\dg^2(y(2) - y(1))\, sgn(r(2) - r(1)) \sg$ in \ref{mu_mubar_brack_1} by 
$\dg^2(y(2) - y(1))\, sgn(r(2) - r(1)) \sg + i \Upsilon(y(1),y(2))$, where $\Upsilon$ is real.
Causality requires that $\Upsilon$ is a singular distribution with support only on 
$y(1) = y(2)$, and covariance under transformations of the $y$ coordinates requires that it 
transforms as $\dg^2(y(2) - y(1))$ does. Some possibilities are 
$\Upsilon = \mbox{constant} \times \dg^2(y(2) - y(1))$ or 
$\Upsilon = \lam(y) \times \dg^2(y(2) - y(1))$. 

The freedom in the bracket $\{\mu(1),\bar{\mu}(2)\}_\circ$ propagates to other brackets. 
Here the simplest bracket, with $\Upsilon = 0$, will be adopted since it seems 
the simplest and most natural seeming choice.

What about the brackets between $\mu_R$ on $\cN_R - S_0$ and $\mu_L$ and $\bar{\mu}_L$ on 
$\cN_L$? Causality and the constraint (\ref{e_consistency}) imply that
\bearr
\{\mu_R(1),\mu_L(2)\}_\circ   & = &  0	\label{muR_muL_brack}\\
\{\mu_R(1),\bar{\mu}_L(2)\}_\circ & = & \left\{\begin{array}{cl} 
[\frac{\di\bar{\mu}_L}{\di\bar{\mu}_R}]_2\{\mu_R(1),\bar{\mu}_R(2))\}_\circ  & 
\mbox{for $2 \in S_0$} \\
0 & \mbox{for $2 \in \cN_L - S_0$} \end{array}\right.	\label{muR_mubarL_brack}
\eearr
The case $2 \in S_0$ requires explanation. In this case the constraint (\ref{e_consistency}) 
implies that $\mu_L(2)$ may be expressed as a function of $\mu_R(2)$, and the coordinate 
transformation from the complex coordinate $z_R = y^1_R + i y^2_R$ to $z_L = y^1_L + i y^2_L$:
\be		\label{mu_transformation2}
\mu_L(\mu_R) = [\frac{\di z_R}{\di \bar{z}_L} + \mu_R\frac{\di \bar{z}_R}{\di \bar{z}_L}]/
		[\frac{\di z_R}{\di z_L} + \mu_R\frac{\di \bar{z}_R}{\di z_L}],
\ee
(see (\ref{mu_transformation})). Similarly $\bar{\mu}_L$ may be expressed in terms of 
$\bar{\mu}_R$, and the 
coordinate transformation, via the formal complex conjugate of (\ref{mu_transformation2}).
Since $\mu_R(1)$ commutes with $\mu_R$, $s_L$, $s_R$ and $\rho_0$
\be
\{\mu_R(1),\mu_L(\mu_R;2)\}_\circ = 0
\ee
and
\be 
\{\mu_R(1),\bar{\mu}_L(\bar{\mu}_R;2)\}_\circ 
= [\frac{\di\bar{\mu}_L}{\di\bar{\mu}_R}]_2\{\mu_R(1),\bar{\mu}_R(2))\}_\circ.
\ee
(Note that we are using the fact that by (\ref{mu_transformation2}) $\mu_L$ depends 
only on $\mu_R$ and the coordinate transformation, and not on $\bar{\mu}_R$.)
The fact that $\{\mu_R(1),\bar{\mu}_L(2)\}_\circ$ does not vanish on all of $\cN_L$
does not contradict the requirement that $\barbox_{L,1}$ vanishes as a distribution,
for if $\barbox_{L,1}$ as defined by (\ref{muR_mubarL_brack}), (\ref{mu_transformation2}), 
and (\ref{mu_mubar_brack_1}) is integrated against a smooth function on $\cN_L$ the result 
vanishes. 

Note that although the variation of $e$ produced by $\{\mu_R(1), \cdot\}_\circ$ is discontinuous
at $S_0$ (\ref{muR_mubarL_brack}) assures that $\{\mu_R(1), \cdot\}_\circ$ respects the 
constraint (\ref{e_consistency}), for the constraint is a requirement of consistency rather than
one of continuity. The constraint requires that $e_{R\,ij}$ and $e_{L\,mn}$ define the same 
$e_{ab}$ on $S_0$, and $\{\mu_R(1),\cdot\}_\circ$ respects this condition.

Instead of parametrizing $e$ by $\mu_R$ on $\cN_R$ and by $\mu_L$ on $\cN_L$, one may 
parametrize $e$ on all of $\cN$ by $\mu$ referred to the $b$ chart. When solving 
(\ref{auxbracket2def}) we wished to use variables with vanishing variation on $\di\cN$ under 
admissible variations of the solution. 
However, now that we have solved (\ref{auxbracket2def}) for the brackets between the 
$\mu$ and $\bar{\mu}$ fields nothing stops us from transforming these brackets into the 
$b$ chart. Indeed this transformation simplifies the 
brackets somewhat. 

To carry out the transformation first note that since the brackets of $\mu$ (and $\bar{\mu}$) 
with $s_R^i$, $s_L^m$, and $\rho_0$ vanish the transformation from the $a$ to the $b$ 
coordinates, although really field dependent, can be treated as field independent when 
transforming the brackets of $\mu$ and $\bar{\mu}$. Second, recall that the only difference 
between (\ref{mu_mubar_brack_1}) and (\ref{mu_mubar_bracket}) for points on $\cN_R$ is that in 
the former the $y_R$ chart is used to label the generators while in the latter these are 
labelled by the $\theta$ coordinates. The same is true for the two versions, 
(\ref{mu_mu_brack_1}) and (\ref{mu_mu_brack}), of the bracket between $\mu(1)$ and 
$\mu(2)$ on $\cN_R$. But we have shown that (\ref{mu_mu_brack}) and (\ref{mu_mubar_bracket}) are
invariant under arbitrary diffeomorphisms of the $\theta$ coordinates, and thus in particular 
under the transformation from the $\theta$ chart to the $y_R$ chart and {\em vice. versa}. 
Transforming (\ref{mu_mu_brack_1}) and (\ref{mu_mubar_brack_1}) - and their analogs on $\cN_L$ 
- to the $\theta$ chart thus yields exactly (\ref{mu_mu_brack}) and (\ref{mu_mubar_bracket}),
\bearr
\{\mu(1),\mu(2)\}_\circ & = & \{\bar{\mu}(1),\bar{\mu}(2)\}_\circ = 0,	\label{mu_mu_brack2}\\
\{\mu(1),\bar{\mu}(2)\}_\circ 
& = & 2\pi G\, \dg^2(\theta(2) - \theta(1))\, sgn(1,2) \nonumber \\
&& [\frac{1 - |\mu|^2}{\sqrt{\rho}}]_1 [\frac{1 - |\mu|^2}{\sqrt{\rho}}]_2\:
e^{\int_1^2 \frac{1}{1 - |\mu|^2}[\bar{\mu}d\mu - \mu d\bar{\mu}]},
\label{mu_mubar_bracket2}
\eearr  
for points $1$ and $2$ belonging to the same branch of $\cN$ with at most one of these lying on 
$S_0$. (\ref{muR_muL_brack}) and (\ref{muR_mubarL_brack}) show that the brackets between these 
fields at points $1$ and $2$ belonging to distinct branches, with neither lying on $S_0$, 
vanishes. If (\ref{mu_mu_brack2}) and (\ref{mu_mubar_bracket2}) hold then the formulae
(\ref{muR_muL_brack}) and (\ref{muR_mubarL_brack}) for the brackets of $\mu_R(1)$ with 
$\mu_L(2)$ and $\bar{\mu}_L(2)$ for $2 \in S_0$ follow as automatic consequences. 

(\ref{mu_mu_brack2}) and (\ref{mu_mubar_bracket2}) also suggest values for the brackets when $1$
and $2$ both belong to $S_0$. They suggest that 
\be	\label{e_e_S0_brack}
0 = \{\mu(1),\mu(2)\}_\circ = \{\mu(1),\bar{\mu}(2)\}_\circ 
= \{\bar{\mu}(1),\bar{\mu}(2)\}_\circ
\ee
for $1, 2 \in S_0$.
In applying (\ref{mu_mu_brack2}) and (\ref{mu_mubar_bracket2}) to the case $1,2 \in S_0$ we are
of course extrapolating. These equations have not yet been shown to be consistent with 
(\ref{auxbracket2def}) in this case. Moreover, (\ref{mu_mubar_bracket2}), which is unambiguous 
when $\bar{\mu}(2)$ is smeared with a smooth function on $\cN$, is ambiguous when $1$ and $2$ 
are restricted to $S_0$ because the discontinuous function $sgn(1,2)$ has not been defined 
except when $1$ and $2$ lie on the same generator. If this function is discontinuous on $S_0$ 
(which $sgn(r(2) - r(1))\sg$ is) then the product $\dg^2(\theta(1) - \theta(2)) sgn(1,2)$ is 
ambiguous. We therefore need to check the compatibility of the proposed brackets 
(\ref{e_e_S0_brack}) with (\ref{auxbracket2def}).

As a first step let us verify that $\hat{\Omega}[\{\mu_R(1),\cdot\}_\circ,\dg] = \dg\mu_R(1)$ 
for $1 \in S_0$ is satisfied for admissible $\dg$ such that $\dg s_R^i = \dg s_L^m = \dg 
\rho_0 = 0$ provided all the expressions (\ref{mu_mu_brack_1}), (\ref{mu_mubar_brack_1}), 
(\ref{muR_muL_brack}), and (\ref{muR_mubarL_brack}) for the brackets among the $\mu$ and 
$\bar{\mu}$ fields obtained so far, as well as the proposed brackets (\ref{e_e_S0_brack}), hold.

To do this we integrate both terms in (\ref{Omega_Box}) by parts according to 
(\ref{parts_int1}). Then 
        \bearr
C[\{\mu_R(1),\cdot\}_\circ,\dg] =
      - \frac{1}{4\pi G}[\!\! &&\! \int_{S_R} dy^1_R \wedge dy^2_R  \int_{r_0}^1 
	\Box_{R\,1} \di_r \barbox_{R\,2} + \barbox_{R\,1} \di_r \Box_{R\,2} dr	\nonumber \\
				& + &\!\int_{S_L} dy^1_L \wedge dy^2_L  \int_{r_0}^1 
	\Box_{L\,1} \di_r \barbox_{L\,2} + \barbox_{L\,1} \di_r \Box_{L\,2} dr 	\nonumber \\
				& + \:\frac{1}{2} &\!\int_{S_0} dy_R^1 \wedge dy_R^2\:
			(\Box_{R\,1}\barbox_{R\,2} + \barbox_{R\,1} \Box_{R\,2}) \nonumber \\
				& + \:\frac{1}{2} &\!\int_{S_0} dy_L^1 \wedge dy_L^2\:(
			\Box_{L\,1} \barbox_{L\,2} + \barbox_{L\,1} \Box_{L\,2})].
        \label{Omega_Box3}
        \eearr
(Recall that $\Box_1$ corresponds to $\dg_1 \equiv \{\mu_R(1),\cdot\}_\circ$ and $\Box_2$ to
$\dg_2 \equiv \dg$.)

By (\ref{e_e_S0_brack}) the $S_0$ integrals vanish. (\ref{mu_mu_brack_1}) and its analog for 
$\cN_L$ imply that
\be	
\Box_{R\,1} = \Box_{L\,1} = 0
\ee
and (\ref{mu_mubar_brack_1}) implies that\footnote{
Recall that $\barbox$ is need not be the complex conjugate of $\Box$.}
\bearr	
\barbox_{R\,1} & = & \left[\frac{\sqrt{\rho_y}\, e^{-\ag}}{1 - |\mu|^2}\right]_R 
\{\mu_R(1),\bar{\mu}_R\}_\circ \\
& = & 2\pi G \left[\frac{1 - |\mu|^2}{\sqrt{\rho_y}}\right]_{1\,R} 
\dg^2(y_R - y_R(1))\label{bboxR1}
\eearr
on $\cN_R - S_0$, and (\ref{e_e_S0_brack}) implies that it is zero on $S_0$ itself. Therefore
\bearr
\lefteqn{-\int_{S_R} dy^1_R \wedge dy^2_R  \int_{r_0}^1 
		\Box_{R\,1} \di_r \barbox_{R\,2} + \barbox_{R\,1} \di_r \Box_{R\,2}\, dr}
\nonumber \\
& = & - 2\pi G \left[\frac{1 - |\mu|^2}{\sqrt{\rho_y}}\right]_{1\,R} [\int_{r_0}^1 \di_r 
\Box_{R\,2}\, dr]_{y = y(1)}	\\
& = &  2\pi G \left[\frac{1 - |\mu|^2}{\sqrt{\rho_y}}\right]_{1\,R} \Box_{R\,2}(r_0)	\\
& = &  2\pi G\:\dg^{a_R} \mu_R(1). \label{C_Rbulk_value}
\eearr
Similarly, at any point of $\cN_L$
\be
\barbox_{L\,1} = \left[\frac{\sqrt{\rho_y}\, e^{-\ag}}{1 - |\mu|^2}\right]_L 
\{\mu_R(1),\bar{\mu}_L\}_\circ
\ee
The bracket $\{\mu_R(1),\bar{\mu}_L(2)\}_\circ$ with $1 \in S_0$ and $2 \in \cN_L - S_0$ may be 
evaluated using the complex conjugate of the analog for $\cN_L$ of (\ref{muR_mubarL_brack}), 
yielding 
\be
\barbox_{L\,1} = \left[\frac{\sqrt{\rho_y}\, e^{-\ag}}{1 - |\mu|^2}\right]_L 
		\left[\frac{\di\mu_R}{\di\mu_L}\right]_1 \{\mu_L(1),\bar{\mu}_L\}_\circ.
\ee
It follows, by the analog of the calculation carried out in 
(\ref{bboxR1} - \ref{C_Rbulk_value}), that
\bearr
- \int_{S_L} dy^1_L \wedge dy^2_L  \int_{r_0}^1 \Box_{L\,1} \di_r \barbox_{L\,2} 
+ \barbox_{L\,1} \di_r \Box_{L\,2}\, dr
& = & 2\pi G\:[\frac{\di\mu_R}{\di\mu_L}]_1 \dg^{a_L} \mu_L(1) \\
& = & 2\pi G\:\dg^{a_L} \mu_R(1).\label{C_Lbulk_value}
\eearr
$\dg^{a_L} \mu_R(1)$ is $\dg^{a_L} \mu$ evaluated in the $a_R$ chart at the values $a_R^\ag(1)$
of the $a_R$ coordinates. By (\ref{mu_transformation}) $\mu$ transforms autonomously under 
change of coordinates, i.e. $\mu$ in the new chart depends only on $\mu$ in the old chart and on
the transformation between the charts - it does not depend on other fields, such as $\bar{\mu}$,
in the old coordinates. $\dg^{a_L} \mu_R(1)$ is thus defined by (\ref{dgwdef}). This definition 
implies that
\be
\dg^{a_L} \mu_R = \di \mu_R/\di \mu_L \dg^{a_L} \mu_L, 
\ee
where in the partial derivative $\mu_R$ is $\mu$ referred to to the {\em fixed} chart $a_{R\,0}$
which agrees with $a_R$ at the fiducial solution, and similarly $\mu_L$ is referred to the 
fixed chart $a_{L\,0}$ defined by to $a_L$ at the fiducial solution. The partial derivative is
of course evaluated at fixed $a_{L\,0}$ (and thus fixed $a_{R\,0}$). 

The sum of all the terms in (\ref{Omega_Box3}) is thus 
\be	\label{C_eval}
\frac{1}{2} \dg^{a_R} \mu_R(1) + \frac{1}{2} \dg^{a_L} \mu_R(1).
\ee
Note that this evaluation of $C[\{\mu_R(1),\cdot\}_\circ,\dg]$ is valid for all admissible $\dg$
- the hypotheses that $\dg s_R^i = \dg s_L^m = \dg \rho_0 = 0$ have not been used.
At present however we are interested in the case that these hypotheses do hold.
Then $\dg^{a_R} = \dg^b = \dg^{a_L}$ and the sum (\ref{C_eval}) of the terms is
$\dg\mu_R(1)$ as required. That is, with the brackets we have adopted 
$\hat{\Omega}[\{\mu_R(1),\cdot\}_\circ,\dg] = \dg\mu_R(1)$ for all $1 \in int\cN$ and all 
admissible $\dg$ satisfying $\dg s_R^i = \dg s_L^m = \dg \rho_0 = 0$.

But of course this condition must hold also for $\dg$ such that $\dg s_R^i$, $\dg s_L^m$, and 
$\dg \rho_0$ are {\em not} all zero. Considering such variations allows us to determine
the brackets between $e$ and $\hat{\omega}_{R\,i}$, $\hat{\omega}_{L\,m}$, and $\lam$.

Let us begin with the case $1 \in S_0$. Instead of working with $\mu_R$ it is somewhat more 
convenient to use the conformal 2-metric in $\theta$ coordinates, $e_{ab}$. Thus we look for 
brackets that verify the relation $\hat{\Omega}[\{e_{ab}(1),\cdot\}_\circ,\dg] = \dg e_{ab}(1)$.

To evaluate $\hat{\Omega}[\{e_{ab}(1),\cdot\}_\circ,\dg]$ we begin by noting that
\be
B[\{e_{ab}(1),\cdot\}_\circ,\dg] = 0
\ee
since $e_{ab}(1)$ on $S_0$ commutes with $e$ on $S_0$ and with $\rho_0$, $s_R$, and $s_L$.
Furthermore
\bearr
\lefteqn{A[\{e_{ab}(1),\cdot\}_\circ,\dg]}\nonumber \\
&& = \frac{1}{16\pi G}\int_{S_0} 2 \{e_{ab}(1),\lam\}_\circ \dg\rho_0
+ \{e_{ab}(1),\hat{\omega}_{R\,i}\}_\circ\dg s_R^i
+ \{e_{ab}(1),\hat{\omega}_{L\,m}\}_\circ\dg s_L^m
\: d\theta^1\wedge d\theta^2 \nonumber\\
&&
\eearr
and we have essentially already evaluated $C[\{e_{ab}(1),\cdot\}_\circ,\dg]$.
The only brackets occurring in $C[\{e_{ab}(1),\cdot\}_\circ,\dg]$ are of the form
$\{e_{ab}(1),e_{A\,kl}\}_\circ$ with $A = L$ or $R$. $e_{ab}$ is a function of $\mu_R$, 
$\bar{\mu}_R$, and $s_R$ and $\rho_0$. Therefore, since $e_{A\,kl}$ commutes with the latter 
two fields,
\be
\{e_{ab}(1),e_{A\,kl}\}_\circ =  [\frac{\di e_{ab}}{\di\mu_R}]_1 \{\mu_R(1),e_{A\,kl}\}_\circ
		  + \mbox{complex conjugate}
\ee
and thus
\be
C[\{e_{ab}(1),\cdot\}_\circ,\dg] = [\frac{\di e_{ab}}{\di\mu_R}]_1 
C[\{\mu_R(1),\cdot\}_\circ,\dg]
		  + \mbox{complex conjugate}.
\ee
Now by (\ref{C_eval})
\be
C[\{\mu_R(1),\cdot\}_\circ,\dg] = \frac{1}{2}[\dg^{a_R}\mu_R 
+ \dg^{a_L}\mu_R]_1.
\ee
Hence
\bearr
C[\{e_{ab}(1),\cdot\}_\circ,\dg] & = & \frac{1}{2}
[\frac{\di e_{ab}}{\di\mu_R}(\dg^{a_R} + \dg^{a_L}) \mu_R]_1 + \mbox{complex conjugate}. \\
				 & = & \frac{1}{2}(\dg^{a_R} + \dg^{a_L}) e_{ab}(1).
\eearr
But 
\bearr
\dg^a e_{ab} & = & \dg^y e_{ab} - \dg^y r \di_r e_{ab}	\label{dga_expansion0}\\
	     & = & \dg^b e_{ab} - {\pounds_{\xi_\perp}} e_{ab} 
- \frac{v}{2\rho_0}[\dg^b - {\pounds_{\xi_\perp}}]\rho_0\: \di_v e_{ab},\label{dga_expansion}
\eearr 
and
\be
{\pounds}_{\xi_\perp}\rho_0 = \di_c[\rho_0\xi^c_\perp],
\ee
with $\xi_\perp^i = \dg s^i$ and $v = r/r_0$ the $v$ coordinate that parametrizes the generators
of the branch under consideration. 
$\hat{\Omega}[\{e_{ab}(1),\cdot\}_\circ,\dg] = \dg e_{ab}(1)$ therefore requires
\bearr
0 & = & [- \frac{1}{2}{\pounds_{\xi_{\perp R}}} e_{ab} 
+ \frac{1}{4}\frac{\di_c[\rho_0 \xi_{\perp R}^c]}{\rho_0} \di_{v^-} e_{ab} 
- \frac{1}{4}\frac{\dg^b \rho_0}{\rho_0} \di_{v^-} e_{ab}]_1 \nonumber\\
&& + \frac{1}{16\pi G}\int_{S_0} \{e_{ab}(1),\lam\}_\circ \dg\rho_0
+ \{e_{ab}(1),\hat{\omega}_{R\,i}\}_\circ\dg s_R^i
\: d\theta^1\wedge d\theta^2 \nonumber\\
&& +\ \mbox{corresponding $L$ branch terms}.
\eearr  
From this one obtains the brackets (\ref{lam_e_0_bracket}) and (\ref{omega_R_e_0_bracket}), and
the $L$ analog of the latter:
\bearr
\{e_{ab}(1),\lam(2)\}_\circ 
& = & 2\pi G \frac{1}{\rho_0}[\di_{v^-} e_{ab} + \di_{v^+} e_{ab}] \dg^2(\theta(2) - \theta(1)),
\label{lam_e_0_bracket1}\\
\{e_{ab}(1), \hat{\omega}_R[f]\}_\circ 
& = & 8\pi G [{\pounds}_f e_{ab} - \frac{1}{2}\di_{v^-} e_{ab}
\frac{{\pounds}_f \rho_0}{\rho_0}]_1,
\label{omega_R_e_0_bracket1}\\
\{e_{ab}(1),\hat{\omega}_L[g]\}_\circ 
& = & 8\pi G [{\pounds}_g e_{ab} - \frac{1}{2}\di_{v^+} e_{ab}
\frac{{\pounds}_g \rho_0}{\rho_0}]_1.
\label{omega_L_e_0_bracket1}
\eearr

Now let us consider the case $1 \in \cN_R - S_0$ again. We require
\be		\label{eR_brackets_def}
\hat{\Omega}[\{e_{R\,ij}(1),\cdot\}_\circ,\dg] = \dg e_{R\,ij}(1)
\ee
for all admissible variations $\dg$. Reviewing the derivation of (\ref{Omega_Box2}) one verifies
that for general admissible $\dg$
\be
C[\{e_{R\,ij}(1),\cdot\}_\circ,\dg] = \dg e_{R\,ij}(1) - \frac{1}{32\pi G} \int_{S_0}\eg
\{e_{R\,ij}(1),e^{ab}\}_\circ [\dg^{a_R} - \dg^{a_L}] e_{ab}.
\ee
The other terms in $\hat{\Omega}$ are
\bearr
\lefteqn{A[\{e_{R\,ij}(1),\cdot\}_\circ,\dg]}\nonumber \\
&& = \frac{1}{16\pi G}\int_{S_0} 2 \{e_{R\,ij}(1),\lam\}_\circ \dg\rho_0
+ \{e_{R\,ij}(1),\hat{\omega}_{R\,i}\}_\circ\dg s_R^i
+ \{e_{R\,ij}(1),\hat{\omega}_{L\,m}\}_\circ\dg s_L^m
\: d\theta^1\wedge d\theta^2 \nonumber\\
&&
\eearr
and  
\bearr
\lefteqn{B[\{e_{R\,ij}(1),\cdot\}_\circ,\dg]}\nonumber \\
&& = -\frac{1}{64\pi G}\int_{S_0} \{e_{R\,ij}(1),e^{ab}\}_\circ [\di_{v^-}e_{ab} \dg^{y_R}
+ \di_{v^+}e_{ab} \dg^{y_L}]\eg. 
\eearr
Thus, by (\ref{dga_expansion}), (\ref{eR_brackets_def}) requires
\bearr
0 = \int_{S_0} d\theta^1\wedge d\theta^2 \!\!\!\!\!&& ( 2 \{e_{R\,ij}(1),\lam\}_\circ 
\dg\rho_0 
\nonumber\\
&& + \{e_{R\,ij}(1),\hat{\omega}_{R\,k}\}_\circ\dg s_R^k
   + \{e_{R\,ij}(1),\hat{\omega}_{L\,m}\}_\circ\dg s_L^m \nonumber \\
&& - \frac{1}{2} \{e_{R\,ij}(1),e^{ab}\}_\circ \di_{v^+}e_{ab} \dg^{y_L}\rho_0 \nonumber \\
&& + \frac{1}{2} \rho_0 \{e_{R\,ij}(1),e^{ab}\}_\circ 
 [{\pounds}_{\xi_{\perp R}} - {\pounds}_{\xi_{\perp L}}] e_{ab}).
\eearr
From which it follows that 
\be	\label{eR_lam_bracket1}
\{e_{R\,ij}(1),\lam(2)\}_\circ = \frac{1}{4} 
\{e_{R\,ij}(1),e^{ab}(2)\}_\circ [\di_{v^+}e_{ab}]_2,
\ee
and
\bearr
\{e_{R\,ij}(1),\hat{\omega}_R[\xi_{\perp R}]\}_\circ & = & 
- \frac{1}{2}\int_{S_0} \eg \{e_{R\,ij}(1),e^{ab}\}_\circ {\pounds}_{\xi_{\perp R}} e_{ab}
\label{eR_omegaR_bracket1}\\
\{e_{R\,ij}(1),\hat{\omega}_L[\xi_{\perp L}]\}_\circ & = & 
  \frac{1}{2}\int_{S_0} \{e_{R\,ij}(1),e^{ab}\}_\circ 
(\eg {\pounds}_{\xi_{\perp L}} e_{ab} - {\pounds}_{\xi_{\perp L}}\eg \di_{v^+}e_{ab}) 
\label{eR_omegaL_bracket1}
\eearr
[Using
\be
{\pounds}_{\xi_{\perp}} e_{ab} = 2 D_{(a}[e_{b)c}\xi_\perp^c] - e_{ab} D_c \xi_\perp^c
\ee
(where $D$ is the covariant derivative defined by the metric $\rho_0 e$ on $S_0$) and the fact
that $\{e_{R\,ij}(1),e^{ab}\}_\circ$ vanishes on $\di S_0$ for all $1 \in int\cN_R$ because 
$\di S_0$ lies outside the causal domain of influence $J[1]$ of the point $1$, the last two 
brackets may be shown to be equivalent to 
\bearr
\{e_{R\,ij}(1),\hat{\omega}_{R\,k}(2)\}_\circ & = & [D_a[\{e_{R\,ij}(1),e^{ab}\}_\circ]\rho_0 
e_{bc}
\frac{\di\theta^c}{\di y_R^k}]_2	\label{eR_omegaR_bracket2}\\
\{e_{R\,ij}(1),\hat{\omega}_{L\,m}(2)\}_\circ & = & [- D_a[\{e_{R\,ij}(1),e^{ab}\}_\circ]\rho_0 
e_{bc}
\frac{\di\theta^c}{\di y_L^m} \nonumber \\
&&+ \frac{1}{2} \di_c[\{e_{R\,ij}(1),e^{ab}\}_\circ \di_{v^+}e_{ab}]\rho_0 
\frac{\di\theta^c}{\di y_L^m}]_2.\ \ ]	\label{eR_omegaL_bracket2}
\eearr

The brackets of $e_{ab}(1)$ and $\mu(1)$ with $\lam(2)$, $\hat{\omega}_R(2)$, and 
$\hat{\omega}_L(2)$ are easily obtained from (\ref{eR_lam_bracket1}), 
(\ref{eR_omegaR_bracket1}), and (\ref{eR_omegaL_bracket1}). They are given in 
(\ref{lam_e_bracket}), (\ref{omegaR_e_bracket}),
and (\ref{omegaL_e_bracket}). But they do not seem to be more 
illuminating than the brackets found above. It is reassuring, given 
the non-obvious expressions for the brackets, that they ensure that $e_{R\,ij}$ commutes with 
$\kappa[v]$, the generator of diffeomorphisms of the $\theta$ coordinates, as it should.

The brackets obtained so far ensure the validity of (\ref{auxbracket2def}) for $\varphi$ equal 
to $\rho_0(1)$, $s_R(1)$, or $s_L(1)$ at any point $1 \in int S_0$ and for $\varphi = e_{ab}(1)$
(or any other representation of the conformal 2-metric) for any $1 \in int \cN$. It remains only
to impose (\ref{auxbracket2def}) for $\varphi$ equal to $\lam$, $\hat{\omega}_R$, and 
$\hat{\omega}_L$.

Let us impose
\be		\label{lam_brackets_def}
\hat{\Omega}[\{\lam(1),\cdot\}_\circ,\dg] = \dg \lam(1).
\ee
The brackets that have already been determined imply that 
\bearr
\lefteqn{A[\{\lam(1),\cdot\}_\circ,\dg] = \dg \lam(1)}\nonumber \\
&& + \frac{1}{16\pi G}\int_{S_0} 2 \{\lam(1),\lam\}_\circ \dg\rho_0
+ \{\lam(1),\hat{\omega}_{R\,i}\}_\circ\dg s_R^i
+ \{\lam(1),\hat{\omega}_{L\,m}\}_\circ\dg s_L^m
\: d\theta^1\wedge d\theta^2 \nonumber\\
&&
\eearr
and  
\bearr
\lefteqn{B[\{\lam(1),\cdot\}_\circ,\dg]}\nonumber \\
& = & - \frac{1}{8} \dg^{y_R} e^{ab} \di_{v^-}e_{ab} - \frac{1}{8} \dg^{y_L} e^{ab} \di_{v^+}e_{ab}
\nonumber \\ 
&& - \frac{1}{64\pi G}\int_{S_0} \{\lam(1), e_{ab}\}_\circ (\di_{v^-}e^{ab} \dg^{y_R}\eg
+ \di_{v^+}e^{ab} \dg^{y_L}\eg)\\
& = & - \frac{1}{8} \dg^{y_R} e^{ab} \di_{v^-}e_{ab} - \frac{1}{8} \dg^{y_L} e^{ab} \di_{v^+}e_{ab}
\nonumber \\ 
&& + \frac{1}{32}(\di_{v^-}e^{ab} + \di_{v^+}e^{ab})
(\di_{v^-}e^{ab} \frac{\dg^{y_R} \rho_0}{\rho_0} + \di_{v^+}e^{ab} \frac{\dg^{y_L} \rho_0}{\rho_0}).
\eearr

Evaluating $C$ requires a little more thought. No derivatives of $\Box_1$ appear in the expression
(\ref{Omega_Box3}) for $C[\dg_1,\dg_2]$. The bulk term in this form of 
$C[\{\lam(1),\cdot\}_\circ,\dg]$ may thus be evaluated using the expression 
(\ref{eR_lam_bracket1}) for $\{\lam(1),e_{R\,ij}(2)\}_\circ$, valid when $2 \in \cN_R - S_0$, and 
the analogous expression corresponding to the $L$ branch - ignoring the jump discontinuity in 
these brackets at $S_0$. Thus
\be
C_{bulk}[\{\lam(1),\cdot\}_\circ,\dg] =
\frac{1}{4}[\di_{v^+} e^{ab}]_1 C_{R\,bulk}[\{e_{ab}(1),\cdot\}_\circ,\dg]
+ \frac{1}{4}[\di_{v^-} e^{ab}]_1 C_{L\,bulk}[\{e_{ab}(1),\cdot\}_\circ,\dg].
\ee
But by (\ref{C_Rbulk_value}) and (\ref{C_Lbulk_value}) 
\bearr
C_{R\,bulk}[\{e_{ab}(1),\cdot\}_\circ,\dg] & = & \frac{1}{2} \dg^{a_R} e_{ab}, \label{C_Rbulk_value2}\\
C_{L\,bulk}[\{e_{ab}(1),\cdot\}_\circ,\dg] & = & \frac{1}{2} \dg^{a_L} e_{ab}, \label{C_Lbulk_value2}
\eearr
so
\be
C_{bulk}[\{\lam(1),\cdot\}_\circ,\dg] =
\frac{1}{8}\di_{v^+} e^{ab} \dg^{a_R} e_{ab}(1)
+ \frac{1}{8}\di_{v^-} e^{ab} \dg^{a_L} e_{ab}(1).
\ee
The $S_0$ surface contribution to $C$ in the form (\ref{Omega_Box3}) is 
\bearr
C_{S_0}[\dg_1,\dg_2] & \equiv & -\frac{1}{8\pi G} \int_{S_0} 
  dy_R^1 \wedge dy_R^2\:(\Box_{R\,1}\barbox_{R\,2} + \barbox_{R\,1} \Box_{R\,2})
+ dy_L^1 \wedge dy_L^2\:(\Box_{L\,1}\barbox_{L\,2} + \barbox_{L\,1} \Box_{L\,2}) \nonumber \\
&& \label{C_S0_def1}\\
	& = & \frac{1}{32\pi G}\int_{S_0} \eg [\dg_1^{a_R} e_{ab}\, \dg_2^{a_R}  e^{ab}
			+ \dg_1^{a_L} e_{ab}\, \dg_2^{a_L} e^{ab}]	\label{C_S0_def2}
\eearr
Now if $\dg_1 = \{\lam(1),\cdot\}_\circ$ then by (\ref{dga_expansion}) 
\be
\dg_1^a e_{ab} = \{\lam(1),e_{ab}\}_\circ - \frac{1}{2\rho_0}\{\lam(1),\rho_0\}_\circ\:\di_v e_{ab}
\ee
(because $\{\lam(1),s^i\}_\circ = 0$). Substituting the expressions (\ref{lam_e_0_bracket}) and 
(\ref{lam_rho_bracket}) for these brackets one obtains
\be
\dg_1^{a_R} e_{ab} = 2\pi G \frac{1}{\rho_0}[\di_{v^-} e_{ab} - \di_{v^+} e_{ab}] 
\dg^2(\theta(2) - \theta(1)) = - \dg_1^{a_L} e_{ab}.
\ee
Hence
\be
C_{S_0}[\{\lam(1),\cdot\}_\circ,\dg] = \frac{1}{16}[\di_{v^-} e_{ab} - \di_{v^+} e_{ab}]
(\dg^{a_R} - \dg^{a_L}) e^{ab}.
\ee
Summing all the contributions, taking into account (\ref{dga_expansion}) and various 
cancellations, one finds
\bearr
\hat{\Omega}[\{\lam(1),\cdot\}_\circ,\dg] & = & \dg \lam(1)
+ \frac{1}{8\pi G}\int_{S_0} \{\lam(1),\lam\}_\circ \dg\rho_0 \: d\theta^1\wedge d\theta^2 
\nonumber\\
&& + \frac{1}{16\pi G}\{\lam(1),\hat{\omega}_R[\xi_{\perp R}]\}_\circ
   + \frac{1}{16\pi G}\{\lam(1),\hat{\omega}_L[\xi_{\perp L}]\}_\circ \nonumber \\ 
&& + \frac{1}{16}[\di_{v^-} e^{ab} - \di_{v^+} e^{ab}]_1
({\pounds}_{\xi_{\perp R}}e_{ab} - {\pounds}_{\xi_{\perp L}}e_{ab})_1  
\eearr
(since $\dg s_R^i = \xi_{\perp R}^i$ and $\dg s_L^m = \xi_{\perp L}^m$).

The requirement (\ref{lam_brackets_def}) thus implies that
\bearr
\{\lam(1),\lam(2)\}_\circ & = & 0 	\label{lam_lam_bracket1}\\
\{\lam(1),\hat{\omega}_R[f]\}_\circ & = & -\pi G [(\di_{v^-} e^{ab} - \di_{v^+} e^{ab}) 
{\pounds}_f e_{ab}]_1 \label{lam_omegaR_bracket1}\\
\{\lam(1),\hat{\omega}_L[g]\}_\circ & = & -\pi G [(\di_{v^+} e^{ab} - \di_{v^-} e^{ab})
{\pounds}_g e_{ab}]_1
\eearr

It remains only to impose 
\be		\label{omega_brackets_def}
\hat{\Omega}[\{\hat{\omega}_R[f],\cdot\}_\circ,\dg] = \dg \hat{\omega}_R[f],
\ee
and the analogous equation for $\hat{\omega}_L[g]$.
Our method will be closely analogous to that used to solve the condition 
(\ref{lam_brackets_def}) corresponding to $\lam(1)$: First note that by 
(\ref{lam_omegaR_bracket1})
\bearr
A[\{\hat{\omega}_R[f],\cdot\}_\circ,\dg] = \dg \hat{\omega}_R[f]\!\!\!\!\!\!\! && 
+ \frac{1}{8}\int_{S_0}{\pounds}_f e_{ab} (\di_{v^-} e^{ab} - \di_{v^+} e^{ab}) \dg\eg	\nonumber\\  
&& - \frac{1}{16\pi G} \{\hat{\omega}_R[f],\hat{\omega}_R[\xi_{\perp R}]\}_\circ
- \frac{1}{16\pi G} \{\hat{\omega}_R[f],\hat{\omega}_L[\xi_{\perp L}]\}_\circ.
\eearr
To evaluate the $B$ contribution recall that $\dg^y  = \dg^b - {\pounds}_{\xi_\perp}$ with
$\xi_\perp^i = \dg s^i$. Thus
\bearr
B_R[\dg_1,\dg_2] & = & -\frac{1}{64\pi G}\int_{S_0} \dg^y_1 e_{ab}\di_{v^-} e^{ab} \dg_2^y \eg 
- (1 \leftrightarrow 2)\\
& = & -\frac{1}{64\pi G}\int_{S_0} \dg_1 e_{ab}\di_{v^-} e^{ab} \dg_2^y \eg  
- {\pounds}_{\xi_{\perp R\,1}} e_{ab}\di_{v^-} e^{ab} \dg_2^y \eg \nonumber\\
&& \ \ \ \ \ \ \ \ \qquad - \dg_1 \eg \di_{v^-}e^{ab} \dg^y_2 e_{ab} + {\pounds}_{\xi_{\perp R\,1}} \eg \di_{v^-}e^{ab} 
\dg^y_2 e_{ab}
\eearr
and, by (\ref{omega_R_e_0_bracket1}), (\ref{omega_y_R_bracket}), and (\ref{dga_expansion}),
\bearr
\lefteqn{B[\{\hat{\omega}_R[f],\cdot\}_\circ,\dg]}\nonumber \\
& = & - \frac{1}{64\pi G}\int_{S_0} \{\hat{\omega}_R[f], e_{ab}\}_\circ 
(\di_{v^-}e^{ab} \dg^{y_R}\eg + \di_{v^+}e^{ab} \dg^{y_L}\eg) \nonumber\\
&& \qquad + \frac{1}{4}\int_{S_0}({\pounds}_f \eg \di_{v^-} e^{ab}\dg^{y_R}e_{ab} 
   - {\pounds}_f e_{ab}\di_{v^-} e^{ab}\dg^{y_R}\eg)\\
& = & - \frac{1}{8}\int_{S_0} 
[{\pounds}_f e_{ab} - \frac{1}{2}\di_{v^-} e_{ab}\frac{{\pounds}_f \rho_0}{\rho_0}]
(\di_{v^-}e^{ab} \dg^{y_R}\eg - \di_{v^+}e^{ab} \dg^{y_L}\eg) \nonumber\\
&& + \frac{1}{4}\int_{S_0}{\pounds}_f \eg \di_{v^-} e^{ab}\dg^{a_R}e_{ab} 
\eearr

The contribution from $C$ is obtained as in the evaluation of $C[\{\lam(1),\cdot\}_\circ,\dg]$,
by expanding $C$ in bulk and surface
terms according to (\ref{Omega_Box3}) and noting that the bracket (\ref{eR_omegaR_bracket1}) and
the analog of (\ref{eR_omegaL_bracket1}) with $L$ and $R$ interchanged imply
\bearr
C_{bulk}[\{\hat{\omega}_R[f],\cdot\}_\circ,\dg] & = & 
- \frac{1}{2} \int_{S_0} \eg {\pounds}_f e_{ab}C_{R\,bulk}[\{e^{ab},\cdot\}_\circ,\dg] \nonumber \\
&& + \frac{1}{2} \int_{S_0} (\eg {\pounds}_f e_{ab} - {\pounds}_f \eg \di_{v^-}e_{ab}) 
C_{L\,bulk}[\{e^{ab},\cdot\}_\circ,\dg],
\eearr
and therefore taking into account 
(\ref{C_Rbulk_value2}) and (\ref{C_Lbulk_value2}), 

\be
C_{bulk}[\{\hat{\omega}_R[f],\cdot\}_\circ,\dg] 
= - \frac{1}{4} \int_{S_0} \eg {\pounds}_f e_{ab} (\dg^{a_R} - \dg^{a_L}) e^{ab}
+ {\pounds}_f \eg \di_{v^-}e_{ab} \dg^{a_L} e^{ab}
\ee

There remains the surface term $C_{S_0}[\{\hat{\omega}_R[f]\cdot\}_\circ,\dg]$ to evaluate.
$C_{S_0}[\dg_1,\dg_2]$ is defined by (\ref{C_S0_def2}). If 
$\dg_1 = \{\hat{\omega}_R[f],\cdot\}_\circ$ then, because $\hat{\omega}_R$ commutes with 
$\rho_0$ and $s_L$, $\dg_1^{a_L} = \dg_1^b$ and thus, by (\ref{omega_R_e_0_bracket1})
\be
\dg_1^{a_L} e_{ab} = \{\hat{\omega}_R[f],e_{ab}\}_\circ 
 = - 8\pi G [{\pounds}_f e_{ab} 
- \frac{1}{2}\frac{{\pounds}_f \rho_0}{\rho_0} \di_{v^-} e_{ab}].
\ee
$\dg_1^{a_R}$, on the other hand, differs from $\dg_1^b$. By (\ref{dga_expansion}),  
(\ref{omega_y_R_bracket}), and (\ref{omega_R_e_0_bracket1})
\bearr
\dg_1^{a_R} e_{ab} & = & \{\hat{\omega}_R[f], e_{ab}\}_\circ 
+ 16 \pi G [{\pounds}_f e_{ab} 
- \frac{1}{2}\frac{{\pounds}_f \rho_0}{\rho_0} \di_{v^-} e_{ab}]\\
& = & - \dg_1^{a_L} e_{ab}.
\eearr
Thus
\be
C_{S_0}[\{\hat{\omega}_R[f],\cdot\}_\circ,\dg] = \frac{1}{4}\int_{S_0} [\eg {\pounds}_f e_{ab} 
- \frac{1}{2}{\pounds}_f \eg \di_{v^-} e_{ab}] (\dg^{a_R} - \dg^{a_L}) e^{ab}.
\ee
Summing this surface term and the bulk term one finds
\be
C[\{\hat{\omega}_{R\,i}(1),\cdot\}_\circ,\dg] = 
- \frac{1}{8}\int_{S_0} {\pounds}_f \eg \di_{v^-} e_{ab}(\dg^{a_R} + \dg^{a_L}) e^{ab}.
\ee

Summing $A$, $B$, and $C$ one is left, after some cancellations, with
\bearr
\hat{\Omega}[\{\hat{\omega}_R[f],\cdot\}_\circ,\dg] & = & \dg \hat{\omega}_R[f]
+ \frac{1}{16\pi G} \{\hat{\omega}_R[f],\hat{\omega}_R[\xi_{\perp R}]\}_\circ
+ \frac{1}{16\pi G} \{\hat{\omega}_R[f],\hat{\omega}_L[\xi_{\perp L}]\}_\circ \nonumber \\
&& + \frac{1}{8}\int_{S_0} {\pounds}_f e_{ab} (\di_{v^-} e^{ab}{\pounds}_{\xi_{\perp R}}\eg 
- \di_{v^+} e^{ab}{\pounds}_{\xi_{\perp L}}\eg) \nonumber\\
&& - \frac{1}{8}\int_{S_0} {\pounds}_f \eg \di_{v^-} e^{ab}({\pounds}_{\xi_{\perp R}}e_{ab} 
- {\pounds}_{\xi_{\perp L}} e_{ab}).
\eearr
Thus
\bearr
\{\hat{\omega}_R[f_1],\hat{\omega}_R[f_2]\}_\circ & = & -2\pi G
\int_{S_0} {\pounds}_{f_1} e_{ab}\,\di_{v^-} e^{ab}{\pounds}_{f_2}\eg - (1 \leftrightarrow 2) 
\label{omega_omega_R_bracket1}\\
\{\hat{\omega}_R[f],\hat{\omega}_L[g]\}_\circ & = &  2\pi G
\int_{S_0} [{\pounds}_f e_{ab} \di_{v^+} e^{ab}{\pounds}_g \eg
			- {\pounds}_g e_{ab} \di_{v^-} e^{ab}{\pounds}_f \eg].
\label{omegaR_omegaL_bracket1}
\eearr
Interchanging the roles of $\cN_R$ and $\cN_L$ gives us the $L$ branch analog of 
(\ref{omega_omega_R_bracket1}) and the same expression (\ref{omegaR_omegaL_bracket1}) for
$\{\hat{\omega}_R[f],\hat{\omega}_L[g]\}_\circ$ again.

With these brackets $\hat{\Omega}[\{\varphi,\cdot\}_\circ,\dg] = \dg \varphi$ for all initial 
data $\varphi$ and all admissible $\dg$. That is $\{\cdot,\cdot\}_\circ$ satisfies
(\ref{auxbracket2def}).

\section{Concluding Remarks}\label{conclusions}

We have found a fairly complete canonical formalism for the gravitational field on the 
domain of dependence of a single pair of intersecting null hypersurfaces. The phase space has 
been defined and a simple representation of this space in terms of metric fields on $\Real^4$
has been given; the Poisson bracket has been defined on a class of ``nice'' phase space 
functions, called ``observables'', and most importantly, a bracket, compatible with the Poisson 
bracket on the observables, has been defined on a set of free initial data variables.

Several things remain to be done. The most important are: 

\begin{itemize}

\item Resolve the problem of the Jacobi relations for the bracket on the initial data.
The bracket on initial data proposed here does not satisfy the Jacobi relations, which precludes
a conventional Dirac quatization of the initial data. The requirement that the bracket on initial
data reproduce the (Peierls) Poisson bracket when applied to the observables leaves some freedom
in the choice of bracket on the initial data. One could try to exploit this freedom to find a 
new bracket on the initial data which does satisfy the Jacobi relations. A hope is that this 
would also lead to brackets that are simpler than the ones found here.

I am currently working along these lines, and at the time of writing it looks like it will work.
If, on the other hand, no such bracket exists, this would also be interesting. That the quantum 
commutators should reproduce the Peierls bracket on the observables seems an eminently 
reasonable requirement. If there is no corresponding bracket satisfying the Jacobi relation on 
the initial data, then this would indicate that the initial data are not quantized in the usual 
sense, and some generalization of Dirac quantization is called for.

\item Define a canonical theory for the whole Universe. The current form of the theory provides 
a canonical description of the domain of dependence of a single hypersurface $\cN$, which never 
constitutes all of an inextendible solution of Einstein's field equations. One way to describe 
all of spacetime is to use an ``atlas'' of interrelated phase spaces corresponding to a covering
of the spacetime by domains of dependence of the type we have considered. A state of this global
theory would consist of points in each of a collection of phase spaces, and a specification
of how the corresponding initial data hypersurfaces, all of the form of $\cN$, intersect.
Of course there would then be many ways to describe the same spacetime corresponding to 
differently positioned initial data hypersurfaces, and one would need to define the action on 
the phase points corresponding to a displacement of one of the initial data hypersurfaces
with respect to the others. This action seems to be generated by a sort of Hamiltonian, which
is currently being studied. 

The multi-phase space formalism outlined seems very attractive and naturally adapted to the
local nature of general relativity. In particular it does not seem to impose the global 
causality restriction that a theory based on global Cauchy surfaces does. This restriction
does correspond to conventional ideas about time, but it is alien to general relativity
and perhaps simply realizes a prejudice we have acquired in our (nearly) Minkowskian environment.

\end{itemize}

Of course there are a number of other, in my opinion lesser, loose ends: 

\begin{itemize}

\item The meaning of gauge invariance it the present canonical formalism should be completely
cleared up. At present it is unclear to me whether diffeomorphisms that map the end surfaces
$S_L, S_R \subset \di\cN$ to themselves, and thus permute the generators of $\cN_L$ and $\cN_R$,
should be treated as gauge or global symmetry transformations.

\item One expects that the causality condition, which we have imposed as an ansatz on the 
bracket on the initial data, really is a consequence of the form of the presymplectic 2-form. 
This should be demonstrated if true.

\item The bracket on initial data has been found assuming that the corresponding solution has no
Killing vectors. The theory should be extended to solutions with Killing vectors, to the extent 
that this is possible. This would close a gap in the theory and make it possible to study the 
theory in concrete exact solutions. If no such extension is possible this would also be an 
interesting fact.

\end{itemize}

Finally, we have assumed at several points that our initial data on $\cN$ define unique maximal
Cauchy developments, and that the metric of this development is differentiable in the parameter 
of any smooth one parameter family of initial data. To my knowledge this has not been proved,
although an heuristic argument has been given by Sachs \cite{Sachs}, and slightly weaker claims 
have been demonstrated by Rendall \cite{Rendall}, and it seems virtually certain to be true. 
To obtain the proofs an extension of Rendall's method seems viable. 

The main motivation for the present work has been the desire to understand the holographic 
principle \cite{Beckenstein}\cite{tHooft}\cite{Susskind}. Susskind \cite{Susskind} has given 
compelling, yet inconclusive, arguments in favour of holography based on semiclassical black 
hole theory. He also gave arguments supporting the conjecture that string theory is holographic,
using a light front formulation of string theory. My aim has been to provide a framework in 
which some of these claims could be stated precisely and proved.

Given the present formalism (and the yet to be found bracket on initial data that satisfies the 
Jacobi relations) one is in a position to try to demonstrate a holographic limit on the 
dimensionality of the Hilbert space quantizing the phase space corresponding to $\cN$ in either 
a full or semi-classical quantization of the initial data.

To see get an idea of how the theory might be quantized one might try to see how (and if) 
the AdS/CFT duality works in terms of a null canonical formulation of the bulk theory. Of course
this requires an extension of the formalism, to supergravity and string theory. Indeed the
present results on gravity on a null hypersurface might give insight into string theory in the 
light front gauge, and thus help to verify Susskind's \cite{Susskind} conjecture that string 
theory is a holographic theory.

\section{Acknowledgements}

The present work has had a long gestation. The key calculations were carried
out during a visit to the Centre de Physique Teorique in Luminy, France, and a visit to the
Albert Einstein Institut in Potsdam, Germany. I would like to thank the CNRS for funding my 
visit to Luminy, and the group at Luminy, especially Carlo Rovelli, for the invitation and many 
discussions, and I would like to thank Thomas Thiemann for hosting me at the AEI.
The work was completed at the Perimeter Institute in Waterloo, Canada and the Universidad de la
Republica Oriental del Uruguay. I thank the group at Perimeter, especially Lee Smolin, 
Fotini Markopolou, Richard Epp, Janet Fesnoux and Howard Burton for inviting me and/or providing
a wonderfully stimulating work environment. At the Universidad de la Republica I would like to 
thank especially Rodolfo Gambini for his constant support and for helpful discussions.
In the course of this work I have benefited from the comments, questions, and encouragement of 
many other colleagues, among them Peter Aichelburg, Alejandro Perez, Jose-Antonio Zapata, 
Laurent Freidel, Jerzy Lewandowski, John Stachel, Alan Rendall, Helmuth Friedrich, 
Gonzalo Aniano, Michael Bradley, and Ingemar Bengtsson. I have used a TeX macro from the 
thesis of Chris Beetle to make some symbols.

This work has been partly supported by Proyecto 8049, Fondo Clemente Estable of the Ministerio 
de Educacion y Cultura, Uruguay.

\appendix

\section{Definitions regarding manifolds and causal structure.}\label{definitions}

Here a number of definitions and a few propositions important to their meaning are gathered 
together for convenience of the reader. The definitions are mostly taken from \cite{Wald} 
or \cite{Hawking_Ellis}, or equivalent to the ones given there, but the definitions of Cauchy 
surfaces and developments are slightly modified. The definitions given here do not form a 
self-contained logical structure. A complete exposition can be found in \cite{Wald} and in 
\cite{Hawking_Ellis}. 

A manifold will be defined as in \cite{Wald} p. 11. A manifold equipped with a non-degenerate 
metric will be called a metric manifold, and it will be called smooth or $C^\infty$ if both the 
atlas and the metric are $C^\infty$. Unless otherwise specified all manifolds will be assumed to
be Hausdorff and paracompact. (see \cite{Hawking_Ellis} and references therein). 

A manifold, according to the (standard) definition of \cite{Wald}, has 
only interior points. However the concept of a manifold can be generalized to that of a 
manifold with boundary. If such a manifold is differentiable one can distinguish in the 
boundary differentiable components as well as edges, corners,
and in general lower dimensional ``strata''.  

\begin{definition}\label{manifold_with_boundary}
A {\em manifold with boundary} is defined in the same way as a manifold without boundary, except
that the charts, instead of being maps to open subsets of
$\Real^n$, are maps to open 
subsets of the half space $\halfR^n$ consisting of the half of $\Real^n$ on which the first 
component, $x^1$, of each n-tuple is $\geq 0$. The boundary $\di Y$ of a manifold with boundary 
$Y$ is composed of the points that are mapped to $x^1 = 0$ by the charts. 
\end{definition}

Differentiable manifolds generalize to {\em stratified manifolds} in which the charts may be 
maps to open subsets of the 1/4 space, 1/8 space, or $1/2^m$ space, consisting of real 
$n-tuples$ of which the first two, three, or $m$ components, as the case may be, are $\geq 0$. 
The $n – m$ dimensional strata consist of the points mapped to $0 = x^1 = x^2 = ... = x^m$ 
in the $1/2^m$ space by these charts. $\cN$ is a stratified manifold.
Note that according to the present definition, the boundary $\di Y$ of a stratified manifold 
includes the union of all the lower dimensional strata.

In the present work ``manifold'', without further qualification, will denote a $C^\infty$ 
manifold with or without boundary (or lower dimensional strata). Note that a manifold 
{\em without} boundary can be the interior of a manifold with boundary. The absence of a 
boundary does not imply that a boundary could not be attached to the manifold, only that it 
is not included in it.  

Now we turn to definitions associated with the causal structure defined by a smooth Lorentzian 
metric on a manifold $X$ without boundary. We shall assume that $X$ is time orientable with 
this metric - there is a continuous choice of the future half of the light cone throughout $X$. 
In fact, unless otherwise specified, we shall always assume that Lorentzian metric manifolds 
representing spacetime are time orientable. 
Furthermore we shall assume that spacetime is 4 dimensional. However most definitions and 
results generalize in an obvious way to spacetimes of any dimensionality.

\begin{definition}\label{causal_future}
The {\em causal future}, $J^+[S;X]$, of a set $S \subset X$ in $X$ is the set of points of $X$ 
that can be reached from some point in the set $S$ via a future directed differentiable causal 
curve in $X$ - a curve with tangent everywhere null or timelike.
\end{definition}

\begin{definition}\label{chronological_future}
The {\em chronological future}, $I^+[S;X]$, of a set $S \subset X$ in $X$ is the set of points 
of $X$ that can be reached from some point in $S$ via a future directed differentiable timelike 
curve in $X$ of non-zero duration - a curve with tangent everywhere non-zero and timelike.
\end{definition}

\begin{definition}\label{future_domain_of_dependence}
The {\em future domain of dependence} $D^+[S;X]$ of $S \subset X$ in $X$ is the set of points 
$p \in X$ such that all inextendible causal curves in $X$ from $p$ intersect $S$ to the past 
of $p$.
\end{definition}

(Often, when the ambient spacetime manifold with respect to which these sets are defined is 
clear from the context, it will not be specified explicitly, and we shall write
$I^+[S]$, $J^+[S]$, and $D^+[S]$.)

The definitions of the causal past $J^-$, chronological past $I^-$, and past domain of 
dependence $D^-$ are the obvious time reversed analogs of those of $J^+$, $I^+$, and $D^+$.

Uniting the past and future parts one may define the full {\em domain of dependence} as
$D[S] = D^+[S]\cup D^-[S]$, the {\em causal domain of influence} as $J[S] = J^+[S]\cup J^-[S]$,
and the {\em chronological domain of influence} as $I[S] = I^+[S]\cup I^-[S]$.
$D[S]$ may be defined directly as the set of points $p \in X$ 
such that all inextendible causal curves in $X$ through $p$ intersect $S$ (to the past or to 
the future of $p$). While $J[S]$ and $I[S]$ are the unions of the causal and timelike curves 
through $S$ respectively.

\begin{definition}\label{achronal}
A subset $S$ of $X$ is {\em achronal} iff no timelike curve in $X$ crosses $S$ more than once, 
i.e. at two or more values of the curve's parameter. 
\end{definition}

Equivalently, $S$ is achronal iff $S$, $I^+[S]$, and $I^-[S]$ are disjoint.

The achronality of a set requires that its interior, suitably defined, be an embedded 
3-manifold: If $S$ is achronal then $S$ is disjoint from $I^+[S]$, but of course 
$S \subset \bar{I}^+[S]$. Thus $S \subset \dot{I}^+[S]$. But by prop. 6.3.1. of 
\cite{Hawking_Ellis} $\dot{I}^+[S]$ is a $C^{1-}$,\footnote{
It satisfies a Lipschitz condition, and {\em a fortiori} is $C^0$. See \cite{Hawking_Ellis}.}
embedded 3-manifold.
If the {\em interior} of $S$, $int S$, is defined to be the topological 
interior of $S$ in $\dot{I}^+[S]$, then $int S$ is also a $C^{1-}$, embedded 3-dimensional 
submanifold (without boundary). Note that $int S$ is {\em not} the interior of $S$ in the 
ambient spacetime $X$, which is empty.

This makes possible a more intrinsic definition of $int S$:

\begin{definition}\label{interior_achronal}
The {\em interior}, $int S$ of an achronal subset $S \subset X$ is the largest subset of 
$S$ that is an embedded 3-submanifold of $X$ without boundary. (Equivalently it is the union
of all such subsets). 
\end{definition}

The equivalence of this definition with the earlier one 
can be deduced from the following topological lemma:
 
\begin{lemma}\label{embedding_lemma0} 
If $A$ is an n-manifold with boundary embedded in a n-manifold without boundary $B$, then 
$A - \di A$, the interior of $A$ according to its structure as a manifold with 
boundary, equals $A - \dot{A}$, the interior of $A$ in the topology of $B$, and 
$\di A \subset \dot{A}$.   
\end{lemma}
{\em Proof}: A point $p \in A$ lies in the domain of
a chart $\psi: V \rightarrow \halfR^n$ of $A$ and a chart $\phi: V' \rightarrow \Real^n$
with $V$ and $V'$ open neighbourhoods of $p$ in $A$ and $B$ respectively. If $U$ is an open 
neighbourhood of $p$ in $A$ which is contained in both $V$ and $V'$ then $\phi \circ \psi^{-1}$ 
is a homeomorphism from $\psi[U]$, an open subset of $\halfR^n$, to the subset 
$\phi[U]$ of $\Real^n$. 

Suppose $p \in  \di A$. $\psi[U] \subset \halfR^n$ cannot then be open in 
$\Real^n$, because it contains $\psi(p)$ which lies on the boundary of $\halfR^n$ in the 
topology of $\Real^n$. On the other hand, since $A$ is an embedded submanifold with boundary 
the open sets of $A$ are the intersections of $A$ with open sets of $B$. Thus there exists
a set $U'$ open in $B$ such that $U = U' \cap A$. $\phi[U']$ is open in $\Real^n$.
Brouwer's invariance of domain theorem \cite{Brouwer} requires that the 
homeomorphic image in $\Real^n$ of an open set in $\Real^n$ be open.\footnote{
A homeomorphism maps the open sets of its domain to open sets in its range. This does not
by itself imply that the range is open in the topology of an ambient space whenever the domain
is open in an ambient space.}
The open set 
$\phi[U']$ thus cannot be homeomorphic to $\psi[U]$, and therefore not to 
$\phi[U] = \phi[U' \cap A]$ either. $U$ must contain a point outside $A$. It 
follows that $p \in \dot{A}$, and hence that $\di A\subset \dot{A}$.

Now suppose that, on the contrary, $p \in A - \di A$. Then $\psi(p)$ does not lie in the
boundary plane of $\halfR^n$ and there exists a neighbourhood $R$ of $\psi(p)$ in 
$\psi[V\cap V'] \subset \halfR^n$ which does not intersect the boundary of $\halfR^n$
and which is open in $\halfR^n$ and also in $\Real^n$.
Define $U = \psi^{-1}[R]$. $U$ is a neighbourhood of $p$ open in $A$ and a subset of $V\cap V'$.
Since $R = \psi[U]$ is open in $\Real^n$, $\phi[U]$ is also. $U$ is thus open in $B$, 
implying that $p$ lies in the interior $A - \dot{A}$ of $A$ according to the topology of $B$. 
Together with the previous result that no point in $\di A$ lies in $A - \dot{A}$ this implies 
that $A - \di A = A - \dot{A}$.\QED
\newline

From the lemma it follows that any subset $U \subset S$ that is a 3-manifold without boundary
($\di U = \emptyset$) must coincide with its interior in the topology of $\dot{I}^+[S]$, and 
thus be open in this topology. This establishes the equivalence of the two definitions of 
$int S$.

A useful notion of boundary for $S$ is the following:

\begin{definition}\label{embedding_boundary}
The {\em embedding boundary} of an achronal subset $S \subset X$ is 
$\ocirc{S} = \bar{S} - int S$, where $\bar{S}$ is the closure of $S$ in the topology of $X$.
\end{definition}
 
$\ocirc{S}$ is simply the topological boundary of $S$ in $\dot{I}^+[S]$. Note that $\ocirc{S}$ 
does not in general coincide with the causal edge of $S$ defined in \cite{Hawking_Ellis} p.202. 
It is easy to show that $\ocirc{S} \subset edge[S]$. However not all points
of the edge need to lie in $\ocirc{S}$.\footnote{$\ocirc{S}$ does equal $edge[S]$ if strong 
causality holds. See \cite{Hawking_Ellis} p. 196-197 on Alexandrov topology.} 

Lemma \ref{embedding_lemma0} provides us with a further useful and obvious corollary.

\begin{proposition}\label{embedding2} 
If $\Sg$ is a 3-manifold with boundary achronally embedded in spacetime then 
$int \Sg = \Sg - \di\Sg$, the interior of $\Sg$ according to its manifold structure, 
and $\di\Sg \subset \ocirc{\Sg}$.
\end{proposition}

Conversely, if the embedding boundary, $\ocirc{S}$, of an achronal set $S \subset X$ is 
embedded in $X$ then $\bar{S}$ is a manifold with boundary, with $\di \bar{S} = \ocirc{S}$. 
[If a manifold is embedded according to our definition (that of \cite{Hawking_Ellis}), which 
is to say {\em regularly} embedded in the definition of \cite{three_muses}, then according to
\cite{three_muses} p. 242 it is a submanifold of the ambient manifold according to the 
definition of \cite{three_muses} p. 239. The result then follows immediately.]
Note that if $\Sg$ is closed prop. \ref{embedding2} implies that $\ocirc{\Sg} = \di \Sg$.

Now we turn to Cauchy surfaces and Cauchy developments.

\begin{definition}\label{Cauchy_surface}
A {\em Cauchy surface} of an open set $A \subset X$ is an achronal subset $C$ of the
closure $\bar{A}$ of $A$ such that any causal curve in $A$ which is inextendible
in $A$ crosses $C$ or has an endpoint on $C$ in $\dot{A}$.
\end{definition}

\begin{definition}\label{globally_hyperbolic}
An open subset $\cH \subset X$ is {\em globally hyperbolic} in $X$ iff $\cH$ contains a Cauchy 
surface of $\cH$.
\end{definition}

Whether $\cH$ is globally hyperbolic in $X$ depends to some extent on 
the properties of the ambient spacetime $X$. Cauchy surfaces must be achronal, and 
whether a given subset is achronal depends on the manifold in which it is embedded.
However several important consequences of global hyperbolicity hold for a subset 
$\cal G$ provided only that it is globally hyperbolic regarded as a manifold by itself, 
i.e. provided that it's Cauchy surfaces are achronal in $\cal G$. For instance, this weaker 
property, which might be called {\em intrinsic global hyperbolicity}, is sufficient 
to establish the existence of a foliation of $\cal G$ by homeomorphic spacelike hypersurfaces
\cite{Hawking_Ellis} prop. 6.6.8.

Def. \ref{globally_hyperbolic} of global hyperbolicity has been adopted because it is 
equivalent to the definition used in \cite{Hawking_Ellis}. There a subset $\cH$ of $X$ is 
globally hyperbolic in $X$ iff strong causality holds on the subset\footnote{
{\em Strong causality} holds at $p$ if every neighbourhood of $p$ contains a neighbourhood
which causal curves cross at most once. See \cite{Hawking_Ellis}.} 
and $J^+(p)\cap J^-(q)$ is compact and contained in $\cH$ for all $p, q \in \cH$. When $\cH$ is 
open this can be shown to be equivalent to the definition \ref{globally_hyperbolic} using 
Geroch's theorem prop. 6.6.8. \cite{Hawking_Ellis} and prop. 6.6.3. of \cite{Hawking_Ellis}.
(See \cite{Wald} p. 209.)

The following result, which extends the relation between Cauchy surfaces and global 
hyperbolicity established by def. \ref{globally_hyperbolic} to Cauchy surfaces outside
the globally hyperbolic set, justifies our definition \ref{Cauchy_surface} of Cauchy surfaces.

\begin{proposition}\label{Cauchy_glob_hyp}
An open subset $\cH$ of $X$ is globally hyperbolic iff $int \bar{\cH} = \cH$, and $\cH$ posses 
a Cauchy surface.
\end{proposition}
{\em Proof}: Let $C \subset \bar{\cH}$ be a Cauchy surface of $\cH$. $\cH$ is therefore a 
subset of $int D[\bar{S}]$ which, by \cite{Hawking_Ellis} Prop 6.6.3, is globally hyperbolic. 
($\bar{S}$ is achronal because $S$ is.) Thus strong causality holds at all points of $\cH$ and 
$J^+(p)\cap J^-(q)$ is compact $\forall p, q \in \cH$. All that remains to show is that 
$J^+(p)\cap J^-(q) \subset \cH$. 

Let $p' \in I^-(p)\cap \cH$ and $q' \in I^+(q)\cap \cH$, then $J^+(p)\cap J^-(q) \subset 
I^+(p')\cap I^-(q')$. It is therefore sufficient to show that $I^+(p')\cap I^-(q') \subset \cH 
\ \ \forall p', q' \in \cH$. 

Suppose this were not so and $I^+(p')\cap I^-(q')$ contains a point $r$ outside $\cH$.
Let $\cg$ be an inextendible timelike curve through $p'$, $r$, and $q'$. Since $q'$ and $p'$
lie in $\cH$ but $r$ does not, $\cg \cap \cH$ consists of at least two segments that are 
inextendible in $\cH$, one before $r$, the other after. Each segment must cross $C$ or have 
an endpoint on $C$ in $\dot{\cH}$, so $\cg$ crosses $C$ at least once at $r$ or later (according
to the parameter of $\cg$), and at least once at $r$ or earlier. On the other hand $C$ is 
achronal, so $\cg$ can in fact cross $C$ only once, implying that the crossing occurs at $r$, 
and therefore that $r \in C \subset \bar{\cH}$. We may conclude that $I^+(p')\cap I^-(q') 
\subset \bar{\cH}$. But $I^+(p')\cap I^-(q')$ is open, so it must actually be contained in 
$int \bar{\cH} = \cH$. 

Conversely, if $\cH$ is globally hyperbolic it contains a Cauchy surface. What must be shown is
that $int \bar{\cH} = \cH$ in this case. Suppose $p \in int \bar{\cH}$, then $\cH$ is dense in 
any sufficiently small open neighbourhood $U$ of $p$. Let $p' \in I^-(p)\cap U \cap\cH$ and 
$q' \in I^+(q)\cap U \cap\cH$, then $p \in I^+(p')\cap I^-(q')$. The argument used to show that 
$I^+(p')\cap I^-(q') \subset \bar{\cH}$ applies, but since the Cauchy surface now is 
contained in $\cH$ one may conclude that $I^+(p')\cap I^-(q') \subset \cH$ and thus 
$p \in \cH$. \QED
\newline

Prop. \ref{Cauchy_glob_hyp} is essentially a generalization of prop. 6.6.3. of 
\cite{Hawking_Ellis} which states that the interior of the domain of dependence of a closed 
achronal set is globally hyperbolic. Here we reproduce this very useful result, and also show 
that the assumption that the achronal set be closed is unnecessary.

\begin{proposition}\label{closed_cauchy_unnecessary}
The interior of the domain of dependence of any achronal subset of $X$ is globally
hyperbolic in $X$.
\end{proposition}
{\em Proof}: 
The claim can be deduced from prop. \ref{Cauchy_glob_hyp}, but it is easier to prove it 
directly.
Suppose $S \subset X$ is achronal and let $\cB = int D[S]$. $\cB$ is a subset of 
$int D[\bar{S}]$ which, by \cite{Hawking_Ellis} Prop 6.6.3, is globally hyperbolic. 
Thus strong causality holds at all points of $\cB$ and $J^+(p)\cap J^-(q)$ is compact 
$\forall p, q \in \cB$. All that remains to show is that $J^+(p)\cap J^-(q) \subset \cB$. 

Let $p' \in I^-(p)\cap \cB$ and $q' \in I^+(q)\cap \cB$, then $J^+(p)\cap J^-(q) \subset 
I^+(p')\cap I^-(q')$. Any point $r \in I^+(p')\cap I^-(q')$ lies between $p'$ and $q'$  
on a timelike curve. If this curve is extended until it is timelike and inextendible then it
must cross $S$, and it must do so precisely once since $S$ is achronal. Any point between $p'$
and the crossing point, or between $q'$ and the crossing point, must lie in $D[S]$ so 
$r \in D[S]$. Since $I^+(p')\cap I^-(q')$ is open it must in fact be contained in the interior,
$\cB$, of $D[S]$.\QED
\newline

A peculiar feature of our definitions is that while the interior of the domain of dependence 
of an achronal set $S$ is globally hyperbolic $S$ is not necessarily a Cauchy surface of 
$int D[S]$, because it may not be contained in $\overline{int D[S]}$. For instance suppose 
$\sg$ is a spacelike disk in Minkowski space, and let $S = \dot{I}^-[\sg]$, 
$S \cap \overline{int D[S]} = \sg$. Nevertheless, it will be proved 
further on (prop. \ref{cN_horizon}) that $\cN$ is a Cauchy surface of $\cD \equiv int D[\cN]$.

\begin{definition}\label{Cauchy_horizon}
The {\em future Cauchy horizon} of a set $S$ is $H^+[S] = \overline{D[S]} - I^-[D[S]]$
\end{definition}

The past Cauchy horizon is defined analogously. 
\newline

We may now give the definition of Cauchy developments used in the present work

\begin{definition}\label{Cauchy_development}
A {\em smooth Cauchy development} of initial data specified on a 3-manifold $\Sg$ is the domain 
of dependence of $\Sg$ embedded achronally in a $C^\infty$,\footnote{
If no hypothesis is made about the degree of differentiability of the 4-metric then it becomes 
difficult to define what is meant by a solution to the field equations.}
time orientable, Lorentzian metric 4-manifold without boundary such that the data specified on 
$\Sg$ match those induced on $\Sg$ by the 4-metric. Moreover the 4-metric is required to satisfy
the field equations in the domain of dependence.
\end{definition} 

In general Cauchy developments defined in this way are not manifolds (with or without boundary).
Nevertheless some work might yield a more intrinsic definition which does not refer to the 
ambient boundaryless manifold in which the domain of dependence exists. On the other hand we 
make use of this ambient manifold on occasion, so the definition seems adequate for our 
purposes.

Given initial data on a given 3-manifold can have many different Cauchy developments. For 
instance if some of the points in the domain of dependence are eliminated from the ambient 
manifold the new domain of dependence in the manifold thus reduced would be a proper subset of 
the original one. 
More generally for any given Cauchy development there are many others that are isometric to a 
proper subset. Moreover the Cauchy development might be isometric to a proper subset of a larger
one. This leads naturally to the concept of a maximal Cauchy development.
 
\begin{definition}\label{maximal_Cauchy_development}
A {\em maximal $C^\infty$ Cauchy development} is a Cauchy development which
is not isometric to a proper subset of any other $C^\infty$ Cauchy development of the same data 
on the same 3-manifold.
\end{definition}

Zorn's lemma asserts that maximal Cauchy developments exist, while
uniqueness theorems assure that they are uniquely determined up to diffeomorphisms by the 
initial data, for the classes of initial data to 
which these theorems apply (see \cite{Wald} p. 263-264).  

\section{Some facts about the null initial data hypersurfaces used in the present work}
\label{nullhypersurfaces}

This appendix collects together results about the topology, geometry, and causal structure 
of the initial data hypersurface $\cN$ (defined in subsection \ref{problem}) and its Cauchy 
developments.

We shall begin with the local aspects of the geometry of $\cN$. It is shown that the tangents 
of the generators of $\cN$ are normal to $\cN$. As a consequence all two dimensional cross 
sections of $\cN$ that are transverse to the generators are spacelike and normal to the 
generators. A converse result is also given that establishes that if the normal vectors of a 
hypersurface are null, making it a null hypersurface, then these normals are tangent to the 
hypersurface and integrate to null geodesics that sweep out the hypersurface.  

Next we turn to the issue of caustics and crossings of generators. Typically the generators,
if extended far enough, will form caustics. But (see fig. \ref{self_intersection}) a generator 
may well cross another generator before reaching a caustic and thus enter the chronological 
future of $\cN$. If this occurs $\cN$ is not achronal. The question therefore arises whether 
it is really true that there always exists a solution in which $\cN$ with given data is 
achronal, or whether the initial data must satisfy some additional conditions in order for 
such a space to exist. Moreover, if there is no such condition, and any data corresponds to 
an achronal embedding, can a solution in which $\cN$ is not achronal be represented by a 
Cauchy development of initial data? 

In a Cauchy development $\cN$ is achronal by definition, but we have at present no proof of 
the existence such developments. There is only a theorem showing the existence of Cauchy 
developments of a neighbourhood of $S_0$ in $\cN$, and we have been {\em assuming} that 
developments of all of $\cN$ exist.  In the following it will be shown, quite independently 
of any existence theorem for Cauchy developments, that if $\cN$ is not achronal in a 
spacetime $X$ then there exists a locally isometric covering manifold $\cV$ of a neighbourhood 
of $\cN$ in $X$ in which the future directed null geodesics normal to $S_0$ sweep out an 
achronal hypersurface covering $\cN$. This answers both questions of the previous paragraph:
Achronality of $\cN$ requires no additional restrictions on the initial data (beyond 
those that ensure the absence of caustics on $\cN$). Furthermore, if we suppose that initial
data determines unique maximal Cauchy developments so that (portions of) solution spacetimes 
can be represented by corresponding initial data, then solutions in which $\cN$ is not achronal
are represented by locally isometric covering manifolds with geometry corresponding to the 
initial data on $\cN$. Therefore such solutions {\em are} represented by initial data, but
the identifications that differentiate the covering manifold from the original solution
spacetime, and make $\cN$ non-achronal, leave no trace in the initial data. 

Adopting the covering manifold as our spacetime we can be certain that $\cN$ is achronal.
Using this we prove a number of further results: Points on $\cN$ are causally related iff they
lie on the same generator; All Cauchy surfaces of $D[\cN]$ share the same boundary $\di\cN$; 
and others.

Finally two useful results regarding Cauchy developments of $\cN$ are established: 
Firstly, that any Cauchy development of $\cN$ may be extended to a boundaryless globally 
hyperbolic manifold containing $\cN$, and secondly, that all Cauchy developments of $\cN$,
with all possible data, are diffeomorphic to each other (though of course not necessarily 
isometric).  

Let us first revisit the definition of $\cN$. Suppose $S_0$ is smoothly embedded as a 
spacelike 2-surface in a 4-manifold $X$ with smooth Lorentzian metric and let $O$ be the normal 
bundle of $S_0$ (i.e. the bundle of normal vectors). This bundle contains the subbundles 
$\bar{N}_L$ and $\bar{N}_R$ of future and inward directed null normal vectors and future and 
outward directed normal null vectors respectively (inward and outward referring to the 
arbitrarily chosen orientation of $S_0$). $\cN_L$ and $\cN_R$ are obtained by suitably 
truncating each ray of $\bar{N}_L$ and $\bar{N}_R$ and applying the exponential map. Here a 
{\em ray} is the set of positive multiples of a vector, and the exponential map is the map 
that takes a point $p \in X$ and a vector $v$ in the tangent space of $X$ at $p$ to the point 
of affine parameter $1$ on the geodesic with tangent $v$ originating at $p$. The exponential 
map is well defined as long as $v$ is not so large that the geodesic leaves $X$ before reaching 
affine parameter $1$, and then it is a smooth map because the metric is smooth (see 
\cite{Hawking_Ellis} p. 33). Restricted to the normal bundle $O$ it defines a smooth map $f$ 
from a neighbourhood of $S_0$ in $O$ into $X$. The image of the neighbourhood is not open in $X$
but a slight modification of our setup makes it so. Since $S_0$ is smoothly embedded
it may be extended to a (boundary-less) spacelike 2-submanifold $S' \supset S_0$ of $X$.
[The smoothness of the embedding and the Whitney extension theorem \cite{Abraham_Robbin} imply 
that the embedding map defined on $S_0$, diffeomorphic to the unit disk in $\Real^2$, may be 
extended to a boundaryless manifold containing $S_0$. An open subset $S' \supset S_0$ of this 
extension will be spacelike.] $f$ extends to the normal bundle $O'$ 
of $S'$ and maps an open neighbourhood of $S_0$ (or indeed of $S'$) in $O'$ to an open 
neighbourhood of the same 2-surface in $X$.

$f$ is not necessarily globally invertible, but if it is restricted to an open subset $V$ of its
domain in $O'$ that excludes all points at which the Jacobian $df$ is degenerate, then by the 
inverse function theorem

\begin{proposition}\label{immersion_thm}
$f: V \rightarrow X$ is a smooth immersion of $V$ in $X$. That is, for each point $P \in V$ 
there is a neighbourhood $U$ of $P$ such that $f$ restricted $U$ is a smooth diffeomorphism of 
$U$ to $f[U]$.
\end{proposition}

Note that the images in $X$ of points in $O$ at which $df$ {\em is} degenerate are caustics of
the congruence of geodesics normal to $S_0$. That is, they are points at which neighbouring 
normal geodesics cross, or nearly cross in the sense that the coordinate separation of points 
of equal affine parameter on a family of nearby normal geodesics is zero to first order in the 
difference between their initial data - in the differences between the coordinates of their 
starting points on $S_0$ and of the components of their initial tangents.

The truncations, $N_L$ and $N_R$, of $\bar{N}_L$ and $\bar{N}_R$ which, when exponentiated,
give rise to $\cN_L$ and $\cN_R$ are chosen so that there are no caustics on $\cN_L$
or $\cN_R$. $V$ may thus be chosen to so that it contains $N_L$ and $N_R$. $V$ will furthermore
be chosen so that any ray of $O'$ leaves $V$ at most once. It is then simply connected.
prop. \ref{immersion_thm} implies that $V$ is a covering manifold of $f[V] \subset X$.
Let us denote by $\cV$ the manifold $V$ equipped with the metric pulled back from $X$ via $f$.
$\cV$ is then a locally isometric covering manifold of $f[V]$. The exponential map 
$\hat{f}:V \rightarrow \cV$ of $V$ into $\cV$ is simply the identity map from $V$ to $\cV$
and thus evidently one to one.

In the following we will mostly study the hypersurface $\cN = \hat{f}[N_L \cup N_R]$ swept out 
by the future directed 
null geodesics normal to $S_0$ in the covering spacetime $\cV$. Since this $\cN$ maps to the 
old $\cN$ in $X$ when the isometric identifications are made that reduce $\cV$ to 
$f[V] \subset X$, any local property of $\cN$ in $\cV$ is shared by $\cN$ in $X$.
In the corresponding Cauchy problem data is given on $N = N_L \cup N_R$ and the embedding
of $N$ in the Cauchy development is diffeomorphic to $\cN$ in $\cV$. 

Note that in our original definition of $\cN$ we only required that the generators of
$\cN$ not form any caustics on $\cN$. Here we require that the congruence of geodesics
normal to $S_0$, which includes also timelike and spacelike geodesics, not form caustics on 
$\cN$. In prop. \ref{no_caustics} the two conditions will be shown to be equivalent.

Now let us establish the key local property of $\cN$, namely that it is a null hypersurface
and the null normal vectors are the tangents to the generators:

\begin{proposition}\label{tangent_is_normal} The tangents to the generators of a branch $\cN_A$
of $\cN$ are normal to $\cN_A$.
\end{proposition}
{\em Proof}: Suppose $\zeta$ is a tangent vector of $\cN_A$ at a point $p$. Let $\lam$ be an 
affine parametrization of the generators of $\cN_A$ such that the corresponding field of 
tangent vectors $l \equiv d/d\lam$ of these is smooth (this requires only that $d/d\lam$ be 
chosen smooth on $S_0$), and define the vector field $\eta$ on the generator through $p$ by 
Lie dragging $\zeta$ along the generator. That is, define $\eta$ to be the field of vectors 
tangent to $\cN_A$ that satisfies the differential equation ${\pounds}_l \eta = 0$ and 
the initial condition $\eta(p) = \zeta$. 

Since $\eta$ is tangent to $\cN_A$ at $S_0$ it can have only components tangent to $S_0$ and
to $l$. This implies that $l \cdot \eta = 0$ there, for the component tangent to $S_0$ 
is orthogonal to $l$ by the definition of the generators, and the component tangent to $l$ is 
orthogonal to $l$ since $l$ is null. But $l \cdot \eta$ is constant along the generator 
through $p$: 
${\pounds}_l \eta = 0$ means that $\nabla_l \eta = \nabla_\eta l$, and, since $l$ is the tangent
to an affinely parametrized geodesic, $\nabla_l l = 0$, so 
\be
\frac{d}{d\lam} [l \cdot \eta] = l \cdot \nabla_l \eta = l \cdot \nabla_\eta l = \frac{1}{2}
\frac{d}{d\lam} l^2 = 0.
\ee
Thus $l \cdot \eta = 0$ at $p$.\QED
\newline

A converse result is worth noting.

\begin{proposition}\label{null_normals_geodesics}
On an arbitrary null hypersurface $\cal K$ (a hypersurface with null normal everywhere)
in a smooth Lorentzian metric manifold $Y$ the normals are tangent to the hypersurface and 

integrate to a congruence of null geodesic curves that sweep out the hypersurface.
\end{proposition}
{\em Proof}:
Let $w$ be a function on $Y$ that is constant on $\cal K$ and has non-zero gradient there. Then 
$k = \vec{d} w$, obtained by acting with the inverse metric on $dw$, is a normal vector field 
on $\cal K$, which by assumption is null and thus has null integral curves. The torsion 
freeness of the connection implies that these integral curves are geodesics:
        \be  \label{normals_to_geodesics}
0 = 2 k^\mu \nabla_{[\mu}\nabla_{\nu]} w = \nabla_k k_\nu - k^\mu \nabla_\nu k_\mu = 
\nabla_k k_\nu - \frac{1}{2} \nabla_\nu k^2, 
        \ee
so $\nabla_k k \propto k$, since $k^2 = 0$ on $\cal K$. That is, $k$ is parallel 
transported, modulo modulus, along it's integral curves which are therefore geodesics.\QED
\newline

Note that {\em any} 2-surface $\sg \subset {\cal K}$ that cuts the integral curves of the 
normals is of course orthogonal to these. (This means that in any local Lorentz frame in 
which an element of $\sg$ is purely spatial the integral curve emerging from this element is 
spatially normal to it.)

Note also that if the level sets of $w$ are null hypersurfaces in a neighbourhood of $\cal K$, 
or indeed if only the gradient of $(dw)^2 \equiv k^2$ vanishes on $\cal K$, then it follows 
from (\ref{normals_to_geodesics}) that the integral curves of $k$ are affinely parametrized 
geodesics. 

We are now ready to show that the restrictions we place on the initial data imply that $df$ is 
non-degenerate on $N$, which is to say, the generators of $\cN$ and neighbouring geodesics 
normal to $S_0$ form no caustics on $\cN$. What are these restrictions? Suppose $\cg$ is a 
generator of a branch $\cN_A$ of $\cN$, and thus the image of a ray of $N$, and suppose $P$
is a point on this ray. Then the area parameter at $P$ is defined to be 
$r(P) = \sqrt{A(P)/\bar{A}}$, the root of the cross sectional area $A$ at $f(P)$ of an 
infinitesimal bundle of generators neighbouring $\cg$, over the cross sectional area 
$\bar{A}$ of the same bundle at the truncating surface $S_A$. (See subsection
\ref{area_parameter}).
In the initial data sets we admit the area parameter along each ray starts at a finite, 
positive, non-zero value $r_0$ at the beginning of the ray on $S_0$ and increases or decreases 
monotonically to a value of $1$ at the truncating surface where the ray leaves $N$. Thus 
a bundle of generators having non-zero cross sectional area on $S_0$ necessarily has non-zero 
cross sectional area everywhere along its trajectory to $S_A$. 

This restriction on the initial data together with the following proposition implies that 
$df$ is non-degenerate on $N$.

\begin{proposition}\label{no_caustics}
$df$ is non-degenerate on $N$ provided $r/r_0$ does not vanish on $\cN$.
\end{proposition}
{\em Proof}: 
If $df$ is degenerate at a point $P \in O$ then $f$ maps any basis of the space $T_P$ of
vectors tangent to $O$ at $P$ to a linearly {\em dependent} set of tangent vectors to $X$
at $p = f(P) \in X$. To show that $df$ is non-degenerate on a branch $N_A$ of $N$ we will 
choose a basis of $T_P$ for all $P \in N_A$ and show that the 4-volume spanned by the image 
of this basis is non-zero, implying that the four image vectors are linearly independent. 

Coordinates can be defined on $N_A$ as follows: Choose coordinates on $S_0$ and 
convect these along the rays to all of $N_A$ (i.e. set them to be constant on the rays), 
then choose a linear parameter $\lam$ on each ray so that $\lam = 0$ on $S_0$ and the 
$\lam = 1$ surface is smooth. The associated coordinate basis consists of $L = d/d\lam$, tangent
to the null rays of $N_A$, and two vectors, $E_1$ and $E_2$, corresponding to the coordinates   
on $S_0$. This can be completed to a basis of tangents to $O$ on $N_A$ by including a timelike 
vector $T$ which is tangent to the fibres of $O$ and constant in each fibre. ($T$ is timelike
according to the metric on the tangent space of $X$ at the base of the fibre on $S_0$ where $T$ 
can be regarded to reside since it is constant on the fibre). 
At each point
$P \in N_A$ $f$ pushes these vectors forward to vectors $l$, $e_1$, $e_2$, and $\tau$ tangent 
to $X$ at $p = f(P)$. $l$ is tangent to the generator through $p$ and thus null and normal to 
$\cN_A$, while $e_1$ and $e_2$ are tangent to $\cN_A$ and thus normal to $l$. Since 
$0 = l \cdot l = l \cdot e_1 = l \cdot e_2$ the $4 \times 4$ matrix of inner products of these
vectors\footnote{
When $l$, $e_1$, $e_2$, and $\tau$ are linearly independent, and thus form a basis, 
the matrix of their inner products is the matrix of covariant components of the metric
referred to this basis. The matrix of inner products is of course defined even when 
these vectors are not linearly independent. The 4-volume is given by the square root of minus 
the determinant of the matrix of inner products in either case.}
falls into a block diagonal form and the 4-volume spanned by them is easily seen to be
$- l \cdot \tau A$, where $A$ is the area spanned by $e_1$ and $e_2$. 
 
Since $L$ and $T$ are non-zero null and timelike vectors respectively 
the inner product $l \cdot \tau$ is necessarily non-zero at $S_0$. 
Less obviously, this inner product is constant along each generator:
The displacement from a point on a ray of $N_A$, generated by $L$, to the corresponding 
point on the timelike ray of $O$ generated by the vector $L + T$ is given by the vector field 
$V(\lam) = \lam T$ on the null ray. $\lam \tau$ is thus a geodesic deviation vector field, or
Jacobi field, and it is subject to the geodesic deviation equation:
$\nabla_l \nabla_l \lam \tau^\mu = \lam R_{\tau l l}^\mu$.
Since $\nabla_l l = 0$ the inner product $\lam l\cdot \tau$ satisfies
\be
\frac{d^2}{d\lam^2} \lam [l \cdot \tau] = l \cdot \nabla_l \nabla_l \lam\tau 
= \lam R_{\tau l l l} = 0,
\ee
which shows that $\lam l\cdot\tau$ is linear in $\lam$, and hence that $l\cdot\tau$ is constant
along the generator. In conclusion $l\cdot\tau$ is non-zero everywhere on the generator. 

Degeneracy of $df$ on $N_A$ would therefore require that the area $A$ spanned
by $e_1$ and $e_2$ vanish somewhere on a generator. But $A = A_0 (r/r_0)^2$, where $A_0$
is the area spanned by these vectors at $S_0$, which is non-zero by definition.
The condition that $r/r_0$ not vanish thus guarantees that $A$ does not vanish and $df$ is 
non-degenerate.\QED
\newline

Now let us show that $\cN$ is achronal in $\cV$. This of course does depend on the fact that
$\cN$ is embedded, and not merely immersed, in $\cV$. The square $s = v\cdot v$ 
(according to the metric on $S'$) of the vectors orthogonal to $S'$ constitutes a smooth 
function on $O'$, and thus on $\cV$. At a point $p \in \cV$ $s$ is minus the proper time 
squared elapsed along the orthogonal geodesic segment $\cg_p$ from $S'$ to $p$, or, if $\cg_p$ 
is spacelike, it is the length squared of $\cg_p$. 

$s$ is also the value on these geodesic segments of the action functional 
$\sg[c] = \int_0^1 v\cdot v\, dt$, when $t$ is an affine parameter. This action functional
is defined for an arbitrary differentiable curve $c:[0,1] \rightarrow \cV$, with $t$ the 
parameter of the curve and $v = d/dt$ the corresponding tangent vector. When the curve is the 
geodesic segment $\cg_p$ normal to $S'$ parametrized affinely so that $c(1) = p$ then, firstly, 
$v(0)$ is the vector at $c(0) \in S'$ that the exponential map associates to $p$, and, 
secondly, $v$ is parallel propagated so $v \cdot v$ is constant on the geodesic. As a 
consequence $\sg = s$.

Using this correspondence we may prove the following useful lemma:
\begin{lemma} \label{level_s_orthogonal}
The level set of $s$ passing through $p$ is orthogonal to $\cg_p$. Indeed the gradient vector
$\vec{ds}$ of $s$ (obtained by acting with the inverse metric on $ds$) is tangent to $\cg_p$ 
and points away from $S'$.
\end{lemma}

{\em Proof}: A direct calculation shows that under a variation of $c$ 
\be    \label{sg_variation}
\dg\sg = 2\left[\dg c\cdot v\right]_0^1 - 2\int_0^1 \dg c\cdot \nabla_v v\,dt.
\ee
($\dg c$ is the vector field on the curve generating its displacement). 
This allows us to evaluate the gradient of $s$. Let us restrict attention to the geodesic 
segments $\cg_p$ and vary $p$. The base point of $\cg_p$ on $S'$ varies, but according to 
(\ref{sg_variation}), since $v(0)$ is orthogonal to $S'$ this variation makes no contribution 
to the variation of $\sg$. Since $\cg_p$ is a geodesic there is no contribution from the 
integral term either. The variation of the action thus is $\dg\sg = 2\dg c(1)\cdot v(1)$
\be   \label{grad_s_v}
\vec{d}s = 2 v(1).
\ee 
From this the claim follows immediately.\QED
\newline

Now let us use this lemma. Let $V^+$ be the subset of $V$ consisting of future directed causal 
vectors. It follows from lemma \ref{level_s_orthogonal} that $s$ is strictly decreasing along 
any future directed timelike curve in $\cV^+ \equiv \hat{f}[V^+]$. From this we see immediately 
that 

\begin{proposition} \label{cN_achronal}
$\cN$ is achronal in $\cV$. 
\end{proposition}

{\em Proof}: If $\cN$ were not achronal then there would exist a future directed timelike curve 
in $\cV$ crossing $\cN$ twice. Moreover, by \cite{Wald} theorem 8.1.2., this curve may be 
chosen to be piecewise geodesic, and thus piecewise differentiable. [An arbitrary timelike 
curve may be covered with a sequence of convex normal neighbourhoods so that successive 
neighbourhoods overlap. Then a sequence of points may be chosen lying on the curve and in the 
overlaps of successive neighbourhoods and successive points in {\em this} sequence may be 
connected by timelike geodesics, which always exist by the theorem.] $s$ is strictly decreasing 
along the curve at any point in $\cV^+ - S_0$. The curve starts on $\cN \subset \cV^+$ where 
$s = 0$, and if it starts on $S_0$ it immediately enters $\cV^+ - S_0$ because it is timelike 
and future directed. Since the curve can leave $\cV^+ - S_0$ only if $s$ returns to $0$, its 
value on the boundary of this set, the curve can in fact not leave $\cV^+ - S_0$ and $s$  
always continues to decrease strictly after the starting point of the curve.
An endpoint of the curve on $\cN$ would therefore have $s < 0$, but $s = 0$ on all of $\cN$. 
\QED
\newline

Prop. \ref{cN_achronal} establishes that the achronality of $\cN$ requires no restrictions on 
the initial data, save that it make $\cN$ caustic free. The absence of caustics implies that 
$\cN$ can always be made achronal by going to the covering manifold $\cV$ of a neighbourhood of 
$\cN$. 

$\cV$ does not necessarily cover all of the domain of dependence of $\cN$, even in
ambient spacetimes in which $\cN$ is achronal. A caustic in the timelike geodesics normal to 
$S_0$ might well form inside $D[\cN]$ without there being any sort of pathology in $D[\cN]$ 
that one would wish to exclude. Thus $\cV$ is not in general large enough to be the ambient 
spacetime of the maximal Cauchy developments of data on $\cN$.  
Nevertheless, the domain of dependence of $\cN$ in $\cV$, while not the {\em maximal} Cauchy 
development is sufficient for our purpose of defining a pre-Poisson bracket on the initial 
data. One may define the observables on these, non-maximal, Cauchy developments, and the Peierls
bracket on these determines an almost unique pre-Poisson bracket on the initial data.

For the sake of completeness, and conceptual simplicity, I define further on a boundaryless 
manifold $M$ which contains both $\cN$ as an achronal submanifold and the interior of the 
maximal Cauchy development of the data on $\cN$. This construction is not necessary to justify 
the main results of the present work.

We are now in a position to show that

\begin{proposition}\label{domain_of_influence_on_N}
In $\cV$ the subset of $\cN$ that is causally connected to a point $p \in \cN$ (i.e. lies in 
$J^+[p] \cup J^-[p]$) consists of the points on the generators through $p$.
\end{proposition}

{\em Proof}: Suppose $q$ and $r$ are points of $\cN$ that are causally connected, with
$q$ to the future of $r$: $q \in J^+(r)$. Let $\cN_e$ be the extension of $\cN$ swept
out by the generators of $\cN$ continued to the future as null geodesics until they leave $\cV$.
$s = 0$ on $\cN_e$ so $\cN_e$ is achronal in $\cV$ by the same argument used to prove the 
achronality of $\cN$ (prop. \ref{cN_achronal}). Now suppose $q'$ lies on the generator through 
$q$ to the future of $q$, then a causal curve $\cg$ from $r$ to $q$ followed by the generator 
segment from $q$ to $q'$ defines a causal curve from $r$ to $q'$ which may be deformed to a 
timelike one unless it is a null geodesic. The achronality of $\cN_e$ thus implies that
$\cg$ is a null geodesic that continues the generator segment from $q$ to $q'$ to the past, 
that is, it implies that $\cg$ coincides with the generator through $q$ and $q'$. It follows
that $r$ and $q$ necessarily lie on the same generator. \QED  
\newline

A further property of $\cN$ that we will use is

\begin{proposition} \label{generators_near_boundary}
If $W$ is a neighbourhood of $\di \cN$ in $\cN$ then there exists a neighbourhood $U$ of the 
boundary in $S_0$ such that all generators of $\cal N$ incident on $U$ lie entirely in $W$.
\end{proposition}
{\em Proof}: 
Consider one branch $\cN_A$ of $\cN$. The complement $C$ of $W$ in $\cN_A$ is a closed subset of
$\cN_A$, and $\cN_A$ is compact. Thus $C$ is compact. The map $\pi:\cN_A \rightarrow S_0$ which 
takes the points on a generator to the origin of that generator on $S_0$ is continuous - it is 
just the composition of the diffeomorphism $\hat{f}^{-1}$ (restricted to $\cN_A$) with the 
projection to the base manifold in $O$. Thus $\pi[C]$ is compact and therefore closed. 
It follows that the complement $U_A$ of $\pi[C]$ is open. Since $C$ does not intersect $\di\cN$ 
and $\pi^{-1}[\di S_0] \subset \di\cN$, $\pi[C]$ does not intersect $\di S_0$. Thus $U_A$ is an 
open neighbourhood of $\di S_0$. The same is of course true of $U = U_R \cap U_L$.\QED
\newline

Now we turn to properties of Cauchy developments of $\cN$. We begin with a pair of lemmas
about achronal sets.

\begin{lemma}\label{boundary_outside} If $S$ is an achronal set then its embedding boundary 
$\ocirc{S}$ lies outside the domain of influence $I[D[S]]$ of the domain of dependence of $S$.
\end{lemma}
{\em Proof}: Since $\dot{I}^+[S]$ is achronal (by \cite{Hawking_Ellis} prop. 6.3.1.) a 
timelike curve can cross $\dot{I}^+[S]$ at most once. On the other hand all inextendible
timelike curves through $D[S]$ must cross $S \subset \dot{I}^+[S]$. They can thus not 
cross $\dot{I}^+[S]$ outside $S$ as well. Indeed no timelike curve through $D[S]$ can 
cross $\dot{I}^+[S] - S$ since any such curve could be extended until inextendible. 
$I[D[S]]\cap \dot{I}^+[S]$ is thus a subset of $S$, and since it is open in the topology
of $\dot{I}^+[S]$ it must be contained in the interior of $S$ and be disjoint from 
$\ocirc{S}$.\QED   
\newline

\begin{lemma}\label{null_geodesic_condition} Suppose $S$ is an achronal set and $p \in int S$, 
then $I^+(p)\cap D[S] = \emptyset$ iff there exists a null geodesic in $S$ passing through $p$ 
without past endpoint in the interior of $S$. 
\end{lemma}
{\em Proof}: If there exists a null geodesic segment $\cg$ in $S$ passing through $p$ without 
past endpoint in $int S$ then either this segment is past inextendible in spacetime or it has a 
past endpoint $s$ in the closure $\bar{S}$ of $S$, and thus in 
$\overline{S} - intS = \ocirc{S}$. In the first case there exists (by \cite{Wald} lemma
8.1.4.) through any $q \in I^+(p)$ a past inextendible timelike curve contained in $I^+[\cg] 
\subset I^+[S]$. Since $S$ is achronal this curve never crosses $S$, implying that
$q \notin D[S]$. In the second case one may remove the end point $s$ from the spacetime
making $\cg$ past inextendible. In the spacetime without $s$ there again exists a past 
inextendible timelike curve through $q$ contained in $I^+[S]$. This curve is either 
inextendible in the full spacetime, indicating that $q \notin D[S]$, or $s \in \dot{S}$ is 
its past endpoint. But by lemma \ref{boundary_outside} no timelike curve through $D[S]$
passes through $\ocirc{S}$, so also in this case $q \notin D[S]$. $I^+(p)$ is thus disjoint
from $D[S]$.

Now to the converse: if $I^+(p)$ is disjoint from $D[S]$ then there exists a sequence of
points $q_i$ in $I^+(p)$ approaching $p$ and causal, past inextendible curves $\cg_i$ through 
the $q_i$ which do not intersect $S$. The portions, $\cg'_i$, of these curves that lie in 
$I^+[S]$ are either past inextendible or have past endpoints on $\dot{I}^+[S]-S$. 
Removing the closed set $\dot{I}^+[S]- intS$ from spacetime leaves a manifold in which the 
$\cg'_i$ are all past inextendible. By lemma 6.2.1. of \cite{Hawking_Ellis} there must exist a 
causal, past inextendible limit curve of the $\cg'_i$ through $p$. Since the 
$\cg'_i$ all lie in the closed set $\bar{I}^+[S]$ the limit curve does so also. But, 
because $S$ is achronal the limit curve cannot enter $I^+[S]$. It is thus confined to
$\bar{I}^+[S] - I^+[S] = \dot{I}^+[S]$, which is just $int S$ in this manifold.
Moreover the limit curve, which is causal, must be a null geodesic, for if it were not then two
points on the curve could be connected by a timelike curve, violating the achronality of $S$. 
\QED
\newline

Note that when $I^+(p)\cap D[S]$ is non empty for some $p \in S$ then in fact all of
$I^+(p)$ that lies within a neighbourhood $U$ of $p$ is contained in $D[S]$: If 
$r \in I^+(p)\cap D[S]$ then we may take $U = I^-(r)$, because $p \in I^-(r)$ and past directed 
timelike curves from $r$ that have not yet reached $S$ must be contained in $D[S]$, 
so $I^-(r)\cap I^+(p) \subset D[S]$.

Lemma \ref{null_geodesic_condition} allows one to demonstrate some intuitively plausible basic 
facts about the domain of dependence of $\cN$.

\begin{proposition}\label{cN_horizon} For all $p \in int \cN$, the part of $I^+(p)$ lying 
in a sufficiently small neighbourhood of $p$ is contained in $D[\cN]$ while $I^-(p)$ does 
not intersect $D[\cN]$. Thus $D[\cN] = D^+[\cN]$, $int D[\cN]$ is non empty, $\cN \subset
\overline{int D[\cN]}$, and $\cN$ is a Cauchy surface of $int D[\cN]$.
\end{proposition}
{\em Proof}: 
By prop. \ref{domain_of_influence_on_N} the only null geodesics in $\cN$ passing through a 
point $p \in int \cN$ are the generators through $p$. These have past endpoints on 
$int S_0 \subset int \cN$, and future endpoints on $S_L$ or $S_R$, both subsets of 
$\di\cN \subset \ocirc{\cN}$. 
Thus by lemma \ref{null_geodesic_condition} $I^+(p)$ must intersect $D[\cN]$,
while $I^-(p)$ must not intersect $D[\cN]$. That $U \cap I^+(p) \subset D[\cN]$ for some 
neighbourhood $U$ of $p$ follows from the remarks after the proof of that lemma. 
$D[\cN] = D^+[\cN]$, $int D[\cN] \neq \emptyset$, and $\cN \subset \overline{int D[\cN]}$
follow immediately. $\cN \subset \overline{int D[\cN]}$ implies that $\cN$ is a Cauchy surface 
of $int D[\cN]$ according to def. \ref{Cauchy_surface}.\QED 
\newline

Since the interior of the domain of dependence of the initial data hypersurface $\cN$ 
we are using does not cover the whole spacetime, and in fact does not even include 
$\cN$ itself, it has been necessary to adopt a wider notion of Cauchy surfaces than is usual
(see def. \ref{Cauchy_surface}). 
In the following we shall see that despite this more ample definition, the Cauchy surfaces 
$int D[\cN]$ are quite simple: They are all embedded manifolds homeomorphic to $\cN$ 
with boundary coinciding with $\di\cN$. 

\begin{proposition}\label{cauchy_boundaries_equal} If $C$ and $C'$ are both Cauchy surfaces
of an open set $\cH \subset X$, then $\ocirc{C} = \ocirc{C'}$.
\end{proposition}
{\em Proof}: $\ocirc{C}$ is the boundary of $\bar{C}$ in $\dot{I}^+[C]$ and also
of its complement in $\dot{I}^+[C]$: $\ocirc{C} = [\dot{I}^+[C] - \bar{C}]^\circ$. But
\be   \label{cbe_proof_eqn}
\dot{I}^+[C] - \bar{C} = \dot{I}^+[\cH] - \bar{\cH},
\ee
the left side being a set depending only on $\cH$ and not on the choice of Cauchy surface.
To justify (\ref{cbe_proof_eqn}) note that $I^+[C] \subset I^+[\bar{\cH}] = I^+[\cH]$,
and conversely $I^+[\cH] - \bar{\cH} \subset I^+[C] - \bar{\cH}$, for if 
$p \in I^+[\cH] - \bar{\cH}$ then there exists a timelike curve from a point 
$q \in \cH \cap I^-(p)$ to $p$ which, when extended to an inextendible timelike curve
necessarily crosses $C$, and does so to the past of $p$ since any point on the curve between
$q$ and the crossing point lies in $\bar{\cH}$. It follows that
\be
I^+[C] - \bar{\cH} = I^+[\cH] - \bar{\cH}.
\ee
Now
\be
\dot{I}^+[\cH] - \bar{\cH} = [I^+[\cH] - \cH]^\cdot - \bar{\cH} 
= [I^+[C] - \cH]^\cdot - \bar{\cH} =\dot{I}^+[C] - \bar{\cH}.
\ee
In the rightmost expression $\bar{\cH}$ may be replaced by $\bar{C}$: $\bar{\cH} \subset
\overline{D[\bar{C}]}$, hence, by \cite{Hawking_Ellis} prop. 6.5.1, for any point 
$r \in \bar{\cH}$ there is a timelike curve passing through both $r$ and a 
point $s \in \bar{C}$. Since $\dot{I}^+[C]$ is 
achronal the curve can intersect $\dot{I}^+[C] \supset \bar{C}$ {\em only} at the one point
$s$, so if $r \in \dot{I}^+[C]$ as well as $\bar{\cH}$ then $r = s \in \bar{C}$. This 
establishes (\ref{cbe_proof_eqn}), and also the claim of the proposition. \QED
\newline

Since $\cN$ is compact, and thus closed in spacetime, $\ocirc{\cN} = \di\cN$, so prop.
\ref{cauchy_boundaries_equal} implies that for any Cauchy surface $C$ of $int D[\cN]$
$\ocirc{C} = \di\cN$. Indeed, we shall see that $\bar{C}$ is a manifold with boundary that
is homeomorphic to $\cN$. It follows that $\di\bar{C} = \ocirc{C} = \di\cN$.

To demonstrate homeomorphism we first establish a lemma:

Let us define the {\em minimal component}, $C^0$, of a Cauchy surface $C$ of an open
hyperbolic set $\cH$, to be $C \cap I[\cH]$.\footnote{
$C^0$ is the only part of $C$ which causal 
curves through $\cH$ actually cross. (Because $\cH$ is open $J[\cH] = I[\cH]$). Thus $C^0$ is 
precisely that part of $C$ that is necessary for it to function as a Cauchy surface for $\cH$.
A Cauchy surface of $\cH$ contained in $\cH$ is equal to its minimal component.
On the other hand $\di\cN$ is not part of the minimal component of $\cN$.
Indeed lemmas \ref{boundary_outside} and \ref{null_geodesic_condition} (and the time 
reverse of lemma \ref{null_geodesic_condition}) indicate that for any Cauchy surface $C^0$
consists of $C$ minus its boundary and any null geodesics without future or past endpoint in
$int C$. In particular, by prop. \ref{cN_horizon} and lemma \ref{boundary_outside} 
$\cN^0 = int \cN$.} 

\begin{proposition}\label{cauchy_boundary} If $C$ is a Cauchy surface of an open set $\cH$
then $\overline{C} - C^0 = \overline{\cH} - I[\cH]$.
\end{proposition}
{\em Proof}: It is evident that $\overline{C} - C^0 \subset \overline{\cH} - I[\cH]$. What must 
be shown is thus that {\em all} points of $\overline{\cH} - I[\cH]$ lie in $\overline{C} - C^0$.
Since $C^0 \subset I[\cH]$ it is sufficient to show that $\overline{\cH} 
- I[\cH]\subset \overline{C}$. The achronality of $C$ implies that
$\overline{C}$ is also achronal. Therefore, since $C$ is a Cauchy surface for $\cH$, 
$\overline{C}$ is as well. As a consequence $\cH \subset D[\overline{C}]$ implying that 
$\overline{\cH} \subset \overline{D[\overline{C}]}$. By \cite{Hawking_Ellis} prop. 6.5.1, 
this implies that all inextendible timelike curves through $\overline{\cH}$ must pass through 
$\overline{C}$.
But if $p \in \overline\cH - I[\cH]$ then $I(p)$ is disjoint from $\cH$, and indeed from 
$\overline{\cH} \supset \overline{C}$, since $I(p)$ is open. The only point on a timelike curve 
through $p$ not in $I(p)$ is $p$ itself, so this is the only point at which the curve can 
intersect $\overline{C}$: $p \in \overline{C}$. \QED
\newline 

\begin{proposition}\label{cauchy_homeo} If $C$ and $C'$ are both Cauchy surfaces
of the open set $\cH \subset X$, then $\bar{C}$ is homeomorphic to  $\bar{C'}$.
\end{proposition}
{\em Proof}: Let $v$ be a smooth vector field on $X$. (By lemma 8.1.1 of \cite{Wald} such a 
field always exists on a paracompact, time orientable, smooth, Lorentzian metric manifold. 
See also \cite{Hawking_Ellis} pp. 38-40.) Each integral curve of $v$ is inextendible in $X$, 
so if it passes through $\cH$ it must
cross both $C$ and $C'$ exactly once. The integral curves thus establish a 1-1 correspondence
between $C^0 \equiv C \cap I[\cH]$ and $C'^0 \equiv C' \cap I[\cH]$. By 
prop. \ref{cauchy_boundary} $\bar{C} - C^0 = \bar{\cH} - I[\cH] = \bar{C'} - C'^0$ so this
1-1 correspondence actually extends to a 1-1 correspondence between $\bar{C}$ and $\bar{C'}$.
Moreover, this correspondence is continuous and has continuous inverse: The smoothness of $v$ 
implies that the points on the integral curves are continuous functions of the curve 
parameter and of the starting point of the curve. Furthermore the achronality of $C$ and $C'$ 
imposes a Lipschitz condition on the curve parameter value of the point $p' \in C'$ that 
corresponds to a point $p \in C$, implying that this parameter is a continuous function on $C$. 
$p'$ is thus a continuous function of $p$. Reversing the roles of $C$ and $C'$ one sees that the
inverse is also continuous
\QED
\newline

The preceding results together establish that

\begin{proposition}\label{cauchy_D} The closure of any Cauchy surface of $\cD = int D[\cN]$ is 
an embedded compact 3-manifold with boundary $\di\cN$ that is homeomorphic to $\cN$, and thus 
to the closed unit cube in $\Real^3$.
\end{proposition}

In an open globally hyperbolic region the wave equation, and similar equations,
such as the linearized Einstein field equation in transverse gauge (\ref{reduced_linearized}), 
have unique advanced and retarded Green's function (see \cite{three_muses} p. 523). It is
therefore relevant when using Green's functions in conjunction with initial data
on $\cN$, as we do, that 

\begin{proposition}\label{globally_hyperbolic_neighbourhood} $\cN$ is contained in 
an open subset $\hat{\cV}$ of $\cV$ which, when regarded as a spacetime in its own right, 
is globally hyperbolic.
\end{proposition}

The proof consists in extending $\cN$ to a larger achronal hypersurface $\Lam$ and showing that
$\cN \subset int D[\Lam] \equiv \hat{\cV}$. In fact it will not be shown that $\Lam$ is 
achronal in $\cV$, but rather in an open subset $\cV'$, which will serve as the spacetime
in which $D[\Lam]$ is defined

{\em Proof part I. Definition of $\cV'$ and $\Lam$}: $\cV'$ is an open subset of $\cV$ 
containing $\cN$, that is, like $\cV$, an exponential of an open subset ($V'$) of $O'$ 
from which any ray of $O'$ exits at most once. $\cV'$ thus has all the properties
that have been demanded of $\cV$, and could be taken in place of $\cV$ in all preceding 
theorems. In addition the closure of $\cV'$ in $\cV$ will be required to be compact. 
[This requirement is easily fulfilled. For instance, define a Riemannian metric on $O'$, and 
let $\cV'$ be a subset of the cylinder formed by vectors in $O'$ that are shorter than twice 
the maximal length of the vectors in the compact subset $N$.] As a consequence any geodesic 
normal to $S'$ crosses the boundary of $\cV'$ before leaving $\cV$.

The construction of $\Lam$ proceeds in two steps. The first step is to extend $\cN$ to a 
hypersurface $\tilde{\cN}$ just like $\cN$ but slightly larger. Like $\cN$, $\tilde{\cN}$ is 
swept out by future directed null geodesic generators normal to a compact spacelike 2-disk that 
are truncated before reaching a caustic. The 2-disk $\tilde{S}_0$ on which the generators of 
$\tilde{\cN}$ originate contains $S_0$ in its interior, and the surfaces $\tilde{S}_L$ and 
$\tilde{S}_R$ on which the generators are truncated are chosen to lie beyond $S_L$ and $S_R$, 
so that $\cN \subset int\tilde{\cN}$.
Recall that $O'$ is the normal bundle of a smoothly embedded, boundaryless, spacelike 2-manifold
$S' \supset S$ (which is defined before prop. \ref{immersion_thm}). $\tilde{S}_0$ will be taken 
to be a compact disk with smooth boundary in $S'$ that contains $S_0$ in its interior. 
Furthermore $\tilde{S}_L$ and $\tilde{S}_R$ will be chosen close enough to $\cN$ so that they 
(and thus all of $\tilde{\cN}$) lie within $\cV'$. Thus $\tilde{\cN} = \hat{f}[\tilde{N}]$ where
$\tilde{N} \supset N$ is a subset of $V'$, and prop. \ref{cN_achronal} implies that 
$\tilde{\cN}$ is achronal. 

Having defined $\tilde{\cN}$ we now define $\Lam$ to be $\dot{I}^-[\tilde{\cN}]$. This is an 
achronal embedded 3-manifold, by \cite{Hawking_Ellis} prop. 6.3.1.   

To show that $\cN \subset int D[\Lam]$ we use the following lemma:

\begin{lemma}\label{generators_not_continued} Suppose $\lam'$ is the maximal extension 
(as a null geodesic) of a generator of $\tilde{\cN}$ in the spacetime $\cV'$, then 
$\lam'\cap\overline{I^-[\tilde{\cN}]} \subset \tilde{\cN}$ in this spacetime.
\end{lemma}
{\em Proof}: Suppose $p \in \lam'\cap\overline{I^-[\tilde{\cN}]} - \tilde{\cN}$. Then there 
exists a sequence of points $\{ p_n\}$ in $I^-[\tilde{\cN}]$ in $\cV'$ approaching $p$, and 
for each $p_n$ a future directed timelike curve $\cg_n$ in $\cV'$ from $p_n$ to $\tilde{\cN}$. 
Now consider the problem in the context of the larger ambient spacetime $\cV \supset \cV'$.
The $\cg_n$ are timelike curves in $\cV$ that are inextendible in $\cV - \tilde{\cN}$ so, 
by \cite{Hawking_Ellis} lemma 6.2.1, they have a causal limit curve $\cg$ through $p$ which 
is also inextendible in $\cV - \tilde{\cN}$, and thus either inextendible in $\cV$ or has an 
endpoint on $\tilde{\cN}$.

Now note that the proof of prop. \ref{cN_achronal} applies equally well to $\tilde{\cN}$ in 
$\cV$
and shows that it is achronal in $\cV$. This implies that $\cg$ coincides with $\lam$, the
maximal extension as a null geodesic of $\lam'$ in $\cV$, to the future of $p$: $\cg$ lies in 
$\overline{I^-[\tilde{\cN}]}$ and thus, since $\tilde{\cN}$ is achronal, outside 
$I^+[\tilde{\cN}]$. But following $\lam$ from a point on $\tilde{\cN}$ to $p$ and then 
continuing along $\cg$ one obtains a causal curve that may be deformed to a timelike one, which 
would imply that $\cg$ enters $I^+[\tilde{\cN}]$, unless $\cg$ coincides with $\lam$ after $p$. 
On the other hand the $\cg_n$ are all contained in $\cV'$, and hence $\cg$ is contained in the 
compact set $\overline{\cV'}$. $\cg$ must thus have an endpoint at $\lam\cap\dot{\cV'}$ 
which does not lie on $\tilde{\cN}$, contained in the open set $\cV'$. A contradiction has 
been reached, implying that $\lam'\cap\overline{I^-[\tilde{\cN}]} - \tilde{\cN}$ is in fact 
empty in $\cV'$. \QED
\newline

{\em proof of prop. \ref{globally_hyperbolic_neighbourhood} part II}: 
Suppose $p \in \cN \subset int \tilde{\cN}$. By prop. \ref{domain_of_influence_on_N} any null 
geodesic through $p$ lying in $\Lam$ is (an extension of) a generator of $\tilde{\cN}$ through 
$p$. Such a generator has a past endpoint on $int \tilde{S}_0\subset int\Lam$ and leaves 
$\tilde{\cN}$ at the boundary of $\tilde{\cN}$. Now recall that 
$\Lam = \dot{I}^-[\tilde{\cN}] \subset \overline{I^-[\tilde{\cN}]}$.
Lemma \ref{generators_not_continued} thus implies that the extension $\lam'$ of a generator as 
a null geodesic cannot remain in $\Lam$ outside $\tilde{\cN}$. $\lam'\cap \tilde{\cN}$ thus has 
a future endpoint on the boundary of $\tilde{\cN}$, which is contained in $int\Lam$ since 
$\tilde{\cN}$ is compact. By prop. \ref{null_geodesic_condition}, and its time reverse, $I^+(p)$
and $I^-(p)$ intersect $D[\Lam]$, so $p \in I^+(q)\cap I^-(r)$ with $q, r \in D[\Lam]$. But 
it is easily shown that the whole open set $I^+(q)\cap I^-(r)$ must be contained in $D[\Lam]$, 
implying that $p \in int D[\Lam]$. \QED
\newline

$\hat{\cV} \equiv int D[\Lam]$ is thus a globally hyperbolic boundaryless manifold which
is a locally isometric covering of the original embedding of $\cN$ in the arbitrary, smooth,
Lorentzian, spacetime $X$. $\hat{\cV}$ is sufficient spacetime to develop our whole theory in 
terms of the domain of dependence of $\cN$. In particular we may define 
the domain of dependence of $\cN$ in $\hat{\cV}$, observables on this domain of dependence,
and a Peierls bracket between these, which in turn induces our auxiliary pre-Poisson bracket
on the initial data. Nevertheless $\hat{\cV}$ does not in general contain the maximal Cauchy
development of data given on $\cN$. Even in a perfectly regular solution to Einstein's field 
equations a caustic may develop in the timelike geodesics normal to $S_0$ within the interior 
of the domain of dependence of $\cN$. It is therefore interesting, though by no means necessary 
for the definition of the phase space or the Poisson structure on it, to show that a boundaryless
manifold $M$ extending $\hat{\cV}$ can always be constructed that is globally hyperbolic 
and contains $\cN$ and the interior of its maximal Cauchy development.

$M$ is constructed by gluing two manifolds: Choose a Cauchy surface $\Sg_0$ of 
$\hat{\cV}$ lying to the future of $\cN$.\footnote{
By \cite{Hawking_Ellis} prop. 6.6.8. there exists a continuous ``time'' function $t$ on 
$\hat{\cV}$ which increases strictly to the future along any causal curve, and ranges from 
$-\infty$ to $\infty$. Since $\cN$ is compact $t$ has a finite maximum on $\cN$. Any level 
set of $t$ with $t$ greater than this maximum will serve as the desired Cauchy surface.}
The chronological past $I^-[\Sg_0]$ of $\Sg_0$ in $\hat{\cV}$, denoted $\cA$, is one of the
components of $M$, which of course contains $\cN$. The other component is obtained as 
follows. Choose a spacelike Cauchy surface $\Sg$ of the interior of the domain of dependence
of $\cN$ in $\cA$, $\cD_0 = int D[\cN;\cA]$. (Note that $\cD_0 = int D[\cN;\hat{\cV}]\cap 
I^-[\Sg_0;\hat{\cV}]$.) If the metric on $\cD_0$ satisfies the field equations then it 
induces admissible, spacelike, initial data on $\Sg$, which in turn define a maximal
Cauchy development $\cD_m$. $\cD_m$ is the other component of $M$.

$\cD_0$ must be isometric to a subset of $\cD_m$. Indeed 

\begin{lemma}\label{D0_embedding} There exists an isometry $\omega: \cD_0 \cup \sg_0
\rightarrow \cD_m$ where $\sg_0 = \Sg_0 \cap int D[\cN;\hat{\cV}]$. Moreover, 
$\sg_0' \equiv \omega[\sg_0]$ is the border $\dot{\omega}[\cD_0]$ in $\cD_m$.
\end{lemma}
{\em Proof}: The definition of the maximal Cauchy development implies that there exists an 
isometry $\omega: \cD_0 \rightarrow \cD_m$. This isometry extends to boundary points of
$\cD'_0 \equiv \omega[\cD_0]$ in $\cD_m$, for all such boundary points are naturally images 
of points on the boundary of $\cD_0$ in $\hat{\cV}$: Suppose $p' \in \cD'_0$, and $\cg'$ is 
a (future or past directed) timelike curve in $\cD'_0$ which is inextendible in $\cD'_0$
but has an endpoint in the direction of increasing curve parameter at $p'$. A timelike curve 
through $\cD_0$ that is inextendible in $\hat{\cV}$ must pass 
through the Cauchy surface $\cN$ of $\cD_0$ to the past and the Cauchy surface $\Sg_0$, of 
$\hat{\cV}$, to the future, it cannot remain in $\cD_0$. Thus $\cg = \omega^{-1}[\cg']$ must 
either have an endpoint in the sense of increasing parameter value inside $\cD_0$ (but it cannot
because $\cg'$ does not end inside $\cD'_0$), or at a point $p$ on the boundary of $\cD_0$ in 
$\hat{\cV}$. 

Now choose a convex normal neighbourhood $U_p$ of $p$ in $\hat{\cV}$ and a convex normal 
neighbourhood $U_{p'}$ of $p'$ in $\cD_m$. $\cg$ is contained $U_p$ for sufficiently large curve 
parameter and $\cg' = \omega[\cg]$ is contained in $U_{p'}$ for sufficiently large parameter 
values. Thus $Y' = U_{p'} \cap \omega[U_p \cap \cD_0]$ contains $\cg'$ beyond a minimum 
parameter value. Let $r' \in \cg' \cap Y'$ and construct Riemann normal coordinates 
$y'$ on $U_{p'}$ based at $r'$. Then define $y$ to be the chart on $U_p$ obtained by first 
pulling back $y'$, restricted to a convex normal neighbourhood of $r'$ contained in $Y'$, to 
$U_p$ via $\omega$, and then extending the resulting Riemann normal coordinates about 
$r = \omega^{-1}(r')$ to all of $U_p$. This defines a diffeomorphism $\phi = y'^{-1} \circ y$ 
from part of $U_p$ to part of $U_{p'}$.

The boundary of $\cD_0$ in $\hat{\cV}$ consists of the achronal hypersurface 
$\overline{\cD_0} - I^-{\cD_0}$ to the future and the achronal hypersurface 
$\overline{\cD_0} - I^+{\cD_0} = \cN$ to the past. Define $I^\circ$ to be $I^+(r;U_p)$ if $p$ 
lies on the future boundary, and $I^-(r;U_p)$ if $p$ lies on the past boundary. Since the
geodesics from $r$ to a point in $I^\circ \cap \cD_0$ are entirely contained in $\cD_0$
$\phi$ coincides with $\omega$ on $I^\circ \cap \cD_0$. $y\circ\cg$ and $y' \circ\cg'$ thus 
define the same curve (for sufficiently large parameter value) in $\Real^4$, and, since the limit
points $p$ and $p'$ lie in the domains of $y$ and $y'$ respectively, $y(p) = y'(p')$.
$p$ therefore lies in the domain of $\phi$ and $\phi(p) = p'$. Indeed a whole neighbourhood $W_p$
of $p$ lies in the domain of $\phi$, which is open. Since $p \in I^\circ$ this neighbourhood
may be taken to lie inside $I^\circ$. On $W_p$ $\phi$ is then a smooth extension of $\omega$. 
Moreover, since the metric is smooth in both of the charts $y$ and $y'$ it is an isometry 
on the closure of $\cD_0$ in $W_p$. 

Covering $\dot{\cD}'_0$ with sets of the form $W_{p'} = \omega[W_p]$ a unique extension of 
$\omega^{-1}$ to $\dot{\cD}'_0$ is obtained by patching together the maps $\phi^{-1} = y^{-1} 
\circ y'$. The extension is well defined and unique because any two diffeomorphisms extending 
$\omega^{-1}$ must agree on $\bar{\cD}'_0$ in the overlap of their domains, since they agree on 
$\cD'_0$ in the overlap of their domains. The extension is an isometry, and its inverse, which 
extends $\omega$ to part of $\dot{\cD}_0$, will be called $\omega$ as well.

What part of $\dot{\cD}_0$ lies in $\omega^{-1}[\dot{\cD}'_0]$?
A boundary point $p'$ of $\cD'_0$ in $\cD_m$ that lies to the past of $\Sg' = \omega[\Sg]$ must
be the image of a point $p \in \cN$. (By prop. \ref{cN_horizon}.) Through $p'$ there must 
therefore pass a null geodesic which is the isometric image of a portion of a generator through 
$p$. This null geodesic would be future inextendible but would not cross $\Sg'$, since the 
generator crosses $\bar{\Sg}$ only at its future endpoint on $\di\cN$, which is not contained 
in $\Sg \subset \cD_0$, by prop. \ref{boundary_outside}. However $\Sg'$ is a Cauchy surface of 
$\cD_m$. Thus there can be no boundary point to the past of $\Sg'$. 

A boundary point $p'$ to the future of $\Sg'$ must be the image of a point $p$ lying 
either on the future Cauchy horizon $H^+[\Sg;\cA]$ of $\Sg$ in $\cA$, or on the future boundary,
$\Sg_0$, of $\cA$ itself. If $p$ lies on $H^+[\Sg;\hat{\cV}] \supset H^+[\Sg;\cA]$ there would 
be a past inextendible null geodesic from $p'$ which does not cross $\Sg'$, for through 
$p$ there must exist, by theorem 8.3.5. of \cite{Wald}, a null geodesic contained in 
$H^+[\Sg;\hat{\cV}]\subset J^+[\Sg;\hat{\cV}]$ which is past inextendible or has past endpoint 
on the causal edge of $\bar{\Sg}$.\footnote{
The causal edge is defined in \cite{Hawking_Ellis} p.202. In $\hat{\cV}$, which is globally 
hyperbolic, and thus strongly causal, $edge[\Sg] = \di\cN$. See \cite{Hawking_Ellis} p. 196-197 
on Alexandrov topology.}
But the causal edge of $\bar{\Sg}$ cannot be included in $\Sg$ since $\Sg'$, being a 
Cauchy surface contained in $\cD_m$, has empty causal edge (by \cite{Wald}, corollary to 
prop. 8.3.6.). $p$ therefore does not lie in $H^+[\Sg;\hat{\cV}]$.

A future directed timelike curve from $\cD_0$ cannot cross $\Sg_0$ without crossing 
$H^+[\Sg;\hat{\cV}]$ - and thus leaving $int D[\cN;\hat{\cV}]$ - first, except by passing 
through $\sg_0 = \Sg_0 \cap int D[\cN;\hat{\cV}]$. $\sg'_0 = \omega[\sg_0]$ thus contains the 
boundary of $\cD'_0$ in $\cD_m$. Moreover, all points of $\sg_0$ are in fact mapped into 
$\cD_m$: Consider using a different Cauchy surface, $\tilde{\Sg_0}$, of $\hat{\cV}$, lying to 
the future of $\Sg_0$, in place of $\Sg_0$ in our construction. The set $\tilde{\cD_0}
= int D[\cN;\hat{\cV}]\cap I^-[\tilde{\Sg_0};\hat{\cV}]$, which contains $\cD_0$ and in 
particular $\sg_0$, would then be mapped by an isometry $\tilde{\omega}$ into $\cD_m$. The 
$\tilde{\omega}$ may be chosen so that it agrees with $\omega$ on the overlap of their 
domains, including $\sg_0$\footnote{
$\tilde{\omega}\circ \omega^{-1}$ defines an isometry of $\cD'_0$ to another open subset 
of $\cD_m$. Since $\cD'_0$ contains $\Sg'$ the unicity of the maximal Cauchy 
development implies that this isometry extends to an isometry of $\cD_m$ to $\cD_m$. 
If this isometry is non-trivial one may simply compose its inverse with $\tilde{\omega}$ to 
obtain a new $\tilde{\omega}$, satisfying all requirements, which agrees with $\omega$ on 
$\cD_0$.}
$\omega$ thus maps all of $\sg_0$ into $\cD_m$.
\QED
\newline 

$M$ is the union of $\cA$ and $\cD_m$ with points in $\cD_0$ and 
$\cD_m\cap I^-[\Sg'_0]$ identified according to the isometry $\omega$. The open sets of 
$M$ are those sets which have open intersections with $\cA$ and $\cD_m$ in the topologies
of those manifolds. (This is the standard ``gluing topology'').

\begin{proposition}\label{Mhat_Hausdorff} $M$ is Hausdorff
\end{proposition}
{\em Proof}: Suppose $M$ is not Hausdorff. Then there must exist points $p$ and $q$
in $M$ such that all neighbourhoods of $q$ intersect all neighbourhoods of $p$. If both
$p$ and $q$ lie in $\cD_m$, or both lie in $\cA$ this cannot occur because $\cD_m$ and 
$\cA$ are Hausdorff. Suppose therefore that $q \in \cD_m$ and $p \in \cA$. $\cD_m$
is thus a neighbourhood of $q$ and $\cA$ a neighbourhood of $p$, so in order that the pair
violate the Hausdorff condition it is necessary that $q \in \dot{\cA} = \dot{\cD}'_0 = \sg'_0$ 
(by lemma \ref{D0_embedding}) in $\cD_m$ and $p \in \dot{\cD}_m = \dot{\cD}_0$ in $\cA$.
To show that $M$ is actually Hausdorff it is thus sufficient to find disjoint 
neighbourhoods of $q$ and $p$ in $M$ for such $p$ and $q$.  

Let $q_0 = \omega^{-1}(q)$. By lemma \ref{D0_embedding} 
$q_0 \in \sg_0 \subset int D[\cN;\hat{\cV}]$ which is necessarily distinct from 
$p$ lying in the disjoint set $\dot{\cD}_0 \subset \dot{D[\cN;\hat{\cV}]}$. Since 
$\hat{\cV}$ is Hausdorff there exist disjoint open subsets of $\hat{\cV}$ $U_{q_0} \ni q_0$
and $U_p \ni p$. Indeed we can, and will, require that $U_p \subset \cA$ and 
$U_{q_0} \subset int D[\cN;\hat{\cV}] \cap I^-[\tilde{\Sg}_0,\hat{\cV}]$, where, as in the proof
of lemma \ref{D0_embedding}, $\tilde{\Sg}_0$ is a Cauchy surface of $\hat{\cV}$ lying to the 
future of $\Sg_0$. 

The desired neighbourhoods of $p$ and $q$ in $M$ are constructed from $U_{q_0}$ and $U_p$. 
An open set in $M$ is a pair of open sets, one in $\cA$, one in $\cD_m$, such that
identified pairs of points are either both in the sets or both outside. Let $\tilde{\omega}$ be 
the isometry of lemma \ref{D0_embedding} corresponding to the case in which $\Sg_0$ is replaced 
by $\tilde{\Sg}_0$, and recall from the proof of that lemma that $\tilde{\omega}$ may be chosen 
so that it agrees with $\omega$ on the intersection of their domains. With this choice of 
$\tilde{\omega}$ $\tilde{\omega}[U_{q_0}]$ is an open neighbourhood of $q$ in $\cD_m$, and
$(U_{q_0}\cap \cD_0, \tilde{\omega}[U_{q_0}])$ and $(U_p, \omega[U_p\cap\cD_0])$ are disjoint
open neighbourhoods of $p$ and $q$ are respectively.
\QED


%

The paracompactness and time orientability of $M$ follows from that of $\cA$ and 
$\cD_m$ and the fact that $\cD_m \cap \cA$ has only one connected component. (See 
\cite{Hawking_Ellis} p. 14.)
 
\begin{proposition}\label{Mhat_good} $\Lam$ (and thus $\cN$) is achronal in $M$, and 
$M$ is globally hyperbolic.
\end{proposition}
{\em Proof}: Suppose $p \in \Lam$ and $\cg$ is a future directed timelike curve from $p$.
If $\cg$ remains in $\cA \subset \hat{\cV}$ it cannot cross $\Lam$ again because $\Lam$ 
is achronal in $\hat{\cV}$. If $\cg$ leaves $\cA$ without leaving $M$ it must cross 
the boundary, $\Sg'_0$, of $\cA \cap \cD_m$ in $\cD_m$. If $\cg$ is then to return to $\cA$ 
it must again cross this same boundary, this time entering $\cA$. But $\cg$, which is future 
directed, cannot cross the boundary in both directions since $\Sg_0$ is a level set of a time 
function which increases along any future directed causal curve. Thus $\cg$ cannot cross $\Lam 
\subset \cA$ a second time. This establishes the achronality of $\Lam$ in $M$.

To prove global hyperbolicity it is thus sufficient to show that all of $M$ lies in the
domain of dependence of $\Lam$. Suppose $\cg$ is an inextendible causal curve which passes 
through $\cA$. $\cg$ may be extended to an inextendible causal curve, $\cg'$, in $\hat{\cV}$,
and $\cg'$ must pass through $\Lam$ and through $\Sg_0$. Since $\Sg_0$ is achronal $\cg'$
cannot reenter $\cA$ once it leaves, so $\cg' \cap \cA = \cg$, proving that $\cg$ crosses 
$\Lam$. If on the other hand $\cg$ is an inextendible causal curve passing through $\cD_m$ then
it must cross the Cauchy surface $\Sg$ of $\cD_m$, which lies in $\cA$, implying that 
$\cg$ also passes through $\cA$ and thus through $\Lam$. All inextendible causal curves
in $M$ cross $\Lam$. \QED
\newline

The demonstration that only those points in $\cN$ that lie on the same generator are causally 
related (prop. \ref{domain_of_influence_on_N}) goes through in $M$ precisely as in $\cV$.

Since $M$ (and $\hat{\cV}$) are globally hyperbolic the transverse gauge linearized 
Einstein field equations (\ref{reduced_linearized}) has unique advanced and retarded Green's 
functions $G^\pm(p,q)$ supported (as distributions) at points $q \in J^\pm(p)$. (see 
\cite{three_muses} p. 523)

Using the global hyperbolicity of the ambient spacetime $M$ or $\hat{\cV}$ it may be 
shown that sources with compact support in the interior of $D[\cN]$ do not perturb (to linear 
order) the metric in a spacetime neighbourhood of $\di\cN$:

\begin{proposition}\label{closureJ+_disjoint_N} If $s$ is a compact subset of the interior of 
the domain of dependence of $\cN$ in $M$, then $J^\pm[s]$ is closed and disjoint from an 
open neighbourhood of $\di\cN$. 
\end{proposition}
{\em Proof}: The fact that $J^\pm[s]$ is closed follows from \cite{Hawking_Ellis} prop. 6.6.1.
To prove that $J^\pm[s]$ is disjoint from an open neighbourhood of $\di\cN$ it is therefore
sufficient to show that it is disjoint from $\di\cN$ itself. Now by prop. \ref{cauchy_D},
prop. \ref{embedding2}, and lemma \ref{boundary_outside}
$I[\di\cN]$ is disjoint from $D[\cN]$, and thus $\overline{I[\di\cN]} \supset J[\di\cN]$ is 
disjoint from $int D[\cN]$. It follows at once that $J[s] \supset J^\pm[s]$ is disjoint
from $\di\cN$. \QED
\newline


All Cauchy developments of $\cN$ seem to be diffeomorphic to each other as spacetimes (though 
of course not necessarily isometric). In the following a slightly weaker claim, that they are 
diffeomorphic everywhere except possibly on their future Cauchy horizons (i.e. their future 
boundaries, see \cite{Hawking_Ellis}), will be proved. This result makes possible the concrete 
representation we have used of the set of Cauchy developments as a space of solutions $\cal S$ 
to the field equations on a common, fixed manifold $\Dund_0$ (The union of the interior of the 
hypersurface $\cN_0 = \{ t = |x|, |x|\leq 1, y^2 + z^2 \leq 1\}$ and the interior of its domain 
of dependence in $\Real^4$ equipped with the Minkowski metric 
$ds^2 = -dt^2 + dx^2 + dy^2 + dz^2$ - see Fig. \ref{Mfig}).

This representation makes the set of Cauchy developments easier to think about, 
but it is actually not necessary for our main results, which concern the Peierls bracket, the 
symplectic 2-form, and the auxiliary pre-Poisson bracket on the initial data. These objects are 
defined entirely in terms of first order perturbation theory about a given solution, and thus 
in the context of a single, spacetime manifold, fixed once this solution is chosen. The use of 
a fixed spacetime manifold in developing the main results of the present work is thus justified 
quite independently of the diffeomorphism between Cauchy developments.    

Suppose $\cN$ (defined as on p. \pageref{cN_def}) is smooth and achronal in a smooth, Lorentzian
metric 4-manifold. Let us define $\Dund = int D[\cN] \cup int \cN$. Then 

\begin{proposition} \label{Dund_horizonless} $\Dund$ is the domain of dependence of $\cN$ 
minus its future Cauchy horizon
\end{proposition}
{\em Proof}:
$D[\cN] = D^+[\cN]$ follows from prop. \ref{cN_horizon}, and prop. \ref{cN_horizon} and prop. 
\ref{domain_of_influence_on_N} together imply that $int \cN = I^-[D[\cN]]\cap \cN$. Thus the 
identities $int D[\cN] = I^-[D[\cN]]\cap I^+[D[\cN]]$ and $D^+[\cN] = D^+[\cN]\cap I^+[\cN] 
\cup \cN$ imply $\Dund = D[\cN]\cap I^-[D[\cN]]$, or equivalently $\Dund = D[\cN] - H^+[\cN]$ 
where $H^+[\cN] \equiv D[\cN] - I^-[D[\cN]]$ is the future Cauchy horizon. \QED

\begin{proposition} \label{Cauchys_diffeo} For any two Cauchy developments, 1 and 2, of $\cN$, 
$\Dund_1$ and $\Dund_2$ are diffeomorphic as manifolds (though not necessarily isometric). 
\end{proposition}
{\em Proof}: 
To prove diffeomorphism a smooth chart $\Phi:\Dund \rightarrow \Real^4$ is constructed for each 
Cauchy development, such that the range, $\Phi[\Dund]$, of the coordinates on $\Dund$ is the 
same for all Cauchy developments. Then, if $\Phi_1$ and $\Phi_2$ are the charts for Cauchy 
developments 1 and 2 respectively, $\Phi_1^{-1} \circ \Phi_2$ provides the desired 
diffeomorphism.

The construction of the chart $\Phi$ proceeds in two steps. First the chart $\Phi$ is defined 
on a neighbourhood of $S_0$ and on $\cN$ itself. Then this chart is extended to all of $\Dund$. 
The exponential map $f$ from the normal bundle $O'$ over $S'$ to to the ambient spacetime $X$ 
is a smooth immersion except at degenerate points of $df$. 

Since $df$ is non-degenerate on $S' \supset S_0$, the generalized inverse function theorem
(\cite{Guillemin_Pollack}, p. 19) implies that $f$ maps an open neighbourhood $U$ of $S'$ in 
$O'$ diffeomorphically to a neighbourhood $\cU = f[U]$ of $S'$ in $X$. Moreover, $\cN = f[N]$ 
where $N \subset O'$ is free of degenerate points of $df$. Since $\cN$ is achronal $f$ must 
be one to one on $N$ and thus a diffeomorphism on this set as well.


Coordinates $(x^0, x^1, x^2, x^3)$ are chosen on the normal bundle $O'$ as follows: 
$(x^2, x^3)$ coordinatize the base manifold $S'$, and $(x^0, x^1)$ the fibres, consisting of 
the normal vectors at each point of $S'$. $(x^0, x^1)$ are linear, that is, they are the 
components of the normal vectors with respect to some basis defined at each point of $S'$, and 
$(x^0, x^1, x^2, x^3)$ are chosen so that $int N$ corresponds to the coordinate domain 
$\{(x^0, x^1, x^2, x^3)||x^1| = x^0, |x^1|<1, (x^2)^2 + (x^3)^2 < 1\}$.

$\Phi$ maps each point $p \in \cU \cup \cN$ to the coordinates $(x^0, x^1, x^2, x^3)$ of 
its preimage $P = f|_U^{-1}(p)$ in $U \cup N$.

Note that the definition of the coordinates implies that the vector field $\di_0$ on $S_0$
is timelike and future directed.
 
We turn now to the extension of $\Phi'$. Since the continuous vector field $\di_0$ is timelike 
and future directed on $S_0$ it is also so in an open neighbourhood $\cB_0$ of $S_0$ in $\cal U$.
Let $\cC$ be a compact subset of $\cB_0$ containing $S_0$ in its interior, then $\Phi$ will be 
set equal to $\Phi'$ on a compact subset $\cC$ of $\cB_0$ containing $S_0$ in its interior, and 
on $\cN$ itself . To extend the chart beyond this set $\di_0$ will first be extended to a 
smooth timelike vector field on all of $\Dund$ by splicing it smoothly with a smooth, non-zero, 
future directed, future complete, timelike vector field on $\Dund$. Here ``future complete'' 
means that the supremum of the curve parameter in $\Dund$ is infinite on every integral curve 
of the vector field. 

A field $u$ satisfying these requirements can be found using a suitable 
Riemannian metric on $\Dund$: Choose a partition of unity $\{f_n\}$ of $\Dund$ subordinate to a 
locally finite atlas $\{\phi_n\}$ such that the domain of each chart $\phi_n$ has compact 
closure, and define the Riemannian metric $r = \sum_n f_n \phi_{*\,n} e$, the sum of the 
pullbacks via the various charts of the standard euclidean metric $e$ on $\Real^4$ weighted by 
the partition of unity. Let $\rho = \sum_n n f_n$. It is easy to show that the Riemannian metric
$r' = r (||\nabla \rho||^2 + 1)$ is complete in the sense that any curve without endpoint in 
$\Dund$ has infinite length, whereas a curve from a starting point to an end point, including 
starting and endpoints on $\cN$, has finite length.\footnote{
To show this note that the length of a curve exceeds the absolute value of the integral of 
$\nabla \rho$ along the curve. The claim then follows from the fact that $\rho$ diverges 
along any curve without endpoint.}
$u$ is taken at each point to be the future directed timelike vector of maximal proper time 
among the unit vectors of $r'$. The smoothness of $r'$ and the Lorentzian spacetime metric 
guarantee that this $u$ is smooth.

The splicing is defined using a partition of unity $\varphi_0$, $\varphi_1$
subordinate to the cover $\cB_0$, $\cB_1 = \cU - \cC$. That is, $\varphi_0$ is a smooth function
taking values between $0$ and $1$ such that $\varphi_0 = 1$ on $\cC$ and $0$ outside $\cB_0$, 
and $\varphi_1 = 1 - \varphi_0$. $w = \varphi_0 \di_0 + \varphi_1 u$ is the spliced field, 
defined on all of $\Dund$ (for this it is sufficient that $\di_0$ is defined on the support 
$\varphi_0$). Since both vector fields being added are timelike and future directed, so is 
their sum $w$.

Because $w$ is timelike and non-zero on $\Dund$ the integral curves of $w$ are 
timelike and can have no endpoint in $int \Dund = \Dund - \cN$. Every integral curve must 
therefore cross the Cauchy surface $\cN$, precisely once. Indeed lemma \ref{boundary_outside} 
prop. \ref{cN_horizon} and lemma \ref{boundary_outside} show that the set of points of
$\cN$ at which these integral curves cross is precisely $int \cN$.
 
At a point $p \in \Dund$ the coordinate $x^0$ of the extended chart $\Phi$ is the parameter of 
the integral curve of $w$ through $p$, with zero set so that it agrees with the 
$x^0$ of the chart $\Phi'$ on $\cN$. The coordinates $x^1, x^2, x^3$ assigned to $p$ by $\Phi$
are just the $x^1, x^2, x^3$ assigned by $\Phi'$ to the point at which the integral curve 
through $p$ crosses $\cN$. (That is, $x^1, x^2, x^3$ are held constant along the integral 
curves.)

The smoothness of $w$, of the submanifold $\cN - S_0$ and of the chart $\Phi'$, on $\cU$ and on 
$\cN - S_0$, imply, via standard theorems on the solutions of ODEs, that $\Phi$ is a smooth 
chart of $\Dund$.\footnote{
The definition of $w$, and thus $\Phi$ extends to an open neighbourhood of $\Dund$ in the 
ambient spacetime. The local existence and uniqueness theorem 21.1 of \cite{Abraham_Robbin} 
implies that $\Phi$ is one to one. Theorem 21.2 (flow box theorem) of \cite{Abraham_Robbin} 
shows that $\Phi$ is smooth and $d\Phi$ is invertible. Thus, by \cite{three_muses} p. 93 
theorem 1, $\Phi$ is a diffeomorphism of an open set containing $\Dund$ into $\Real^4$.}
By prop. \ref{cN_horizon} $D[\cN] = D^+[\cN]$ so the value of $x^0$ on $\cN$,
equal to $|x^1|$ is the lower bound of the range of $x^0$ on a given integral curve, while the 
future completeness of $u$ implies that the upper bound is $\infty$.

The range of $\Phi$ on $\Dund$ is therefore 
\be
\Phi[\Dund] = \{(x^0, x^1, x^2, x^3)| |x^1| \leq x^0 < \infty, |x^1|<1, (x^2)^2 + (x^3)^2 < 1\},
\ee
which is independent of the particular Cauchy development in question.
\QED

\section{Solutions to linearized gravity in terms of initial data and the relation
between the variations of observables and the presymplectic 2-form}\label{another_approach}

The key result of subsection \ref{Peierls_symplectic}, equation (\ref{inverse2}), can be proved 
in a less abstract manner, as a corollary to a Kirchhoff type formula for the solutions of the 
linearized 
Einstein equations in terms of initial data. This formula, demonstrated in the sequel, provides 
us with some analytic information about the dependence of the metric, and the observables, on 
initial data.

We begin by examining (\ref{inverse2}). As in subsection \ref{Peierls_symplectic} we chose a 
spacetime region $Q$ bounded by $\cN$ and a Cauchy surface $\Sg_+$ of the future domain of 
dependence of $\cN$, so that the domain of sensitivity of $A$ lies in the interior of $Q$.
Then
\be
\dg A = \int_Q \ag^{\mu\nu}\dg g_{\mu\nu} \sqrt{-g} d^4x.
\ee
The perturbation $\Dg_A g$ of the metric induced by $A$ is determined only modulo a 
diffeomorphism generator, but if we require it to obey the transverse gauge condition
then, by (\ref{Dg_def}) and (\ref{pert1}), it may be written as
\be          
\Dg_A g_{\sg\tau}(y) 
		= -16\pi G\int_Q \ag^{\mu\nu}(x) \Dg_{\mu\nu\,\sg\tau} (x, y) \sqrt{-g} d^4 x.
\ee
(\ref{inverse2}) thus holds for all observables of $\cD = int \Dund$ if, for all solutions
to the linearized vacuum Einstein equations $\dg g$,
\be             \label{Kirch1}
\dg g_{\mu\nu}(x) = -16\pi G\Omega_\cN [\DDg_{\mu\nu} (x), \dg] + 2\nabla_{(\mu} \xi_{\mu)}
\ee
at all points $x$ of $\cD$, for some vector field $\xi$. Here $\DDg_{\mu\nu} (x)$ is the 
tensor field (depending on $y$) obtained from the two point tensor 
$\Dg_{\mu\nu\,\sg\tau} (x, y)$ when $x$, $\mu$, and $\nu$ are held fixed.  
The $\xi$ term reflects the 
freedom to change $\dg g$ by a linearized diffeomorphism. When both sides of (\ref{Kirch1}) 
are integrated against $\ag$ to recover (\ref{inverse2}) it disappears because $\ag$ is 
divergenceless and vanishes at the boundary of integration. (Note that (\ref{Kirch1}) implies
that (\ref{inverse2}) holds even when $\Dg_A$ is not in transverse gauge, because, as 
(\ref{boundary_diff}) will show, when $\dg$ is a solution to the linearized field equations 
that vanishes near $\di\cN$ $\bar{\Omega}[{\pounds_\xi}, \dg] = 0$ for any diffeomorphism 
generator ${\pounds_\xi}$.)

Equation (\ref{Kirch1}) is in fact valid: Suppose $\cg_1 \equiv \dg_1 g$ is a solution to the 
linearized field equations that satisfies the transverse gauge condition, 
$\chi_\nu \equiv\nabla^\mu \bar{\cg}_{\mu\nu} = 0$ with $\bar{\cg}_{\mu\nu} = \cg_{\mu\nu} - 
\frac{1}{2} g_{\mu\nu} g^{\sg\rho}\cg_{\sg\rho}$, and let $\cg_2 \equiv \dg_2 g$ be the 
(unique) advanced solution to the linear, diagonal, second order hyperbolic system
\be             \label{reduced_linearized2}
\nabla^\sg\nabla_\sg \bar{\cg}_{\mu\nu} - 2 R^\sg{}_{\mu\nu}{}^\rho \bar{\cg}_{\sg\rho} 
= \theta_{\mu\nu},
\ee
with $\theta$ a smooth, symmetric tensor field of compact support contained in $\cD$.
(This equation is just (\ref{reduced_linearized}) with the source $32\pi G \ag$ replaced by
$\theta$.) $\cg_2$ may be expressed in terms of the advanced Green's function as
\be          \label{advanced_transverse_solution}
\cg_{2\,\sg\tau}(y) 
		= -\frac{1}{2}\int_Q \theta^{\mu\nu}(x) G^-_{\mu\nu\,\sg\tau} (x, y) 
		\sqrt{-g} d^4 x.
\ee
Here $Q$ is defined as earlier, but with future boundary chosen so that $Q$ contains the 
support of $\theta$. Note that $G^-$ is {\em regular} in the sense that when it is integrated 
against a smooth source $\theta$ the result is a smooth solution \cite{three_muses}. 

Now, for any variation $\dg$ the variation of the Einstein tensor is
\be
\dg G_{\mu\nu}  =  \nabla_{(\mu}\chi_{\nu)} - 1/2 g_{\mu\nu}\nabla_\sg\chi^\sg
- 1/2 (\nabla^\sg\nabla_\sg \bar{\cg}_{\mu\nu} - 2R^\sg{}_{\mu\nu}{}^\rho \bar{\cg}_{\sg\rho}),
\ee
so $\cg_2$ would be a solution to the linearized Einstein equations, with source $-\frac{1}{2}
\theta$, if $\chi_{2\,\nu}$ were zero.

However, (\ref{reduced_linearized2}) and the Bianchi identity imply that 
\be             \label{transverse_gauge_preservation2}
\nabla^\mu\nabla_\mu \chi_{2\,\nu} = \nabla^\mu \theta_{\mu\nu},
\ee
implying that $\chi_2$ does not vanish unless $\theta$ is divergenceless - which is not assumed 
here. (\ref{transverse_gauge_preservation2}) also has unique, regular retarded and advanced  
Green's functions $H^\pm$. Hence
\bearr          
\chi_{2\,\tau}(y) 
	& = & \int_Q \nabla_\mu\theta^{\mu\nu}(x) H^-_{\nu\,\tau} (x, y) 
		\sqrt{-g} d^4 x\\
	& = & -\int_Q \theta^{\mu\nu}(x) \nabla_\mu H^-_{\nu\,\tau} (x, y) 
		\sqrt{-g} d^4 x
\eearr

Now consider the identity used in (\ref{hypersurface_independence}) to prove the hypersurface 
independence of $\bar{\Omega}$:
\bearr
0 & = & \dg_1\dg_2 I - \dg_2\dg_1 I - [\dg_1,\dg_2] I \\
  & = & \dg_1 \phi[\dg_2] - \dg_2 \phi[\dg_1] - \phi[[\dg_1,\dg_2]]\\
     &&   + \frac{1}{16\pi G}\int_Q [\dg_2 g_{\mu\nu}\dg_1 G^{\mu\nu}
- \dg_1 g_{\mu\nu}\dg_2 G^{\mu\nu} ]\sqrt{-g}\, d^4 x\\
  & = & - \bar{\Omega}_{\cN}[\dg_1,\dg_2] 
	- \frac{1}{16\pi G}\int_Q \dg_1 g_{\mu\nu}[\nabla^\mu\chi_2^\nu 
	- \frac{1}{2} g^{\mu\nu}\nabla_\sg\chi_2^\sg 
	- \frac{1}{2} \theta^{\mu\nu}]\sqrt{-g}\, d^4 x. \label{dggOmega0}                   
\eearr
In (\ref{dggOmega0}) the facts that $\dg_1 G^{\mu\nu} = 0$, that $\dg_2 g$ vanishes in a 
neighbourhood of the future boundary of $Q$, and that the orientation of $\cN$ is opposite that 
of $\di Q$ have been taken into account. 
It is easy to obtain (\ref{inverse2}) directly from this last expression, by setting 
$\theta = 32\pi G \ag$, but it is illuminating to press on to obtain an expression for 
$\dg_1 g$ in terms of data on $\cN$.

The last integral of (\ref{dggOmega0}) can be written as
\be \label{ddg_int_simp}
\int_Q \bar{\cg}_{1\,\mu\nu}\nabla^\mu\chi^\nu \sqrt{-g}\, d^4 x 
-\frac{1}{2}\int_Q \theta^{\mu\nu}\dg g_{\mu\nu}\sqrt{-g}\, d^4 x.
\ee
Since $\cg_1$ satisfies the transverse gauge condition, integrating the first term by parts 
reduces it to a boundary integral, in fact an integral over $\cN$ only:
\be
-\int_\cN \bar{\cg}_1^{\sg\tau}\chi_{2\,\tau} \sqrt{-g}\, d\Sg_\sg
= \int_Q \theta^{\mu\nu}(x) \nabla_\mu \xi_\nu \sqrt{-g}\, d^4 x,
\ee
where
\be  \label{ddg_int_simp2}
\xi_\nu(x) = \int_\cN \bar{\cg}_1^{\sg\tau} H^-_{\nu\,\tau} (x, y) \sqrt{-g}\, d\Sg_\sg. 
\ee
Now $\theta$ is an arbitrary smooth symmetric tensor field with support contained in $\cD$
so (\ref{dggOmega0}), (\ref{ddg_int_simp}), and (\ref{ddg_int_simp2}) imply that at any point 
$p$ of $\cD$
\be             \label{Kirch2}
\dg_1 g_{\mu\nu}(p) = 16\pi G\,\Omega_\cN [{\bf G}^-_{\mu\nu} (p), \dg_1] + 
2\nabla_{(\mu} \xi_{\nu)},
\ee
where ${\bf G}^-_{\mu\nu} (p)$ is the tensor field obtained by fixing the first argument point
in the advanced Green's function to be $p$ (and the pair of indices associated with the first
point to be $\mu$ and $\nu$). 
(\ref{Kirch1}) follows immediately if one notes that the retarded Green's function 
$G^+(p,q)$ vanishes for $q \in \cN$ since $\cN$ lies outside $J^+(p)$.

Solutions to the linearized field equations that are {\em not} in transverse gauge only differ 
from transverse gauge solutions by a linearized diffeomorphism. Such solutions can thus
also be expressed in terms of initial data via (\ref{Kirch2}) or (\ref{Kirch1}), by using $\xi$ 
other than that defined by (\ref{ddg_int_simp2}).

\section{$G_{rr} = 0$}\label{Grrzero}

In this appendix it is shown that on $\cN_R$ the Einstein field equation $G_{rr} = 0$,
in the adapted coordinates $a^\ag = (u,r,y^1,y^2)$, is equivalent to 
\be
\Gamma^r_{rr} = - \frac{r}{8}\,\di_r e_{ij}\,\di_r e^{ij}.
\ee

We proceed by direct calculation of $G_{rr}$. The form (\ref{a_line_element}) that the metric 
takes in the adapted coordinates implies that in these coordinates the Christoffel symbols,
\be
\Gamma^\ag_{\bg\cg} = \frac{1}{2} g^{\ag\eg}\{\di_\cg g_{\eg\bg} + \di_\bg g_{\eg\cg} 
- \di_\eg g_{\bg\cg}\},
\ee 
on $\cN_R$ satisfy
\bearr
&&\Gamma_{\ag rr} = \dg^u_\ag\Gamma_{urr},\ \ \ \ \Gamma^\ag_{rr} = \dg^\ag_r \Gamma^r_{rr}\\
&&\Gamma^r_{rr} = g^{ur}\Gamma_{urr} = \frac{1}{2}\di_u g_{rr} \\
&&\Gamma_{r\bg r} = \dg^u_\bg \Gamma_{rur},\ \ \ \ \Gamma^u_{\bg r} = \dg^u_\bg\Gamma^u_{ur}\\
&&\Gamma^u_{u r} = g^{ur}\Gamma_{rur} = -\frac{1}{2}\di_u g_{rr} = - \Gamma^r_{rr}\\
&&\Gamma^i_{jr} = \frac{1}{2}g^{ik}\di_r g_{jk} = \dg^i_j\,\frac{1}{r}
+ \frac{1}{2}e^{ik}\di_r e_{jk}, \ \ \ \Gamma^i_{ir} = \frac{2}{r}\\
&&\di_u\Gamma_{rrr} - \di_r\Gamma_{rur} = \frac{1}{2}\{\di_u\di_r g_{rr} - \di_r\di_u g_{rr}\} = 0.
\eearr
Using these facts, and the adapted coordinate form (\ref{a_line_element}) of the metric, $G_{rr}$
can be simplified as follows:
\bearr
G_{rr} = R_{rr} & \equiv & \di_\ag\Gamma^\ag_{rr} - \di_r\Gamma^\ag_{\ag r} 
+ \Gamma^\bg_{rr}\Gamma^\ag_{\ag\bg} - \Gamma^\bg_{\ag r}\Gamma^\ag_{\bg r} \\
& = & \di_u\Gamma^u_{rr} - \di_r\Gamma^u_{ur} - \di_r\Gamma^i_{ir} 
+ \Gamma^r_{rr}\Gamma^\ag_{\ag r} - \Gamma^u_{ur}\Gamma^u_{ur} - \Gamma^r_{rr}\Gamma^r_{rr}\\
&& - \Gamma^i_{jr}\Gamma^j_{ir} \\
& = & [\di_u g^{uu}]\Gamma_{urr} + g^{ur}\di_u\Gamma_{rrr} - [\di_r g^{ur}]\Gamma_{rur} 
- g^{ur}\di_r\Gamma_{rur} + \frac{2}{r^2}\\
&& + \Gamma^r_{rr}\Gamma^u_{ur} + \frac{2}{r}\Gamma^r_{rr} - \Gamma^u_{ur}\Gamma^u_{ur}  
- \frac{2}{r^2} - \frac{1}{4}e^{ik}e^{jl}\di_r e_{jk}\di_r e_{il}\\
& = & - g^{ur}[\di_u g_{rr}]\Gamma^r_{rr} + g^{ur}[\di_r g_{ru}]\Gamma^u_{ur} 
+ \Gamma^r_{rr}\Gamma^u_{ur} - \Gamma^u_{ur}\Gamma^u_{ur}\\
&& + \frac{2}{r}\Gamma^r_{rr} + \frac{1}{4}\di_r e_{ij}\di_r e^{ij}\\
& = & \frac{2}{r}\Gamma^r_{rr} + \frac{1}{4}\di_r e_{ij}\di_r e^{ij},
\eearr
which implies the result.

\end{document}